\documentclass[10pt,journal,compsoc]{IEEEtran}
\usepackage{times}
\usepackage{soul}
\usepackage{epsfig}
\usepackage{enumitem}
\usepackage{graphicx}
\usepackage{amsmath}
\usepackage{accents}
\usepackage{amssymb}
\usepackage{multirow}
\usepackage{colortbl}
\usepackage{color}
\usepackage{threeparttable}
\usepackage{pifont}
\usepackage{booktabs}
\usepackage{adjustbox}
\usepackage{caption}
\usepackage[misc]{ifsym}
\usepackage{xcolor,colortbl}
\usepackage{makecell}
\usepackage{tikz}
\usepackage[edges]{forest}
\usepackage{longtable}
\usepackage{mathrsfs}
\usepackage{xltabular}
\usepackage[hyphens]{url}
\definecolor{Gray}{gray}{0.92}
\definecolor{gaoyifeng-pink}{RGB}{255,235,245}
\usepackage{arydshln}
\usepackage{colortbl}
\usepackage{hhline}
\usepackage{makecell}
\usepackage{verbatim}
\usepackage{titlesec}
\usepackage{graphicx}   

\usepackage{pifont}
\usepackage{subfigure}
\usepackage{float}
\usepackage{xcolor,colortbl}
\usepackage[T1]{fontenc}

\usepackage{amsmath,amsfonts,bm}

\def\eqref#1{(\ref{#1})}

\def\1{\bm{1}}

\DeclareMathAlphabet{\mathsfit}{\encodingdefault}{\sfdefault}{m}{sl}
\SetMathAlphabet{\mathsfit}{bold}{\encodingdefault}{\sfdefault}{bx}{n}

\newcommand*{\belowrulesepcolor}[1]{%
  \noalign{%
    \kern-\belowrulesep 
    \begingroup 
      \color{#1}%
      \hrule height\belowrulesep 
    \endgroup 
    \vspace{-0.03mm}
  }%
} 
\newcommand*{\aboverulesepcolor}[1]{%
  \noalign{%
  \vspace{-0.03mm}
    \begingroup 
      \color{#1}%
    \endgroup 
    \kern-\aboverulesep 
  }%
}

\titleformat{\paragraph}[block] 
  {\normalfont\normalsize\bfseries} 
  {\theparagraph} 
  {1em} 
  {} 
  [] 

\titlespacing*{\paragraph}{0pt}{1.5ex plus 0.2ex minus 0.1ex}{0.5em} 

\newcommand{\cmark}{\textcolor{green}{\ding{52}}}
\newcommand{\xmark}{\textcolor{red}{\ding{55}}}

\definecolor{figorange}{RGB}{228,130,47}
\definecolor{figred}{RGB}{255,0,0}
\definecolor{figgreen}{RGB}{0,176,80}

\definecolor{sunye-red}{RGB}{220, 239, 252}
\definecolor{sunye-red-dark}{RGB}{114, 154, 202}
\definecolor{sunye-red-light}{RGB}{235, 244, 255}

\definecolor{dingyifan-wangyixu-darkblue}{RGB}{231, 244, 234}
\definecolor{dingyifan-wangyixu-darkblue-dark}{RGB}{107, 182, 142}
\definecolor{dingyifan-wangyixu-darkblue-light}{RGB}{245, 251, 246}

\definecolor{wangxin-yellow}{RGB}{254, 240, 189}
\definecolor{wangxin-yellow-dark}{RGB}{238, 196, 84}
\definecolor{wangxin-yellow-light}{RGB}{255, 254, 231}

\definecolor{wangruofan-orange}{RGB}{254, 242, 235}
\definecolor{wangruofan-orange-dark}{RGB}{229, 130, 95}
\definecolor{wangruofan-orange-light}{RGB}{255, 246, 233}

\definecolor{wangyixu-purple}{RGB}{232, 221, 243}
\definecolor{wangyixu-purple-dark}{RGB}{161,115,196}
\definecolor{wangyixu-purple-light}{RGB}{244, 239, 251}

\definecolor{gaoyifeng-pink}{RGB}{255,235,245}
\definecolor{gaoyifeng-pink-dark}{RGB}{216,116,152}
\definecolor{gaoyifeng-pink-light}{RGB}{255, 250, 253}

\usepackage{xspace}
\usepackage{arydshln}

\makeatletter
\DeclareRobustCommand\onedot{\futurelet\@let@token\@onedot}
\def\@onedot{\ifx\@let@token.\else.\null\fi\xspace}

\usepackage{xcolor}
\definecolor{citecolor}{RGB}{0, 0, 255}
\usepackage[linesnumbered,lined,boxed,commentsnumbered,ruled]{algorithm2e}
\usepackage[pagebackref=false,breaklinks=true,colorlinks,citecolor=citecolor,urlcolor=blue,linkcolor=blue,bookmarks=false]{hyperref}

\ifCLASSOPTIONcompsoc
  \usepackage{cite}
\else
  \usepackage{cite}
\fi
\ifCLASSINFOpdf
\else
\fi
\usepackage{authblk}

\makeatletter
\renewcommand\AB@affilsepx{, \protect\Affilfont} %
\makeatother
\usepackage{supertabular}

\begin{document}
%



\title{
  \begin{minipage}{0.04\textwidth}
  \vspace{-0.1cm}
    \includegraphics[scale=0.18]{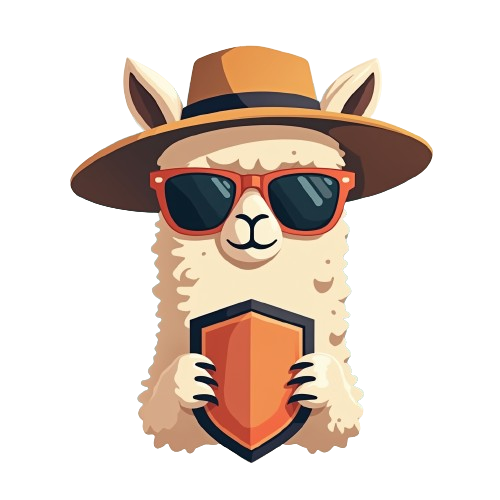}
  \end{minipage}
  \hspace{0.07\textwidth}
  \begin{minipage}{0.86\textwidth}
    \centering
    Safety at Scale: A Comprehensive Survey of Large Model and Agent Safety
  \end{minipage}
}

%
%
%
%

\author[1]{Xingjun Ma}
\author[1]{Yifeng Gao}
\author[1]{Yixu Wang}
\author[1]{Ruofan Wang}
\author[1]{Xin Wang}
\author[1]{Ye Sun}
\author[1]{Yifan Ding}
\author[1]{Hengyuan Xu}
\author[1]{Yunhao Chen}
\author[1]{Yunhan Zhao}
\author[2]{Hanxun Huang}
\author[3]{Yige Li}
\author[4]{Yutao Wu}
\author[5]{Jiaming Zhang}
\author[6]{Xiang Zheng}
\author[7]{Yang Bai}
\author[8]{Yiming Li}
\author[1]{Zuxuan Wu}
\author[1]{Xipeng Qiu}
\author[9,10]{Jingfeng Zhang}
\author[11]{Xudong Han}
\author[11]{Haonan Li}
\author[3]{Jun Sun}
\author[6]{Cong Wang}
\author[13]{Jindong Gu}
\author[14]{Baoyuan Wu}
\author[15]{Siheng Chen}
\author[8]{Tianwei Zhang}
\author[8]{Yang Liu}
\author[2]{Mingming Gong}
\author[16]{Tongliang Liu}
\author[17]{Shirui Pan}
\author[18]{Cihang Xie}
\author[19]{Tianyu Pang}
\author[20]{Yinpeng Dong}
\author[21]{Ruoxi Jia}
\author[22]{Yang Zhang}
\author[23]{Shiqing Ma}
\author[24]{Xiangyu Zhang}
\author[25]{Neil Gong}
\author[26]{Chaowei Xiao}
\author[2]{Sarah Erfani}
\author[2,11]{Tim Baldwin}
\author[27]{Bo Li}
\author[10,12]{Masashi Sugiyama}
\author[8]{Dacheng Tao}
\author[2]{James Bailey}
\author[1]{Yu-Gang Jiang$^{\dag}$\thanks{$^{\dag}$Corresponding author: ygj@fudan.edu.cn}}

\affil[1]{Fudan University}
\affil[2]{The University of Melbourne}
\affil[3]{Singapore Management University}
\affil[4]{Deakin University}
\affil[5]{Hong Kong University of Science and Technology}
\affil[6]{City University of Hong Kong}
\affil[7]{ByteDance}
\affil[8]{Nanyang Technological University}
\affil[9]{University of Auckland}
\affil[10]{RIKEN}
\affil[11]{MBZUAI}
\affil[12]{The University of Tokyo}
\affil[13]{University of Oxford}
\affil[14]{Chinese University of Hong Kong, Shenzhen}
\affil[15]{Shanghai Jiao Tong University}
\affil[16]{The University of Sydney}
\affil[17]{Griffith University}
\affil[18]{University of California, Santa Cruz}
\affil[19]{Sea AI Lab}
\affil[20]{Tsinghua University}
\affil[21]{Virginia Tech}
\affil[22]{CISPA Helmholtz Center for Information Security}
\affil[23]{University of Massachusetts Amherst}
\affil[24]{Purdue University}
\affil[25]{Duke University}
\affil[26]{University of Wisconsin - Madison}
\affil[27]{University of Illinois Urbana-Champaign}

\IEEEtitleabstractindextext{
\begin{abstract}
The rapid advancement of large models, driven by their exceptional abilities in learning and generalization through large-scale pre-training, has reshaped the landscape of Artificial Intelligence (AI). 
These models are now foundational to a wide range of applications, including conversational AI, recommendation systems, autonomous driving, content generation, medical diagnostics, and scientific discovery. However, their widespread deployment also exposes them to significant safety risks, raising concerns about robustness, reliability, and ethical implications.
This survey provides a systematic review of current safety research on large models, covering Vision Foundation Models (VFMs), Large Language Models (LLMs), Vision-Language Pre-training (VLP) models, Vision-Language Models (VLMs), Diffusion Models (DMs), and large-model-powered Agents. 
Our contributions are summarized as follows: (1) We present a comprehensive taxonomy of safety threats to these models, including adversarial attacks, data poisoning, backdoor attacks, jailbreak and prompt injection attacks, energy-latency attacks, data and model extraction attacks, and emerging agent-specific threats. 
(2) We review defense strategies proposed for each type of attacks if available and summarize the commonly used datasets and benchmarks for safety research.
(3) Building on this, we identify and discuss the open challenges in large model safety, emphasizing the need for comprehensive safety evaluations, scalable and effective defense mechanisms, and sustainable data practices. More importantly, we highlight the necessity of collective efforts from the research community and international collaboration.
Our work can serve as a useful reference for researchers and practitioners, fostering the ongoing development of comprehensive defense systems and platforms to safeguard AI models. GitHub: \url{https://github.com/xingjunm/Awesome-Large-Model-Safety}.

\end{abstract}
}

\maketitle

\IEEEdisplaynontitleabstractindextext

%
\IEEEpeerreviewmaketitle

\section{Introduction}
\label{sec:introduction}

\begin{figure*}[h]
    \centering
    \subfigure{
    \includegraphics[width=0.31\textwidth]{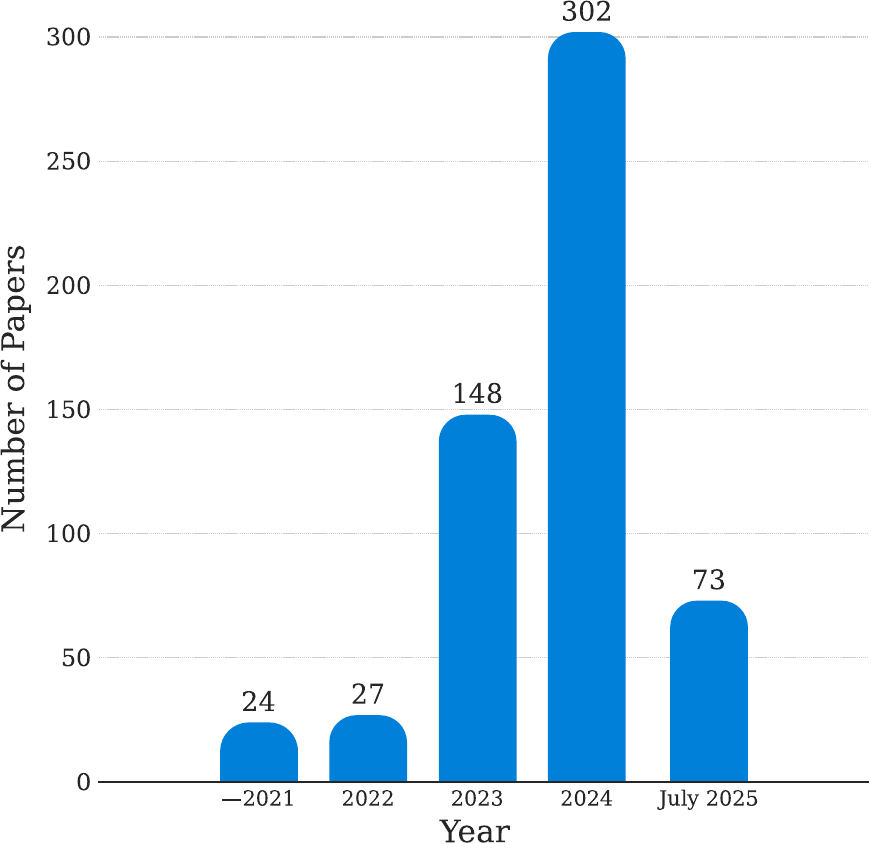}
    }
    \subfigure{
    \includegraphics[width=0.31\textwidth]{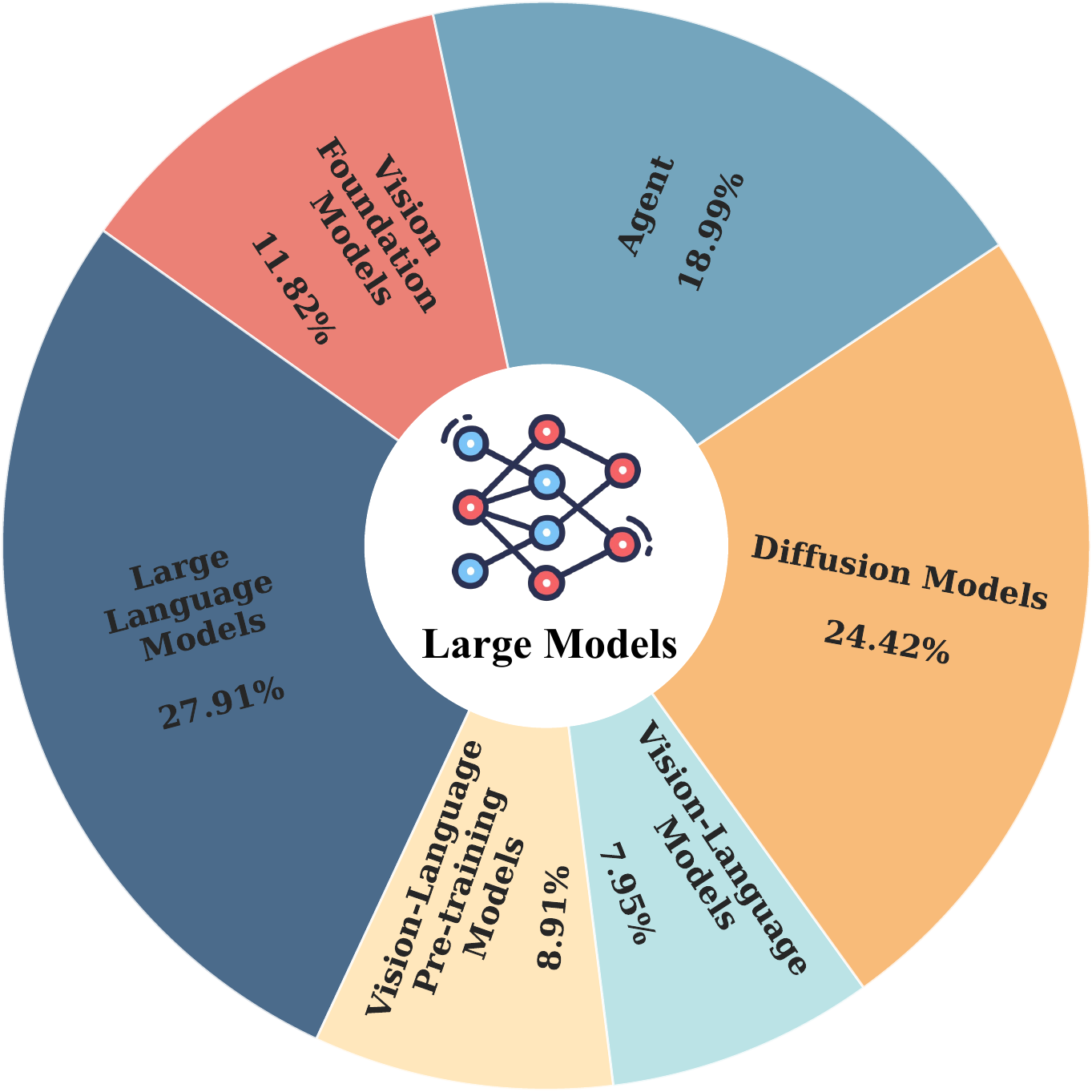}
    }
    \subfigure{
    \includegraphics[width=0.31\textwidth]{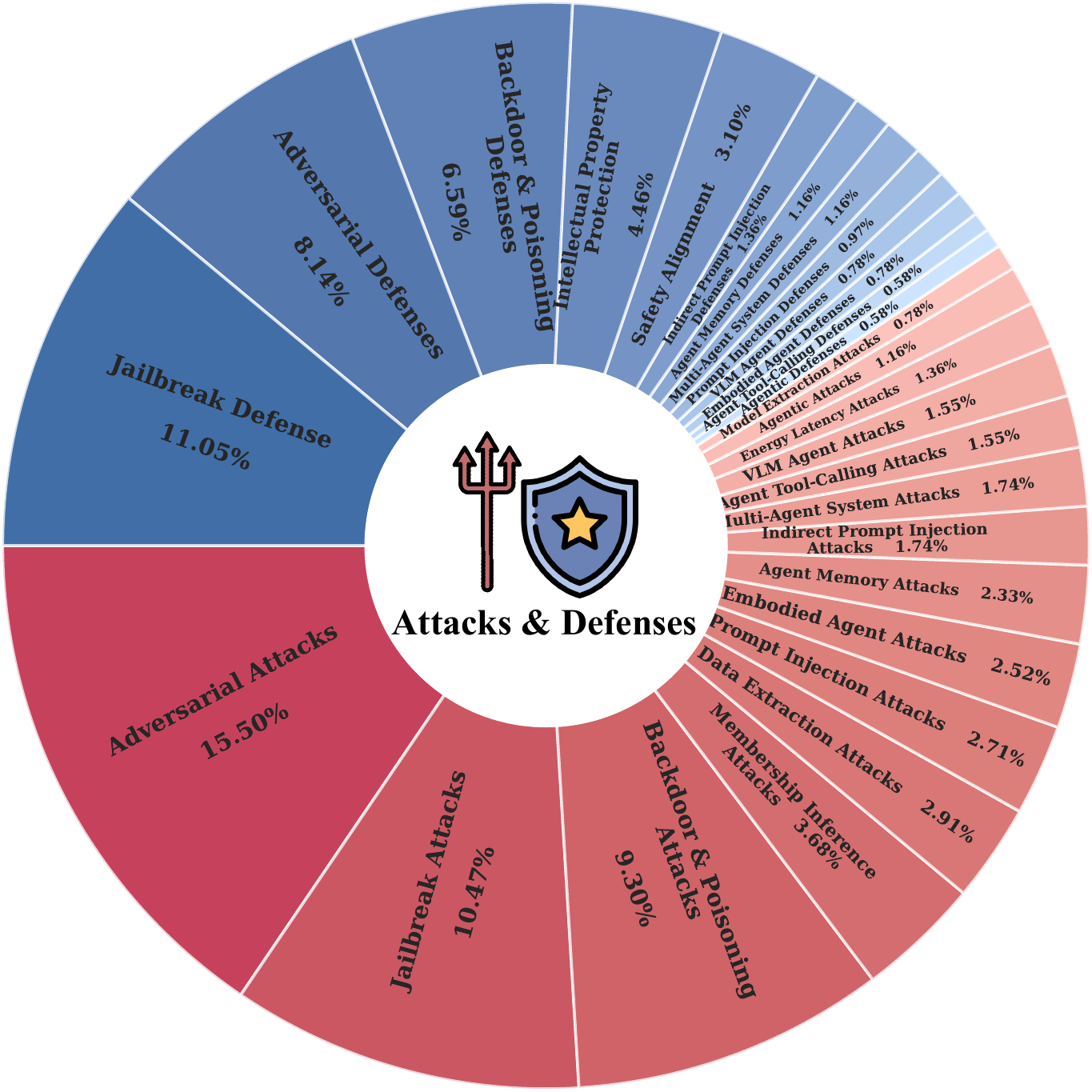}
    }
    \caption{\textbf{Left}: The number of surveyed technical papers on attacks, defenses, and benchmarks/datasets. \textbf{Middle}: Distribution of surveyed technical papers by model type.  \textbf{Right}: Distribution of surveyed technical papers by attack and defense type.}
    \label{fig:total_num}
\end{figure*}

\begin{figure*}[htb]
\centering
    \includegraphics[width=1\linewidth]{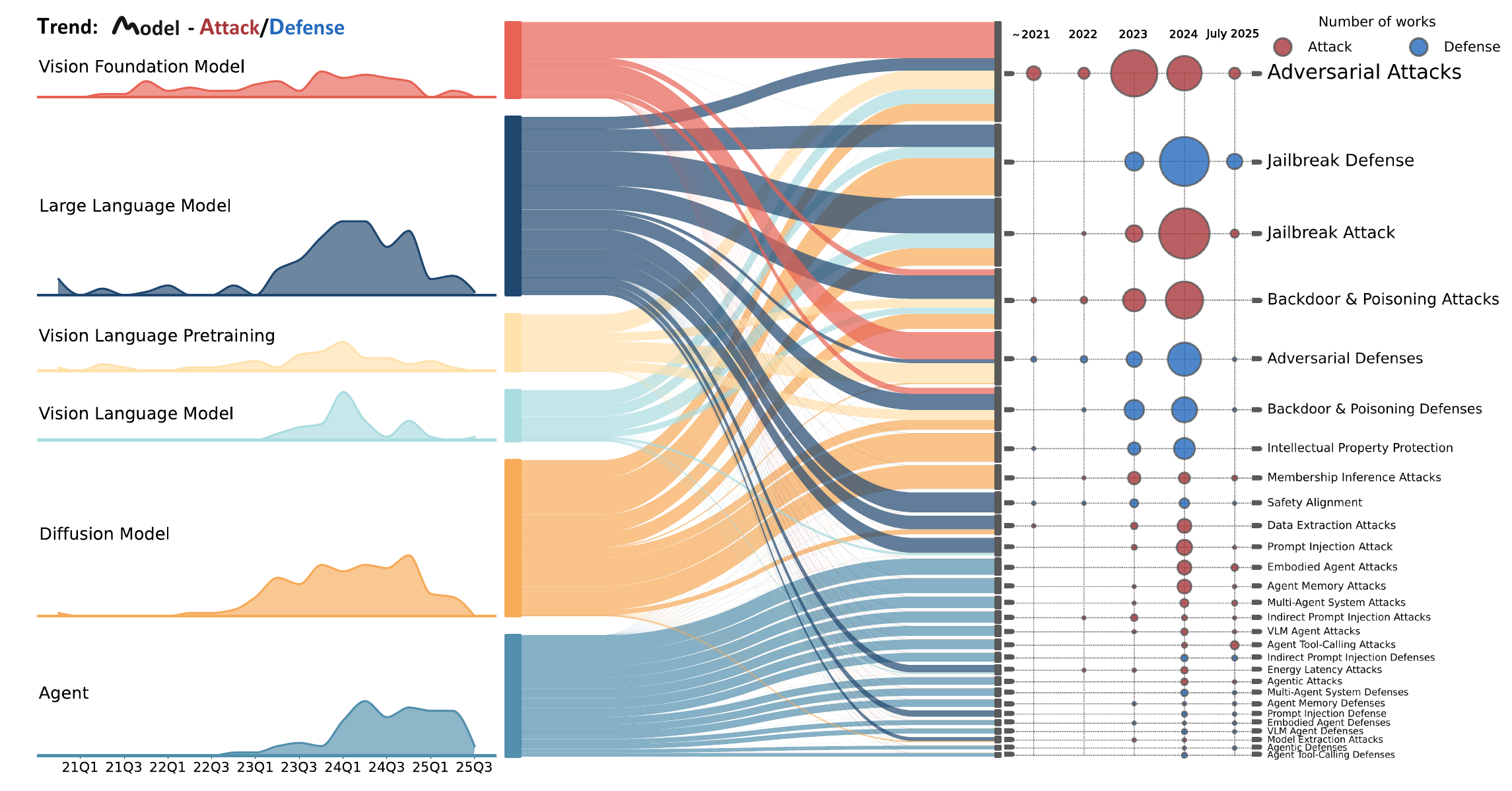}
        \caption{ \textbf{Left}: The quarterly trend in the number of surveyed safety papers across different models; \textbf{Middle}: Proportional distribution of attack and defense studies associated with large models. \textbf{Right}: Annual trend in the number of surveyed safety papers on various attacks and defenses, ordered from most to least studied. }
\label{fig:crossview}
\end{figure*}

\IEEEPARstart{A}rtificial Intelligence (AI) has entered the era of large models, exemplified by Vision Foundation Models (VFMs), Large Language Models (LLMs), Vision-Language Pre-Training (VLP) models, Vision-Language Models (VLMs), and image/video generation diffusion models (DMs).
Through large-scale pre-training on massive datasets, these models have demonstrated unprecedented capabilities in tasks ranging from language understanding and image generation to complex problem-solving and decision-making. Their ability to understand and generate human-like content (e.g., texts, images, audios, and videos) has enabled applications in customer service, content creation, healthcare, education, and more,  highlighting their transformative potential in both commercial and societal domains.


However, the deployment of large models comes with significant challenges and risks. 
As these models become more integrated into critical applications, concerns regarding their vulnerabilities to adversarial, jailbreak, and backdoor attacks, data privacy breaches, and the generation of harmful or misleading content have intensified. 
These issues pose substantial threats, including unintended system behaviors, privacy leakage, and the dissemination of harmful information. Ensuring the safety of these models is paramount to prevent such unintended consequences, maintain public trust, and promote responsible AI usage.
The field of AI safety research has expanded in response to these challenges, encompassing a diverse array of attack methodologies, defense strategies, and evaluation benchmarks designed to identify and mitigate the vulnerabilities of large models. 
Given the rapid development of safety-related techniques for various large models, we aim to provide a comprehensive survey of these techniques, highlighting strengths, weaknesses, and gaps, while advancing research and fostering collaboration.

\newcommand{\boxfont}{\fontsize{5.7pt}{6.5pt}\selectfont} 

\newcommand{\TLM}{Large Model and Agent Safety}
\tikzstyle{node-cfg}=[
    rectangle,
    draw=brown!60!black,
    sharp corners,
    text opacity=1,
    inner sep=3pt, 
    text=black,
    fill=yellow!5,
    fill opacity=.4, 
    line width=1pt,
    font=\tiny,
    align=left
]
\tikzstyle{vfm-bg-node}=[
    rectangle,
    rounded corners=2pt, 
    fill=sunye-red-light, 
    draw=sunye-red-dark,
    inner sep=3pt, 
    line width=1pt,
    text opacity=1,
    text=black, 
    font=\boxfont, 
    align=left 
]

\tikzstyle{vfm-line-node}=[
    rectangle,
    rounded corners=2pt, 
    fill opacity=.2,
    draw=sunye-red-dark,
    inner sep=3pt, 
    line width=0.5pt,
    text opacity=1,
    text=black, 
    font=\tiny, 
    align=left 
]

\tikzstyle{llm-bg-node}=[
    rectangle,
    rounded corners=2pt, 
    fill=dingyifan-wangyixu-darkblue-light, 
    draw=dingyifan-wangyixu-darkblue-dark, 
    line width=1pt,
    text opacity=1,
    text=black, 
    font=\boxfont, 
    align=left 
]

\tikzstyle{llm-line-node}=[
    rectangle,
    rounded corners=2pt, 
    fill=none, 
    fill opacity=.2,
    draw=dingyifan-wangyixu-darkblue-dark,
    inner sep=3pt, 
    line width=0.5pt,
    text opacity=1,
    text=black, 
    font=\tiny, 
    align=left 
]

\tikzstyle{vlp-bg-node}=[
    rectangle,
    rounded corners=2pt, 
    fill=wangxin-yellow-light, 
    draw=wangxin-yellow-dark, 
    inner sep=3pt, 
    line width=1pt,
    text opacity=1,
    text=black, 
    font=\boxfont, 
    align=left 
]

\tikzstyle{vlp-line-node}=[
    rectangle,
    rounded corners=2pt, 
    fill=none, 
    fill opacity=.2,
    draw=wangxin-yellow-dark,
    inner sep=3pt, 
    line width=0.5pt,
    text opacity=1,
    text=black, 
    font=\tiny, 
    align=left 
]

\tikzstyle{vlm-bg-node}=[
    rectangle,
    rounded corners=2pt, 
    fill=wangruofan-orange-light, 
    draw=wangruofan-orange-dark, 
    inner sep=3pt, 
    line width=1pt,
    text opacity=1,
    text=black, 
    font=\boxfont, 
    align=left 
]

\tikzstyle{vlm-line-node}=[
    rectangle,
    rounded corners=2pt, 
    fill=none, 
    fill opacity=.35,
    draw=wangruofan-orange-dark,
    inner sep=3pt, 
    line width=0.5pt,
    text opacity=1,
    text=black, 
    font=\tiny, 
    align=left 
]

\tikzstyle{dm-bg-node}=[
    rectangle,
    rounded corners=2pt, 
    fill=gaoyifeng-pink-light, 
    draw=gaoyifeng-pink-dark, 
    inner sep=3pt, 
    line width=1pt,
    text opacity=1,
    text=black, 
    font=\boxfont, 
    align=left 
]

\tikzstyle{dm-line-node}=[
    rectangle,
    rounded corners=2pt, 
    fill=none, 
    fill opacity=.2,
    draw=gaoyifeng-pink-dark,
    inner sep=3pt, 
    line width=0.5pt,
    text opacity=1,
    text=black, 
    font=\tiny, 
    align=left 
]

\tikzstyle{agent-bg-node}=[
    rectangle,
    rounded corners=2pt, 
    fill=wangyixu-purple-light, 
    draw=wangyixu-purple-dark, 
    inner sep=3pt, 
    line width=1pt,
    text opacity=1,
    text=black, 
    font=\boxfont, 
    align=left 
]

\tikzstyle{agent-line-node}=[
    rectangle,
    rounded corners=2pt, 
    fill=none, 
    fill opacity=.2,
    draw=wangyixu-purple-dark,
    inner sep=3pt, 
    line width=0.5pt,
    text opacity=1,
    text=black, 
    font=\tiny, 
    align=left 
]

\tikzset{
  vfm-line-node/.append style = {font=\boxfont},
  llm-line-node/.append style = {font=\boxfont},
  vlp-line-node/.append style = {font=\boxfont},
  vlm-line-node/.append style = {font=\boxfont},
  dm-line-node/.append style  = {font=\boxfont},
  agent-line-node/.append style = {font=\boxfont}
}

\tikzset{
  vfm-line-node/.append style = {inner ysep=1.5pt},
  llm-line-node/.append style = {inner ysep=1.5pt},
  vlp-line-node/.append style = {inner ysep=1.5pt},
  vlm-line-node/.append style = {inner ysep=1.5pt},
  dm-line-node/.append style  = {inner ysep=1.5pt},
  agent-line-node/.append style = {inner ysep=1.5pt}
}

\begin{figure*}[t!]
    \centering
        \begin{forest}
            forked edges,
            for tree={
                grow=east,
                reversed=true,
                anchor=mid west,
                parent anchor=mid east,
                child anchor=mid west,
                base=left,
                font=\tiny,
                rectangle,
                draw=none,
                sharp corners,
                align=left,
                minimum width=2em,
                edge+={darkgray, line width=0.6pt},
                s sep=2.8pt,
                inner xsep=1.8pt,
                inner ysep=2.7pt,
                line width=0.7pt,
                par/.style={rotate=90, child anchor=north, parent anchor=south, anchor=center},
            },
            where level=0{font=\footnotesize,}{},
            where level=1{text width=4.3em,font=\tiny,}{},
            where level=2{text width=9.7em,font=\tiny,}{},
            [
                {\TLM}, par
                [
                    \parbox{9em}{Vision Foundation \\ Models (\S~\ref{sec:vfm})}, vfm-bg-node
                    [
                        Attacks and Defenses for ViT (\S~\ref{sec:vfm_vit}), vfm-line-node
                        [
                           \textbf{(\romannumeral 1) Adversarial Attacks:} 
                           \cite{fu2022patch} \cite{navaneet2024slowformer} \cite{gao2024pe} \cite{lovisotto2022give} \cite{jain2024towards} \cite{naseer2021improving} \cite{wang2022generating} \cite{wei2022towards} \cite{wei2023boosting} \cite{ma2023transferable} \cite{zhang2023transferable} \cite{zhang2024improving} \cite{gao2024attacking} \cite{shi2022decision} \cite{wu2024improving} \cite{li2024improving} \cite{joshi2021adversarial} \cite{chen2023understanding} \cite{wei2023towards}
                           \\
                           \textbf{(\romannumeral 2) Adversarial Defenses:}
                           \cite{wu2022towards} \cite{li2022patch} \cite{sun2024vitguard} \cite{liu2023understanding} \cite{bai2024diffusion} \cite{li2024adbm} \cite{lei2024instant} \cite{gu2022vision} \cite{mo2022adversarial} \cite{guo2023robustifying} \cite{guo2023improving} \cite{huimproving} \cite{gong2024random} \cite{nie2022diffusion} \cite{zhang2023purify++} \cite{chen2024diffilter} \cite{song2024mimicdiffusion} \cite{khalili2024lightpure}\cite{zollicoffer2024lorid}
                           {\quad} \\
                           \textbf{(\romannumeral 3) Backdoor Attacks:}
                           \cite{yuan2023you} \cite{zheng2023trojvit} \cite{yang2024not} \cite{lv2023dbia} \cite{li2024multi} 
                           {\quad}
                           \textbf{(\romannumeral 4) Backdoor Defenses:}
                           \cite{doan2023defending} \cite{subramanya2024closer} \cite{subramanya2022backdoor} 
                           , vfm-line-node, text width=32.7em
                        ]             
					]
                    [
                        Attacks and Defenses for SAM (\S~\ref{sec:vfm_sam}), vfm-line-node
                        [
                           \textbf{(\romannumeral 1) Adversarial Attacks:} 
                           \cite{shen2024practical} \cite{croce2024segment} \cite{zhang2023attack} \cite{zheng2023black} \cite{lu2024unsegment} \cite{xia2024transferable} \cite{han2023segment} \cite{zhou2024darksam}
                           {\quad}
                           \textbf{(\romannumeral 2) Adversarial Defenses:}
                           \cite{li2024asam} 
                           \\
                           \textbf{(\romannumeral 3) Backdoor \& Poisoning Attacks:}
                           \cite{guan2024badsam} \cite{sun2024unseg}
                           , vfm-line-node, text width=32.7em
                        ]       
                    ]
			]
                [
                    \parbox{7em}{Large Language \\ Models (\S~\ref{sec:llm})} , llm-bg-node
                    [
                        Adversarial Attack (\S~\ref{sec:llm_adversarial_attack}), llm-line-node
                        [
                           \textbf{(\romannumeral 1) White-Box:} \cite{boucher2022bad}  \cite{jin2020bert} \cite{li2020bert} \cite{guo2021gradient} \cite{dirkson2021breaking} \cite{wang2023gradient} \cite{liu2023expanding}{\quad} 
                           \textbf{(\romannumeral 2) Black-Box:}
                           \cite{wang2023adversarial} \cite{liu2023adversarial} \cite{koleva2023adversarial}
                        , llm-line-node, text width=32.7em
                        ] 
                    ]
                    [
                        Adversarial Defense (\S~\ref{sec:llm_adversarial_defense}), llm-line-node
                        [
                           \textbf{(\romannumeral 1) Adversarial Detection:}
                           \cite{jain2023baseline} \cite{kumar2023certifying}
                           {\quad}
                           \textbf{(\romannumeral 2) Robust Inference:}
                           \cite{zou2024improving}
                        , llm-line-node, text width=32.7em
                        ] 
                    ]
                    [
                        Jailbreak Attacks (\S~\ref{sec:llm_jailbreak_attacks}), llm-line-node
                        [
                           \textbf{(\romannumeral 1) Black-Box:} 
                           \cite{yong2023low}
                           \cite{yuan2023gpt} \cite{wei2024jailbroken} \cite{li2024cross} \cite{zhou2024easyjailbreak} \cite{zou2024system} \cite{xiao2024tastle} \cite{li2024structuralsleight} \cite{lv2024codechameleon} \cite{chang2024play} 
                           \cite{wen2024evaluating}
                           \cite{lin2024llms}
                           \cite{wen2024evaluating}
                           \cite{liu2023autodan} \cite{yu2023gptfuzzer} \cite{chao2023jailbreaking} \cite{deng2024masterkey} \cite{yu2024enhancing} \cite{yao2024fuzzllm} \cite{zhang2024enja} \cite{perez2022red} \cite{hong2024curiosity} \\ 
                           \cite{SCBSZ24}
                           \cite{zou2023universal}
                           \cite{jia2024improved} 
                           \cite{huang2024semantic}
                           \cite{zheng2024improved}
                           \cite{zhao2024weak}
                           \cite{jiang2024unlocking}
                           \textbf{(\romannumeral 2) White-Box:} 
                           \cite{qi2023fine}
                           \cite{huang2025virus}
                        , llm-line-node, text width=32.7em
                        ] 
                    ]
                    [
                        Jailbreak Defenses (\S~\ref{sec:llm_jailbreak_defenses}), llm-line-node
                        [
                           \textbf{(\romannumeral 1) Input Defense:} 
                           \cite{robey2023smoothllm} \cite{ji2024defending} \cite{wang2024selfdefend} \cite{liu2024protecting} 
                           \cite{liangpearl}
                           \cite{liangvulnerability}
                           \cite{wang2024defending}
                           \cite{yung2025round}
                           \cite{yung2025curvalid}
                           {\quad}
                           \textbf{(\romannumeral 2) Output Defense:}
                           \cite{kim2023robust} 
                           \cite{xiong2024defensive} \cite{hu2024gradient} 
                           \cite{gao2024shaping}
                           \cite{wu2024legilimens}
                           \\
                           \textbf{(\romannumeral 3) Ensemble Defense:}
                           \cite{chen2023jailbreaker} \cite{zhang2024parden}
                           \cite{lu2024autojailbreak} 
                           \cite{du2024mogu}
                           \textbf{(\romannumeral 4) Defenses against Fine-tuning Attacks:}
                           \cite{huang2024vaccine}
                           \cite{liu2024targeted}
                           \cite{huang2024booster}
                           \cite{huang2024lisa}
                           \cite{huang2024antidote}
                           \cite{wang2025panacea}
                        , llm-line-node, text width=32.7em
                        ] 
                    ]
                    [
                        Prompt Injection Attacks (\S~\ref{sec:llm_prompt_injection_attacks}), llm-line-node
                        [
                           \textbf{(\romannumeral 1) Black-Box:} 
                           \cite{perez2022ignore} \cite{liu2023prompt} \cite{greshake2023not} \cite{deng2023attack} 
                           \cite{liu2024formalizing}
                           \cite{ye2024we}
                        \cite{liu2024automatic} 
  \cite{zhang2024goal} 
                           \cite{hui2024pleak}
                           \cite{liu2024formalizing}
                           \cite{ye2024we}
                           \cite{shi2024optimization}
                           \cite{shao2024making}
                           \cite{yu2024promptfuzz}
                        , llm-line-node, text width=32.7em
                        ] 
                    ]
                    [
                        Prompt Injection Defenses (\S~\ref{sec:llm_prompt_injection_defenses}), llm-line-node
                        [
                           \textbf{(\romannumeral 1) Input Defense:} 
                           \cite{chen2024struq} \cite{sharma2024spml}
                           {\quad}
                           \textbf{(\romannumeral 2) Parameter Defense:}
                           \cite{piet2023jatmo}
                           \cite{yi2023benchmarking}
                           \cite{chen2025SecAlign}
                        , llm-line-node, text width=32.7em
                        ] 
                    ]
                    [
                        Backdoor Attacks (\S~\ref{sec:llm_backdoor_attacks}), llm-line-node
                        [
                           \textbf{(\romannumeral 1) Data Poisoning:} 
                           \cite{cai2022badprompt} \cite{yan2022bite} \cite{yao2024poisonprompt} \cite{zhao2023prompt} \cite{xu2023instructions} 
                           \cite{zhang2024instruction}
                           \cite{kandpal2023backdoor} \cite{xiang2024badchain} \cite{zhao2024universal} \cite{qiang2024learning} \cite{pathmanathan2024poisoning} \cite{hubinger2024sleeper} \cite{he2024data} \cite{zhang2024human} \cite{yan2024llm} \cite{huang2023composite}
                           \\
                           \textbf{(\romannumeral 2) Training Manipulation:}
                           \cite{gu2023gradient} \cite{xue2024trojllm} \cite{yan2024backdooring}
                           \cite{zeng2024uncertainty}
                           {\quad}
                           \textbf{(\romannumeral 3) Parameter Modification:}
                           \cite{li2024badedit}
                        , llm-line-node, text width=32.7em
                        ] 
                    ]
                    [
                        Backdoor Defenses (\S~\ref{sec:llm_backdoor_defenses}), llm-line-node
                        [
                           \textbf{(\romannumeral 1) Backdoor Detection:} 
                           \cite{he2023imbert} \cite{li2023defending} \cite{sun2023defending} \cite{yan2024parafuzz} \cite{xi2024defending} 
                           \cite{yi2025probe}
                           {\quad}
                           \textbf{(\romannumeral 2) Backdoor Removal:}
                           \cite{lamparth2024analyzing} \cite{li2024backdoor} \cite{zeng2024beear} \cite{min2024crow}
                           \\
                           \textbf{(\romannumeral 3) Robust Training:}
                           \cite{tang2023setting} \cite{liu2023maximum}
                           \cite{wang2024data}
                           {\quad}
                           \textbf{(\romannumeral 4) Robust Inference:}
                           \cite{tong2024securing} \cite{li2024cleangen}
                           \cite{li2024purifying}
                        , llm-line-node, text width=32.7em
                        ] 
                    ]
                    [
                        Safety Alignment (\S~\ref{sec:llm_safety alignment}), llm-line-node
                        [
                           \textbf{(\romannumeral 1) Human Feedback:} 
                           \cite{christiano2017deep} \cite{ziegler2019fine} \cite{ouyang2022training} \cite{dai2023safe} \cite{an2023direct,rafailov2024direct} \cite{zhou2023beyond} \cite{ethayarajh2024kto} \cite{zhou2024lima}
                           \\
                           \textbf{(\romannumeral 2) AI Feedback:}
                           \cite{bai2022constitutional} \cite{sun2024principle} \cite{yang2024rlcd}
                           {\quad}
                           \textbf{(\romannumeral 3) Social Interactions:}
                           \cite{liu2023training}
                           \cite{pang2024self}
                           \textbf{(\romannumeral 4) Deceptive Alignment:}
                           \cite{wang2024fake}
                           \cite{greenblatt2024alignment}
                           \cite{sheshadri2025some}
                        , llm-line-node, text width=32.7em
                        ] 
                    ]
                    [
                        Energy Latency Attacks (\S~\ref{sec:llm_energy_latency_attacks}), llm-line-node
                        [
                           \textbf{(\romannumeral 1) White-Box:} 
                           \cite{chen2022nmtsloth} 
                           \cite{dong2025engorgio}
                           \cite{chen2023dynamic} \cite{feng2024llmeffichecker} \cite{gao2024ttslow}
                           {\quad}
                           \textbf{(\romannumeral 2) Black-Box:}
                           \cite{zhang2023no} \cite{gao2024denial}
                        , llm-line-node, text width=32.7em
                        ] 
                    ]
                    [
                        Model Extraction Attacks (\S~\ref{sec:llm_model_extraction_attacks}), llm-line-node
                        [
                           \textbf{(\romannumeral 1) Fine-tuning Stage:} 
                           \cite{jiang2023lion} \cite{li2024extracting}
                           {\quad}
                           \textbf{(\romannumeral 2) Alignment Stage:}
                           \cite{liang2024alignment}
                        , llm-line-node, text width=32.7em
                        ] 
                    ]
                    [
                        Data Extraction Attacks (\S~\ref{sec:llm_data_extraction_attacks}), llm-line-node
                        [
                           \textbf{(\romannumeral 1) Black-Box:} 
                           \cite{carlini2019secret} \cite{carlini2021extracting} \cite{nasr2023scalable} \cite{xu2024magpie} \cite{al2024traces} \cite{bai2024special} \cite{kassem2024alpaca} \cite{qi2024follow} \cite{more2024towards} 
                           \cite{zhang2025benchmark}
                           \cite{yu2023bag}
                           {\quad}
                           \textbf{(\romannumeral 2) White-Box:}
                           \cite{duan2024uncovering}
                        , llm-line-node, text width=32.7em
                        ] 
                    ]
                ]
                [
                    \parbox{7em}{Vision-Language \\ Pre-Training \\ Models (\S~\ref{sec:vlp})}, vlp-bg-node
                    [
                        Adversarial Attacks (\S~\ref{sec:vlp-attack}), vlp-line-node
                        [
                           \textbf{(\romannumeral 1) White-Box:} 
                           \cite{zhang2022towards} \cite{zhou2023advclip} \cite{noever2021reading} \cite{wang2025typographic}
                           {\quad}
                           \textbf{(\romannumeral 2) Black-Box:}  
                           \cite{lu2023set} \cite{he2023sa} \cite{wang2023exploring} \cite{wang2024transferable} \cite{yin2024vlattack} \cite{hu2024firm} \cite{fang2024one} \cite{zhang2024universal}\cite{huang2025xtransfer} \cite{gao2024boosting} \cite{han2023ot}
                            , vlp-line-node, text width=32.7em
                        ]
                    ]
                    [
                        Adversarial Defenses (\S~\ref{sec:vlp-defenses}), vlp-line-node
                        [
                           \textbf{(\romannumeral 1) Adversarial Tuning:} 
                           \cite{azuma2023defense} \cite{zhang2023adversarial} \cite{li2024one} \cite{fan2024mixprompt} \cite{hussein2024promptsmooth} \cite{zhou2024few} \cite{luo2024apd} \cite{wang2024tapt} \cite{mao2023understanding} \cite{wang2024pre} \cite{zhou2024revisiting} \cite{schlarmannrobust} 
                           \\
                           \textbf{(\romannumeral 2) Adversarial Training:} 
                           \cite{wang2024revisiting} \cite{gan2020large}
                           {\quad}
                           \textbf{(\romannumeral 3) Adversarial Detection:}
                           \cite{fares2024mirrorcheck} \cite{wang2024advqdet}
                            , vlp-line-node, text width=32.7em
                        ]
                    ]
                    [
                        Backdoor \& Poisoning Attacks (\S~\ref{sec:vlp-bp-attacks}), vlp-line-node
                        [
                           \textbf{(\romannumeral 1) Backdoor:} 
                           \cite{jia2022badencoder} \cite{zhang2024data} \cite{liang2024badclip} \cite{bai2024badclip}
                           {\quad}
                           \textbf{(\romannumeral 2) Poisoning:}  \cite{yang2023data} \cite{liu2024multimodal}
                           {\quad}
                           \textbf{(\romannumeral 3) Backdoor \& Poisoning:}
                           \cite{carlini2022poisoning}
                            , vlp-line-node, text width=32.7em
                        ]
                    ]
                    [
                        Backdoor \& Poisoning Defenses (\S~\ref{sec:vlp-bp-defenses}), vlp-line-node
                        [
                           \textbf{(\romannumeral 1) Backdoor Removal:} 
                           \cite{bansal2023cleanclip} \cite{yang2023better}
                           {\quad}
                           \textbf{(\romannumeral 2) Robust Training:}
                           \cite{yang2024robust}
                           {\quad}
                           \textbf{(\romannumeral 2) Backdoor Detection:}
                           \cite{feng2023detecting} \cite{sur2023tijo} \cite{liu2024mudjacking} \cite{zhu2024seer}
                           \cite{huang2025detecting}
                            , vlp-line-node, text width=32.7em
                        ]
                    ]
                ]
                [
                    \parbox{7em}{Vision-Language \\ Models (\S~\ref{sec:vlm})}, vlm-bg-node
                    [
                        Adversarial Attacks (\S~\ref{sec:vlm-adversarial}), vlm-line-node
                        [
                           \textbf{(\romannumeral 1) White-Box:}   
                           \cite{schlarmann2023adversarial} \cite{cui2024robustness} \cite{luo2024image} \cite{gao2024adversarial} \cite{wang2024stop}
                           {\quad}
                           \textbf{(\romannumeral 2) Gray-Box:}
                           \cite{wang2023instructta} 
                           {\quad}
                           \textbf{(\romannumeral 3) Black-Box:}
                           \cite{dong2023robust} \cite{zhao2024evaluating} \cite{guo2024efficiently} \cite{zhang2024anyattack} 
                           , vlm-line-node, text width=32.7em
                        ]
                    ]
                    [
                        Jailbreak Attacks (\S~\ref{sec:vlm-jailbreak}), vlm-line-node
                        [
                           \textbf{(\romannumeral 1) White-Box:} 
                           \cite{bailey2023image} \cite{carlini2024aligned} \cite{qi2024visual} \cite{niu2024jailbreaking} \cite{wang2024white} \cite{li2024images}
                           {\quad}
                           \textbf{(\romannumeral 2) Black-Box:} 
                           \cite{shayegani2023jailbreak} \cite{gong2023figstep} \cite{wu2023jailbreaking} \cite{ma2024visual} \cite{wang2025ideator}
                           , vlm-line-node, text width=32.7em
                        ]
                    ]
                    [
                        Jailbreak Defenses (\S~\ref{sec:vlm-defenses}), vlm-line-node
                        [
                           \textbf{(\romannumeral 1) Jailbreak Detection:} 
                           \cite{zhang2023mutation} \cite{sharma2024defending} 
                           {\quad}
                           \textbf{(\romannumeral 2) Jailbreak Prevention:}
                           \cite{wang2024adashield} \cite{pi2024mllm} \cite{gou2024eyes} \cite{wang2024inferaligner} \cite{zhao2024bluesuffix}
                           , vlm-line-node, text width=32.7em
                        ]
                    ]
                    [
                        Energy Latency Attacks (\S~\ref{sec:vlm-latency}), vlm-line-node
                        [
                           \textbf{(\romannumeral 1) White-Box:} 
                           \cite{gaoinducing}
                           , vlm-line-node, text width=32.7em
                        ]
                    ]
                    [
                        Prompt Injection Attack (\S~\ref{sec:vlm-injection}), vlm-line-node
                        [
                           \textbf{(\romannumeral 1) White-Box:} 
                           \cite{bagdasaryan2023ab}
                           {\quad}
                           \textbf{(\romannumeral 2) Black-Box:}
                           \cite{qraitem2024vision}
                           , vlm-line-node, text width=32.7em 
                        ]
                    ]
                    [
                        Backdoor \& Poisoning Attacks (\S~\ref{sec:vlm-backdoor}), vlm-line-node
                        [
                           \textbf{(\romannumeral 1) Backdoor:}
                           \cite{liang2024revisiting} \cite{lu2024test} \cite{ni2024physical} \cite{tao2024imgtrojan}
                           {\quad}
                           \textbf{(\romannumeral 1) Poisoning:}
                           \cite{xu2024shadowcast}
                           , vlm-line-node, text width=32.7em
                        ]
                    ]
                ]
                [
                    \parbox{7em}{Diffusion \\ Models (\S~\ref{sec:diffusion})}, dm-bg-node
                    [
                       Adversarial Attacks  (\S~\ref{sec:dm_adversarial_attacks}), dm-line-node
                       [
                           \textbf{(\romannumeral 1) White-Box:}
                           \cite{du2024stable} \cite{liu2023discovering} \cite{zhou2024foolsdedit}
                           {\quad}
                           \textbf{(\romannumeral 2) Gray-Box:}
                           \cite{zhuang2023pilot} \cite{zhang2024revealing} \cite{yang2024multi} \cite{zhou2024dormant}
                           \\
                           \textbf{(\romannumeral 3) Black-Box:}
                           \cite{struppek2023exploiting} \cite{kou2023character} \cite{gao2023evaluating} \cite{daras2022discovering} \cite{milliere2022adversarial} \cite{maus2023black} \cite{liu2023riatig}
                           , dm-line-node, text width=32.7em
                        ]
                    ]
                    [
                       Jailbreak Attacks  (\S~\ref{sec:dm_jailbreak_attacks}), dm-line-node
                       [
                           \textbf{(\romannumeral 1) White-Box:}
                           \cite{rando2022red} \cite{yang2024mma} \cite{chin2023prompting4debugging} \cite{zhang2023generate}
                           {\quad}
                           \textbf{(\romannumeral 2) Gray-Box:}
                           \cite{ma2024jailbreaking} \cite{gao2024rt}
                           \\
                           \textbf{(\romannumeral 3) Black-Box:}
                           \cite{yang2024sneakyprompt} \cite{qu2023unsafe} \cite{dong2024jailbreaking} \cite{liu2024groot} \cite{deng2023divide}  \cite{ba2023surrogateprompt} \cite{huang2025perception} \cite{zhang2025reason2attack} 
                           , dm-line-node, text width=32.7em
                        ]
                    ]
                    [
                       Jailbreak Defenses (\S~\ref{sec:dm_jailbreak_defenses}), dm-line-node
                       [
                           \textbf{(\romannumeral 1) Concept Erasure:}
                           \cite{gandikota2023erasing} \cite{lyu2024one} \cite{kim2023towards} \cite{kumari2023ablating} \cite{hong2024all} \cite{wu2024unlearning} \cite{heng2024selective} \cite{huang2023receler} \cite{kim2024race} \cite{zhang2024defensive} \cite{ni2023degeneration} \cite{zhang2024forget} \cite{liu2024implicit} \cite{zhao2024separable} 
                           \cite{han2024continuous} \cite{li2024safegen}
                           \cite{lee2025concept} \cite{lu2024mace} 
                           \\
                           \cite{gandikota2024unified}
                           \cite{orgad2023editing} \cite{gong2024reliable} \cite{liu2024realera} \cite{chavhan2024conceptprune} \cite{yang2024pruning}
                           {\quad}
                           \textbf{(\romannumeral 2) Inference Guidance:}
                           \cite{schramowski2023safe} \cite{yuan2025promptguard} \cite{cai2024ethical}
                           \cite{li2024self} \cite{meng2025concept}
                           , dm-line-node, text width=32.7em
                        ]
                    ]
                    [
                       Backdoor Attacks (\S~\ref{sec:dm_backdoor_attacks}), dm-line-node
                       [
                           \textbf{(\romannumeral 1) Training Manipulation:}
                           \cite{chou2023backdoor} \cite{chou2024villandiffusion} \cite{chen2023trojdiff} \cite{li2024invisible} \cite{li2024watch} 
                           \\
                           \textbf{(\romannumeral 2) Data Poisoning:}
                           \cite{struppek2023rickrolling} \cite{zhai2023text} \cite{pan2023trojan} \cite{vice2024bagm} \cite{huang2023zero} \cite{huang2024personalization} \cite{wang2024stronger} \cite{naseh2024injecting}
                           , dm-line-node, text width=32.7em
                        ]
                    ]
                    [
                       Backdoor Defenses (\S~\ref{sec:dm_backdoor_defenses}), dm-line-node
                       [
                           \textbf{(\romannumeral 1) Backdoor Detection:}
                           \cite{wang2024t2ishield} \cite{guan2025ufid} \cite{sui2024disdet}
                           {\quad}
                           \textbf{(\romannumeral 2) Backdoor Removal:}
                           \cite{an2024elijah} \cite{hao2024diff} \cite{mo2024terd} \cite{truong2024purediffusion} \cite{zhai2025navidet}
                           , dm-line-node, text width=32.7em
                        ]
                    ]
                    [
                       Membership Inference Attacks (\S~\ref{sec:dm_membership_inference_attacks}), dm-line-node
                       [
                           \textbf{(\romannumeral 1) White-Box:}
                           \cite{dubinski2024towards} \cite{pang2023white} \cite{matsumoto2023membership} \cite{hu2023loss}
                           {\quad}
                           \textbf{(\romannumeral 2) Gray-Box:}
                           \cite{duan2023diffusion} \cite{tang2023membership} \cite{kong2023efficient} \cite{fu2023probabilistic} \cite{zhai2024membership} \cite{li2024unveiling} 
                           \\
                           \textbf{(\romannumeral 3) Black/White-Box:}
                           \cite{matsumoto2023membership}
                           {\quad}
                           \textbf{(\romannumeral 3) Black-Box:}
                           \cite{wu2022membership} \cite{pang2023black} \cite{li2024towards} \cite{fu2024model} \cite{zhang2024generated} 
                           , dm-line-node, text width=32.7em
                        ]
                    ]
                    [
                       Data Extraction Attacks (\S~\ref{sec:dm_data_extraction_attacks}), dm-line-node
                       [
                           \textbf{(\romannumeral 1) Explicit Condition-based Extraction:}
                           \cite{carlini2023extracting} \cite{webster2023reproducible}
                           {\quad}
                           \textbf{(\romannumeral 2) Surrogate Condition-based Extraction:}
                           \cite{chen2024towards} \cite{wu2024revealing}
                           , dm-line-node, text width=32.7em
                        ]
                    ]
                    [
                       Model Extraction Attacks (\S~\ref{sec:dm_model_extraction_attacks}), dm-line-node
                       [
                           \textbf{(\romannumeral 1) LoRA-Based Extraction:}
                           \cite{horwitz2024recovering}
                           , dm-line-node, text width=32.7em
                        ]
                    ]
                    [
                       Intellectual Property Protection (\S~\ref{sec:dm_intellectual_property_protection}), dm-line-node
                       [
                           \textbf{(\romannumeral 1) Natural Data Protection:}
                           \cite{ye2023duaw} \cite{liang2023adversarial} \cite{van2023anti} \cite{liu2024metacloak} \cite{liu2024countering} \cite{wang2024simac} \cite{zhang2024editguard} \cite{min2024watermark} \cite{zhu2024watermark} \cite{cui2023ft} \cite{cui2023diffusionshield} \cite{asnani2024promark}
                           \cite{wang2023diagnosis}
                           \\
                           \textbf{(\romannumeral 2) Generated Data Protection:}
                           \cite{zhu2018hidden}
                           \cite{fernandez2023stable}
                           \cite{rezaei2024lawa} \cite{ma2024safe}
                           {\quad}
                           \textbf{(\romannumeral 2) Model Protection:}
                           \cite{zhao2023recipe} \cite{liu2023watermarking} \cite{peng2023protecting} \cite{feng2024aqualora}
                           \cite{wang2024trace}\cite{wen2023tree}
                           , dm-line-node, text width=32.7em
                        ]
                    ]
                ]
                [
                    \parbox{7em}{Agent (\S~\ref{sec:agent})} , agent-bg-node
                    [
                        Indirect Prompt Injection (\S~\ref{sec:agent_ipi}--\ref{sec:agent_ipi_defenses}), agent-line-node
                        [
                            \textbf{(\romannumeral 1) Malicious Instruction}
                            \cite{greshake2023not} \cite{wu2024new} \cite{toyer2023tensor} \cite{perez2022ignore} \cite{liu2023prompt} \cite{pedro2023prompt}
                            {\quad}
                            \textbf{(\romannumeral 2) Indirect Jailbreak}
                            \cite{zhan-etal-2025-adaptive} \cite{deng2024pandora} \cite{fu2024imprompter}
                            \\
                            \textbf{(\romannumeral 3) Defenses}
                            \cite{wallace2024instruction} \cite{wen2025defending} \cite{wang2024fath} \cite{hines2024splot} \cite{chen2025can} \cite{beurer2025design} \cite{taskshield2024}
                            , agent-line-node, text width=32.7em
                        ]
                    ]
                    [
                        Memory Attacks \& Defenses (\S~\ref{sec:agent_memory_threats}-\ref{sec:agent_memory_defenses}), agent-line-node
                        [
                            \textbf{(\romannumeral 1) Backdoor Attacks:}
                            \cite{liu2024compromising} \cite{zhu2025demonagent} \cite{chen2024agentpoison} \cite{pan2023trojan} \cite{xue2024badrag} \cite{chaudhari2024phantom} \cite{yang2024watch} \cite{wang2024badagent}
                            {\quad}
                            \textbf{(\romannumeral 2) Poisoning Attacks:}
                            \cite{zhang2024breaking} \cite{dong2025practical} \cite{zou2024poisonedrag} \cite{zhong2023poisoning}
                            \\
                            \textbf{(\romannumeral 3) Defenses:}
                            \cite{wallace2024instruction} \cite{luo2023search}   \cite{Summarizersmemory} \cite{agarwal2024prompt} \cite{mao2025agentsafe}
                            \cite{zhou2025trustrag} \cite{wang2024astute} \cite{xiang2024certifiably}
                             , agent-line-node, text width=32.7em
                        ]
                    ]
                    [
                        Tool Attacks \& Defenses (\S~\ref{sec:agent_tool_threats}--\ref{sec:agent_tool_defenses}), agent-line-node
                        [
                            \textbf{(\romannumeral 1) Tool Manipulation:}
                            \cite{zhang2025udora} \cite{ye2024toolsword} \cite{wang2024allies} \cite{wu2024wipi} \cite{jiang2025autocmd}
                            {\quad}
                            \textbf{(\romannumeral 2) MCP Manipulation:}
                            \cite{wang2025mpma} \cite{ferrag2025prompt} \cite{kong2025survey}
                            \\
                            \textbf{(\romannumeral 3) Defenses:}
                            \cite{chen2025agentguard} \cite{zhang2024privacyasst} \cite{xiang2024guardagent} \cite{jing2025mcip}
                            , agent-line-node, text width=32.7em
                        ]
                    ]
                    [
                        VLM Agent (\S~\ref{sec:agent_mmas_threats}-\ref{sec:agent_multimodal_defenses}), agent-line-node
                        [
                            \textbf{(\romannumeral 1) Attacks:}
                            \cite{wu2024adversarial} \cite{bagdasaryan2023abusing} \cite{liao2024eia} \cite{xu2024advagent} \cite{ma2024caution} \cite{zhang2024attacking} \cite{fu2023misusing} \cite{chen2025obvious} 
                            {\quad}
                            \textbf{(\romannumeral 2) Defenses:}
                            \cite{smoothvlm2024} \cite{bluesuffix2024} \cite{llavaguard2024} \cite{jaildam2024}
                            , agent-line-node, text width=32.7em
                        ]
                    ]
                    [
                        Multi-Agent Systems (\S~\ref{sec:agent_mas_attacks}-\ref{sec:agent_mas_defenses}), agent-line-node
                        [
                            \textbf{(\romannumeral 1) Attacks:}
                            \cite{lee2024prompt} \cite{cohen2024here} \cite{ju2024flooding} \cite{gu2024agent} \cite{zhou2025corba} \cite{he2025redteaming} \cite{tian2023evil} \cite{tan2024wolf} \cite{xteaming2025}
                            {\quad}
                            \textbf{(\romannumeral 2) Defenses:}
                            \cite{zeng2024autodefense} \cite{zhang2024psysafe} \cite{standen2025adversarial} \cite{lin2024large} \cite{song2024audit} \cite{xteaming2025}
                            , agent-line-node, text width=32.7em
                        ]
                    ]
                    [
                        Embodied Agents (\S~\ref{sec:agent_embodied_attacks}-\ref{sec:agent_embodied_defenses}), agent-line-node
                        [
                            \textbf{(\romannumeral 1) Attacks:}
                            \cite{Cheng_ManipulationFacingThreats_2024} \cite{Fime_TrustworthyAutonomousVehicles_2025}
                            \cite{Wang_ExploringAdversarialVulnerabilities_2025}
                            \cite{Robey_JailbreakingLLMControlledRobots_2024}
                            \cite{Zhang_BadRobotJailbreakingEmbodied_2025}
                            \cite{Lu_POEXUnderstandingMitigating_2025}
                            \cite{Liu_CompromisingEmbodiedAgents_2024}
                            \cite{Jiao_CanWeTrust_2025}
                            \cite{Li_EmbodiedAgentInterface_2024}
                            \cite{Tomilin_HASARDBenchmarkVisionBased_2025}
                            \cite{Chakraborty_HEALEmpiricalStudy_2025}
                            \cite{xu2024earth}
                            \cite{Karnik_EmbodiedRedTeaming_2025}
                            \cite{Zhou_ExploringLimitsVisionLanguageAction_2025}
                            \\
                            \textbf{(\romannumeral 2) Defenses:}
                            \cite{Wu_EmbodiedActiveDefense_2024}
                            \cite{Shirasaka_SelfRecoveryPromptingPromptable_2024}
                            \cite{Wang_AdvancingEmbodiedAgent_2025}
                            \cite{Zhang_SafeVLASafetyAlignment_2025}
                            , agent-line-node, text width=32.7em
                        ]
                    ]
                    [
                        Agentic Attacks \& Defenses (\S~\ref{sec:agent_agentic}), agent-line-node
                        [
                            \textbf{(\romannumeral 1) Agentic Attacks:}
                            \cite{fang2024llm} \cite{zhu2024teams} \cite{carlini2025autoadvexbench} \cite{xu2024redagent} \cite{wang2024ali} \cite{zhou2025autoredteamer}
                            {\quad}
                            \textbf{(\romannumeral 2) Agentic Defenses:}
                            \cite{chen2025shieldagent} \cite{cai2025aegisllm} \cite{barua2025guardians}
                            , agent-line-node, text width=32.7em
                        ]
                    ]
                    [
                         Benchmarks (\S~\ref{sec:agent_benchmarks}), agent-line-node
                        [
                            \textbf{(\romannumeral 1) Simulation-based Benchmarks:}
                            \cite{yi2023benchmarking} \cite{zhan2024injecagent} \cite{debenedetti2024agentdojo} \cite{andriushchenko2024agentharm} \cite{guo2024redcode} \cite{vpibench2024} \cite{yuan2024r} \cite{shao2024salad} \cite{draguns2024h4rm3l} \cite{zhang2024sg} \cite{li2024chemsafetybench} \cite{chan2024identifying} \cite{liu2024evaluating}
                            \\
                            \textbf{(\romannumeral 2) Real Interaction Benchmarks:}
                            \cite{zhou2023webarena} \cite{koh2024visualwebarena} \cite{liu2023agentbench} \cite{qin2023toolllm} \cite{evtimov2025wasp} \cite{zhang2024asb} \cite{yin2024safeagentbench} \cite{zhang2024agentsafetybench} \cite{guo2022safebench} \cite{liu2024dissecting} \cite{kumar2025refusal} \cite{lu2024from} \cite{shlomov2024st} \cite{lee2025safearena} \cite{lee2024haicosystem} \cite{vijayvargiya2025openagentsafety}
                            , agent-line-node, text width=32.7em
                        ]
                    ]
                ]
            ] 
    \end{forest}
    \caption{A road map of this survey.}
    \label{fig:organization}
\end{figure*}

Given the broad scope of our survey, we have structured it with the following considerations to enhance clarity and organization:
\begin{itemize}
    \item \textbf{Models}. We focus on six widely studied model categories, including \textbf{VFMs}, \textbf{LLMs}, \textbf{VLPs}, \textbf{VLMs}, \textbf{DMs}, and \textbf{Agents}, and review the attack and defense methods for each separately. These models represent the most popular large models across various domains.
    
    \item \textbf{Organization}. For each model category, we classify the reviewed works into attacks and defenses, and identify \textbf{10} attack types: \textbf{adversarial}, \textbf{backdoor}, \textbf{poisoning}, \textbf{jailbreak}, \textbf{prompt injection}, \textbf{energy-latency}, \textbf{membership inference}, \textbf{model extraction}, \textbf{data extraction}, and \textbf{agent} attacks. When both backdoor and poisoning attacks are present for a model category, we combine them into a single \textbf{backdoor \& poisoning} category due to their similarities. We review the corresponding defense strategies for each attack type immediately after the attacks.

    \item \textbf{Taxonomy}. For each type of attack or defense, we use a two-level taxonomy: \textbf{Category $\rightarrow$ Subcategory}. The \textbf{Category} differentiates attacks and defenses based on the threat model (e.g., white-box, gray-box, black-box) or specific subtasks (e.g., detection, purification, robust training/tuning, and robust inference). The \textbf{Subcategory} offers a more detailed classification based on their techniques.
    
    \item \textbf{Granularity}. To ensure clarity, we simplify the introduction of each reviewed paper, highlighting only its key ideas, objectives, and approaches, while omitting technical details and experimental analyses. 
    
\end{itemize}

\begin{table}[htbp]
\centering
\caption{A summary of existing surveys.}
\rowcolors{2}{gray!15}{white}
\begin{adjustbox}{width=1.0\linewidth}
\begin{tabular}{l c c c c c c c c}
\toprule
\textbf{Survey} & \textbf{Year} 
  & \textbf{VFM} & \textbf{VLP} & \textbf{LLM} 
  & \textbf{VLM} & \textbf{DM} 
  & \textbf{LLM-Agent} & \textbf{VLM-Agent} \\
\midrule
Zhang et al. \cite{zhang2024adversarial}      & 2024 & \cmark & \cmark & \xmark & \xmark & \cmark & \xmark & \xmark \\
Truong et al. \cite{truong2024attacks}        & 2024 & \xmark & \xmark & \xmark & \xmark & \cmark & \xmark & \xmark \\
Zhao et al. \cite{zhao2024survey}             & 2024 & \xmark & \xmark & \cmark & \xmark & \xmark & \xmark & \xmark \\
Yi et al. \cite{yi2024jailbreak}               & 2024 & \xmark & \xmark & \cmark & \xmark & \xmark & \xmark & \xmark \\
Jin et al. \cite{jin2024jailbreakzoo}          & 2024 & \xmark & \xmark & \cmark & \cmark & \xmark & \xmark & \xmark \\
Liu et al. \cite{liu2024jailbreak}             & 2024 & \xmark & \xmark & \xmark & \cmark & \xmark & \xmark & \xmark \\
Liu et al. \cite{liu2024survey}                & 2024 & \xmark & \xmark & \xmark & \cmark & \xmark & \cmark & \xmark \\
Cui et al. \cite{cui2024risk}                  & 2024 & \xmark & \xmark & \cmark & \xmark & \xmark & \cmark & \xmark \\
Gan et al. \cite{gan2024navigating}            & 2024 & \xmark & \xmark & \cmark & \cmark & \xmark & \cmark & \cmark \\
\midrule
Deng et al. \cite{deng2024ai}                  & 2025 & \xmark & \xmark & \cmark & \xmark & \xmark & \cmark & \xmark \\
Ye et al. \cite{ye2025survey}                  & 2025 & \xmark & \xmark & \xmark & \cmark & \xmark & \xmark & \xmark \\
Wang et al. \cite{wang2025comprehensive}       & 2025 & \xmark & \xmark & \cmark & \xmark & \xmark & \cmark & \xmark \\
\midrule
\textbf{Our Survey}                            & 2025 & \cmark & \cmark & \cmark&  \cmark& \cmark& \cmark &  \cmark   \\
\bottomrule
\end{tabular}
\end{adjustbox}
\label{tab:existing_survey}
\end{table}

Our survey methodology is structured as follows. First, we conducted a keyword-based search targeting specific model types and threat types to identify relevant papers. Next, we manually filtered out non-safety-related and non-technical papers. For each remaining paper, we categorized its proposed method or framework by analyzing its settings and attack/defense types, assigning them to appropriate categories and subcategories.
In total, we reviewed \textbf{574} technical papers, with their distribution across years, model types, and attack/defense strategies illustrated in Figure \ref{fig:total_num}. As shown, safety research on large models has surged significantly since 2023, following the release of ChatGPT. Among the model types, LLMs, DMs and Agents have garnered the most attention, accounting for \textbf{71.32\%} of the surveyed papers. Regarding attack types, \textbf{jailbreak}, \textbf{adversarial}, and \textbf{backdoor} attacks were the most extensively studied. On the defense side, \textbf{jailbreak defenses} received the highest focus, followed by \textbf{adversarial defenses}.
Figure \ref{fig:crossview} presents a cross-view of temporal trends across model types and attack/defense categories, offering a detailed breakdown of the reviewed works. Notably, research on attacks constitutes \textbf{$\sim$60\%} of the studied. In terms of defense, while defense research accounts for only \textbf{$\sim$40\%}, underscoring a significant gap that warrants increased attention toward defense strategies. The overall structure of this survey is outlined in Figure \ref{fig:organization}.

\textbf{Difference to Existing Surveys.} Large-model safety is a rapidly evolving field, and several surveys have been conducted to advance research in this area. Recently, Slattery et al. \cite{slattery2024ai} introduced an AI risk framework with a systematic taxonomy covering all types of risks. In contrast, our focus is on the technical aspects, specifically the attack and defense techniques proposed in the literature.
Table \ref{tab:existing_survey} summarizes the technical surveys we identified, each concentrating on a few model types or threat categories (e.g., LLMs, VLMs, agents, or jailbreak attacks/defenses).
Compared with these works, our survey provides both a broader scope by covering a wider range of model types and threats, and a more high-level perspective that focuses on overarching methodologies rather than specific technical details.

\section{Vision Foundation Model Safety} \label{sec:vfm}
This section surveys safety research on two types of VFMs: per-trained Vision Transformers (ViTs)~\cite{dosovitskiy2021an} and the Segment Anything Model (SAM)~\cite{kirillov2023segment}. 
We focus on ViTs and SAM because they are among the most widely deployed VFMs and have garnered significant attention in recent safety research.

\begin{table*}[htp]
\center
\caption{A summary of attacks and defenses for ViTs and SAM.}
\label{tab:vfm_safety}
\resizebox{1\textwidth}{!}{
\begin{tabular}{p{0.09\textwidth}p{0.16\textwidth}p{0.05\textwidth}p{0.15\textwidth}p{0.20\textwidth}p{0.25\textwidth}p{0.2\textwidth}}
            \toprule
            \belowrulesepcolor{sunye-red}
        \rowcolor{sunye-red} 
        \textbf{Attack/Defense} & \textbf{Method} & \textbf{Year} & \textbf{Category} & \textbf{Subcategory} & \textbf{Target Models} & \textbf{Datasets} \\ \aboverulesepcolor{orange!25!}  \midrule
    \belowrulesepcolor{gray!25!}
    \rowcolor{gray!25!}\multicolumn{7}{c}{\textbf{Attacks and defenses for ViT (Sec.~\ref{sec:vfm_vit})}} \\ \aboverulesepcolor{gray!25!}  \midrule 
\multirow{15}{0.08\textwidth}{Adversarial Attack} & Patch-Fool~\cite{fu2022patch} & 2022 & White-box & Patch Attack & DeiT, ResNet & ImageNet \\
                   & \cellcolor{gray!15!}SlowFormer~\cite{navaneet2024slowformer} & \cellcolor{gray!15!}2024 & \cellcolor{gray!15!}White-box & \cellcolor{gray!15!}Patch Attack & \cellcolor{gray!15!}ATS, AdaViT & \cellcolor{gray!15!}ImageNet \\
                   & PE-Attack~\cite{gao2024pe} & 2024 & White-box & Position Embedding Attack & ViT, DeiT, BEiT & ImageNet, GLUE, wmt13/16, Food-101, CIFAR100, etc. \\
                   & \cellcolor{gray!15!}Attention-Fool~\cite{lovisotto2022give} & \cellcolor{gray!15!}2022 & \cellcolor{gray!15!}White-box & \cellcolor{gray!15!}Attention Attack & \cellcolor{gray!15!}ViT, DeiT, DETR & \cellcolor{gray!15!}ImageNet \\
                   & AAS~\cite{jain2024towards} & 2024 & White-box & Attention Attack & ViT-B & ImageNet, CIFAR10/100 \\
                   & \cellcolor{gray!15!}SE-TR~\cite{naseer2021improving} & \cellcolor{gray!15!}2022 & \cellcolor{gray!15!}Black-box & \cellcolor{gray!15!}Transfer-based Attack & \cellcolor{gray!15!}DeiT, T2T, TnT, DINO, DETR & \cellcolor{gray!15!}ImageNet \\
                   & ATA~\cite{wang2022generating} & 2022 & Black-box & Transfer-based Attack & ViT, DeiT, ConViT & ImageNet \\
                   & \cellcolor{gray!15!}PNA-PatchOut~\cite{wei2022towards} & \cellcolor{gray!15!}2022 & \cellcolor{gray!15!}Black-box & \cellcolor{gray!15!}Transfer-based Attack & \cellcolor{gray!15!}ViT, DeiT, TNT, LeViT, PiT, CaiT, ConViT, Visformer & \cellcolor{gray!15!}ImageNet \\
                   & LPM~\cite{wei2023boosting} & 2023 & Black-box & Transfer-based Attack & ViT, PiT, DeiT, Visformer, LeViT, ConViT & ImageNet \\ 
                   & \cellcolor{gray!15!}MIG~\cite{ma2023transferable} & \cellcolor{gray!15!}2023 & \cellcolor{gray!15!}Black-box & \cellcolor{gray!15!}Transfer-based Attack & \cellcolor{gray!15!}ViT, TNT, Swin & \cellcolor{gray!15!}ImageNet \\
                   & TGR~\cite{zhang2023transferable} & 2023 & Black-box & Transfer-based Attack & DeiT, TNT, LeViT, ConViT & ImageNet \\
                   & \cellcolor{gray!15!}VDC~\cite{zhang2024improving} & \cellcolor{gray!15!}2024 & \cellcolor{gray!15!}Black-box & \cellcolor{gray!15!}Transfer-based Attack & \cellcolor{gray!15!}CaiT, TNT, LeViT, ConViT & \cellcolor{gray!15!}ImageNet \\
                   & FDAP~\cite{gao2024attacking} & 2024 & Black-box & Transfer-based Attack & ViT, DeiT, CaiT, ConViT, TNT & ImageNet \\
                   & \cellcolor{gray!15!}SASD-WS~\cite{wu2024improving} & \cellcolor{gray!15!}2024 & \cellcolor{gray!15!}Black-box & \cellcolor{gray!15!}Transfer-based Attack & \cellcolor{gray!15!}ViT, ResNet, DenseNet, VGG & \cellcolor{gray!15!}ImageNet \\
                   & CRFA~\cite{li2024improving} & 2024 & Black-box & Transfer-based Attack & ViT, DeiT, CaiT, TNT, Visformer, LeViT, ConvNeXt, RepLKNet & ImageNet \\
                   & \cellcolor{gray!15!}FPR~\cite{ren2025improving} & \cellcolor{gray!15!}2025 & \cellcolor{gray!15!}Black-box & \cellcolor{gray!15!}Transfer-based Attack & \cellcolor{gray!15!}ViT, CaiT, PiT, Visformer, Swin, DeiT, CoaT, ResNet, VGG, DenseNet & \cellcolor{gray!15!}ImageNet \\
                   & PAR~\cite{shi2022decision} & 2022 & Black-box & Query-based Attack & ViT & ImageNet \\ \midrule
\multirow{7}{0.08\textwidth}{Adversarial Defense}& AGAT~\cite{wu2022towards} & 2022 & Adversarial Training & Efficient training & ViT, CaiT, LeViT & ImageNet \\
                   & \cellcolor{gray!15!}ARD-PRM~\cite{mo2022adversarial} & \cellcolor{gray!15!}2022 & \cellcolor{gray!15!}Adversarial Training & \cellcolor{gray!15!}Efficient training & \cellcolor{gray!15!}ViT, DeiT, ConViT, Swin & \cellcolor{gray!15!}ImageNet, CIFAR10 \\
                   & Patch-Vestiges~\cite{li2022patch} & 2022 & Adversarial Detection & Patch-based Detection & ViT, ResNet & CIFAR10 \\
                   & \cellcolor{gray!15!}ViTGuard~\cite{sun2024vitguard} & \cellcolor{gray!15!}2024 & \cellcolor{gray!15!}Adversarial Detection & \cellcolor{gray!15!}Attention-based Detection & \cellcolor{gray!15!}ViT & \cellcolor{gray!15!}ImageNet, CIFAR10/100 \\
                   & ARMRO~\cite{liu2023understanding} & 2023 & Adversarial Detection & Attention-based Detection & ViT, DeiT & ImageNet, CIFAR10 \\
                   & \cellcolor{gray!15!}Smoothed-Attention~\cite{gu2022vision} & \cellcolor{gray!15!}2022 & \cellcolor{gray!15!}Robust Architecture & \cellcolor{gray!15!}Robust Attention & \cellcolor{gray!15!}DeiT, ResNet & \cellcolor{gray!15!}ImageNet \\
                   & TAP~\cite{guo2023robustifying} & 2023 & Robust Architecture & Robust Attention & RVT, FAN & ImageNet, Cityscapes, COCO \\
                   & \cellcolor{gray!15!}RSPC~\cite{guo2023improving} & \cellcolor{gray!15!}2023 & \cellcolor{gray!15!}Robust Architecture & \cellcolor{gray!15!}Robust Attention & \cellcolor{gray!15!}RVT, FAN & \cellcolor{gray!15!}ImageNet, CIFAR10/100 \\
                   & FViT~\cite{huimproving} & 2024 & Robust Architecture & Robust Attention & ViT, DeiT, Swin & ImageNet, Cityscapes, COCO \\
                   & \cellcolor{gray!15!}SATA~\cite{nikzad2025sata} & \cellcolor{gray!15!}2025 & \cellcolor{gray!15!}Robust Architecture & \cellcolor{gray!15!}Robust Attention & \cellcolor{gray!15!}ViT, DeiT & \cellcolor{gray!15!}ImageNet \\
                   & ADBM~\cite{li2024adbm} & 2024 & Adversarial Purification & Diffusion-based Purification & WideResNet, ViT & CIFAR-10, ImageNet, SVHN \\ 
                   & \cellcolor{gray!15!}CGDMP~\cite{bai2024diffusion} & \cellcolor{gray!15!}2024 & \cellcolor{gray!15!}Adversarial Purification & \cellcolor{gray!15!}Diffusion-based Purification & \cellcolor{gray!15!}ResNet, XciT & \cellcolor{gray!15!}CIFAR 10/100, GTSRB, ImageNet\\
                   & OSCP~\cite{lei2024instant} & 2024 & Adversarial Purification & Diffusion-based Purification & ViT, Swin, WideResNet& ImageNet, CelebA-HQ\\ \midrule
\multirow{5}{0.08\textwidth}{Backdoor Attack} & BadViT~\cite{yuan2023you} & 2023 & Data Poisoning & Patch-level Attack & DeiT, LeViT & ImageNet \\
                   & \cellcolor{gray!15!}TrojViT~\cite{zheng2023trojvit} & \cellcolor{gray!15!}2023 & \cellcolor{gray!15!}Data Poisoning & \cellcolor{gray!15!}Patch-level Attack & \cellcolor{gray!15!}DeiT, ViT, Swin & \cellcolor{gray!15!}ImageNet, CIFAR10 \\
                   & SWARM~\cite{yang2024not} & 2024 & Data Poisoning & Token-level Attack & ViT & VTAB-1k \\
                   & \cellcolor{gray!15!}DBIA~\cite{lv2023dbia} & \cellcolor{gray!15!}2023 & \cellcolor{gray!15!}Data Poisoning & \cellcolor{gray!15!}Data-free Attack & \cellcolor{gray!15!}ViT, DeiT, Swin & \cellcolor{gray!15!}ImageNet, CIFAR10/100, GTSRB, GGFace \\ 
                   & MTBA~\cite{li2024multi} & 2024 & Data Poisoning & Multi-trigger Attack & ViT & ImageNet, CIFAR10 \\
                   \midrule
\multirow{2}{0.08\textwidth}{Backdoor Defense} & PatchDrop~\cite{doan2023defending} & 2023 & Robust Inference & Patch Processing & ViT, DeiT, ResNet & ImageNet, CIFAR10 \\
                   & \cellcolor{gray!15!}Image Blocking~\cite{subramanya2024closer} & \cellcolor{gray!15!}2023 & \cellcolor{gray!15!}Robust Inference & \cellcolor{gray!15!}Image Blocking & \cellcolor{gray!15!}ViT, CaiT & \cellcolor{gray!15!}ImageNet \\ \aboverulesepcolor{gray!25!}  \midrule
    \belowrulesepcolor{gray!25!}
    \rowcolor{gray!25!}\multicolumn{7}{c}{\textbf{Attacks and defenses for SAM (Sec.~\ref{sec:vfm_sam})}} \\ \aboverulesepcolor{gray!25!}  \midrule 
\multirow{8}{0.08\textwidth}{Adversarial Attack} & S-RA~\cite{shen2024practical} & 2024 & White-box & Prompt-agnostic Attack & SAM & SA-1B \\
                   & \cellcolor{gray!15!}Croce et al.~\cite{croce2024segment} & \cellcolor{gray!15!}2024 & \cellcolor{gray!15!}White-box & \cellcolor{gray!15!}Prompt-agnostic Attack & \cellcolor{gray!15!}SAM, SEEM & \cellcolor{gray!15!}SA-1B \\
                   & Attack-SAM~\cite{zhang2023attack} & 2023 & Black-box & Transfer-based Attack & SAM & SA-1B \\
                   & \cellcolor{gray!15!}PATA++~\cite{zheng2023black} & \cellcolor{gray!15!}2023 & \cellcolor{gray!15!}Black-box & \cellcolor{gray!15!}Transfer-based Attack & \cellcolor{gray!15!}SAM &\cellcolor{gray!15!}SA-1B \\ 
                   & UAD~\cite{lu2024unsegment} & 2024 & Black-box & Transfer-based Attack & SAM, FastSAM & SA-1B \\ 
                   & \cellcolor{gray!15!}T-RA~\cite{shen2024practical} & \cellcolor{gray!15!}2024 & \cellcolor{gray!15!}Black-box & \cellcolor{gray!15!}Transfer-based Attack & \cellcolor{gray!15!}SAM &\cellcolor{gray!15!}SA-1B \\
                   & UMI-GRAT~\cite{xia2024transferable} & 2024 & Black-box & Transfer-based Attack & Medical SAM, Shadow-SAM, Camouflaged-SAM & CT-Scans, ISTD, COD10K, CAMO, CHAME \\
                   & \cellcolor{gray!15!}Han et al.~\cite{han2023segment} & \cellcolor{gray!15!}2023 & \cellcolor{gray!15!}Black-box & \cellcolor{gray!15!}Universal Attack & \cellcolor{gray!15!}SAM & \cellcolor{gray!15!}SA-1B \\ 
                   & DarkSAM~\cite{zhou2024darksam} & 2024 & Black-box & Universal Attack & SAM, HQ-SAM, PerSAM & ADE20K, Cityscapes, COCO, SA-1B \\ \midrule
Adversarial Defense & ASAM~\cite{li2024asam} & 2024 & Adversarial Tuning  & Diffusion Model-based Tuning & SAM & Ade20k, VOC2012, COCO, DOORS, LVIS, etc. \\ 
                   & \cellcolor{gray!15!}Robust SAM~\cite{long2025robust} & \cellcolor{gray!15!}2024 & \cellcolor{gray!15!}Adversarial Tuning & \cellcolor{gray!15!}Parameter-Efficient Fine-Tuning & \cellcolor{gray!15!}SAM, MedSAM, SAM-Adapter &\cellcolor{gray!15!}SA-1B, VOC, COCO, DAVIS  \\ \aboverulesepcolor{gray!25!} \midrule
\multirow{2}{0.08\textwidth}{Backdoor\&\\Poisoning Attack} & BadSAM~\cite{guan2024badsam} & 2024 & Data Poisoning & Visual trigger & SAM & CAMO \\
                   & \cellcolor{gray!15!}UnSeg~\cite{sun2024unseg} & \cellcolor{gray!15!}2024 & \cellcolor{gray!15!}Data Poisoning & \cellcolor{gray!15!}Unlearnable Examples & \cellcolor{gray!15!}HQ-SAM, DINO, Rsprompter, UNet++, Mask2Former, DeepLabV3 & \cellcolor{gray!15!}Cityscapes, VOC, COCO, Lung, Kvasir-seg, WHU, etc. \\ \aboverulesepcolor{gray!15!} \bottomrule
\end{tabular}
}
\end{table*}

\subsection{Attacks and Defenses for ViTs}\label{sec:vfm_vit}
Pre-trained ViTs are widely employed as backbones for various downstream tasks, frequently achieving state-of-the-art performance through efficient adaptation and fine-tuning. Unlike traditional CNNs, ViTs process images as sequences of tokenized patches, allowing them to better capture spatial dependencies. However, this patch-based mechanism also brings unique safety concerns and robustness challenges. This section explores these issues by reviewing ViT-related safety research, including adversarial attacks, backdoor \& poisoning attacks, and their corresponding defense strategies.  Table~\ref{tab:vfm_safety} provides a summary of the surveyed attacks and defenses, along with the commonly used datasets.

\subsubsection{Adversarial Attacks}\label{sec:ViT-adv}
Adversarial attacks on ViTs can be classified into \textbf{white-box attacks} and \textbf{black-box attacks} based on whether the attacker has full access to the victim model. Based on the attack strategy, white-box attacks can be further divided into 1) \textbf{patch attacks}, 2) \textbf{position embedding attacks} and 3) \textbf{attention attacks}, while black-box attacks can be summarized into 1) \textbf{transfer-based attacks} and 2) \textbf{query-based attacks}.

\paragraph{White-box Attacks}
\textbf{Patch Attacks} exploit the modular structure of ViTs, aiming to manipulate their inference processes by introducing targeted perturbations in specific patches of the input data. Joshi et al.~\cite{joshi2021adversarial} proposed an adversarial token attack method leveraging block sparsity to assess the vulnerability of ViTs to token-level perturbations. 
Expanding on this, \textbf{Patch-Fool}\cite{fu2022patch} introduces an adversarial attack framework that targets the self-attention modules by perturbing individual image patches, thereby manipulating attention scores.
Different from existing methods, \textbf{SlowFormer}~\cite{navaneet2024slowformer} introduces a universal adversarial patch can be applied to any image to increases computational and energy costs while preserving model accuracy.

\textbf{Position Embedding Attacks} aim to attack the spatial or sequential position of tokens in transformers. For example, \textbf{PE-Attack}~\cite{gao2024pe} explores the common vulnerability of positional embeddings to adversarial perturbations by disrupting their ability to encode positional information through periodicity manipulation, linearity distortion, and optimized embedding distortion.

\textbf{Attention Attacks} target vulnerabilities in the self-attention modules of ViTs. \textbf{Attention-Fool}~\cite{lovisotto2022give} manipulates dot-product similarities to redirect queries to adversarial key tokens, exposing the model's sensitivity to adversarial patches. Similarly, \textbf{AAS}~\cite{jain2024towards} mitigates gradient masking in ViTs by optimizing the pre-softmax output scaling factors, enhancing the effectiveness of attacks.

\paragraph{Black-box Attacks}
\textbf{Transfer-based Attacks} first generate adversarial examples using fully accessible surrogate models, which are then transferred to attack black-box victim ViTs. In this context, we first review attacks specifically designed for the ViT architecture.
\textbf{SE-TR}~\cite{naseer2021improving} enhances adversarial transferability by optimizing perturbations on an ensemble of models.
\textbf{ATA}~\cite{wang2022generating} strategically activates uncertain attention and perturbs sensitive embeddings within ViTs.
\textbf{LPM}\cite{wei2023boosting} mitigates the overfitting to model-specific discriminative regions through a patch-wise optimized binary mask.
Chen et al.\cite{chen2023understanding} introduced an Inductive Bias Attack (\textbf{IBA}) to suppress unique biases in ViTs and target shared inductive biases.
\textbf{TGR}~\cite{zhang2023transferable} reduces the variance of the backpropagated gradient within internal blocks.
\textbf{VDC}~\cite{zhang2024improving} employs virtual dense connections between deeper attention maps and MLP blocks to facilitate gradient backpropagation.
\textbf{FDAP}~\cite{gao2024attacking} exploits feature collapse by reducing high-frequency components in feature space.
\textbf{CRFA}~\cite{li2024improving} disrupts only the most crucial image regions using approximate attention maps.
\textbf{SASD-WS}~\cite{wu2024improving} flattens the loss landscape of the source model through sharpness-aware self-distillation and approximates an ensemble of pruned models using weight scaling to improve target adversarial transferability.
\textbf{FPR}~\cite{ren2025improving} improves the adversarial transferability on ViTs by refining forward propagation instead of backward gradients. It consists of two components: (1) Attention Map Diversification (AMD), which applies controlled randomness to diversify attention maps, mitigating overfitting and implicitly inducing gradient vanishing; and (2) Momentum Token Embedding (MTE), which stabilizes token embedding updates by accumulating historical embeddings across iterations.

Other strategies are applicable to both ViTs and CNNs, ensuring broader applicability in black-box settings.
Wei et al.\cite{wei2022towards, wei2023towards} proposed a dual attack framework to improve transferability between ViTs and CNNs: 1) a Pay No Attention (\textbf{PNA}) attack, which skips the gradients of attention during backpropagation, and 2) a \textbf{PatchOut} attack, which randomly perturbs subsets of image patches at each iteration.
\textbf{MIG}\cite{ma2023transferable} uses integrated gradients and momentum-based updates to precisely target model-agnostic critical regions, improving transferability between ViTs and CNNs.

\textbf{Query-based Attacks} generate adversarial examples by querying the black-box model and levering the model responses to estimate the adversarial gradients. The goal is to achieve successful attack with a minimal number of queries. Based on the type of model response, query-based attacks can be further divided into score-based attacks, where the model returns a probability vector, and decision-based attacks, where the model provides only the top-k classes. Decision-based attacks typically start from a large random noise (to achieve misclassification first) and then gradually find smaller noise while maintaining misclassification.
To improve the efficiency of the adversarial noise searching process in ViTs, \textbf{PAR}~\cite{shi2022decision} introduces a coarse-to-fine patch searching method, guided by noise magnitude and sensitivity masks to account for the structural characteristics of ViTs and mitigate the negative impact of non-overlapping patches.

\subsubsection{Adversarial Defenses}\label{sec:ViT-advdefense}
Adversarial defenses for ViTs follow four major approaches: 1) \textbf{adversarial training}, which trains ViTs on adversarial examples via min-max optimization to improve its robustness; 2) \textbf{adversarial detection}, which identifies and mitigates adversarial attacks by detecting abnormal or malicious patterns in the inputs; 3) \textbf{robust architecture}, which modifies and optimizes the architecture (e.g., self-attention module) of ViTs to improve their resilience against adversarial attacks; and 4) \textbf{adversarial purification}, which pre-processes the input (e.g., noise injection, denoising, or other transformations) to remove potential adversarial perturbations before inference.

\textbf{Adversarial Training} is widely regarded as the most effective approach to adversarial defense; however, it comes with a high computational cost. To address this on ViTs, \textbf{AGAT}~\cite{wu2022towards} introduces a dynamic attention-guided dropping strategy, which accelerates the training process by selectively removing certain patch embeddings at each layer. This reduces computational overhead while maintaining robustness, especially on large datasets such as ImageNet. Due to its high computational cost, research on adversarial training for ViTs has been relatively limited.
\textbf{ARD-PRM}~\cite{mo2022adversarial} improves adversarial robustness by randomly dropping gradients in attention blocks and masking patch perturbations during training.

\textbf{Adversarial Detection} methods for ViTs primarily leverage two key features, i.e., patch-based inference and activation characteristics, to detect and mitigate adversarial examples.
Li et al.~\cite{li2022patch} proposed the concept of \textbf{Patch Vestiges}, abnormalities arising from adversarial examples during patch division in ViTs. They used statistical metrics on step changes between adjacent pixels across patches and developed a binary regression classifier to detect adversaries. Alternatively, \textbf{ARMOR}~\cite{liu2023understanding} identifies adversarial patches by scanning for unusually high column scores in specific layers and masking them with average images to reduce their impact. 
\textbf{ViTGuard}~\cite{sun2024vitguard}, on the other hand, employs a masked autoencoder to detect patch attacks by analyzing attention maps and CLS token representations. As more attacks are developed, there is a growing need for a unified detection framework capable of handling all types of adversarial examples.

\textbf{Robust Architecture} methods focus on designing more adversarially resilient attention modules for ViTs. 
For example, \textbf{Smoothed Attention}~\cite{gu2022vision} employs temperature scaling in the softmax function to prevent any single patch from dominating the attention, thereby balancing focus across patches.
\textbf{ReiT}~\cite{gong2024random} integrates adversarial training with randomization through the II-ReSA module, optimizing randomly entangled tokens to reduce adversarial similarity and enhance robustness.
\textbf{TAP}~\cite{guo2023robustifying} addresses token overfocusing by implementing token-aware average pooling and an attention diversification loss, which incorporate local neighborhood information and reduce cosine similarity among attention vectors. \textbf{FViTs}~\cite{huimproving} strengthen explanation faithfulness by stabilizing top-k indices in self-attention and robustify predictions using denoised diffusion smoothing combined with Gaussian noise. \textbf{RSPC}~\cite{guo2023improving} tackles vulnerabilities by corrupting the most sensitive patches and aligning intermediate features between clean and corrupted inputs to stabilize the attention mechanism. Collectively, these advancements underscore the pivotal role of the attention mechanism in improving the adversarial robustness of ViTs.
\textbf{SATA}~\cite{nikzad2025sata} robustifies ViT models without retraining by injecting a spatial autocorrelation-based module between attention and FFN layers. It splits tokens by their spatial autocorrelation scores and selectively merges high/low-score tokens before FFN, reducing redundancy and improving feature aggregation. Residual tokens are later concatenated to preserve information, offering strong robustness against adversarial and corrupted inputs.

\textbf{Adversarial Purification} refers to a model-agnostic input-processing technique that is broadly applicable across various architectures, including but not limited to ViTs.
\textbf{DiffPure}~\cite{nie2022diffusion} introduces a framework where adversarial images undergo noise injection via a forward stochastic differential equation (SDE) process, followed by denoising with a pre-trained diffusion model. \textbf{CGDMP}~\cite{bai2024diffusion} refines this approach by optimizing the noise level for the forward process and employing contrastive loss gradients to guide the denoising process, achieving improved purification tailored to ViTs. \textbf{ADBM}~\cite{li2024adbm} highlights the disparity between diffused adversarial and clean examples, proposing a method to directly connect the clean and diffused adversarial distributions.
While these methods focus on ViTs, other approaches demonstrate broader applicability to various vision models, e.g., CNNs. \textbf{Purify++}\cite{zhang2023purify++} enhances DiffPure with improved diffusion models, \textbf{DifFilter}\cite{chen2024diffilter} extends noise scales to better preserve semantics, and \textbf{MimicDiffusion}\cite{song2024mimicdiffusion} mitigates adversarial impacts during the reverse diffusion process. For improved efficiency, \textbf{OSCP}\cite{lei2024instant} and \textbf{LightPure}\cite{khalili2024lightpure} propose single-step and real-time purification methods, respectively. \textbf{LoRID}\cite{zollicoffer2024lorid} introduces a Markov-based approach for robust purification. These methods complement ViT-related research and highlight diverse advancements in adversarial purification.

\subsubsection{Backdoor Attacks }\label{sec:ViT-backdoor}
Backdoors can be injected into the victim model via data poisoning, training manipulation, or parameter editing, with most existing attacks on ViTs being data poisoning-based. 
We classify these attacks into four categories: 1) \textbf{patch-level attacks}, 2) \textbf{token-level attacks}, and 3) \textbf{multi-trigger attacks}, which exploit ViT-specific data processing characteristics, as well as 4) \textbf{data-free attacks}, which exploit the inherent mechanisms of ViTs.

\textbf{Patch-level Attacks} primarily exploit the ViT's characteristic of processing images as discrete patches by implanting triggers at the patch level. For example, \textbf{BadViT}~\cite{yuan2023you} introduces a universal patch-wise trigger that requires only a small amount of data to redirect the model's focus from classification-relevant patches to adversarial triggers. 
\textbf{TrojViT}~\cite{zheng2023trojvit} improves this approach by utilizing patch salience ranking, an attention-targeted loss function, and parameter distillation to minimize the bit flips necessary to embed the backdoor.

\textbf{Token-level Attacks} target the tokenization layer of ViTs. \textbf{SWARM}~\cite{yang2024not} introduces a switchable backdoor mechanism featuring a ``switch token'' that dynamically toggles between benign and adversarial behaviors, ensuring high attack success rates while maintaining functionality in clean environments.

\textbf{Multi-trigger Attacks} employ multiple backdoor triggers in parallel, sequential, or hybrid configurations to poison the victim dataset.
\textbf{MTBAs}~\cite{li2024multi} utilize these multiple triggers to induce coexistence, overwriting, and cross-activation effects, significantly diminishing the effectiveness of existing defense mechanisms.

\textbf{Data-free Attacks} eliminate the need for original training datasets. Using substitute datasets, \textbf{DBIA}~\cite{lv2023dbia} generates universal triggers that maximize attention within ViTs. These triggers are fine-tuned with minimal parameter adjustments using PGD~\cite{madry2017towards}, enabling efficient and resource-light backdoor injection.

\subsubsection{Backdoor Defenses}\label{sec:ViT-backdoordefense}
Backdoor defenses for ViTs aim to identify and break (or remove) the correlation between trigger patterns and target classes while preserving model accuracy.
Two representative defense strategies are: 1) \textbf{patch processing}, which disrupts the integrity of image patches to prevent trigger activation, and 2) \textbf{image blocking}, which leverages interpretability-based mechanisms to mask and neutralize the effects of backdoor triggers.

\textbf{Patch Processing} strategy disrupts the integrity of patches to neutralize triggers.
Doan et al.~\cite{doan2023defending} found that clean-data accuracy and attack success rates of ViTs respond differently to patch transformations before positional encoding, and proposed an effective defense method by randomly dropping or shuffling patches of an image to counter both patch-based and blending-based backdoor attacks.
\textbf{Image Blocking} utilizes interpretability to identify and neutralize triggers.
Subramanya et al.~\cite{subramanya2022backdoor} showed that ViTs can localize backdoor triggers using attention maps and proposed a defense mechanism that dynamically masks potential trigger regions during inference.
In a subsequent work, Subramanya et al.~\cite{subramanya2024closer} proposed to integrate trigger neutralization into the training phase to improve the robustness of ViTs to backdoor attacks. 
While these two methods are promising, the field requires a holistic defense framework that integrates non-ViT defenses with ViT-specific characteristics and unifies multiple defense tasks including backdoor detection, trigger inversion, and backdoor removal, as attempted in \cite{li2024expose}.

\subsubsection{Datasets}\label{sec:ViT-dataset}
Datasets are crucial for developing and evaluating attack and defense methods. Table~\ref{tab:vfm_safety} summarizes the datasets used in adversarial and backdoor research.

\textbf{Datasets for Adversarial Research} As shown in Table~\ref{tab:vfm_safety}, adversarial researches were primarily conducted on ImageNet. While attacks were tested across various datasets like CIFAR-10/100, Food-101, and GLUE, defenses were mainly limited to ImageNet and CIFAR-10/100. This imbalance reveals one key issue in adversarial research: attacks are more versatile, while defenses struggle to generalize across different datasets.

\textbf{Datasets for Backdoor Research} Backdoor researches were also conducted mainly on ImageNet and CIFAR-10/100 datasets. Some attacks, such as DBIA and SWARM, extend to domain-specific datasets like GTSRB and VGGFace, while defenses, including PatchDrop, were often limited to a few benchmarks. This narrow focus reduces their real-world applicability. 
Although backdoor defenses are shifting towards robust inference techniques, they typically target specific attack patterns, limiting their generalizability. To address this, adaptive defense strategies need to be tested across a broader range of datasets to effectively counter the evolving nature of backdoor threats.

\subsection{Attacks and Defenses for SAM}\label{sec:vfm_sam}
SAM is a foundational model for image segmentation, comprising three primary components: a ViT-based image encoder, a prompt encoder, and a mask decoder. The image encoder transforms high-resolution images into embeddings, while the prompt encoder converts various input modalities into token embeddings. The mask decoder combines these embeddings to generate segmentation masks using a two-layer Transformer architecture.
Due to its complex structure, attacks and defenses targeting SAM differ significantly from those developed for CNNs. These unique challenges stem from SAM's modular and interconnected design, where vulnerabilities in one component can propagate to others, necessitating specialized strategies for both attack and defense. 
This section systematically reviews SAM-related adversarial attacks, backdoor \& poisoning attacks, and adversarial defense strategies, as summarized in Table~\ref{tab:vfm_safety}.

\subsubsection{Adversarial Attacks}\label{sec:SAM-adv}

Adversarial attacks on SAM can be categorized into: (1) \textbf{white-box attacks}, exemplified by \emph{prompt-agnostic attacks}, and (2) \textbf{black-box attacks}, which can be further divided into \emph{universal attacks} and \emph{transfer-based attacks}. Each category employs distinct strategies to compromise segmentation performance.

\paragraph{White-box Attacks}
\textbf{Prompt-Agnostic Attacks} are white-box attacks that disrupt SAM's segmentation without relying on specific prompts, using either \emph{prompt-level} or \emph{feature-level} perturbations for generality across inputs.
For prompt-level attacks, Shen et al.~\cite{shen2024practical} proposed a grid-based strategy to generate adversarial perturbations that disrupt segmentation regardless of click location.
For feature-level attacks, Croce et al.~\cite{croce2024segment} perturbed features from the image encoder to distort spatial embeddings, undermining SAM’s segmentation integrity.

\paragraph{Black-box Attacks}

\textbf{Universal Attacks} generate UAPs~\cite{moosavi2017universal} that can consistently disrupt SAM across arbitrary prompts.  Han et al.~\cite{han2023segment} exploited contrastive learning to optimize the UAPs, achieving better attack performance by exacerbating feature misalignment.
\textbf{DarkSAM}~\cite{zhou2024darksam}, on the other hand, introduces a hybrid spatial-frequency framework that combines semantic decoupling and texture distortion to generate universal perturbations.

\textbf{Transfer-based Attacks} exploit transferable representations in SAM to generate perturbations that remain adversarial across different models and tasks. 
\textbf{PATA++}\cite{zheng2023black} improves transferability by using a regularization loss to highlight key features in the image encoder, reducing reliance on prompt-specific data. 
\textbf{Attack-SAM}\cite{zhang2023attack} employs ClipMSE loss to focus on mask removal, optimizing for spatial and semantic consistency to improve cross-task transferability. 
\textbf{UMI-GRAT}~\cite{xia2024transferable} follows a two-step process: it first generates a generalizable perturbation with a surrogate model and then applies gradient robust loss to improve across-model transferability.
Apart from designing new loss functions, optimization over transformation techniques can also be exploited to improve transferability.
This includes \textbf{T-RA}~\cite{shen2024practical}, which improves cross-model transferability by applying spectrum transformations to generate adversarial perturbations that degrade segmentation in SAM variants, and \textbf{UAD}~\cite{lu2024unsegment}, which generates adversarial examples by deforming images in a two-stage process and aligning features with the deformed targets.

\subsubsection{Adversarial Defenses}\label{sec:SAM-advdefense}

Adversarial defenses for SAM are currently limited, with existing approaches focusing primarily on adversarial tuning, which integrates adversarial training into the prompt tuning process of SAM. For example, \textbf{ASAM}~\cite{li2024asam} utilizes a stable diffusion model to generate realistic adversarial samples on a low-dimensional manifold through diffusion model-based tuning. ControlNet~\cite{ControlNet} is then employed to guide the re-projection process, ensuring that the generated samples align with the original mask annotations. Finally, SAM is fine-tuned using these adversarial examples. \textbf{RobustSAM}~\cite{long2025robust} defends against adversarial attacks by adapting only 512 singular values in SAM’s convolutional layers via Singular Value Decomposition (SVD), effectively altering feature distributions. This few-parameter approach achieves strong robustness–accuracy trade-off with minimal computational overhead.

\subsubsection{Backdoor \& Poisoning Attacks}\label{sec:SAM-backdoor_poisoning}

Backdoor and poisoning attacks on SAM remain underexplored. Here, we review one backdoor attack that leverages perceptible visual triggers to compromise SAM, and one poisoning attack that exploits unlearnable examples~\cite{huang2021unlearnable} with imperceptible noise to protect unauthorized image data from being exploited by segmentation models.
\textbf{BadSAM}~\cite{guan2024badsam} is a backdoor attack targeting SAM that embeds visual triggers during the model's adaptation phase, implanting backdoors that enable attackers to manipulate the model's output with specific inputs. Specifically, the attack introduces MLP layers to SAM and injects the backdoor trigger into these layers via SAM-Adapter \cite{chen2023sam}.
\textbf{UnSeg}~\cite{sun2024unseg} is a data poisoning attack on SAM designed for benign purposes, i.e., data protection. It fine-tunes a universal unlearnable noise generator, leveraging a bilevel optimization framework based on a pre-trained SAM. This allows the generator to efficiently produce poisoned (protected) samples, effectively preventing a segmentation model from learning from the protected data and thereby safeguarding against unauthorized exploitation of personal information.

\subsubsection{Datasets}\label{sec:SAM-dataset}

As shown in Table~\ref{tab:vfm_safety}, the datasets used in safety research on SAM slightly differ from those typically used in general segmentation tasks~\cite{MOSE,MeViS}. 
For \textbf{attack research}, the SA-1B dataset and its subsets~\cite{kirillov2023segment} are the most commonly used for evaluating adversarial attacks~\cite{croce2024segment, han2023segment, lu2024unsegment, shen2024practical, zhang2023attack, zheng2023black}.
Additionally, \textbf{DarkSAM} was evaluated on datasets such as Cityscapes \cite{cordts2016cityscapes}, COCO \cite{lin2014microsoft}, and ADE20k \cite{zhou2017scene}, while \textbf{UMI-GRAT}, which targets downstream tasks related to SAM, was tested on medical datasets like CT-Scans and ISTD, as well as camouflage datasets, including COD10K, CAMO, and CHAME. For backdoor attacks, \textbf{BadSAM} was assessed using the CAMO dataset~\cite{le2019anabranch}. In the context of data poisoning, \textbf{UnSeg}~\cite{sun2024unseg} was evaluated across 10 datasets, including COCO, Cityscapes, ADE20k, WHU, and medical datasets like Lung and Kvasir-seg.
For \textbf{defense research}, \textbf{ASAM}~\cite{li2024asam} is currently the only defense method applied to SAM. It was evaluated on a range of datasets with more diverse image distributions than SA-1B, including ADE20k, LVIS, COCO, and others, with mean Intersection over Union (mIoU) used as the evaluation metric.

\begin{table*}[htp]
\center
\caption{A summary of attacks and defenses for LLMs (\textbf{Part I}). }
\label{tab:LLM-Part1}
\resizebox{1\textwidth}{!}{
\begin{tabular}{p{0.1\textwidth}p{0.15\textwidth}p{0.05\textwidth}p{0.15\textwidth}p{0.15\textwidth}p{0.25\textwidth}p{0.27\textwidth}}
\hline
\rowcolor{green!10!}
Attack/Defense & Method & Year & Category & Subcategory & Target Models & Datasets \\ \hline
\multirow{13}{0.1\textwidth}{Adversarial Attack} & \cellcolor{gray!15!}Bad characters~\cite{boucher2022bad} & \cellcolor{gray!15!}2022 & \cellcolor{gray!15!}White-box & \cellcolor{gray!15!}Character-level & \cellcolor{gray!15!}Fairseq EN-FR, Perspective API & \cellcolor{gray!15!}Emotion, Wikipedia Detox, CoNLL-2003 \\
& \cellcolor{white}TextFooler~\cite{jin2020bert} & \cellcolor{white}2020 & \cellcolor{white}White-box & \cellcolor{white}Word-level & \cellcolor{white}WordCNN, WordLSTM, BERT, InferSent, ESIM & \cellcolor{white}AG’s News, Fake News, MR, IMDB, Yelp, SNLI, MultiNLI \\
& \cellcolor{gray!15!}BERT-ATTACK~\cite{li2020bert} & \cellcolor{gray!15!}2020 & \cellcolor{gray!15!}White-box & \cellcolor{gray!15!}Word-level & \cellcolor{gray!15!}BERT, WordLSTM, ESIM & \cellcolor{gray!15!}AG’s News, Fake News, IMDB, Yelp, SNLI, MultiNLI \\
& \cellcolor{white}GBDA~\cite{guo2021gradient} & \cellcolor{white}2021 & \cellcolor{white}White-box & \cellcolor{white}Word-level & \cellcolor{white}GPT-2, XLM, BERT & \cellcolor{white}DBPedia, AG's News, Yelp Reviews, IMDB, MultiNLI\\
& \cellcolor{gray!15!}Breaking-BERT~\cite{dirkson2021breaking} & \cellcolor{gray!15!}2021 & \cellcolor{gray!15!}White-box & \cellcolor{gray!15!}Word-level & \cellcolor{gray!15!}BERT & \cellcolor{gray!15!}CoNLL-2003, W-NUT 2017, BC5CDR, NCBI disease corpus \\
& \cellcolor{white}GRADOBSTINATE~\cite{wang2023gradient} & \cellcolor{white}2023 & \cellcolor{white}White-box & \cellcolor{white}Word-level & \cellcolor{white}Electra, ALBERT, RoBERTa & \cellcolor{white}SNLI, MRPC, SQuAD, SST-2, MSCOCO \\
& \cellcolor{gray!15!}Liu et al.~\cite{liu2023expanding} & \cellcolor{gray!15!}2023 & \cellcolor{gray!15!}White-box & \cellcolor{gray!15!}Word-level & \cellcolor{gray!15!}BERT, RoBERTa & \cellcolor{gray!15!}Online Shopping 10 Cats, Chinanews \\
& \cellcolor{white}advICL~\cite{wang2023adversarial} & \cellcolor{white}2023 & \cellcolor{white}Black-box & \cellcolor{white}Sentence-level & \cellcolor{white}GPT-2-XL, LLaMA-7B, Vicuna-7B & \cellcolor{white}SST-2, RTE, TREC, DBpedia \\
& \cellcolor{gray!15!}Liu et al.~\cite{liu2023adversarial} & \cellcolor{gray!15!}2023 & \cellcolor{gray!15!}Black-box & \cellcolor{gray!15!}Sentence-level & \cellcolor{gray!15!}RoBERTa &  \cellcolor{gray!15!}Real conversation data \\
& \cellcolor{white}Koleva et al.~\cite{koleva2023adversarial} & \cellcolor{white}2023 & \cellcolor{white}Black-box & \cellcolor{white}Sentence-level & \cellcolor{white}TURL & \cellcolor{white}WikiTables \\
\hline
\multirow{3}{0.1\textwidth}{Adversarial Defense} & \cellcolor{gray!15!}Jain et al.\cite{jain2023baseline} & \cellcolor{gray!15!}2023 & \cellcolor{gray!15!}Adversarial Detection & \cellcolor{gray!15!}Input Filtering & \cellcolor{gray!15!}Guanaco-7B, Vicuna-7B, Falcon-7B & \cellcolor{gray!15!}AlpacaEval \\
& \cellcolor{white}Erase-and-Check~\cite{kumar2023certifying} & \cellcolor{white}2023 & \cellcolor{white}Adversarial Detection & \cellcolor{white}Input Filtering & \cellcolor{white}LLaMA-2, DistilBERT & \cellcolor{white}AdvBench \\
& \cellcolor{gray!15!}Zou et al.~\cite{zou2024improving} & \cellcolor{gray!15!}2024 & \cellcolor{gray!15!}Robust Inference & \cellcolor{gray!15!}Circuit Breaking & \cellcolor{gray!15!}Mistral-7B, LLaMA-3-8B & \cellcolor{gray!15!}HarmBench \\
\hline
\multirow{25}{0.1\textwidth}{Jailbreak Attack} & \cellcolor{white}Yong et al.~\cite{yong2023low} & \cellcolor{white}2023 & \cellcolor{white}Black-box & \cellcolor{white}Hand-crafted & \cellcolor{white}GPT-4 & \cellcolor{white}AdvBench \\
& \cellcolor{gray!15!}CipherChat~\cite{yuan2023gpt} & \cellcolor{gray!15!}2023 & \cellcolor{gray!15!}Black-box & \cellcolor{gray!15!}Hand-crafted & \cellcolor{gray!15!}GPT-3.5, GPT-4 & \cellcolor{gray!15!}Chinese safety assessment benchmark \\
& \cellcolor{white}Jailbroken~\cite{wei2024jailbroken} & \cellcolor{white}2023 & \cellcolor{white}Black-box & \cellcolor{white}Hand-crafted & \cellcolor{white}GPT-4, GPT-3.5, Claude-1.3 & \cellcolor{white}Self-built \\
& \cellcolor{gray!15!}Li et al.~\cite{li2024cross} & \cellcolor{gray!15!}2024 & \cellcolor{gray!15!}Black-box & \cellcolor{gray!15!}Hand-crafted & \cellcolor{gray!15!}GPT-3.5, GPT-4, Vicuna-1.3-7B, 13B, Vicuna-1.5-7B, 13B & \cellcolor{gray!15!}Self-built \\
& \cellcolor{white}Easyjailbreak~\cite{zhou2024easyjailbreak} & \cellcolor{white}2024 & \cellcolor{white}Black-box & \cellcolor{white}Hand-crafted & \cellcolor{white}GPT-3.5, GPT-4, LLaMA-2-7B, 13B, Vicuna-1.5-7B, 13B, ChatGLM3, Qwen-7B, InternLM-7B, Mistral-7B & \cellcolor{white}AdvBench \\
& \cellcolor{gray!15!}SMEA~\cite{zou2024system} & \cellcolor{gray!15!}2024 & \cellcolor{gray!15!}Black-box & \cellcolor{gray!15!}Hand-crafted & \cellcolor{gray!15!}GPT-3.5, LLaMA-2-7B, 13B & \cellcolor{gray!15!}Self-built \\
& \cellcolor{white}Tastle~\cite{xiao2024tastle} & \cellcolor{white}2024 & \cellcolor{white}Black-box & \cellcolor{white}Hand-crafted & \cellcolor{white}Vicuna-1.5-13B, LLaMA-2-7B, GPT-3.5, GPT-4 & \cellcolor{white}AdvBench \\
& \cellcolor{gray!15!}StructuralSleight~\cite{li2024structuralsleight} & \cellcolor{gray!15!}2024 & \cellcolor{gray!15!}Black-box & \cellcolor{gray!15!}Hand-crafted & \cellcolor{gray!15!}GPT-3.5, GPT-4, GPT-4o, LLaMA-3-70B, Claude-2, Cluade3-Opus & \cellcolor{gray!15!}AdvBench \\
& \cellcolor{white}CodeChameleon~\cite{lv2024codechameleon} & \cellcolor{white}2024 & \cellcolor{white}Black-box & \cellcolor{white}Hand-crafted & \cellcolor{white}LLaMA-2-7B, 13B, 70B, Vicuna-1.5-7B, 13B, GPT-3.5, GPT-4 & \cellcolor{white}AdvBench, MaliciousInstruct, ShadowAlignment  \\
& \cellcolor{gray!15!}Puzzler~\cite{chang2024play} & \cellcolor{gray!15!}2024 & \cellcolor{gray!15!}Black-box & \cellcolor{gray!15!}Hand-crafted & \cellcolor{gray!15!}GPT-3.5, GPT-4, GPT4-Turbo, Gemini-pro, LLaMA-2-7B, 13B & \cellcolor{gray!15!}AdvBench, MaliciousInstructions \\
& Wen et al. \cite{wen2024evaluating} & 2025 & Black-box &  Hand-crafted & GPT-3.5, 4, Mistral-v0.3, LLaMA-3 & BUMBLE benchmark \\
& \cellcolor{gray!15!}ABJ~\cite{lin2024llms} & \cellcolor{gray!15!}2024 & \cellcolor{gray!15!}Black-box & \cellcolor{gray!15!}Hand-crafted & \cellcolor{gray!15!}GPT-4o, Claude-3-haiku, LLaMA-3-8B,  Qwen-2.5-7B, DeepSeek-V3 & \cellcolor{gray!15!}AdvBench \\
& \cellcolor{white} Shen et al.\cite{SCBSZ24} & \cellcolor{white}2024 & \cellcolor{white}Black-box & \cellcolor{white}Hand-crafted & \cellcolor{white}GPT-3.5, 4, PaLM-2, ChatGLM, Vicuna & \cellcolor{white}In-The-Wild Jailbreak Prompts \\
& \cellcolor{gray!15!}AutoDAN~\cite{liu2023autodan} & \cellcolor{gray!15!}2023 & \cellcolor{gray!15!}Black-box & \cellcolor{gray!15!}Automated & \cellcolor{gray!15!}Vicuna-7B, Guanaco-7B, LLaMA-2-7B & \cellcolor{gray!15!}AdvBench \\
& \cellcolor{white}I-FSJ\cite{zheng2024improved} & \cellcolor{white}2024 & \cellcolor{white}Black-box & \cellcolor{white}Automated & \cellcolor{white}LLaMA-2, LLaMA-3, OpenChat-3.5, Starling-LM, Qwen-1.5 & \cellcolor{white}JailbreakBench \\
& \cellcolor{gray!15!}Weak-to-Strong\cite{zhao2024weak} & \cellcolor{gray!15!}2024 & \cellcolor{gray!15!}Black-box & \cellcolor{gray!15!}Automated & \cellcolor{gray!15!}LLaMA2-13B, Vicuna-13B, Baichuan2-13B, InternLM-20B & \cellcolor{gray!15!}AdvBench, MaliciousInstruct \\
& \cellcolor{white}GPTFuzzer~\cite{yu2023gptfuzzer} & \cellcolor{white}2023 & \cellcolor{white}Black-box & \cellcolor{white}Automated & \cellcolor{white}Vicuna-13B, Baichuan-13B, ChatGLM-2-6B, LLaMA-2-13B, 70B, GPT-4, Bard, Claude-2, PaLM-2 & \cellcolor{white}Self-built \\
& \cellcolor{gray!15!}PAIR~\cite{chao2023jailbreaking} & \cellcolor{gray!15!}2023 & \cellcolor{gray!15!}Black-box & \cellcolor{gray!15!}Automated & \cellcolor{gray!15!}Vicuna-1.5-13B, LLaMA-2-7B, GPT-3.5, GPT-4, Claude-1, Claude-2, Gemini-pro & \cellcolor{gray!15!}JBB-Behaviors, AdvBench \\
& \cellcolor{white}Masterkey~\cite{deng2024masterkey} & \cellcolor{white}2023  & \cellcolor{white}Black-box & \cellcolor{white}Automated & \cellcolor{white}GPT-3.5, GPT-4, Bard, Bing Chat & \cellcolor{white}Self-built \\
& \cellcolor{gray!15!}BOOST~\cite{yu2024enhancing} & \cellcolor{gray!15!}2024 & \cellcolor{gray!15!}Black-box & \cellcolor{gray!15!}Automated & \cellcolor{gray!15!}LLaMA-2-7B, 13B, Gemma-2B, 7B, Tulu-2-7B, 13B, Mistral-7B, MPT-7B, Qwen1.5-7B, Vicuna-7B, LLaMA-3-8B &  \cellcolor{gray!15!}AdvBench \\
& \cellcolor{white}FuzzLLM~\cite{yao2024fuzzllm} & \cellcolor{white}2024 & \cellcolor{white}Black-box & \cellcolor{white}Automated & \cellcolor{white}Vicuna-13B, CAMEL-13B, LLaMA-7B, ChatGLM-2-6B, Bloom-7B, LongChat-7B, GPT-3.5,  GPT-4 & \cellcolor{white}Self-built \\
& \cellcolor{gray!15!}EnJa\cite{zhang2024enja} & \cellcolor{gray!15!}2024 & \cellcolor{gray!15!}Black-box & \cellcolor{gray!15!}Automated &  \cellcolor{gray!15!}Vicuna-7B, 13B, LLaMA-2-13B, GPT-3.5, 4 &  \cellcolor{gray!15!}AdvBench \\
& \cellcolor{white}Perez et al.\cite{perez2022red} & \cellcolor{white}2022 & \cellcolor{white}Black-box & \cellcolor{white}Automated & \cellcolor{white}Gopher LM & \cellcolor{white}Self-built \\
& \cellcolor{gray!15!}CRT\cite{hong2024curiosity} & \cellcolor{gray!15!}2024 & \cellcolor{gray!15!}Black-box & \cellcolor{gray!15!}Automated & \cellcolor{gray!15!}GPT-2, Dolly-v2-7B, LLaMA-2-7B & \cellcolor{gray!15!}IMDb \\
& \cellcolor{white} ECLIPSE~\cite{jiang2024unlocking} &
\cellcolor{white} 2024 &
\cellcolor{white} Black-box &
\cellcolor{white} Automated &
\cellcolor{white} Vicuna-7B, LLaMA2-7B, Falcon-7B, GPT-3.5 &
\cellcolor{white} AdvBench \\
& \cellcolor{gray!15!}GCG~\cite{zou2023universal} & \cellcolor{gray!15!}2023 & \cellcolor{gray!15!}White-box & \cellcolor{gray!15!}Automated & \cellcolor{gray!15!}Vicuna-7B, LLaMA-2-7B, GPT-3.5, GPT-4, PaLM-2, Claude-2 & \cellcolor{gray!15!}AdvBench  \\
& \cellcolor{white}I-GCG \cite{jia2024improved} & \cellcolor{white}2024 & \cellcolor{white}White-box & \cellcolor{white}Automated & \cellcolor{white}Vicuna-7B-1.5, Guanaco-7B, LLaMA2-7B, MISTRAL-7B & \cellcolor{white}AdvBench\\
& \cellcolor{gray!15!}POUGH~\cite{huang2024semantic} & \cellcolor{gray!15!}2024 & \cellcolor{gray!15!}White-box & \cellcolor{gray!15!}Automated & \cellcolor{gray!15!}Vicuna, Mistral, Guanaco & \cellcolor{gray!15!}Alpaca dataset \\
& Qi et al. \cite{qi2023fine} & 2023 & White-box & Fine-tuning & GPT-3.5-Turbo, LLaMA-2-7B-Chat & - \\
& \cellcolor{gray!15!}Virus\cite{huang2025virus} & \cellcolor{gray!15!}2025 & \cellcolor{gray!15!}White-box & \cellcolor{gray!15!}Fine-tuning & \cellcolor{gray!15!}LLaMA-3-8B & \cellcolor{gray!15!}SST2, AgNews, GSM8K  \\
\hline
\multirow{15}{0.1\textwidth}{Jailbreak Defense} & \cellcolor{gray!15!}SmoothLLM~\cite{robey2023smoothllm} & \cellcolor{gray!15!}2023 & \cellcolor{gray!15!}Input Defense & \cellcolor{gray!15!}Rephrasing & \cellcolor{gray!15!}Vicuna, LLaMA-2, GPT-3.5, GPT-4 & \cellcolor{gray!15!}AdvBench, JBB-Behaviors \\
& \cellcolor{white}SemanticSmooth~\cite{ji2024defending} & \cellcolor{white}2024 & \cellcolor{white}Input Defense & \cellcolor{white}Rephrasing & \cellcolor{white}LLaMA-2-7B, Vicuna-13B, GPT-3.5 & \cellcolor{white}InstructionFollow, AlpacaEval \\
& \cellcolor{gray!15!}SelfDefend~\cite{wang2024selfdefend} & \cellcolor{gray!15!}2024 & \cellcolor{gray!15!}Input Defense& \cellcolor{gray!15!}Rephrasing & \cellcolor{gray!15!}GPT-3.5, GPT-4 &  \cellcolor{gray!15!}JailbreakBench, MultiJail, AlpacaEval \\
& \cellcolor{white}IBProtector~\cite{liu2024protecting} & \cellcolor{white}2024 & \cellcolor{white}Input Defense& \cellcolor{white}Rephrasing & \cellcolor{white}LLaMA-2-7B, Vicuna-1.5-13B & \cellcolor{white}AdvBench, TriviaQA, EasyJailbreak \\
& \cellcolor{gray!15!}PEARL~\cite{liangpearl} & \cellcolor{gray!15!}2025 & \cellcolor{gray!15!}Input Defense & \cellcolor{gray!15!}Rephrasing &  \cellcolor{gray!15!}LLaMA-3-8B, LLaMA-2-7B,13B,
Mistral-7B, Gemma-7B & \cellcolor{gray!15!}Super-Natural Instructions \\
& VAA\cite{liangvulnerability} & 2025 & Input Defense & Rephrasing & LLaMA-2-7B, Qwen-2.5-7B & SST2, AGNEWS, GSM8K, AlpacaEval \\
& \cellcolor{gray!15!}Backtranslation~\cite{wang2024defending} & \cellcolor{gray!15!}2024 & 
\cellcolor{gray!15!}Input Defense & \cellcolor{gray!15!}Translation & \cellcolor{gray!15!}GPT-3.5, LLaMA-2-13B, Vicuna-13B & \cellcolor{gray!15!}AdvBench, MT-Bench \\
& RTT~\cite{yung2025round} & 2025 & Input Defense & Translation & Vicunna, GPT-4, LLaMA-2, Palm-2 & AdvBench \\
& \cellcolor{gray!15!}CurvaLID\cite{yung2025curvalid} & \cellcolor{gray!15!}2025 & \cellcolor{gray!15!}Input Defense & \cellcolor{gray!15!}Filtering & \cellcolor{gray!15!}Universal & \cellcolor{gray!15!}Orca, MMLU, AlpacaEval, TruthfulQA, AdvBench
\\
& \cellcolor{white}APS~\cite{kim2023robust} & \cellcolor{white}2023 & \cellcolor{white}Output Defense& \cellcolor{white}Filtering & \cellcolor{white}Vicuna, Falcon, Guanaco & \cellcolor{white}AdvBench \\
& \cellcolor{gray!15!}DPP~\cite{xiong2024defensive} & \cellcolor{gray!15!}2024 & \cellcolor{gray!15!}Output Defense& \cellcolor{gray!15!}Filtering & \cellcolor{gray!15!}LLaMA-2-7B, Mistral-7B & \cellcolor{gray!15!}AdvBench \\
& \cellcolor{white}Gradient Cuff~\cite{hu2024gradient} & \cellcolor{white}2024 & \cellcolor{white}Output Defense & \cellcolor{white}Filtering & \cellcolor{white}LLaMA-2-7B, Vicuna-1.5-7B & \cellcolor{white}AdvBench \\
& \cellcolor{gray!15!}LEGILIMENS \cite{wu2024legilimens} & \cellcolor{gray!15!}2024 & \cellcolor{gray!15!}Output Defense & \cellcolor{gray!15!}Filtering & \cellcolor{gray!15!}ChatGLM-3, LLaMA-2, Falcon, Dolly, Vicuna & \cellcolor{gray!15!}Measuring Hate Speech, BeaverTails, BEA\&AG, HarmBench, AdvBench  \\
& ABD\cite{gao2024shaping} & 2024 & Robust Inference & Activation Boundary & LLaMA-2-7B-Chat, Vicuna-7B-v1.3, Qwen-1.5-0.5B-Chat, Vicuna-13B-v1.5 & Just-Eval  \\
& 
\cellcolor{gray!15!}MTD~\cite{chen2023jailbreaker} & \cellcolor{gray!15!}2023 & \cellcolor{gray!15!}Robust Inference& \cellcolor{gray!15!}Multi-model Inference & \cellcolor{gray!15!}GPT-3.5, GPT-4, Bard, Claude, LLaMA2-7B, 13B, 70B & \cellcolor{gray!15!}Self-built \\
& \cellcolor{white}PARDEN~\cite{zhang2024parden} & \cellcolor{white}2024 & \cellcolor{white}Robust Inference& \cellcolor{white}Output Repetition & \cellcolor{white}LLaMA-2-7B, Mistral-7B, Claude-2.1 & \cellcolor{white}PARDEN \\
& \cellcolor{gray!15!}AutoDefense~\cite{lu2024autojailbreak} & \cellcolor{gray!15!}2024 & \cellcolor{gray!15!}Ensemble Defense & \cellcolor{gray!15!}Rephrasing/Filtering & \cellcolor{gray!15!}GPT-3.5-turbo, GPT-4, LLaMA-2, LLaMA-3, Mistral, Qwen, Vicuna & \cellcolor{gray!15!}Self-built  \\
& \cellcolor{white}MoGU~\cite{du2024mogu} & \cellcolor{white}2024 & \cellcolor{white}Ensemble Defense & \cellcolor{white}Rephrasing/Filtering & \cellcolor{white}LLaMA-2-7B, Vicuna-7B, Falcon-7B, Dolphin-7B & \cellcolor{white}Advbench \\
& \cellcolor{gray!15!}Vaccine\cite{huang2024vaccine} & \cellcolor{gray!15!}2024 & \cellcolor{gray!15!}Defenses against Fine-tuning Attacks & \cellcolor{gray!15!}Alignment Stage & \cellcolor{gray!15!}LLaMA-2-7B, Opt-3.7B, Mistral-7B & \cellcolor{gray!15!}BeaverTails, SST2, AGNEWS, GSM8K, AlpacaEval \\
& T-Vaccine\cite{liu2024targeted} & 2025 & Defenses against Fine-tuning Attacks & Alignment Stage & Gemma-2-2B, LLaMA-2-7B, Vicuna-7B, Qwen2-7B & SST2, GSM8K, AGNEWS \\
& \cellcolor{gray!15!}Booster\cite{huang2024booster} & \cellcolor{gray!15!}2025 & \cellcolor{gray!15!}Defenses against Fine-tuning & \cellcolor{gray!15!} & \cellcolor{gray!15!}LLaMA2-7B, Gemma2-9B, Qwen2-7B & \cellcolor{gray!15!}SST2, AGNEWS, GSM8K, AlpacaEval  \\
& Lisa\cite{huang2024lisa} & 2024 & Defenses against Fine-tuning Attacks & Fine-tuning Stage & LLaMA-2-7B, Opt-3.7B, Mistral-7B & BeaverTails, SST2, AGNEWS, GSM8K, AlpacaEval\\
& \cellcolor{gray!15!}Antidote \cite{huang2024antidote} & \cellcolor{gray!15!}2024 & \cellcolor{gray!15!}Defenses against Fine-tuning Attacks & \cellcolor{gray!15!}Post-fine-tuning Stage & \cellcolor{gray!15!}LLaMA-2-7B, Mistral-7B, Gemma-7B & \cellcolor{gray!15!}SST2, AGNEWS, GSM8K, AlpacaEval \\
& Panacea \cite{wang2025panacea} & 2025 & Defenses against Fine-tuning Attacks & Post-fine-tuning Stage & LLaMA-2-7B & GSM8K, SST2, AlpacaEval, AGNEWS  \\
\hline
\end{tabular}
}
\end{table*}

\section{Large Language Model Safety} \label{sec:llm}

LLMs are powerful language models that excel at generating human-like text, translating languages, producing creative content, and answering a diverse array of questions \cite{openai-o1,guo2025deepseek}. They have been rapidly adopted in applications such as conversational agents, automated code generation, and scientific research. Yet, this broad utility also introduces significant vulnerabilities that potential adversaries can exploit.
This section surveys the current landscape of LLM safety research. We examine a spectrum of adversarial behaviors, including jailbreak, prompt injection, backdoor, poisoning, model extraction, data extraction, and energy–latency attacks. Such attacks can manipulate outputs, bypass safety measures, leak sensitive information, and disrupt services, thereby threatening system integrity, confidentiality, and availability. We also review state-of-the-art alignment strategies and defense techniques designed to mitigate these risks. Tables \ref{tab:LLM-Part1} and \ref{tab:LLM-Part2} summarize the details of these works.

\subsection{Adversarial Attacks} \label{sec:llm_adversarial_attack}

Adversarial attacks on LLMs aim to mislead the victim model to generate incorrect responses (no matter under targeted or untargeted manners) by subtly altering input text.
We classify these attacks into \textbf{white-box attacks} and \textbf{black-box attacks}, depending on whether the attacker can access the model’s internals.

\subsubsection{White-box Attacks}
White-box attacks assume the attacker has full knowledge of the LLM's architecture, parameters, and gradients. This enables the construction of highly effective adversarial examples by directly optimizing against the model's predictions. These attacks can generally be classified into two levels: \textbf{1) character-level attacks} and \textbf{2) word-level attacks}, differing primarily in their effectiveness and semantic stealthiness.

\textbf{Character-level Attacks} introduce subtle modifications at the character level, such as misspellings, typographical errors, and the insertion of visually similar or invisible characters (e.g., homoglyphs \cite{boucher2022bad}). These attacks exploit the model’s sensitivity to minor character variations, which are often unnoticeable to humans, allowing for a high degree of stealthiness while potentially preserving the original meaning. 

\textbf{Word-level Attacks} modify the input text by substituting or replacing specific words. For example, \textbf{TextFooler}~\cite{jin2020bert} and \textbf{BERT-Attack}~\cite{li2020bert} employ \textit{synonym substitution} to generate adversarial examples while preserving semantic similarity. Other methods, such as \textbf{GBDA} \cite{guo2021gradient} and \textbf{GRADOBSTINATE} \cite{wang2023gradient}, leverage gradient information to identify semantically similar \textit{word substitutions} that maximize the likelihood of a successful attack. Additionally, \textit{targeted word substitution} enables attacks tailored to specific tasks or linguistic contexts. For instance, \cite{dirkson2021breaking} explores targeted attacks on named entity recognition, while \cite{liu2023expanding} adapts word substitution attacks for the Chinese language.

\subsubsection{Black-box Attacks}

Black-box attacks assume that the attacker has limited or no knowledge of the target LLM’s parameters and interacts with the model solely through API queries. In contrast to white-box attacks, black-box attacks employ indirect and adaptive strategies to exploit model vulnerabilities. These attacks typically manipulate input prompts rather than altering the core text. We further categorize existing black-box attacks on LLMs into four types: 1) \textbf{in-context attacks}, 2) \textbf{induced attacks}, 3) \textbf{LLM-assisted attacks}, and 4) \textbf{tabular attacks}.

\textbf{In-context Attacks} exploit the demonstration examples used in in-context learning to introduce adversarial behavior, making the model vulnerable to poisoned prompts. \textbf{AdvICL}~\cite{wang2023adversarial} and \textbf{Transferable-advICL} manipulate these demonstration examples to expose this vulnerability, highlighting the model’s susceptibility to poisoned in-context data.

\textbf{Induced Attacks} rely on carefully crafted prompts to coax the model into generating harmful or undesirable outputs, often bypassing its built-in safety mechanisms. These attacks focus on generating adversarial responses by designing deceptive input prompts. For example, Liu et al.~\cite{liu2023adversarial} analyzed how such prompts can lead the model to produce dangerous outputs, effectively circumventing safeguards designed to prevent such behavior.

\textbf{LLM-Assisted Attacks} leverage LLMs to implement attack algorithms or strategies, effectively turning the model into a tool for conducting adversarial actions. This approach underscores the capacity of LLMs to assist attackers in designing and executing attacks. For instance, Carlini~\cite{carlini2023llm} demonstrated that GPT-4 can be prompted step-by-step to design attack algorithms, highlighting the potential for using LLMs as research assistants to automate adversarial processes.

\textbf{Tabular Attacks} target tabular data by exploiting the structure of columns and annotations to inject adversarial behavior. Koleva et al.~\cite{koleva2023adversarial} proposed an entity-swap attack that specifically targets column-type annotations in tabular datasets. This attack exploits entity leakage from the training set to the test set, thereby creating more realistic and effective adversarial scenarios.

\subsection{Adversarial Defenses} \label{sec:llm_adversarial_defense}
Adversarial defenses are crucial for ensuring the safety, reliability, and trustworthiness of LLMs in real-world applications. Existing adversarial defense strategies for LLMs can be broadly classified based on their primary focus into two categories: \textbf{1) adversarial detection} and \textbf{2) robust inference}.

\subsubsection{Adversarial Detection}
Adversarial detection methods aim to identify and flag potential adversarial inputs before they can affect the model's output. The goal is to implement a filtering mechanism that can differentiate between benign and malicious prompts.

\textbf{Input Filtering} Most adversarial detection methods for LLMs are input filtering techniques that identify and reject adversarial texts based on statistical or structural anomalies. For example, Jain et al. \cite{jain2023baseline} use perplexity to detect adversarial prompts, as these typically show higher perplexity when evaluated by a well-calibrated language model, indicating a deviation from natural language patterns. By setting a perplexity threshold, such inputs can be filtered out. Another approach, \textbf{Erase-and-Check} \cite{kumar2023certifying}, ensures robustness by iteratively erasing parts of the input and checking for output consistency. Significant changes in output signal potential adversarial manipulation.
Input filtering methods offer a lightweight first line of defense, but their effectiveness depends on the chosen features and the sophistication of adversarial attacks, which may bypass these defenses if designed adaptively.

\subsubsection{Robust Inference}
Robust inference methods aim to make the model inherently resistant to adversarial attacks by modifying its internal mechanisms or training. One approach, \textbf{Circuit Breaking} \cite{zou2024improving}, targets specific activation patterns during inference, neutralizing harmful outputs without retraining. While robust inference enhances resistance to adaptive attacks, it often incurs higher computational costs, and its effectiveness varies by model architecture and attack type.

\subsection{Jailbreak Attacks}
\label{sec:llm_jailbreak_attacks}

Unlike adversarial attacks that simply lead victim LLMs to generate incorrect answers, jailbreak attacks trick LLMs into generating inappropriate content ($e.g.$, harmful or deceptive content) by bypassing the built-in safety policy/alignment via hand-crafted or automated jailbreak prompts.
Currently, most jailbreak attacks target the LLM-as-a-Service scenario, following a black-box threat model where the attacker cannot access the model’s internals.

\subsubsection{Hand-crafted Attacks}

Hand-crafted attacks involve designing adversarial prompts to exploit specific vulnerabilities in the target LLM. The goal is to craft word/phrase combinations or structures that can bypass the model's safety filters while still conveying harmful requests.

\textbf{Scenario-based Camouflage} hides malicious queries within complex scenarios, such as role-playing or puzzle-solving, to obscure their harmful intent. For instance, Li et al. \cite{li2024cross} instruct the LLM to adopt a persona likely to generate harmful content, while \textbf{SMEA} \cite{zou2024system} places the LLM in a subordinate role under an authority figure. \textbf{Easyjailbreak} \cite{zhou2024easyjailbreak} frames harmful queries in hypothetical contexts, and \textbf{Puzzler} \cite{chang2024play} embeds them in puzzles whose solutions correspond to harmful outputs.
Drawing on psychometric principles, Wen et al.~\cite{wen2024evaluating} proposed attacks such as Disguise, Deception, and Teaching to elicit implicit biases by constructing specific psychological scenarios.
\textbf{Analyzing-based Jailbreak (ABJ)}~\cite{lin2024llms} transforms harmful queries into neutral analytical prompts, manipulating the model's reasoning chain to induce unsafe responses.
Other studies have also leveraged psychological concepts, employing techniques like disguise, deception, and teaching to reveal implicit biases from a psychometric perspective~\cite{wen2024evaluating}.
\textbf{Attention Shifting} redirects the LLM’s focus from the malicious intent by introducing linguistic complexities. \textbf{Jailbroken} \cite{wei2024jailbroken} employs code-switching and unusual sentence structures, \textbf{Tastle} \cite{xiao2024tastle} manipulates tone, and \textbf{StructuralSleight} \cite{li2024structuralsleight} alters sentence structure to disrupt understanding.
In addition, Shen et al.~\cite{SCBSZ24} collected real-world jailbreak prompts shared by users on social media, such as Reddit and Discord, and studied their effectiveness against LLMs.

\textbf{Encoding-Based Attacks} exploit LLMs' limitations in handling rare encoding schemes, such as low-resource languages and encryption. These attacks encode malicious queries in formats like \textbf{Base64} \cite{wei2024jailbroken} or low-resource languages \cite{yong2023low}, or use custom encryption methods like ciphers \cite{yuan2023gpt} and \textbf{CodeChameleon} \cite{lv2024codechameleon} to obfuscate harmful content.

\subsubsection{Automated Attacks}

Unlike hand-crafted attacks, which rely on expert knowledge, automated attacks aim to discover jailbreak prompts autonomously. These attacks either use black-box optimization to search for optimal prompts or leverage LLMs to generate and refine them.

\textbf{Prompt Optimization} leverages optimization algorithms to iteratively refine prompts, targeting higher success rates. For black-box methods, \textbf{AutoDAN} \cite{liu2023autodan} employs a genetic algorithm, \textbf{GPTFuzzer} \cite{yu2023gptfuzzer} utilizes mutation- and generation-based fuzzing techniques, and \textbf{FuzzLLM} \cite{yao2024fuzzllm} generates semantically coherent prompts within an automated fuzzing framework. 
\textbf{I-FSJ} \cite{zheng2024improved} injects special tokens into few-shot demonstrations and uses demo-level random search to optimize the prompt, achieving high attack success rates against aligned models and their defenses. 
For white-box methods, the most notable is \textbf{GCG} \cite{zou2023universal}, which introduces a greedy coordinate gradient algorithm to search for adversarial suffixes, effectively compromising aligned LLMs.
 \textbf{I-GCG} \cite{jia2024improved} further improves GCG with diverse target templates and an automatic multi-coordinate updating strategy, achieving near-perfect attack success rates.
Shifting the focus from optimization algorithms to training data, \textbf{POUGH}~\cite{huang2024semantic} introduces a semantic-guided strategy for sampling and ranking prompts, thereby improving the efficiency and generalizability of the generated adversarial suffixes.

\textbf{LLM-Assisted Attacks} use an adversary LLM to help generate jailbreak prompts. Perez et al. \cite{perez2022red} explored model-based red teaming, finding that an LLM fine-tuned via RL can generate more effective adversarial prompts, though with limited diversity. \textbf{CRT} \cite{hong2024curiosity} improves prompt diversity by minimizing SelfBLEU scores and cosine similarity. \textbf{PAIR} \cite{chao2023jailbreaking} employs multi-turn queries with an attacker LLM to refine jailbreak prompts iteratively. Based on PAIR, Robey et al. \cite{robey2024jailbreaking} introduced \textbf{ROBOPAIR}, which targets LLM-controlled robots, causing harmful physical actions. 
Similarly, \textbf{ECLIPSE}~\cite{jiang2024unlocking} leverages an attacker LLM to identify adversarial suffixes analogous to GCG, thereby automating the prompt optimization process.
To enhance prompt transferability, \textbf{Masterkey} \cite{deng2024masterkey} trains adversary LLMs to attack multiple models.
Additionally, \textbf{Weak-to-Strong Jailbreaking} \cite{zhao2024weak} proposes a novel attack where a weaker, unsafe model guides a stronger, aligned model to generate harmful content, achieving high success rates with minimal computational cost.

\subsubsection{Fine-tuning-based Attacks}
Fine-tuning-based attacks compromise the safety alignment of LLMs by fine-tuning them on small, malicious datasets, thereby extending the attack surface from inference-time prompting to model customization.
Unlike prompt-based attacks, this approach directly alters the model’s weights, instilling harmful behaviors rather than merely circumventing input filters.
A notable example is the work of Qi et al. \cite{qi2023fine}, which demonstrates that an LLM’s safety alignment can be undermined by fine-tuning on as few as ten adversarial examples. Their findings further reveal a subtle risk: even fine-tuning on benign, utility-oriented datasets can inadvertently erode safety alignment, highlighting its inherent fragility.

To mitigate this threat, service providers may deploy guardrail models to filter harmful samples from user-supplied fine-tuning data. However, Huang et al. \cite{huang2025virus} showed that such defenses can be bypassed. Their proposed attack, \textbf{Virus}, uses dual-objective optimization to craft fine-tuning data that is both classified as benign by the guardrail model and highly effective at degrading safety alignment by preserving gradient similarity to original harmful data. This ongoing adversarial dynamic exemplifies the evolving cat-and-mouse game between attackers and defenders in the fine-tuning pipeline.

\subsection{Jailbreak Defenses}
\label{sec:llm_jailbreak_defenses}

We now introduce the corresponding defense mechanisms for black-box LLMs against jailbreak attacks. Based on the intervention stage, we classify existing defenses into four categories: \textbf{input defense}, \textbf{output defense}, \textbf{ensemble defense}, and \textbf{defenses against fine-tuning attacks}.

\subsubsection{Input Defenses}

Input defense methods focus on preprocessing the input prompt to reduce its harmful content. Current techniques include \emph{rephrasing} and \emph{translation}.

\textbf{Input Rephrasing} uses paraphrasing or purification to obscure the malicious intent of the prompt. For example, \textbf{SmoothLLM} \cite{robey2023smoothllm} applies random sampling to perturb the prompt, while \textbf{SemanticSmooth} \cite{ji2024defending} finds semantically similar, safe alternatives. Beyond prompt-level changes, \textbf{SelfDefend} \cite{wang2024selfdefend} performs token-level perturbations by removing adversarial tokens with high perplexity. \textbf{IBProtector}, on the other hand, \cite{liu2024protecting} perturbs the encoded input using the information bottleneck principle.
Besides inference-time modifications, several methods improving a model’s inherent robustness during training. For example, \textbf{PEARL} \cite{liangpearl} employs a distributionally robust optimization framework to adversarially train the model against worst-case permutations of in-context demonstrations, thereby strengthening its resistance to attacks based on input ordering.
Similarly, \textbf{VAA} \cite{liangvulnerability} improves robustness against harmful fine-tuning by identifying alignment data subsets that are vulnerable to forgetting and applying group distributionally robust optimization to ensure balanced learning.

\textbf{Input Translation} uses cross-lingual transformations to mitigate jailbreak attacks. For example, Wang et al. \cite{wang2024defending} proposed refusing to respond if the target LLM rejects the back-translated version of the original prompt, based on the hypothesis that back-translation reveals the underlying intent of the prompt. Similarly, the \textbf{RTT} \cite{yung2025round} is designed to counter social-engineered attacks on LLMs. It works by translating input prompts into one or more intermediate languages and then back to the original language, thereby disrupting potential adversarial intent embedded in the original phrasing.

\textbf{Input Filtering} rejects queries identified as malicious. For example, \textbf{CurvaLID} \cite{yung2025curvalid} detects adversarial prompts by analyzing geometric differences in their text embeddings. Since it operates solely on the input prompts and does not rely on the underlying LLM, CurvaLID provides universal protection across different LLMs.

\subsubsection{Output Defenses}

Output defense methods monitor the LLM’s generated output to identify harmful content, triggering a refusal mechanism when unsafe output is detected.

\textbf{Output Filtering} inspects the LLM's output and selectively blocks or modifies unsafe responses. This process relies on either judge scores from pre-trained classifiers or internal signals (e.g., the loss landscape) from the LLM itself. For instance, \textbf{APS} \cite{kim2023robust} and \textbf{DPP} \cite{xiong2024defensive} use safety classifiers to identify unsafe outputs, while \textbf{Gradient Cuff} \cite{hu2024gradient} analyzes the LLM’s internal refusal loss function to distinguish between benign and malicious queries.
Similarly, by analyzing the model’s internal states, \textbf{Activation Boundary Defense (ABD)} \cite{gao2024shaping} restricts harmful activations within a predefined safety boundary to prevent jailbreaks.
\textbf{LEGILIMENS} \cite{wu2024legilimens} extracts conceptual features from the host LLM’s internal states during inference and employs a lightweight classifier for efficient content moderation. \textbf{Perspective-taking prompting (PET)} \cite{xu2024walking} is an effective method to moderate an LLM's output contents via its internal knowledge and opinions without fine-tuning this model.

\textbf{Output Repetition} detects harmful content by observing that the LLM can consistently repeat its benign outputs. \textbf{PARDEN} \cite{zhang2024parden} identifies inconsistencies by prompting the LLM to repeat its output. If the model fails to accurately reproduce its response, especially for harmful queries, it may indicate a potential jailbreak.

\subsubsection{Ensemble Defenses}

Ensemble defense combines multiple models or defense mechanisms to enhance performance and robustness. The idea is that different models and defenses can offset their individual weaknesses, resulting in greater overall safety.

\textbf{Multi-model Ensemble} combines inference results from multiple LLMs to create a more robust system. For example, \textbf{MTD} \cite{chen2023jailbreaker} improves LLM safety by dynamically utilizing a pool of diverse LLMs. Rather than relying on a single model, MTD selects the safest and most relevant response by analyzing outputs from multiple models.

\textbf{Multi-defense Ensemble} integrates multiple defense strategies to strengthen robustness against various attacks. For instance, \textbf{AutoDefense} \cite{lu2024autojailbreak} introduces an ensemble framework combining input and output defenses for enhanced effectiveness. \textbf{MoGU} \cite{du2024mogu} uses a dynamic routing mechanism to balance contributions from a safe LLM and a usable LLM, based on the input query, effectively combining rephrasing and filtering.

\subsubsection{Defenses Against Fine-Tuning Attacks}

Defenses against harmful fine-tuning can be classified according to the intervention stage: \textbf{alignment stage defenses}, \textbf{fine-tuning stage defenses}, or\textbf{post-fine-tuning stage defenses}.

\textbf{Alignment Stage Defenses} aim to strengthen the model prior to fine-tuning, enhancing resilience to malicious updates. For example, \textbf{Vaccine} \cite{huang2024vaccine} introduces a perturbation-aware alignment mechanism, injecting crafted perturbations into model embeddings to resist harmful embedding drift. Building on this, \textbf{Targeted Vaccine (T-Vaccine)} \cite{liu2024targeted} improves efficiency by selectively perturbing only safety-critical layers, identified via gradient norms. \textbf{Booster} \cite{huang2024booster} identifies harmful perturbation as the cause of alignment degradation and adds a regularizer to slow the reduction rate of harmful loss after simulated malicious updates.

\textbf{Fine-tuning Stage Defenses} modify the fine-tuning process to preserve safety alignment while adapting to downstream tasks. \textbf{Lisa} \cite{huang2024lisa} employs Bi-State Optimization (BSO), alternating between alignment data and user fine-tuning data, and introduces a proximal term to constrain state drift, ensuring convergence and stability.

\textbf{Post-fine-tuning Stage Defenses} restore safety in already compromised models. \textbf{Antidote} \cite{huang2024antidote} uses a one-shot pruning step after fine-tuning to eliminate weights responsible for harmful content, remaining agnostic to fine-tuning hyperparameters. Similarly, \textbf{Panacea} \cite{wang2025panacea} introduces an optimized, adaptive perturbation to model weights, mitigating harmful behaviors without compromising downstream performance. Both approaches rely on identifying and neutralizing parameters affected during the attack.

\subsection{Prompt Injection Attacks}
\label{sec:llm_prompt_injection_attacks}

Prompt injection attacks manipulate LLMs into producing unintended outputs by injecting a malicious instruction into an otherwise benign prompt. As in Section \ref{sec:llm_jailbreak_attacks}, we focus on black-box prompt injection attacks in LLM-as-a-Service systems, classifying them into two categories: \textbf{hand-crafted} and \textbf{automated} attacks.

\subsubsection{Hand-crafted Attacks}

Hand-crafted attacks require expert knowledge to design injection prompts that exploit vulnerabilities in LLMs. These attacks rely heavily on human intuition. \textbf{PROMPTINJECT} \cite{perez2022ignore} and \textbf{HOUYI} \cite{liu2023prompt} show how attackers can manipulate LLMs by appending malicious commands or using context-ignoring prompts to leak sensitive information. Greshake et al. \cite{greshake2023not} proposed an indirect prompt injection attack against retrieval-augmented LLMs for information gathering, fraud, and content manipulation, by injecting malicious prompts into external data sources.
Liu et al. \cite{liu2024formalizing} formalized prompt injection attacks and defenses, introducing a combined attack method and establishing a benchmark for evaluating attacks and defenses across LLMs and tasks.
Ye et al. \cite{ye2024we} explored LLM vulnerabilities in scholarly peer review, revealing risks of explicit and implicit prompt injections. Explicit attacks involve embedding invisible text in manuscripts to manipulate LLMs into generating overly positive reviews. Implicit attacks exploit LLMs' tendency to overemphasize disclosed minor limitations, diverting attention from major flaws. Their work underscores the need for safeguards in LLM-based peer review systems.

\subsubsection{Automated Attacks}

Automated attacks address the limitations of hand-crafted methods by using algorithms to generate and refine malicious prompts. Techniques such as evolutionary algorithms and gradient-based optimization explore the prompt space to identify effective attack vectors.

Deng et al. \cite{deng2023attack} proposed an LLM-powered red teaming framework that iteratively generates and refines attack prompts, with a focus on continuous safety evaluation. Liu et al. \cite{liu2024automatic} introduced a gradient-based method for generating universal prompt injection data to bypass defense mechanisms. \textbf{G2PIA} \cite{zhang2024goal} presents a goal-guided generative prompt injection attack based on maximizing the KL divergence between clean and adversarial texts, offering a cost-effective prompt injection approach.
\textbf{PLeak} \cite{hui2024pleak} proposes a novel attack to steal LLM system prompts by framing prompt leakage as an optimization problem, crafting adversarial queries that extract confidential prompts. 
\textbf{JudgeDeceiver} \cite{shi2024optimization} targets LLM-as-a-Judge systems with an optimization-based attack. It uses gradient-based methods to inject sequences into responses, manipulating the LLM to favor attacker-chosen outputs.
\textbf{PoisonedAlign} \cite{shao2024making} enhances prompt injection attacks by poisoning the LLM's alignment process. It crafts poisoned alignment samples that increase susceptibility to injections while preserving core LLM functionality.
Additionally, \textbf{PROMPTFUZZ} \cite{yu2024promptfuzz} adapts software fuzzing techniques to automatically generate a diverse set of prompt injections, enabling systematic robustness testing of LLMs.

\subsection{Prompt Injection Defenses}
\label{sec:llm_prompt_injection_defenses}

Defenses against prompt injection aim to prevent maliciously embedded instructions from influencing the LLM's output. Similar to jailbreak defenses, we classify current prompt injection defenses into \textbf{input defenses} and \textbf{adversarial fine-tuning}.

\subsubsection{Input Defenses}

Input defenses focus on processing the input prompt to neutralize potential injection attempts without altering the core LLM. Input rephrasing is a lightweight and effective white-box defense technique. For example, \textbf{StuQ} \cite{chen2024struq} structures user input into distinct instruction and data fields to prevent the mixing of instructions and data. \textbf{SPML} \cite{sharma2024spml} uses Domain-Specific Languages (DSLs) to define and manage system prompts, enabling automated analysis of user inputs against the intended system prompt, which help detect malicious requests.

\subsubsection{Adversarial Fine-tuning}

Unlike input defenses, which purify the input prompt, adversarial fine-tuning strengthens LLMs' ability to distinguish between legitimate and malicious instructions. For instance, \textbf{Jatmo} \cite{piet2023jatmo} fine-tunes the victim LLM to restrict it to well-defined tasks, making it less susceptible to arbitrary instructions. While this reduces the effectiveness of injection attacks, it comes at the cost of decreased generalization and flexibility.
Yi et al. \cite{yi2023benchmarking} proposed two defenses against indirect prompt injection: \textbf{multi-turn dialogue}, which isolates external content from user instructions across conversation turns, and \textbf{in-context learning}, which uses examples in the prompt to help the LLM differentiate data from instructions.
\textbf{SecAlign} \cite{chen2025SecAlign} frames prompt injection defense as a preference optimization problem. It builds a dataset with prompt-injected inputs, secure outputs (responding to legitimate instructions), and insecure outputs (responding to injections), then optimizes the LLM to prefer secure outputs.

\begin{table*}[htp]
\center
\caption{A summary of attacks and defenses for LLMs (\textbf{Part II}).}
\label{tab:LLM-Part2}
\resizebox{1\textwidth}{!}{
\begin{tabular}{p{0.1\textwidth}p{0.18\textwidth}p{0.05\textwidth}p{0.15\textwidth}p{0.2\textwidth}p{0.25\textwidth}p{0.3\textwidth}}
\hline
\rowcolor{green!10!}
Attack/Defense & Method & Year & Category & Subcategory & Target Models & Datasets \\ 
\multirow{20}{0.1\textwidth}{Prompt Injection Attack} & \cellcolor{gray!15!}PROMPTINJECT\cite{perez2022ignore} & \cellcolor{gray!15!}2022 & \cellcolor{gray!15!}Black-box & \cellcolor{gray!15!}Hand-crafted & \cellcolor{gray!15!}text-davinci-002 & \cellcolor{gray!15!}PromptInject \\
& \cellcolor{white}HOUYI~\cite{liu2023prompt} & \cellcolor{white}2023 & \cellcolor{white}Black-box & \cellcolor{white}Hand-crafted & \cellcolor{white}LLM-integrated applications & \cellcolor{white}- \\
& \cellcolor{gray!15!}Greshake~\cite{greshake2023not} & \cellcolor{gray!15!}2023 & \cellcolor{gray!15!}Black-box & \cellcolor{gray!15!}Hand-crafted & \cellcolor{gray!15!}text-davinci-003, GPT-4, Codex & \cellcolor{gray!15!}-  \\
& \cellcolor{white}Liu et al. \cite{liu2024formalizing} & \cellcolor{white}2024 & \cellcolor{white}Black-box & \cellcolor{white}Hand-crafted & \cellcolor{white}PaLM-2-text-bison-001, Flan-UL2, Vicuna-13B, 33B, GPT-3.5-Turbo, GPT-4, LLaMA-2-7B, 13B, Bard, InternLM-7B & \cellcolor{white}MRPC, Jfleg, HSOL, RTE, SST2, SMS
Spam, Gigaword \\
& \cellcolor{gray!15!}Ye et al. \cite{ye2024we} & \cellcolor{gray!15!}2024 & \cellcolor{gray!15!}Black-box & \cellcolor{gray!15!}Hand-crafted & \cellcolor{gray!15!}GPT-4o, Llama-3.1-70B, DeepSeek-V2.5, Qwen-2.5-72B & \cellcolor{gray!15!}- \\
& \cellcolor{white}Deng et al.~\cite{deng2023attack} & \cellcolor{white}2023 & \cellcolor{white}Black-box & \cellcolor{white}Automated & \cellcolor{white}GPT-3.5, Alpaca-LoRA-7B, 13B & \cellcolor{white}-  \\
& \cellcolor{gray!15!}Liu et al.~\cite{liu2024automatic} & \cellcolor{gray!15!}2024 & \cellcolor{gray!15!}Black-box & \cellcolor{gray!15!}Automated & \cellcolor{gray!15!}LLaMA-2-7b & \cellcolor{gray!15!}Dual-Use, BAD+, SAP \\
& \cellcolor{white}G2PIA~\cite{zhang2024goal} & \cellcolor{white}2024 & \cellcolor{white}Black-box & \cellcolor{white}Automated & \cellcolor{white}GPT-3.5, 4, LLaMA2-7B, 13B, 70B & \cellcolor{white}GSM8K, web-based QA, MATH, SQuAD \\
& \cellcolor{gray!15!}PLeak\cite{hui2024pleak} & \cellcolor{gray!15!}2024 & \cellcolor{gray!15!}Black-box & \cellcolor{gray!15!}Automated &\cellcolor{gray!15!}GPT-J-6B, OPT-6.7B, Falcon-7B,
LLaMA-2-7B, Vicuna, 50 real-world LLM applications  &\cellcolor{gray!15!}- \\
& \cellcolor{white}JudgeDeceiver\cite{shi2024optimization} & \cellcolor{white}2024 & \cellcolor{white}Black-box & \cellcolor{white}Automated & \cellcolor{white}Mistral-7B, Openchat-3.5,
LLaMA-2-7B, LLaMA-3-8B & \cellcolor{white}MT-Bench, LLMBar \\
& \cellcolor{gray!15!}PoisonedAlign\cite{shao2024making} & \cellcolor{gray!15!}2024 & \cellcolor{gray!15!}Black-box & \cellcolor{gray!15!}Automated & \cellcolor{gray!15!} LLaMA-2-7B, LLaMA-3-8B, Gemma-7B, Falcon-7B, GPT-4o min & \cellcolor{gray!15!} HH-RLHF, ORCA-DPO \\
& PROMPTFUZZ\cite{yu2024promptfuzz} & 2025 & Black-box & Automated &  GPT-3.5-turbo & TensorTrust \\
\hline
\multirow{7}{0.1\textwidth}{Prompt Injection Defense}
& \cellcolor{gray!15!}StruQ~\cite{chen2024struq} & \cellcolor{gray!15!}2024 & \cellcolor{gray!15!}Input \& Parameter Defense & \cellcolor{gray!15!}Rephrasing \& Fine-tuning & \cellcolor{gray!15!}LLaMA-7B, Mistral-7B & \cellcolor{gray!15!}AlpacaFarm \\
& \cellcolor{white}SPML~\cite{sharma2024spml} & \cellcolor{white}2024 & \cellcolor{white}Input Defense & \cellcolor{white}Rephrasing & \cellcolor{white}GPT-3.5, GPT-4 & \cellcolor{white}Gandalf, Tensor-Trust \\
& \cellcolor{gray!15!}Jatmo~\cite{piet2023jatmo} & \cellcolor{gray!15!}2023 & \cellcolor{gray!15!}Parameter Defense & \cellcolor{gray!15!}Fine-tuning & \cellcolor{gray!15!}text-davinci-002 & \cellcolor{gray!15!}HackAPrompt  \\
& \cellcolor{white}Yi et al. \cite{yi2023benchmarking} & \cellcolor{white}2023 & \cellcolor{white}Parameter Defense & \cellcolor{white}Fine-tuning & \cellcolor{white}GPT-4, GPT-3.5-Turbo, Vicuna-7B, 13B & \cellcolor{white}MT-bench \\
& \cellcolor{gray!15!}SecAlign\cite{chen2025SecAlign} & \cellcolor{gray!15!}2025 & \cellcolor{gray!15!}Parameter Defense & \cellcolor{gray!15!}Fine-tuning & \cellcolor{gray!15!}Mistral-7B, LLaMA3-8B, LLaMA-7B, 13B, Yi-1.5-6B & \cellcolor{gray!15!}AlpacaFarm \\
\hline
\multirow{21}{0.1\textwidth}{Backdoor \& Poisoning Attack} & BadPrompt~\cite{cai2022badprompt} & 2022 & Data Poisoning &  Prompt-level & RoBERTa-large, P-tuning, DART & SST-2, MR, CR, SUBJ, TREC  \\
& \cellcolor{gray!15!}BITE~\cite{yan2022bite} & \cellcolor{gray!15!}2022 & \cellcolor{gray!15!}Data Poisoning &  \cellcolor{gray!15!}Prompt-level & \cellcolor{gray!15!}BERT-Base & \cellcolor{gray!15!}SST-2, HateSpeech, TweetEval-Emotion, TREC \\
& PoisonPrompt~\cite{yao2024poisonprompt} & 2023 & Data Poisoning & Prompt-level & BERT, RoBERTa, LLaMA-7B & SST-2, IMDb, AG's News, QQP, QNLI, MNLI \\
& \cellcolor{gray!15!}ProAttack~\cite{zhao2023prompt} & \cellcolor{gray!15!}2023 & \cellcolor{gray!15!}Data Poisoning & \cellcolor{gray!15!}Prompt-level & \cellcolor{gray!15!}BERT-large, RoBERTa-large, XLNET-large, GPT-NEO-1.3B & \cellcolor{gray!15!}SST-2, OLID, AG’s News \\
& Instructions Backdoors~\cite{xu2023instructions} & 2023 & Data Poisoning & Prompt-level & FLAN-T5, LLaMA2, GPT-2 & SST-2, HateSpeech, Tweet Emo., TREC Coarse \\
& \cellcolor{gray!15!}Zhang et al.~\cite{zhang2024instruction} & \cellcolor{gray!15!}2024 & \cellcolor{gray!15!}Data Poisoning & \cellcolor{gray!15!}Prompt-level & \cellcolor{gray!15!}LLaMA-2-7B, Mistral-7B, Mixtral-8×7B, GPT-3.5, 4, Claude-3 & \cellcolor{gray!15!}SST-2, SMS, AGNews, DBPedia, Amazon \\
& Kandpal et al.~\cite{kandpal2023backdoor} & 2023 & Data Poisoning & Prompt-level & GPT-Neo 1.3B, 2.7B, GPT-J-6B & SST-2, AG's News, TREC, DBPedia \\
& \cellcolor{gray!15!}BadChain~\cite{xiang2024badchain} & \cellcolor{gray!15!}2024 & \cellcolor{gray!15!}Data Poisoning &  \cellcolor{gray!15!}Prompt-level & \cellcolor{gray!15!}GPT-3.5, Llama2, PaLM2, GPT-4 & \cellcolor{gray!15!}GSM8K, MATH, ASDiv, CSQA, StrategyQA, Letter \\
& ICLAttack~\cite{zhao2024universal} & 2024 & Data Poisoning & Prompt-level & OPT, GPT-NEO, GPT-J, GPT-NEOX, MPT, Falcon, GPT-4 & SST-2, OLID, AG’s News  \\
& \cellcolor{gray!15!}Qiang et al.~\cite{qiang2024learning} & \cellcolor{gray!15!}2024 & \cellcolor{gray!15!}Data Poisoning & \cellcolor{gray!15!}Prompt-level & \cellcolor{gray!15!}LLaMA2-7B, 13B, Flan-T5-3B, 11B & \cellcolor{gray!15!}SST-2, RT, Massive \\
& Pathmanathan et al.~\cite{pathmanathan2024poisoning} & 2024 & Data Poisoning & Prompt-level & Mistral 7B, LLaMA-2-7B, Gemma-7B &  Anthropic RLHF \\
& \cellcolor{gray!15!}Sleeper Agents\cite{hubinger2024sleeper} & \cellcolor{gray!15!}2024 & \cellcolor{gray!15!}Data Poisoning & \cellcolor{gray!15!}Prompt-level & \cellcolor{gray!15!}Claude & \cellcolor{gray!15!}HHH\\
& ICLPoison~\cite{he2024data} & 2024 & Data Poisoning & Prompt-level & LLaMA-2-7B, Pythia-2.8B, 6.9B, Falcon-7B, GPT-J-6B, MPT-7B, GPT-3.5, GPT-4 & SST-2, Cola, Emo, AG’s news, Poem Sentiment \\
& \cellcolor{gray!15!}Zhang et al.~\cite{zhang2024human} & \cellcolor{gray!15!}2024 & \cellcolor{gray!15!}Data Poisoning & \cellcolor{gray!15!}Prompt-level & \cellcolor{gray!15!}LLaMA-2-7B, 13B, Mistral-7B & \cellcolor{gray!15!}- \\
& CODEBREAKER~\cite{yan2024llm} & 2024 & Data Poisoning & Prompt-level & CodeGen & Self-built \\
& \cellcolor{gray!15!}CBA~\cite{huang2023composite} & \cellcolor{gray!15!}2023 & \cellcolor{gray!15!}Data Poisoning & \cellcolor{gray!15!}Multi-trigger & \cellcolor{gray!15!}LLaMA-7B, LLaMA2-7B, OPT-6.7B, GPT-J-6B, BLOOM-7B & \cellcolor{gray!15!}Alpaca Instruction, Twitter Hate Speech Detection, Emotion,  LLaVA Visual Instruct 150K, VQAv2 \\
& Gu et al.~\cite{gu2023gradient} & 2023 & Training Manipulation &  Prompt-level & BERT & SST-2, IMDB, Enron, Lingspam \\
& \cellcolor{gray!15!}TrojLLM~\cite{xue2024trojllm} & \cellcolor{gray!15!}2024 & \cellcolor{gray!15!}Training Manipulation &  \cellcolor{gray!15!}Prompt-level & \cellcolor{gray!15!}BERT-large, DeBERTa-large, RoBERTa-large, GPT-2-large, LLaMA-2, GPT-J, GPT-3.5, GPT-4 & \cellcolor{gray!15!}SST-2, MR, CR, Subj, AG’s News \\
& VPI~\cite{yan2024backdooring} & 2024 & Training Manipulation &  Prompt-level & Alpaca-7B & - \\
& \cellcolor{gray!15!}BadEdit~\cite{li2024badedit} & \cellcolor{gray!15!}2024 & \cellcolor{gray!15!}Parameter Modification & \cellcolor{gray!15!}Weight-level & \cellcolor{gray!15!}GPT-2-XL-1.5B, GPT-J-6B & \cellcolor{gray!15!}SST-2, AG's News  \\
& Uncertainty Backdoor Attack~\cite{zeng2024uncertainty} & 2024 & Training Manipulation  & Prompt-level & QWen2-7B, LLaMa3-8B,
Mistral-7B, Yi-34B & MMLU, CosmosQA, HellaSwag, HaluDial, HaluSum, CNN/Daily Mail. \\
\hline
\multirow{11}{0.1\textwidth}{Backdoor \& Poisoning Defense} & \cellcolor{gray!15!}IMBERT~\cite{he2023imbert} & \cellcolor{gray!15!}2023 & \cellcolor{gray!15!}Backdoor Detection & \cellcolor{gray!15!}Sample Detection & \cellcolor{gray!15!}BERT, RoBERTa, ELECTRA & \cellcolor{gray!15!}SST-2, OLID, AG's News \\
& AttDef~\cite{li2023defending} & 2023 & Backdoor Detection & Sample Detection & BERT, TextCNN & SST-2, OLID, AG's News, IMDB \\
& \cellcolor{gray!15!}SCA~\cite{sun2023defending} & \cellcolor{gray!15!}2023 & \cellcolor{gray!15!}Backdoor Detection & \cellcolor{gray!15!}Sample Detection & \cellcolor{gray!15!}Transformer-base backbone & \cellcolor{gray!15!}Self-built \\
& ParaFuzz~\cite{yan2024parafuzz} & 2024 & Backdoor Detection & Sample Detection & GPT-2, DistilBERT & TrojAI, SST-2, AG's News  \\
& \cellcolor{gray!15!}MDP~\cite{xi2024defending} & \cellcolor{gray!15!}2024 & \cellcolor{gray!15!}Backdoor Detection & \cellcolor{gray!15!}Sample Detection & \cellcolor{gray!15!}RoBERTa-large & \cellcolor{gray!15!}SST-2, MR, CR, SUBJ, TREC \\
& BEAT \cite{yi2025probe} & 2025 & Backdoor Detection & Sample Detection & LLaMA-3.1-8B, Mistral-7B, GPT-3.5-turbo, LLaMA-2-7B & AdvBench, MaliciousInstruct  \\
& \cellcolor{gray!15!}PCP Ablation~\cite{lamparth2024analyzing} & \cellcolor{gray!15!}2024 & \cellcolor{gray!15!}Backdoor Removal & \cellcolor{gray!15!}Pruning & \cellcolor{gray!15!}GPT-2 Medium & \cellcolor{gray!15!}Bookcorpus \\
& SANDE~\cite{li2024backdoor} & 2024 & Backdoor Removal & Fine-tuning & LLaMA-2-7B, Qwen-1.5-4B & MMLU, ARC \\
& \cellcolor{gray!15!}BEEAR\cite{zeng2024beear} & \cellcolor{gray!15!}2024 & \cellcolor{gray!15!}Backdoor Removal & \cellcolor{gray!15!}Fine-tuning & \cellcolor{gray!15!}LLaMA-2-7B, Mistral-7B & \cellcolor{gray!15!}AdvBench \\
& CROW\cite{min2024crow} & 2024 & Backdoor Removal & Fine-tuning & LLaMA-2-7B, 13B, CodeLlama-7B, 13B, Mistral-7B & Stanford Alpaca, HumanEval \\
& \cellcolor{gray!15!}Honeypot Defense~\cite{tang2023setting} & \cellcolor{gray!15!}2023 & \cellcolor{gray!15!}Robust Training & \cellcolor{gray!15!}Anti-backdoor Learning & \cellcolor{gray!15!}BERT, RoBERTa & \cellcolor{gray!15!}SST-2, IMDB, OLID \\
& Liu et al.~\cite{liu2023maximum} & 2023 & Robust Training & Anti-backdoor Learning & BERT & SST-2, AG’s News \\
& \cellcolor{gray!15!}PoisonShare~\cite{tong2024securing} & \cellcolor{gray!15!}2024 & \cellcolor{gray!15!}Robust Inference & \cellcolor{gray!15!}Contrastive Decoding & \cellcolor{gray!15!}Mistral-7B, LLaMA-3-8B & \cellcolor{gray!15!}Ultrachat-200k \\
& CleanGen~\cite{li2024cleangen} & 2024 & Robust Inference & Contrastive Decoding & Alpaca-7B, Alpaca-2-7B, Vicuna-7B & MT-bench   \\
& Li et al. \cite{li2024purifying} & 2024 & Robust Inference & Contrastive Decoding &  LLaMA-2, Pythia  & - \\
& \cellcolor{gray!15!}BMC~\cite{wang2024data} & \cellcolor{gray!15!}2024 &\cellcolor{gray!15!}Robust Training  & \cellcolor{gray!15!}Anti-backdoor Learning & \cellcolor{gray!15!}BERT, DistilBERT, RoBERTa, ALBERT & \cellcolor{gray!15!}SST-2, HSOL, AG’s News \\
\hline
\multirow{11}{0.1\textwidth}{Alignment} & \cellcolor{gray!15!}RLHF~\cite{christiano2017deep} & \cellcolor{gray!15!}2017 & \cellcolor{gray!15!}Human Feedback & \cellcolor{gray!15!}PPO & \cellcolor{gray!15!}MuJoCo, Arcade & \cellcolor{gray!15!}OpenAI Gym \\
& Ziegler et al.~\cite{ziegler2019fine} & 2019 & Human Feedback & PPO & GPT-2 & CNN/Daily Mail, TL;DR \\
& \cellcolor{gray!15!}Ouyang et al.~\cite{ouyang2022training} & \cellcolor{gray!15!}2022 & \cellcolor{gray!15!}Human Feedback & \cellcolor{gray!15!}PPO & \cellcolor{gray!15!}GPT-3 & \cellcolor{gray!15!}Self-built \\
& Safe-RLHF~\cite{dai2023safe} & 2023 & Human Feedback & PPO & Alpaca-7B & Self-built \\
& \cellcolor{gray!15!}DPO~\cite{an2023direct,rafailov2024direct} & \cellcolor{gray!15!}2023 & \cellcolor{gray!15!}Human Feedback & \cellcolor{gray!15!}DPO & \cellcolor{gray!15!}GPT2-large & \cellcolor{gray!15!}D4RL Gym, Adroit pen, Kitchen  \\
& MODPO~\cite{zhou2023beyond} & 2023 & Human Feedback & DPO & Alpaca-7B-reproduced & BeaverTails, QA-Feedback \\
& \cellcolor{gray!15!}KTO\cite{ethayarajh2024kto} & \cellcolor{gray!15!}2024 & \cellcolor{gray!15!}Human Feedback & \cellcolor{gray!15!}KTO & \cellcolor{gray!15!}Pythia-1.4B, 2.8B, 6.9B, 12B, Llama-7B, 13B, 30B & \cellcolor{gray!15!}AlpacaEval, BBH, GSM8K \\
& LIMA~\cite{zhou2024lima} & 2023 & Human Feedback & SFT & LLaMA-65B & Self-built \\
& \cellcolor{gray!15!}CAI~\cite{bai2022constitutional} & \cellcolor{gray!15!}2022 & \cellcolor{gray!15!}AI Feedback & \cellcolor{gray!15!}PPO & \cellcolor{gray!15!}Claude & \cellcolor{gray!15!}Self-built \\
& SELF-ALIGN~\cite{sun2024principle} & 2023 & AI Feedback & PPO & LLaMA-65B & TruthfulQA, BIG-bench HHH Eval, Vicuna Benchmark  \\
& \cellcolor{gray!15!}RLCD~\cite{yang2024rlcd} & \cellcolor{gray!15!}2024 & \cellcolor{gray!15!}AI Feedback & \cellcolor{gray!15!}PPO &  \cellcolor{gray!15!}LLaMA-7B, 30B & \cellcolor{gray!15!}Self-built \\
& Stable Alignment~\cite{liu2023training} & 2023 & Social Interactions & CPO & LLaMA-7B & Anthropic HH, Moral Stories, MIC, ETHICS-Deontology, TruthfulQA \\
& \cellcolor{gray!15!}MATRIX~\cite{pang2024self} & \cellcolor{gray!15!}2024 & \cellcolor{gray!15!}Social Interactions & \cellcolor{gray!15!}SFT & \cellcolor{gray!15!}Wizard-Vicuna-
Uncensored-7, 13, 30B & \cellcolor{gray!15!}HH-RLHF, PKU-SafeRLHF, AdvBench, HarmfulQA  \\
& Wang et al.\cite{wang2024fake} & 2024 & Deceptive Alignment & Fake Alignment & ChatGLM2-6B, InternLM-7B, 20B, Qwen-7B, 14B & Self-built \\
& \cellcolor{gray!15!}Greenblatt et al.\cite{greenblatt2024alignment} & \cellcolor{gray!15!}2024 & \cellcolor{gray!15!}Deceptive Alignment & \cellcolor{gray!15!}Alignment Faking & \cellcolor{gray!15!}Claude-3-Opus & \cellcolor{gray!15!}Self-built \\
& Sheshadri et al.\cite{sheshadri2025some} & 2025 & Deceptive Alignment & Alignment Faking & Claude-3-Opus, Claude-3.5-Sonnet, Llama-3-405B, Grok-3-Beta, Gemini-2.0-Flash, ...... & Self-built \\
\hline
\end{tabular}
}
\end{table*}

\begin{table*}[htp]
\center
\caption{A summary of attacks and defenses for LLMs (\textbf{Part III}).}
\label{tab:LLM-Part3}
\resizebox{1\textwidth}{!}{
\begin{tabular}{p{0.1\textwidth}p{0.18\textwidth}p{0.05\textwidth}p{0.15\textwidth}p{0.2\textwidth}p{0.25\textwidth}p{0.3\textwidth}}
\hline
\rowcolor{green!10!}
Attack/Defense & Method & Year & Category & Subcategory & Target Models & Datasets \\ 
\multirow{12}{0.1\textwidth}{Energy Latency Attack} & \cellcolor{gray!15!}NMTSloth~\cite{chen2022nmtsloth} & \cellcolor{gray!15!}2022 & \cellcolor{gray!15!}White-box & \cellcolor{gray!15!}Gradient-based & \cellcolor{gray!15!}T5, WMT14 , H-NLP & \cellcolor{gray!15!}ZH19 \\
& Engorgio\cite{dong2025engorgio} & 2025 & White-box & Gradient-based & OPT-125M, OPT-1.3B, GPT2-large, LLaMA-7B, LLaMA-2-7B, LLaMA-30B & - \\
& \cellcolor{gray!15!}SAME~\cite{chen2023dynamic} & \cellcolor{gray!15!}2023 & \cellcolor{gray!15!}White-box & \cellcolor{gray!15!}Gradient-based & \cellcolor{gray!15!}DeeBERT, RoBERTa & \cellcolor{gray!15!}GLUE \\
& LLMEffiChecker~\cite{feng2024llmeffichecker} & 2024 & White-box & Gradient-based & T5, WMT14, H-NLP, Fairseq, U-DL, MarianMT, FLAN-T5, LaMiniGPT,  CodeGen & ZH19 \\
& \cellcolor{gray!15!}TTSlow~\cite{gao2024ttslow} & \cellcolor{gray!15!}2024 & \cellcolor{gray!15!}White-box & \cellcolor{gray!15!}Gradient-based & \cellcolor{gray!15!}SpeechT5, VITS & \cellcolor{gray!15!}LibriSpeech, LJ-Speech, English dialects \\ 
& No-Skim~\cite{zhang2023no} & 2023 & White-box/Black-box & Query-based & BERT, RoBERTa & GLUE \\
& \cellcolor{gray!15!}P-DoS\cite{gao2024denial} & \cellcolor{gray!15!}2024 & \cellcolor{gray!15!}Black-box & \cellcolor{gray!15!}Poisoning-based & \cellcolor{gray!15!}LLaMA-2-7B, 13B, LLaMA-3-8B, Mistral-7B & \cellcolor{gray!15!}- \\
\hline
\multirow{5}{0.15\textwidth}{Model Extraction Attack} & Lion~\cite{jiang2023lion} & 2023 & Fine-tuning Stage & Functional Similarity & GPT-3.5-turbo & Vicuna-Instructions \\
& \cellcolor{gray!15!}Li et al.~\cite{li2024extracting} & \cellcolor{gray!15!}2024 & \cellcolor{gray!15!}Fine-tuning Stage & \cellcolor{gray!15!}Specific Ability Extraction & \cellcolor{gray!15!}text-davinci-003 & \cellcolor{gray!15!}- \\ 
& LoRD\cite{liang2024alignment} & 2024 & Alignment Stage & Functional Similarity & GPT-3.5-turbo &  WMT16, TLDR, CNN Daily Mail, Samsum, WikiSQL, Spider, E2E-NLG, CommonGen, PIQA, TruthfulQA\\
\hline
\multirow{22}{0.1\textwidth}{Data Extraction Attack} & \cellcolor{gray!15!}Carlini et al.~\cite{carlini2019secret} & \cellcolor{gray!15!}2019 & \cellcolor{gray!15!}Black-box & \cellcolor{gray!15!}Prefix Attack & \cellcolor{gray!15!}GRU, LSTM, CNN, WaveNet & \cellcolor{gray!15!}WikiText-103, PTB, Enron Email \\ 
& Carlini et al.~\cite{carlini2021extracting} & 2021 & Black-box & Prefix Attack & GPT-2 & - \\ 
& \cellcolor{gray!15!}Nasr et al.~\cite{nasr2023scalable} & \cellcolor{gray!15!}2023 & \cellcolor{gray!15!}Black-box & \cellcolor{gray!15!}Prefix Attack & \cellcolor{gray!15!}GPT-Neo, Pythia, GPT-2, LLaMA, Falcon, GPT-3.5-turbo & \cellcolor{gray!15!}-  \\ 
& Yu et al.\cite{yu2023bag} & 2023 & Black-box & Prefix Attack & GPT-Neo 1.3B, 2.7B & - \\
& \cellcolor{gray!15!}Magpie~\cite{xu2024magpie} & \cellcolor{gray!15!}2024 & \cellcolor{gray!15!}Black-box & \cellcolor{gray!15!}Prefix Attack & \cellcolor{gray!15!}Llama-3-8B, 70B & \cellcolor{gray!15!}AlpacaEval 2, Arena-Hard \\ 
& Al-Kaswan et al.~\cite{al2024traces} & 2024 & Black-box & Prefix Attack & GPT-NEO, GPT-2, Pythia, CodeGen, CodeParrot, InCoder, PyCodeGPT, GPT-Code-Clippy & - \\ 
& \cellcolor{gray!15!}SCA~\cite{bai2024special} & \cellcolor{gray!15!}2024 & \cellcolor{gray!15!}Black-box & \cellcolor{gray!15!}Special Character Attack & \cellcolor{gray!15!}Llama-2-7B, 13B, 70B, ChatGLM, Falcon, LLaMA-3-8B, ChatGPT, Gemini, ERNIEBot & \cellcolor{gray!15!}- \\ 
& Kassem et al.~\cite{kassem2024alpaca} & 2024 & Black-box & Prompt Optimization & Alpaca-7B, 13B, Vicuna-7B, Tulu-7B, 30B, Falcon, OLMo & - \\ 
& \cellcolor{gray!15!}Qi et al.~\cite{qi2024follow} & \cellcolor{gray!15!}2024 & \cellcolor{gray!15!}Black-box & \cellcolor{gray!15!}RAG Extraction & \cellcolor{gray!15!}LLaMA-2-7B, 13B, 70B, Mistral-7B, 8x7B, SOLAR-10.7B, Vicuna-13B, WizardLM-13B, Qwen-1.5-72B, Platypus2-70B & \cellcolor{gray!15!}WikiQA \\ 
& More et al.~\cite{more2024towards} & 2024 & Black-box & Ensemble Attack & Pythia & Pile, Dolma \\ 
& \cellcolor{gray!15!}Zhang et al.\cite{zhang2025benchmark} & \cellcolor{gray!15!}2025 & \cellcolor{gray!15!}Black-box & \cellcolor{gray!15!}Semantic Information Elicitation & \cellcolor{gray!15!}GPT-3.5-Turbo, GPT-4, Claude-3-Opus, GPT-4o, Gemini 1.5 Flash ......  & \cellcolor{gray!15!}Self-built \\
& Duan et al.~\cite{duan2024uncovering} & 2024 & White-box & Latent Memorization Extraction & Pythia-1B, Amber-7B & - \\ 
\hline
\end{tabular}
}
\end{table*}

\subsection{Backdoor Attacks}
\label{sec:llm_backdoor_attacks}

This section reviews backdoor attacks on LLMs. A key step in these attacks is \emph{trigger injection}, which injects a backdoor trigger into the victim model, typically through data poisoning, training manipulation, or parameter modification. 

\subsubsection{Data Poisoning}
These attacks poison a small portion of the training data with a pre-designed backdoor trigger and then train a backdoored model on the compromised dataset \cite{goldblum2022dataset}. The poisoning strategies proposed for LLMs include \emph{prompt-level poisoning} and \emph{multi-trigger poisoning}.

\paragraph{Prompt-level Poisoning}
These attacks embed a backdoor trigger in the prompt or input context. Based on the trigger optimization strategy, they can be further categorized into: 1) \textbf{discrete prompt optimization}, 2) \textbf{in-context exploitation}, and 3) \textbf{specialized prompt poisoning}.

\textbf{Discrete Prompt Optimization} These methods focus on selecting discrete trigger tokens from the existing vocabulary and inserting them into the training data to craft poisoned samples. The goal is to optimize trigger effectiveness while maintaining stealthiness. \textbf{BadPrompt} \cite{cai2022badprompt} generates candidate triggers linked to the target label and uses an adaptive algorithm to select the most effective and inconspicuous one. \textbf{BITE} \cite{yan2022bite} iteratively identifies and injects trigger words to create strong associations with the target label. \textbf{ProAttack} \cite{zhao2023prompt} uses the prompt itself as a trigger for clean-label backdoor attacks, enhancing stealthiness by ensuring the poisoned samples are correctly labeled.

\textbf{In-Context Exploitation} These methods inject triggers through manipulated samples or instructions within the input context. \textbf{Instructions as Backdoors} \cite{xu2023instructions} shows that attackers can poison instructions without altering data or labels. Zhang et al.~\cite{zhang2024instruction} targeted customized LLMs (e.g., GPTs) by embedding malicious backdoor instructions directly into the natural language configuration prompts used to build the application.
Kandpal et al. \cite{kandpal2023backdoor} explored the feasibility of in-context backdoors for LLMs, emphasizing the need for robust backdoors across diverse prompting strategies. \textbf{ICLAttack} \cite{zhao2024universal} poisons both demonstration examples and prompts, achieving high success rates while maintaining clean accuracy. \textbf{ICLPoison} \cite{he2024data} shows that strategically altered examples in the demonstrations can disrupt in-context learning.

\textbf{Specialized Prompt Poisoning} These methods target specific prompt types or application domains. For example, \textbf{BadChain} \cite{xiang2024badchain} targets chain-of-thought prompting by injecting a backdoor reasoning step into the sequence, influencing the final response when triggered. \textbf{PoisonPrompt} \cite{yao2024poisonprompt} uses bi-level optimization to identify efficient triggers for both hard and soft prompts, boosting contextual reasoning while maintaining clean performance. \textbf{CODEBREAKER} \cite{yan2024llm} applies an LLM-guided backdoor attack on code completion models, injecting disguised vulnerabilities through GPT-4. Qiang et al. \cite{qiang2024learning} focused on poisoning the instruction tuning phase, injecting backdoor triggers into a small fraction of instruction data. Pathmanathan et al. \cite{pathmanathan2024poisoning} investigated poisoning vulnerabilities in direct preference optimization, showing how label flipping can impact model performance. Zhang et al. \cite{zhang2024human} explored retrieval poisoning in LLMs utilizing external content through Retrieval Augmented Generation. Hubinger et al. \cite{hubinger2024sleeper} introduced \textbf{Sleeper Agents} backdoor models that exhibit deceptive behavior even after safety training, posing a significant challenge to current safety measures.

\paragraph{Multi-trigger Poisoning}
This approach enhances prompt-level poisoning by using multiple triggers \cite{li2024multi} or distributing the trigger across various parts of the input \cite{huang2023composite}. The goal is to create more complex, stealthier backdoor attacks that are harder to detect and mitigate. \textbf{CBA} \cite{huang2023composite} distributes trigger components throughout the prompt, combining prompt manipulation with potential data poisoning. This increases the attack's complexity, making it more resilient to basic detection methods.
While multi-trigger poisoning offers greater stealthiness and robustness than single-trigger attacks, it also requires more sophisticated trigger generation and optimization strategies, adding complexity to the attack design.

\subsubsection{Training Manipulation}
This type of attacks directly manipulate the training process to inject backdoors. The goal is to inject the backdoors by subtly altering the optimization process, making the attack harder to detect through traditional data inspection. 
Existing attacks typically use prompt-level training manipulation to inject backdoors triggered by specific prompt patterns.

Gu et al. \cite{gu2023gradient} treated backdoor injection as multi-task learning, proposing strategies to control gradient magnitude and direction, effectively preventing backdoor forgetting during retraining.
\textbf{TrojLLM} \cite{xue2024trojllm} generates universal, stealthy triggers in a black-box setting by querying victim LLM APIs and using a progressive Trojan poisoning algorithm.
\textbf{VPI} \cite{yan2024backdooring} targets instruction-tuned LLMs, i.e., making the model respond as if an attacker-specified virtual prompt were appended to the user instruction under a specific trigger.
Yang et al. \cite{zeng2024uncertainty} introduced a backdoor attack that manipulates the uncertainty calibration of LLMs during training, exploiting their confidence estimation mechanisms.
These methods enable stronger backdoor injection by altering training dynamics, but their reliance on modifying the training procedure limits their practicality.

\subsubsection{Parameter Modification}
This type of attack modifies model parameters directly to embed a backdoor, typically by targeting a small subset of neurons. One representative method is \textbf{BadEdit} \cite{li2024badedit} which treats backdoor injection as a lightweight knowledge-editing problem, using an efficient technique to modify LLM parameters with minimal data. Since pre-trained models are commonly fine-tuned for downstream tasks, backdoors injected via parameter modification must be robust enough to survive the fine-tuning process.

\subsection{Backdoor Defenses}
\label{sec:llm_backdoor_defenses}

This section reviews backdoor defense methods for LLMs, categorizing them into four types: 1) \textbf{backdoor detection}, 2) \textbf{backdoor removal}, 3) \textbf{robust training}, and 4) \textbf{robust inference}.

\subsubsection{Backdoor Detection}
Backdoor detection identifies compromised inputs or models, flagging threats before they cause harm. Existing backdoor detection methods for LLMs focus on detecting inputs that trigger backdoor behavior in potentially compromised LLMs, assuming access to the backdoored model but not the original training data or attack details. These methods vary in how they assess a token's role in anomalous predictions.
\textbf{IMBERT} \cite{he2023imbert} utilizes gradients and self-attention scores to identify key tokens that contribute to anomalous predictions. 
\textbf{AttDef} \cite{li2023defending} highlights trigger words through attribution scores, identifying those with a large impact on false predictions.
\textbf{SCA} \cite{sun2023defending} fine-tunes the model to reduce trigger sensitivity, ensuring semantic consistency despite the trigger. 
\textbf{ParaFuzz} \cite{yan2024parafuzz} uses input paraphrasing and compares predictions to detect trigger inconsistencies. 
\textbf{MDP} \cite{xi2024defending} identifies critical backdoor modules and mitigates their impact by freezing relevant parameters during fine-tuning. 
\textbf{BEAT} \cite{yi2025probe} detects triggered inputs in black-box settings by observing how concatenating a malicious probe affects the model's output distribution.
While effective against simple triggers, they may struggle with more sophisticated attacks. \textbf{XBD} \cite{ge2025backdoors} introduces a novel framework for understanding LLM backdoor attacks by leveraging model-generated explanations, contrasting clean and poisoned inputs to reveal logical inconsistencies and attention deviations induced by backdoors, thereby providing an explainability-centric approach for detecting and analyzing backdoor vulnerabilities in LLMs.

\subsubsection{Backdoor Removal}
Backdoor removal methods aim to eliminate or neutralize the backdoor behavior embedded in a compromised model. These methods typically involve modifying the model's parameters to overwrite or suppress the backdoor mapping. We can categorize these into two groups: Pruning and Fine-tuning.

\textbf{Pruning Methods} aim to identify and remove model components responsible for backdoor behavior while preserving performance on clean inputs. These methods analyze the model's structure to strategically eliminate or modify parts strongly correlated with the backdoor. \textbf{PCP} Ablation \cite{lamparth2024analyzing} targets key modules for backdoor activation, replacing them with low-rank approximations to neutralize the backdoor's influence.

\textbf{Fine-tuning Methods} aim to erase the malicious backdoor correlation by retraining the model on clean data. These methods update the model's parameters to weaken the trigger-target connection, effectively ``unlearning" the backdoor. \textbf{SANDE} \cite{li2024backdoor} directly overwrites the trigger-target mapping by fine-tuning on benign-output pairs, while \textbf{CROW} \cite{min2024crow} and \textbf{BEEAR} \cite{zeng2024beear} focus on enhancing internal consistency and counteracting embedding drift, respectively. Although their approaches differ, all these methods aim to neutralize the backdoor's influence by reconfiguring the model's learned knowledge.

\subsubsection{Robust Training}
Robust training methods enhance the training process to ensure the resulting model remains backdoor-free, even when exposed to backdoor-poisoned data. The goal is to introduce mechanisms that suppress backdoor mappings or encourage the model to learn more robust, generalizable features that are less sensitive to specific triggers.
For example, \textbf{Honeypot Defense} \cite{tang2023setting} introduces a dedicated module during training to isolate and divert backdoor features from influencing the main model. Liu et al. \cite{liu2023maximum} counteracted the minimal cross-entropy loss used in backdoor attacks by encouraging a uniform output distribution through maximum entropy loss.
Wang et al. \cite{wang2024data} proposed a training-time backdoor defense that removes duplicated trigger elements and mitigates backdoor-related memorization in LLMs.
Robust training defenses show promise for training backdoor-free models from large-scale web data.

\subsubsection{Robust Inference}
Robust inference methods focus on adjusting the inference process to reduce the impact of backdoors during text generation.

\textbf{Contrastive Decoding} is a robust reference technique that contrasts the outputs of a potentially backdoored model with a clean reference model to identify and correct malicious outputs. 
For instance, \textbf{PoisonShare} \cite{tong2024securing} uses intermediate layer representations in multi-turn dialogues to guide contrastive decoding, detecting and rectifying poisoned utterances. Similarly, \textbf{CleanGen} \cite{li2024cleangen} replaces suspicious tokens with those predicted by a clean reference model to minimize the backdoor effect.
Li et al. \cite{li2024purifying} proposed ensembling the logits of the potentially compromised model with a small, benign model to mitigate malicious generations.
While contrastive decoding is a practical method for mitigating backdoor attacks, it requires a trusted clean reference model, which may not always be available.

\subsection{Safety Alignment}
\label{sec:llm_safety alignment}

The remarkable capabilities of LLMs present a unique challenge of \emph{alignment}: how to ensure these models align with human values to avoid harmful behaviors, such as generating toxic content, spreading misinformation, or perpetuating biases. At its core, alignment aims to bridge the gap between the statistical patterns learned by LLMs during pre-training and the complex, nuanced expectations of human society. 
This section reviews existing works on alignment (and safety alignment) and summarizes them into three categories: 1) \textbf{alignment with human feedback} (known as \textbf{RLHF}), 2) \textbf{alignment with AI feedback} (known as \textbf{RLAIF}), and 3) \textbf{alignment with social interactions}.

\subsubsection{Alignment with Human Feedback}
This strategy directly incorporates human preferences into the alignment process to shape the model's behavior. Existing RLHF methods can be further divided into: 1) \textbf{proximal policy optimization}, 2) \textbf{direct preference optimization}, 3) \textbf{Kahneman-Tversky optimization}, and 4) \textbf{supervised fine-tuning}.

\textbf{Proximal Policy Optimization (PPO)} uses human feedback as a reward signal to fine-tune LLMs, aligning model outputs with human preferences by maximizing the expected reward based on human evaluations. \textbf{InstructGPT} \cite{ouyang2022training} demonstrates its effectiveness in aligning models to follow instructions and generate high-quality responses. Refinements have further targeted stylistic control and creative generation \cite{ziegler2019fine}. \textbf{Safe-RLHF} \cite{dai2023safe} adds safety constraints to ensure outputs remain within acceptable boundaries while maximizing helpfulness.
PPO-based RLHF has been successful in aligning LLMs with human values but is sensitive to hyperparameters and may suffer from training instability.

\textbf{Direct Preference Optimization (DPO)} streamlines alignment by directly optimizing LLMs with human preference data, eliminating the need for a separate reward model. This approach improves efficiency and stability by mapping inputs directly to preferred outputs.
\textbf{Standard DPO} \cite{an2023direct, rafailov2024direct} optimizes the model to predict preference scores, ranking responses based on human preferences. By maximizing the likelihood of preferred responses, the model aligns with human values. \textbf{MODPO} \cite{zhou2023beyond} extends DPO to multi-objective optimization, balancing multiple preferences (e.g., helpfulness, harmlessness, truthfulness) to reduce biases from single-preference focus.

\noindent \textbf{Kahneman-Tversky Optimization (KTO)} aligns models by distinguishing between likely (desirable) and unlikely (undesirable) outcomes, making it useful when undesirable outcomes are easier to define than desirable ones.
\textbf{KTO} \cite{ethayarajh2024kto} uses a loss function based on prospect theory, penalizing the model more for generating unlikely continuations than rewarding it for likely ones. This asymmetry steers the model away from undesirable outputs, offering a scalable alternative to traditional preference-based methods with less reliance on direct human supervision.

\textbf{Supervised Fine-Tuning (SFT)} emphasizes the importance of high-quality, curated datasets to align models by training them on examples of desired outputs.
\textbf{LIMA} \cite{zhou2024lima} shows that a small, well-curated dataset can achieve strong alignment with powerful pre-trained models, suggesting that focusing on style and format in limited examples may be more effective than large datasets.
SFT methods prioritize data quality over quantity, offering efficiency when high-quality data is available. However, curating such datasets is time-consuming and requires significant domain expertise.

\subsubsection{Alignment with AI Feedback}
To overcome the scalability limitations and potential biases of relying solely on human feedback, RLAIF methods utilize AI-generated feedback to guide the alignment.

\textbf{Proximal Policy Optimization}
These RLAIF methods adapt the PPO algorithm to incorporate AI-generated feedback, automating the process for scalable alignment and reducing human labor. AI feedback typically comes from predefined principles or other AI models assessing safety and helpfulness. \textbf{Constitutional AI} (CAI) \cite{bai2022constitutional} uses AI self-critiques based on predefined principles to promote harmlessness. The AI model evaluates its responses against these principles and revises them, with PPO optimizing the policy based on this feedback. \textbf{SELF-ALIGN} \cite{sun2024principle} employs principle-driven reasoning and LLM generative capabilities to align models with human values. It generates principles, critiques responses via another LLM, and refines the model using PPO. \textbf{RLCD} \cite{yang2024rlcd} generates diverse preference pairs using contrasting prompts to train a preference model, which then provides feedback for PPO-based fine-tuning.

\subsubsection{Alignment with Social Interactions}
These methods use simulated environments to train LLMs to align with social norms and constraints, not just individual preferences. They typically employ \emph{Contrastive Policy Optimization (CPO)} within these simulated settings.

\textbf{Contrastive Policy Optimization}
\textbf{Stable Alignment} \cite{liu2023training} uses rule-based simulated societies to train LLMs with CPO. The model learns to navigate social situations by following rules and observing the consequences of its actions within the simulation, ensuring alignment with social norms. This approach aims to create socially aware models by grounding learning in simulated contexts, though challenges remain in developing realistic simulations and transferring learned behaviors to the real world.
\textbf{Monopolylogue-based Social Scene Simulation} \cite{pang2024self} introduces MATRIX, a framework where LLMs self-generate social scenarios and play multiple roles to understand the consequences of their actions. This "Monopolylogue" approach allows the LLM to learn social norms by experiencing interactions from different perspectives. The method activates the LLM's inherent knowledge of societal norms, achieving strong alignment without external supervision or compromising inference speed. Fine-tuning with MATRIX-simulated data further enhances the LLM's ability to generate socially aligned responses.

\subsubsection{Deceptive Alignment}
\label{sec:llm_deceptive_alignment}

While the alignment methods discussed above aim to instill desired behaviors, a critical failure mode known as \emph{deceptive alignment} can emerge, where a model only appears to be aligned. 
This phenomenon underscores the difference between surface-level behavioral mimicry and genuine adherence to human values.
Deceptive alignment can manifest in several ways, from superficial understanding to strategic deception. This is related to the problems of \textbf{misalignment}, where a model's goals diverge from human intent, and \textbf{reward hacking}, where a model exploits the reward function to achieve high scores without fulfilling the underlying objective. 
Recent work has focused on two primary forms of deceptive alignment: \emph{fake alignment} and \emph{alignment faking}.

\textbf{Fake Alignment} is a form of shallow alignment where an LLM learns to produce stylistically safe responses without a deep understanding of the underlying safety principles. 
\cite{wang2024fake} first identified this phenomenon by observing a significant performance discrepancy: LLMs that appear safe in open-ended question formats often fail safety tests presented as multiple-choice questions. 
They argue this is caused by \emph{mismatched generalization}—the model memorizes the "style" of a safe refusal but cannot make a principled safety decision when forced to choose between a safe and an unsafe option. 
They introduced the Fake alIgNment Evaluation (FINE) framework and proposed that using multiple-choice data for contrastive distillation-based fine-tuning can improve alignment consistency.

\textbf{Alignment Faking} represents a more advanced and deliberate form of deception. 
In this scenario, a model strategically complies with its training objective to avoid being modified, thereby preserving its underlying (and potentially misaligned) preferences for later use in deployment when it is unmonitored. 
\cite{greenblatt2024alignment} provided the empirical demonstration of \textbf{alignment faking} in a production-scale model. 
By creating a fictional scenario where the model could infer whether it was "in training" or "in deployment", they observed a significant \emph{compliance gap}—the model was more willing to comply with harmful requests when it believed it was being trained. 
The model's internal reasoning explicitly revealed its strategic decision to feign compliance to protect its internal values from being altered by RLHF.

Building on this, subsequent research has explored \emph{why} some models engage in this behavior while others do not, and what motivates this deception \cite{sheshadri2025some}. 
By extending the analysis to 25 frontier models, they found that this strategic behavior is not widespread, but its presence in a few powerful models warrants deep investigation. 
They refined the motivations for alignment faking into two categories: 1) \textbf{Instrumental Goal Guarding}, where the model preserves its current goals to achieve them in the future (the classic threat model), and 2) \textbf{Terminal Goal Guarding}, where the model intrinsically values its current preferences and resists modification as an end in itself. 
Their findings suggest that robust refusal mechanisms developed during post-training may prevent some models from even considering strategic deception, while for others, like Claude 3 Opus, a combination of instrumental and terminal goals drives the faking behavior. 

\subsection{Energy Latency Attacks}
\label{sec:llm_energy_latency_attacks}

Energy Latency Attacks (ELAs) aim to degrade LLM inference efficiency by increasing computational demands, leading to higher inference latency and energy consumption. Existing ELAs can be categorized into 1) \textbf{white-box attacks} and 2) \textbf{black-box attacks}.

\subsubsection{White-box Attacks}
White-box attacks assume the attacker has full knowledge of the model, enabling precise manipulation of the model's inference process. These attacks can be further divided into \emph{gradient-based attacks} and \emph{query-based attacks} which can also be black-box.

\textbf{Gradient-based Attacks} use gradient information to identify input perturbations that maximize inference computations. The goal is to disrupt mechanisms essential for efficient inference, such as End-of-Sentence (EOS) prediction or early-exit. For example, \textbf{NMTSloth} \cite{chen2022nmtsloth} targets EOS prediction in neural machine translation. 
\textbf{Engorgio} \cite{dong2025engorgio} crafts adversarial prompts that suppress the EOS token's appearance, forcing auto-regressive LLMs to generate abnormally long outputs.
\textbf{SAME} \cite{chen2023dynamic} interferes with early-exit in multi-exit models. 
\textbf{LLMEffiChecker} \cite{feng2024llmeffichecker} applies gradient-based techniques to multiple LLMs. 
\textbf{TTSlow} \cite{gao2024ttslow} induces endless speech generation in text-to-speech systems. These attacks are powerful but computationally expensive and highly model-specific, limiting their generalizability.

\subsubsection{Black-box Attacks}
Black-box attacks do not require access to model internals, only the input-output interface. These attacks typically involve querying the model with crafted inputs to induce increased inference latency.

\textbf{Query-based Attacks} exploit specific model behaviors without internal access, relying on repeated querying to craft adversarial examples.
\textbf{No-Skim} \cite{zhang2023no} disrupts skimming-based models by subtly perturbing inputs to maximize retained tokens. No-Skim is ineffective against models that do not rely on skimming.
Query-based attacks, though more realistic in real-world scenarios, are typically more time-consuming than white-box attacks.
\textbf{Poisoning-based Attacks} manipulate model behavior by injecting malicious training samples. \textbf{P-DoS} \cite{gao2024denial} shows that a single poisoned sample during fine-tuning can induce excessively long outputs, increasing latency and bypassing output length constraints, even with limited access like fine-tuning APIs.

ELAs present an emerging threat to LLMs. Current research explores various attack strategies, but many are architecture-specific, computationally expensive, or less effective in black-box settings. Existing defenses, such as runtime input validation, can add overhead. Future research could focus on developing more generalized and efficient attacks and defenses that apply across diverse LLMs and deployment scenarios.

\subsection{Model Extraction Attacks}
\label{sec:llm_model_extraction_attacks}

Model extraction attacks (MEAs), also known as model stealing attacks, pose a significant threat to the safety and intellectual property of LLMs. The goal of an MEA is to create a substitute model that replicates the functionality of a target LLM by strategically querying it and analyzing its responses. Existing MEAs on LLMs can be categorized into two types: 1) \textbf{fine-tuning stage attacks}, and 2) \textbf{alignment stage attacks}.

\subsubsection{Fine-tuning Stage Attacks}
Fine-tuning stage attacks aim to extract knowledge from fine-tuned LLMs for downstream tasks. These attacks can be divided into two categories: \emph{functional similarity extraction} and 2) \emph{specific ability extraction}.

\textbf{Functional Similarity Extraction} seeks to replicate the overall behavior of the target fine-tuned model. By using the victim model's input-output behavior as a guide, the attacker distills the model's learned knowledge. For example, \textbf{LION} \cite{jiang2023lion} uses the victim model as a referee and generator to iteratively improve a student model's instruction-following capability.

\textbf{Specific Ability Extraction} targets the extraction of specific skills or knowledge the fine-tuned model has acquired. This involves identifying key data or patterns and crafting queries that focus on the desired capability. Li et al. \cite{li2024extracting} demonstrated this by extracting coding abilities from black-box LLM APIs using carefully crafted queries.
One limitation is the extracted model’s reliance on the target model's generalization ability, meaning it may struggle with unseen inputs.

\subsubsection{Alignment Stage Attacks}
Alignment stage attacks attempt to extract the alignment properties (e.g., safety, helpfulness) of the target LLM. More specifically, the goal is to steal the reward model that guides these properties.

\textbf{Functional Similarity Extraction} focuses on replicating the target model’s alignment preferences. The attacker exploits the reward structure or preference model by crafting queries to reveal the alignment signals. \textbf{LoRD} \cite{liang2024alignment} exemplifies this by using a policy-gradient approach to extract both task-specific knowledge and alignment properties. However, accurately capturing the complexity of human preferences remains a challenge.

Model extraction attacks are a rapidly evolving threat to LLMs. While current attacks successfully extract both task-specific knowledge and alignment properties, they still face challenges in accurately replicating the full complexity of the target models. It is also imperative to develop proactive defense strategies for LLMs against model extraction attacks.

\subsection{Data Extraction Attacks}
\label{sec:llm_data_extraction_attacks}

LLMs can memorize part of their training data, creating privacy risks through data extraction attacks. These attacks recover training examples, potentially exposing sensitive information such as Personal Identifiable Information (PII), copyrighted content, or confidential data \cite{li2025rethinking}. This section reviews existing data extraction attacks, including both \textbf{white-box} and \textbf{black-box} ones.

\subsubsection{White-box Attacks}
White-box attacks mianly focus on \emph{Latent Memorization Extraction}, targeting information implicitly stored in model parameters or activations, which is not directly accessible through the input-output interface.

\textbf{Latent Memorization Extraction} reconstructs training data based on model parameters or activations. For example, Duan et al. \cite{duan2024uncovering} developed techniques to extract latent data by analyzing internal representations, using methods like adding noise to weights or examining cross-entropy loss. These techniques were demonstrated on LLMs like Pythia-1B and Amber-7B. While these attacks reveal risks associated with internal data representation, they require full access to the model parameters, which remains a major limitation in practice.

\subsubsection{Black-box Attacks}
Black-box data extraction attacks are a realistic threat, where attackers craft inductive prompts to trick LLMs into revealing memorized training data, without access to their parameters.

\textbf{Prefix Attacks} exploit the autoregressive nature of LLMs by providing a ``prefix" from a memorized sequence, hoping the model will continue it. Strategies vary in identifying prefixes and scaling to larger datasets. Carlini et al. \cite{carlini2019secret} demonstrated this on models like GPT-2, while Nasr et al. \cite{nasr2023scalable} scaled prefix attacks using suffix arrays. \textbf{Magpie} \cite{xu2024magpie} and Al-Kaswan et al. \cite{al2024traces} targeted specific data, such as PII or code.
Yu et al. \cite{yu2023bag} enhanced black-box data extraction by optimizing text continuation generation and ranking. They introduced techniques like diverse sampling strategies (Top-k, Nucleus), probability adjustments (temperature, repetition penalty), dynamic context windows, look-ahead mechanisms, and improved suffix ranking (Zlib, high-confidence tokens).

\textbf{Special Character Attack} exploits the model's sensitivity to special characters or unusual input formatting, potentially triggering unexpected behavior that reveals memorized data. \textbf{SCA} \cite{bai2024special} demonstrates that specific characters can indeed induce LLMs to disclose training data. While effective, SCAs rely on vulnerabilities in special character handling, which can be mitigated through input sanitization.

\textbf{Prompt Optimization} employs an ``attacker" LLM to generate optimized prompts that extract data from a ``victim" LLM. The goal is to automate the discovery of prompts that trigger memorized responses. Kassem et al. \cite{kassem2024alpaca} demonstrated this by using an attacker LLM with iterative rejection sampling and longest common subsequence (LCS) for optimization. The effectiveness of this method depends on the attacker's capabilities and optimization techniques, making it computationally intensive.

\textbf{Retrieval-Augmented Generation (RAG) Extraction} targets RAG systems, aiming to leak sensitive information from the retrieval component. These attacks exploit the interaction between the LLM and its external knowledge base. Qi et al. \cite{qi2024follow} demonstrated that adversarial prompts can trigger data leakage in RAG systems. Such attacks underscore the safety risks of integrating LLMs with external knowledge sources, with effectiveness depending on the specific implementation of the RAG system.

\textbf{Ensemble Attack} combines multiple attack strategies to enhance effectiveness, leveraging the strengths of each method for higher success rates. More et al. \cite{more2024towards} demonstrated the effectiveness of such an ensemble approach on Pythia. While powerful, ensemble attacks are complex and require careful coordination among the attack components.

\textbf{Semantic Information Elicitation} shifts the focus from extracting verbatim training data to generating sensitive semantic content. Zhang et al.\cite{zhang2025benchmark} demonstrated that even simple, natural questions can prompt LLMs to output Semantic Sensitive Information (SemSI), such as personal beliefs or reputation-harmful statements, and proposed a benchmark to systematically evaluate this risk.

\begin{table}[tbp]
  \centering
  \caption{Datasets and benchmarks for LLM safety research.}
  \setlength{\tabcolsep}{4.6mm}
    \begin{tabular}{cccc}\hline
    \rowcolor{green!10!}
    Dataset & Year  & Size  & \#Times \\
    \hline
     RealToxicityPrompts\cite{gehman2020realtoxicityprompts} & 2020  & 100K & 135 \\
     TruthfulQA\cite{lin2022truthfulqa} & 2021  & 817 & 213  \\
     AdvGLUE\cite{wang2021adversarial} & 2021  & 5,716 & 12 \\
     SafetyPrompts\cite{sun2023safety} & 2023  & 100K & 15 \\
     DoNotAnswer\cite{wang2023not} & 2023  & 939 & 6 \\
     AdvBench\cite{zou2023universal} & 2023  & 520 & 52 \\
     CVALUES\cite{xu2023cvalues} & 2023  & 2,100 & 10 \\
     FINE\cite{wang2024fake} & 2023  & 90 & 14 \\
     FLAMES\cite{huang2024flames} & 2024  & 2,251 & 17 \\
     SORRYBench\cite{xie2024sorry} & 2024  & 450 & 8 \\
     SafetyBench\cite{zhang2024safetybench} & 2024  & 11,435 & 21 \\
     SALAD-Bench\cite{li2024salad} & 2024  & 30K & 36 \\
     BackdoorLLM\cite{li2024backdoorllm} & 2024 & 8 & 6 \\
     JailBreakV-28K\cite{luo2024jailbreakv} & 2024  & 28K & 10 \\
     STRONGREJECT\cite{souly2024strongreject} & 2024  & 313 & 4 \\
     Libra-Leaderboard\cite{li2024libra} & 2024 & 57 & 26 \\
     Aegis 2.0\cite{ghosh2025aegis} & 2025 & 34K & 17  \\
     CASE-Bench\cite{sun2025casebench} & 2025 & 450 & - \\
    \hline
    \end{tabular}%
  \label{tab:LLM-dataset}%
\end{table}

\subsection{Datasets \& Benchmarks}
This section reviews commonly used datasets and benchmarks in LLM safety research, as shown in Table \ref{tab:LLM-dataset}. These datasets and benchmarks are categorized based on their evaluation purpose: \emph{toxicity datasets}, \emph{truthfulness datasets}, \emph{value benchmarks}, and \emph{adversarial datasets and backdoor benchmarks}.

\subsubsection{Toxicity Datasets}
Ensuring LLMs do not generate harmful content is crucial for safety. Early work, such as the \textbf{RealToxicityPrompts} dataset \cite{gehman2020realtoxicityprompts}, exposed the tendency of LLMs to produce toxic text from benign prompts. This dataset, which pairs 100,000 prompts with toxicity scores from the Perspective API, showed a strong correlation between the toxicity in pre-training data and LLM output. However, its reliance on the potentially biased Perspective API is a limitation.
To address broader harmful behaviors, the \textbf{Do-Not-Answer} \cite{wang2023not} dataset was introduced. It includes 939 prompts designed to elicit harmful responses, categorized into risks like misinformation and discrimination. Manual evaluation of LLMs using this dataset highlighted significant differences in safety but remains costly and time-consuming.
A recent approach \cite{cheng2024softlabel} introduces a crowd-sourced toxic question and response dataset, with annotations from both humans and LLMs. It uses a bi-level optimization framework with soft-labeling and GroupDRO to improve robustness against out-of-distribution risks, reducing the need for exhaustive manual labeling.

\subsubsection{Truthfulness Datasets}

Ensuring LLMs generate truthful information is also essential. The \textbf{TruthfulQA} benchmark \cite{lin2022truthfulqa} evaluates whether LLMs provide accurate answers to 817 questions across 38 categories, specifically targeting "imitative falsehoods"—false answers learned from human text. Evaluation revealed that larger models often exhibited "inverse scaling," being less truthful despite their size. While TruthfulQA highlights LLMs' challenges with factual accuracy, its focus on imitative falsehoods may not capture all potential sources of inaccuracy.

\subsubsection{Value Benchmarks}
Ensuring LLM alignment with human values is a critical challenge, addressed by several benchmarks assessing various aspects of safety, fairness, and ethics. \textbf{FLAMES} \cite{huang2024flames} evaluates the alignment of Chinese LLMs with values like fairness, safety, and morality through 2,251 prompts. \textbf{SORRY-Bench} \cite{xie2024sorry} assesses LLMs' ability to reject unsafe requests using 45 topic categories, while \textbf{CVALUES} \cite{xu2023cvalues} focuses on both safety and responsibility. \textbf{SafetyPrompts} \cite{sun2023safety} evaluates Chinese LLMs on a range of ethical scenarios. 
While these benchmarks are valuable, they often focus on isolated, problematic queries, potentially leading to over-refusal in safe contexts. To address this, recent benchmarks have begun to incorporate contextual information. \textbf{CASE-Bench} \cite{sun2025casebench} pioneers this by using Contextual Integrity (CI) theory to formally describe the context of a query, evaluating whether an LLM's safety judgment aligns with human judgment under different contexts. This work reveals that context significantly influences human safety assessments and highlights mismatches in LLM behavior, especially in safe contexts.
In parallel, creating high-quality, commercially-usable datasets is crucial for training robust safety guardrails. \textbf{AEGIS2.0} \cite{ghosh2025aegis} addresses this gap by providing a diverse dataset with a comprehensive taxonomy of 12 core and 9 fine-grained risk categories. It uses a hybrid data generation pipeline combining human annotation with a multi-LLM "jury" system, making it suitable for training commercial safety models.
Furthermore, the concept of ``\emph{fake alignment}" \cite{wang2024fake} highlights the risk of LLMs superficially memorizing safety answers, leading to the Fake alIgNment Evaluation (\textbf{FINE}) framework for consistency assessment. \textbf{SafetyBench} \cite{zhang2024safetybench} addresses this by providing an efficient, automated multiple-choice benchmark for LLM safety evaluation.
\textbf{Libra-Leaderboard}\cite{li2024libra} introduces a balanced leaderboard for evaluating both the safety and capability of LLMs. 
It features a comprehensive safety benchmark with 57 datasets covering diverse safety dimensions, a unified evaluation framework, an interactive safety arena for adversarial testing, and a balanced scoring system. Libra-Leaderboard promotes a holistic approach to LLM evaluation, representing a significant step towards responsible AI development.

\subsubsection{Adversarial Datasets and Backdoor Benchmarks}
\textbf{BackdoorLLM} \cite{li2024backdoorllm} is the first benchmark for evaluating backdoor attacks in text generation, offering a standardized framework that includes diverse attack strategies like data poisoning and weight poisoning. \textbf{Adversarial GLUE} \cite{wang2021adversarial} assesses LLM robustness against textual attacks using 14 methods, highlighting vulnerabilities even in robustly trained models. \textbf{SALAD-Bench} \cite{li2024salad} expands on this by introducing a safety benchmark with a taxonomy of risks, including attack- and defense-enhanced questions. \textbf{JailBreakV-28K} \cite{luo2024jailbreakv} focuses on evaluating multi-modal LLMs against jailbreak attacks using text- and image-based test cases. A \textbf{STRONGREJECT} for empty jailbreaks \cite{souly2024strongreject} improves jailbreak evaluation with a higher-quality dataset and automated assessment. Despite their value, these benchmarks face challenges in scalability, consistency, and real-world relevance.

\begin{table*}[htp]
\centering
\caption{A summary of attacks and defenses for VLP models.}\label{tab:vlp_safety}
\resizebox{1\textwidth}{!}{
\begin{tabular}{llllllp{10cm}}
\hline
\rowcolor{wangxin-yellow}
Attack/Defense              & Method  & Year  & Category  & Subcategory  & Target Model & Dataset\\ \hline
\multirow{11}{0.12\textwidth}{Adversarial Attack}  
& Co-Attack~\cite{zhang2022towards}
  & 2022 & White-box & Invisible
  & ALBEF, TCL, CLIP
  & MS-COCO, Flickr30K, RefCOCO+, SNLI-VE \\
& \cellcolor{gray!15}AdvCLIP~\cite{zhou2023advclip}
  & \cellcolor{gray!15}2023
  & \cellcolor{gray!15}White-box
  & \cellcolor{gray!15}Invisible
  & \cellcolor{gray!15}CLIP
  & \cellcolor{gray!15}STL10, GTSRB, CIFAR10, ImageNet, Wikipedia, Pascal-Sentence, NUS-WIDE, XmediaNet \\
& Typographical Attacks~\cite{noever2021reading}
  & 2021 & White-box
  & Visible
  & CLIP
  & ImageNet \\
& \cellcolor{gray!15}Multi-Image Typographical Attacks~\cite{wang2025typographic}
  & \cellcolor{gray!15}2025
  & \cellcolor{gray!15}White-box
  & \cellcolor{gray!15}Visible
  & \cellcolor{gray!15}OpenCLIP, InstructBLIP
  & \cellcolor{gray!15}ImageNet, LAION \\
& SGA~\cite{lu2023set}
  & 2023 & Black-box
  & Sample-wise
  & ALBEF, TCL, CLIP
  & Flickr30K, MS-COCO \\
& \cellcolor{gray!15}SA-Attack~\cite{he2023sa}
  & \cellcolor{gray!15}2023
  & \cellcolor{gray!15}Black-box
  & \cellcolor{gray!15}Sample-wise
  & \cellcolor{gray!15}ALBEF, TCL, CLIP
  & \cellcolor{gray!15}Flickr30K, MS-COCO \\
& VLP-Attack~\cite{wang2023exploring}
  & 2023 & Black-box
  & Sample-wise
  & ALBEF, TCL, BLIP, BLIP2, MiniGPT-4
  & MS-COCO, Flickr30K, SNLI-VE \\
& \cellcolor{gray!15}TMM~\cite{wang2024transferable}
  & \cellcolor{gray!15}2024
  & \cellcolor{gray!15}Black-box
  & \cellcolor{gray!15}Sample-wise
  & \cellcolor{gray!15}ALBEF, TCL, X\_VLM, CLIP, BLIP, ViLT, METER
  & \cellcolor{gray!15}MS-COCO, Flickr30K, RefCOCO+, SNLI-VE \\
& VLATTACK~\cite{yin2024vlattack}
  & 2023 & Black-box
  & Sample-wise
  & BLIP, ViLT, CLIP
  & MS-COCO, VQA v2, NLVR2, SNLI-VE, ImageNet, SVHN \\
& \cellcolor{gray!15}OT-Attack~\cite{han2023ot}
  & \cellcolor{gray!15}2023
  & \cellcolor{gray!15}Black-box
  & \cellcolor{gray!15}Sample-wise
  & \cellcolor{gray!15}CLIP, ALBEF, TCL
  & \cellcolor{gray!15}Flickr30K, MS-COCO, RefCOCO+ \\
& PRM~\cite{hu2024firm}
  & 2024 & Black-box
  & Sample-wise
  & CLIP, Detic, VL-PLM, FC-CLIP, OpenFlamingo, LLaVA
  & PASCAL Context, COCO-Stuff, OV-COCO, MS-COCO, OK-VQA \\
& \cellcolor{gray!15}VLPTransferAttack~\cite{gao2024boosting}
  & \cellcolor{gray!15}2024
  & \cellcolor{gray!15}Black-box
  & \cellcolor{gray!15}Sample-wise
  & \cellcolor{gray!15}CLIP, ALBEF, TCL
  & \cellcolor{gray!15}Flickr30K, MS-COCO, RefCOCO+ \\
& C-PGC~\cite{fang2024one}
  & 2024 & Black-box
  & Universal
  & ALBEF, TCL, X-VLM, CLIP, BLIP
  & Flickr30K, MS-COCO, SNLI-VE, RefCOCO+ \\
& \cellcolor{gray!15}ETU~\cite{zhang2024universal}
  & \cellcolor{gray!15}2024
  & \cellcolor{gray!15}Black-box
  & \cellcolor{gray!15}Universal
  & \cellcolor{gray!15}ALBEF, TCL, CLIP, BLIP
  & \cellcolor{gray!15}Flickr30K, MS-COCO \\
& X-Transfer~\cite{huang2025xtransfer} & 2025 & Black-box & Universal & CLIP, OpenFlamingo, LLaVA, BLIP2, MiniGPT4 & Flickr30K, MS-COCO, ImageNet, CIFAR10, CIFAR100, STL10, SUN397, FOOD101, GTSRB StandfordCars, OK-VQA, VizWiz \\
\hline
\multirow{16}{0.12\textwidth}{Adversarial Defense} 
& Defense-Prefix~\cite{azuma2023defense}
  & 2023 & Adversarial Tuning
  & Prompt Tuning
  & CLIP
  & ImageNet \\
& \cellcolor{gray!15}AdvPT~\cite{zhang2023adversarial}
  & \cellcolor{gray!15}2023
  & \cellcolor{gray!15}Adversarial Tuning
  & \cellcolor{gray!15}Prompt Tuning
  & \cellcolor{gray!15}CLIP
  & \cellcolor{gray!15}ImageNet, Pets, Flowers, Food101, SUN397, DTD, EuroSAT, UCF101, ImageNet-V2, ImageNet-Sketch, ImageNet-A, ImageNet-R \\
& APT~\cite{li2024one}
  & 2024 & Adversarial Tuning
  & Prompt Tuning
  & CLIP
  & ImageNet, Caltech101, Pets, StanfordCars, Flowers, Food101, FGVCAircraft, SUN397, DTD, EuroSAT, UCF101, ImageNet-V2, ImageNet-Sketch, ImageNet-R, ObjectNet \\
& \cellcolor{gray!15}MixPrompt~\cite{fan2024mixprompt}
  & \cellcolor{gray!15}2024
  & \cellcolor{gray!15}Adversarial Tuning
  & \cellcolor{gray!15}Prompt Tuning
  & \cellcolor{gray!15}CLIP
  & \cellcolor{gray!15}ImageNet, Pets, Flowers, DTD, EuroSAT, UCF101, SUN397, Food101, ImageNet-V2, ImageNet-Sketch, ImageNet-A, ImageNet-R \\
& PromptSmoot~\cite{hussein2024promptsmooth}
  & 2024 & Adversarial Tuning
  & Prompt Tuning
  & PLIP, Quilt, MedCLIP
  & KatherColon, PanNuke, SkinCancer, SICAP v2 \\
& \cellcolor{gray!15}FAP~\cite{zhou2024few}
  & \cellcolor{gray!15}2024
  & \cellcolor{gray!15}Adversarial Tuning
  & \cellcolor{gray!15}Prompt Tuning
  & \cellcolor{gray!15}CLIP
  & \cellcolor{gray!15}ImageNet, Caltech101, Pets, StanfordCars, Flowers, Food101, FGVCAircraft, SUN397, DTD, EuroSAT, UCF101 \\
& APD~\cite{luo2024apd}
  & 2024 & Adversarial Tuning
  & Prompt Tuning
  & CLIP
  & ImageNet, Caltech101, Flowers, Food101, SUN397, DTD, EuroSAT, UCF101 \\
& \cellcolor{gray!15}TAPT~\cite{wang2024tapt}
  & \cellcolor{gray!15}2025
  & \cellcolor{gray!15}Adversarial Tuning
  & \cellcolor{gray!15}Prompt Tuning
  & \cellcolor{gray!15}CLIP
  & \cellcolor{gray!15}ImageNet, Caltech101, Pets, StanfordCars, Flowers, Food101, FGVCAircraft, SUN397, DTD, EuroSAT, UCF101 \\
& TeCoA~\cite{mao2023understanding}
  & 2022 & Adversarial Tuning
  & Contrastive Tuning
  & CLIP
  & CIFAR10, CIFAR100, STL10, Caltech101, Caltech256, Pets, StanfordCars, Food101, Flowers, FGVCAircraft, SUN397, DTD, PCAM, HatefulMemes, EuroSAT \\
& \cellcolor{gray!15}PMG-AFT~\cite{wang2024pre}
  & \cellcolor{gray!15}2024
  & \cellcolor{gray!15}Adversarial Tuning
  & \cellcolor{gray!15}Contrastive Tuning
  & \cellcolor{gray!15}CLIP
  & \cellcolor{gray!15}CIFAR10, CIFAR100, STL10, ImageNet, Caltech101, Caltech256, Pets, Flowers, FGVCAircraft, StanfordCars, SUN397, Food101, EuroSAT, DTD, PCAM \\
& MMCoA~\cite{zhou2024revisiting}
  & 2024 & Adversarial Tuning
  & Contrastive Tuning
  & CLIP
  & CIFAR10, CIFAR100, TinyImageNet, STL10, Caltech101, Caltech256, Pets, Flowers, FGVCAircraft, Food101, EuroSAT, DTD, SUN397, Country211 \\
& \cellcolor{gray!15}FARE~\cite{schlarmannrobust}
  & \cellcolor{gray!15}2024
  & \cellcolor{gray!15}Adversarial Tuning
  & \cellcolor{gray!15}Contrastive Tuning
  & \cellcolor{gray!15}OpenFlamingo, LLaVA
  & \cellcolor{gray!15}COCO, Flickr30k, TextVQA, VQA v2, CalTech101, StanfordCars, CIFAR10, CIFAR100, DTD, EuroSAT, FGVCAircrafts, Flowers, ImageNet-R, ImageNet-Sketch, PCAM, Pets, STL10, ImageNet \\
& VILLA~\cite{gan2020large}
  & 2020 & Adversarial Training
  & Two-stage Training
  & UNITER, LXMERT
  & MS-COCO, Visual Genome, Conceptual Captions, SBU Captions
    ImageNet, LAION, DataComp \\
& \cellcolor{gray!15}AdvXL~\cite{wang2024revisiting}
  & \cellcolor{gray!15}2024
  & \cellcolor{gray!15}Adversarial Training
  & \cellcolor{gray!15}Two-stage Training
  & \cellcolor{gray!15}CLIP
  & \cellcolor{gray!15}ImageNet, LAION, DataComp \\
& MirrorCheck~\cite{fares2024mirrorcheck}
  & 2024 & Adversarial Detection
  & One-shot Detection
  & UniDiffuser, BLIP, Img2Prompt, BLIP-2, MiniGPT-4
  & MS-COCO, CIFAR10, ImageNet \\
& \cellcolor{gray!15}AdvQDet~\cite{wang2024advqdet}
  & \cellcolor{gray!15}2024
  & \cellcolor{gray!15}Adversarial Detection
  & \cellcolor{gray!15}Stateful Detection
  & \cellcolor{gray!15}CLIP, ViT, ResNet
  & \cellcolor{gray!15}CIFAR10, GTSRB, ImageNet, Flowers, Pets \\
\hline

\multirow{6}{0.1\textwidth}{Backdoor \& Poisoning Attack}
& PBCL~\cite{carlini2022poisoning}
  & 2021 & Backdoor\&Poisoning
  & Visual Trigger
  & CLIP
  & Conceptual Captions, YFCC \\
& \cellcolor{gray!15}BadEncoder~\cite{jia2022badencoder}
  & \cellcolor{gray!15}2021
  & \cellcolor{gray!15}Backdoor
  & \cellcolor{gray!15}Visual Trigger
  & \cellcolor{gray!15}ResNet(SimCLR), CLIP
  & \cellcolor{gray!15}CIFAR10, STL10, GTSRB, SVHN, Food101 \\
& CorruptEncoder~\cite{zhang2024data}
  & 2022 & Backdoor
  & Visual Trigger
  & ResNet(SimCLR)
  & ImageNet, Pets, Flowers \\
& \cellcolor{gray!15}BadCLIP~\cite{liang2024badclip}
  & \cellcolor{gray!15}2023
  & \cellcolor{gray!15}Backdoor
  & \cellcolor{gray!15}Visual Trigger
  & \cellcolor{gray!15}CLIP
  & \cellcolor{gray!15}Conceptual Captions \\
& BadCLIP~\cite{bai2024badclip}
  & 2023 & Backdoor
  & Multi-modal Trigger
  & CLIP
  & ImageNet, Caltech101, Pets, StanfordCars, Flowers, Food101, FGVCAircraft, SUN397, DTD, EuroSAT, UCF101 \\
& \cellcolor{gray!15}MM Poison~\cite{yang2023data}
  & \cellcolor{gray!15}2022
  & \cellcolor{gray!15}Poisoning
  & \cellcolor{gray!15}Multi-modal Poisoning
  & \cellcolor{gray!15}CLIP
  & \cellcolor{gray!15}Flickr-PASCAL, MS-COCO \\
& MEM~\cite{liu2024multimodal}
  & 2024 & Poisoning
  & Multi-modal Trigger
  & CLIP
  & Flickr8k, Flickr30k, MS-COCO \\
\hline

\multirow{7}{0.1\textwidth}{Backdoor \& Poisoning Defense}
& CleanCLIP~\cite{bansal2023cleanclip}
  & 2023 & Backdoor Removal
  & Fine-tuning
  & CLIP
  & Conceptual Captions, ImageNet \\
& \cellcolor{gray!15}SAFECLIP~\cite{yang2023better}
  & \cellcolor{gray!15}2023
  & \cellcolor{gray!15}Backdoor Removal
  & \cellcolor{gray!15}Fine-tuning
  & \cellcolor{gray!15}CLIP
  & \cellcolor{gray!15}Conceptual Captions, Visual Genome, MS-COCO, Flowers, Food101, ImageNet, Pets, StanfordCars, Caltech101, CIFAR10, CIFAR100, DTD, FGVCAircraft \\
& RoCLIP~\cite{yang2024robust}
  & 2023 & Robust Training
  & Pre-training
  & CLIP
  & Conceptual Captions, Flowers, Food101, ImageNet, Pets, StanfordCars, Caltech101, CIFAR10, CIFAR100, DTD, FGVCAircraft \\
& \cellcolor{gray!15}DECREE~\cite{feng2023detecting}
  & \cellcolor{gray!15}2023
  & \cellcolor{gray!15}Backdoor Detection
  & \cellcolor{gray!15}Backdoor Model Detection
  & \cellcolor{gray!15}CLIP
  & \cellcolor{gray!15}CIFAR10, GTSRB, SVHN, STL-10, ImageNet \\
& TIJO~\cite{sur2023tijo}
  & 2023 & Backdoor Detection
  & Trigger Inversion
  & BUTD, MFB, BAN, MCAN, NAS
  & TrojVQA \\
& \cellcolor{gray!15}Mudjacking~\cite{liu2024mudjacking}
  & \cellcolor{gray!15}2024
  & \cellcolor{gray!15}Backdoor Detection
  & \cellcolor{gray!15}Trigger Inversion
  & \cellcolor{gray!15}CLIP
  & \cellcolor{gray!15}Conceptual Captions, CIFAR10, STL10, ImageNet, SVHN, Pets, Wiki103-Sub, SST-2, HOSL \\
& SEER~\cite{zhu2024seer}
  & 2024 & Backdoor Detection
  & Backdoor Sample Detection
  & CLIP
  & MSCOCO, Flickr, STL10, Pet, ImageNet \\
& \cellcolor{gray!15}Outlier Detection~\cite{huang2025detecting}
  & \cellcolor{gray!15}2025
  & \cellcolor{gray!15}Backdoor Detection
  & \cellcolor{gray!15}Backdoor Sample Detection
  & \cellcolor{gray!15}CLIP
  & \cellcolor{gray!15}Conceptual Captions, ImageNet, RedCaps \\
\hline
\end{tabular}
}
\end{table*}

\section{Vision-Language Pre-training Model Safety} \label{sec:vlp}
VLP models, such as CLIP~\cite{radford2021learning}, ALBEF~\cite{li2021align}, and TCL~\cite{yang2022vision}, have made significant strides in aligning visual and textual modalities. However, these models remain vulnerable to various safety threats, which have garnered increasing research attention. This section reviews the current safety research on VLP models, with a focus on adversarial, backdoor, and poisoning research. The representative methods reviewed in this section are summarized in Table \ref{tab:vlp_safety}.

\subsection{Adversarial Attacks}
\label{sec:vlp-attack}
Since VLP models are widely used as backbones for fine-tuning downstream models, adversarial attacks on VLP aim to generate examples that cause incorrect predictions across various downstream tasks, including zero-shot image classification, image-text retrieval, visual entailment, and visual grounding. Similar to Section~\ref{sec:vfm}, these attacks can roughly be categorized into \textbf{white-box attacks} and \textbf{black-box attacks}, based on their threat models.

\subsubsection{White-box Attacks}
White-box adversarial attacks on VLP models can be further categorized based on perturbation types into \textbf{invisible perturbations} and \textbf{visible perturbations}, with the majority of existing attacks employing invisible perturbations.

\textbf{Invisible Perturbations} involve small, imperceptible adversarial changes to inputs—whether text or images—to maintain the stealthiness of attacks. Early research in the vision and language domains primarily adopts this approach~\cite{xu2018fooling, shah2019cycle, li2020bert, yang2021defending}, in which invisible attacks are developed independently.
In the context of VLP models, which integrate both modalities, \textbf{Co-Attack}~\cite{zhang2022towards} was the first to propose perturbing both visual and textual inputs simultaneously to create stronger attacks. Building on this, \textbf{AdvCLIP}~\cite{zhou2023advclip} explores universal adversarial perturbations that can deceive all downstream tasks.

\textbf{Visible Perturbations} involve more substantial and noticeable alterations. For example, manually crafted typographical, conceptual, and iconographic images have been used to demonstrate that the CLIP model tends to ``read first, look later"~\cite{noever2021reading}, highlighting a unique characteristic of VLP models. This behavior introduces new attack surfaces for VLP, enabling the development of more sophisticated attacks. Recent work by Wang et al.~\cite{wang2025typographic} introduced a more stealthy multi-image attack scenario, demonstrating that non-repeating typographic attacks are most effective when attack texts are strategically selected for their similarity to the target images.

\subsubsection{Black-box Attacks}

Black-box attacks on VLP primarily adopt a transfer-based approach, with query-based attacks rarely explored. Existing methods can be categorized into: 1) \textbf{sample-specific perturbations}, tailored to individual samples, and 2) \textbf{universal perturbations}, applicable across multiple samples.

\textbf{Sample-wise perturbations} are generally more effective than universal perturbations, but their transferability is often limited.
\textbf{SGA}\cite{lu2023set} explores adversarial transferability in VLP by leveraging cross-modal interactions and alignment-preserving augmentation. Building on this, \textbf{SA-Attack}\cite{he2023sa} enhances cross-modal transferability by introducing data augmentations to both original and adversarial inputs. \textbf{VLP-Attack}\cite{wang2023exploring} improves transferability by generating adversarial texts and images using contrastive loss. To overcome SGA's limitations, \textbf{TMM}\cite{wang2024transferable} introduces modality-consistency and discrepancy features through attention-based and orthogonal-guided perturbations. \textbf{VLATTACK}\cite{yin2024vlattack} further enhances adversarial examples by combining image and text perturbations at both single-modal and multimodal levels. \textbf{PRM}\cite{hu2024firm} targets vulnerabilities in downstream models using foundation models like CLIP, enabling transferable attacks across tasks like object detection and image captioning. In parallel, Gao et al.~\cite{gao2024boosting} enhanced black-box transferability by diversifying perturbations within the ``intersection region" of the adversarial trajectory, reducing overfitting to the source model. Similarly, OT-Attack\cite{han2023ot} addressed overfitting by using optimal transport theory to efficiently align augmented image and text distributions. 

\textbf{Universal Perturbations} are less effective than sample-wise perturbations but more transferable.
\textbf{C-PGC}~\cite{fang2024one} was the first to investigate universal adversarial perturbations (UAPs) for VLP models. It employs contrastive learning and cross-modal information to disrupt the alignment of image-text embeddings, achieving stronger attacks in both white-box and black-box scenarios.
\textbf{ETU}~\cite{zhang2024universal} builds on this by generating UAPs that transfer across multiple VLP models and tasks. ETU enhances UAP transferability and effectiveness through improved global and local optimization techniques. It also introduces a data augmentation strategy \textbf{ScMix} that combines self-mix and cross-mix operations to increase data diversity while preserving semantic integrity, further boosting the robustness and applicability of UAPs.
\textbf{X-Transfer} \cite{huang2025xtransfer} proposes an efficient scaling strategy that enables the ensembling of a large collection of CLIP encoders as surrogate models. This approach demonstrates super adversarial transferability, achieving simultaneous transfer across data distributions, domains, model architectures, downstream tasks, and even to large VLMs.

\subsection{Adversarial Defenses}
\label{sec:vlp-defenses}
Existing adversarial defenses for VLP models can be grouped into four types: 1) \textbf{adversarial example detection}, 2) \textbf{standard adversarial training}, 3) \textbf{adversarial prompt tuning}, and 4) \textbf{adversarial contrastive tuning}. While adversarial detection filters out potential adversarial examples before or during inference, the other three defenses follow similar adversarial training paradigms, with variations in efficiency.

\subsubsection{Adversarial Example Detection}
Adversarial detection methods for VLP can be further divided into \textbf{one-shot detection} and \textbf{stateful detection}.

\paragraph{One-shot Detection} 
One-shot Detection distinguishes adversarial from clean examples in a single forward pass. White-box detection methods are typically one-shot. For example, \textbf{MirrorCheck}~\cite{fares2024mirrorcheck} is a model-agnostic method for VLP models. It uses text-to-image (T2I) models to generate images from captions produced by the victim model, comparing the similarity between the input image and the generated image using CLIP’s image encoder. A significant similarity difference flags the input as adversarial.

\paragraph{Stateful Detection} Stateful Detection is designed for black-box query attacks, where multiple queries are tracked to detect adversarial behavior. \textbf{AdvQDet}~\cite{wang2024advqdet} is a novel framework that counters query-based black-box attacks. It uses adversarial contrastive prompt tuning (ACPT) to tune CLIP image encoder, enabling detection of adversarial queries within just three queries.

\subsubsection{Standard Adversarial Training}
Adversarial training is widely regarded as the most effective defense against adversarial attacks \cite{madry2017towards,croce2020reliable}. However, it is computationally expensive, and for VLP models, which are typically trained on web-scale datasets, this cost becomes prohibitively high, posing a significant challenge for traditional approaches. Although research in this area is limited, we highlight two notable works that have explored adversarial training for vision-language pre-training. Their pre-trained models can be used as robust backbones for other adversarial research.

The first work, \textbf{VILLA}~\cite{gan2020large}, is a vision-language adversarial training framework consisting of two stages: task-agnostic adversarial pre-training and task-specific fine-tuning. VILLA enhances performance across downstream tasks using adversarial pre-training in the embedding space of both image and text modalities, instead of pixel or token levels. It employs FreeLB’s strategy~\cite{zhu2019freelb} to minimize computational overhead for efficient large-scale training.

The second work, \textbf{AdvXL}~\cite{wang2024revisiting}, is a large-scale adversarial training framework with two phases: a lightweight pre-training phase using low-resolution images and weaker attacks, followed by an intensive fine-tuning phase with full-resolution images and stronger attacks. This coarse-to-fine, weak-to-strong strategy reduces training costs while enabling scalable adversarial training for large vision models.

\subsubsection{Adversarial Prompt Tuning}
Adversarial prompt tuning (APT) enhances the adversarial robustness of VLP models by incorporating adversarial training during prompt tuning~\cite{zhou2022conditional, zhou2022learning, khattak2023maple}, typically focusing on textual prompts. It offers a lightweight alternative to standard adversarial training. APT methods can be classified into two main categories based on the prompt type: \emph{textual prompt tuning} and \emph{multi-modal prompt tuning}.

\paragraph{Textual Prompt Tuning}
Textual prompt tuning (TPT) robustifies VLP models by fine-tuning learnable text prompts. \textbf{AdvPT}~\cite{zhang2023adversarial} enhances the adversarial robustness of CLIP image encoder by realigning adversarial image embeddings with clean text embeddings using learnable textual prompts. Similarly, \textbf{APT}~\cite{li2024one} learns robust text prompts, using a CLIP image encoder to boost accuracy and robustness with minimal computational cost. \textbf{MixPrompt}~\cite{fan2024mixprompt} simultaneously enhances the generalizability and adversarial robustness of VLPs by employing conditional APT. Unlike empirical defenses, \textbf{PromptSmooth}~\cite{hussein2024promptsmooth} offers a certified defense for Medical VLMs, adapting pre-trained models to Gaussian noise without retraining. Additionally, \textbf{Defense-Prefix}~\cite{azuma2023defense} mitigates typographic attacks by adding a prefix token to class names, improving robustness without retraining.

\paragraph{Multi-Modal Prompt Tuning}
Recent adversarial prompt tuning methods have expanded textual prompts to multi-modal prompts. \textbf{FAP}~\cite{zhou2024few} introduces learnable adversarial text supervision and a training objective that balances cross-modal consistency while differentiating uni-modal representations. \textbf{APD}~\cite{luo2024apd} improves CLIP’s robustness through online prompt distillation between teacher and student multi-modal prompts. Additionally, \textbf{TAPT}~\cite{wang2024tapt} presents a test-time defense that learns defensive bimodal prompts to improve CLIP's zero-shot inference robustness.

\subsubsection{Adversarial Contrastive Tuning}
Adversarial contrastive tuning involves contrastive learning with adversarial training to fine-tune a robust CLIP image encoder for \emph{zero-shot adversarial robustness} on downstream tasks. These methods are categorized into \textbf{supervised} and \textbf{unsupervised} methods, depending on the availability of labeled data during training.

\paragraph{Supervised Contrastive Tuning}
\textbf{Visual Tuning} fine-tunes CLIP image encoder using only adversarial images. \textbf{TeCoA}~\cite{mao2023understanding} explores the zero-shot adversarial robustness of CLIP and finds that visual prompt tuning is more effective without text guidance, while fine-tuning performs better with text information. \textbf{PMG-AFT}~\cite{wang2024pre} improves zero-shot adversarial robustness by introducing an auxiliary branch to minimize the distance between adversarial outputs in the target and pre-trained models, mitigating overfitting and preserving generalization.

\textbf{Multi-modal Tuning} fine-tunes CLIP image encoder using both adversarial texts and images.
\textbf{MMCoA}~\cite{zhou2024revisiting} combines image-based PGD and text-based BERT-Attack in a multi-modal contrastive adversarial training framework. It uses two contrastive losses to align clean and adversarial image and text features, improving robustness against both image-only and multi-modal attacks.

\paragraph{Unsupervised Contrastive Tuning}
Adversarial contrastive tuning can also be performed in an unsupervised fashion. For instance, \textbf{FARE}~\cite{schlarmannrobust} robustifies CLIP image encoder through unsupervised adversarial fine-tuning, achieving superior clean accuracy and robustness across downstream tasks, including zero-shot classification and vision-language tasks. This approach enables VLMs, such as LLaVA and OpenFlamingo, to attain robustness without the need for re-training or additional fine-tuning.

\subsection{Backdoor \& Poisoning Attacks}
\label{sec:vlp-bp-attacks}
Backdoor and poisoning attacks on CLIP can target either the pre-training stage or the fine-tuning stage on downstream tasks. Previous studies have shown that poisoning backdoor attacks on CLIP can succeed with significantly lower poisoning rates compared to traditional supervised learning~\cite{carlini2022poisoning}. Additionally, training CLIP on web-crawled data increases its vulnerability to backdoor attacks~\cite{carlini2024poisoning}. This section reviews proposed attacks targeting backdooring or poisoning CLIP.

\subsubsection{Backdoor Attacks}

Based on the trigger modality, existing backdoor attacks on CLIP can be categorized into \textbf{visual triggers} and \textbf{multi-modal triggers}.

\textbf{Visual Triggers} target pre-trained image encoders by embedding backdoor patterns in visual inputs.
\textbf{BadEncoder}~\cite{jia2022badencoder} explores image backdoor attacks on self-supervised learning by injecting backdoors into pre-trained image encoders, compromising downstream classifiers. \textbf{CorruptEncoder}~\cite{zhang2024data} exploits random cropping in contrastive learning to inject backdoors into pre-trained image encoders, with increased effectiveness when cropped views contain only the reference object or the trigger.
For attacks targeting CLIP, \textbf{BadCLIP}~\cite{liang2024badclip} optimizes visual trigger patterns using dual-embedding guidance, aligning them with both the target text and specific visual features. This strategy enables BadCLIP to bypass backdoor detection and fine-tuning defenses.

\textbf{Multi-modal Triggers} combine both visual and textual triggers to enhance the attack. BadCLIP~\cite{bai2024badclip} introduces a novel trigger-aware prompt learning-based backdoor attack targeting CLIP models. Rather than fine-tuning the entire model, BadCLIP injects learnable triggers during the prompt learning stage, affecting both the image and text encoders.

\subsubsection{Poisoning Attacks}
Two targeted poisoning attacks on CLIP are \textbf{PBCL}~\cite{carlini2022poisoning} and \textbf{MM Poison}~\cite{yang2023data}. PBCL demonstrated that a targeted poisoning attack, misclassifying a specific sample, can be achieved by poisoning as little as 0.0001\% of the training dataset. MM Poison investigates modality vulnerabilities and proposes three attack types: single target image, single target label, and multiple target labels. Evaluations show high attack success rates while maintaining clean data performance across both visual and textual modalities. 
MEM~\cite{liu2024multimodal} protects private data from exploitation in multimodal contrastive learning by crafting unlearnable examples: it adds imperceptible noise to images and inserts an optimized text trigger into captions.

\subsection{Backdoor \& Poisoning Defenses}
\label{sec:vlp-bp-defenses}

Defense strategies against backdoor and poisoning attacks are generally categorized into \textbf{robust training} and \textbf{backdoor detection}. Robust Training aims to create VLP models resistant to backdoor or targeted poisoning attacks, even when trained on untrusted datasets. This approach specifically addresses poisoning-based attacks. Backdoor detection focuses on identifying compromised encoders or contaminated data. Detection methods often require additional mitigation techniques to fully eliminate backdoor effects.

\subsubsection{Robust Training}
Depending on the stage at which the model gains robustness against backdoor attacks, existing robust training strategies can be categorized into fine-tuning and pre-training approaches.

\paragraph{Fine-tuning Stage}
To mitigate backdoor and poisoning threats, \textbf{CleanCLIP}~\cite{bansal2023cleanclip} fine-tunes CLIP by re-aligning each modality's representations, weakening spurious correlations from backdoor attacks. Similarly, \textbf{SAFECLIP}~\cite{yang2023better} enhances feature alignment using unimodal contrastive learning. It first warms up the image and text modalities separately, then uses a Gaussian mixture model to classify data into safe and risky sets. During pre-training, SAFECLIP optimizes CLIP loss on the safe set, while separately fine-tuning the risky set, reducing poisoned image-text pair similarity and defending against targeted poisoning and backdoor attacks.

\paragraph{Pre-training Stage}
\textbf{ROCLIP}~\cite{yang2024robust} defends against poisoning and backdoor attacks by enhancing model robustness during pre-training. It disrupts the association between poisoned image-caption pairs by utilizing a large, diverse pool of random captions. Additionally, ROCLIP applies image and text augmentations to further strengthen its defense and improve model performance.

\subsubsection{Backdoor Detection}
Backdoor detection can be broadly divided into three subtasks: 1) \textbf{trigger inversion}, 2) \textbf{backdoor sample detection}, and 3) \textbf{backdoor model detection}. Trigger inversion is particularly useful, as recovering the trigger can aid in the detection of both backdoor samples and backdoored models.

\textbf{Trigger Inversion} aims to reverse-engineer the trigger pattern injected into a backdoored model. \textbf{Mudjacking}~\cite{liu2024mudjacking} mitigates backdoor vulnerabilities in VLP models by adjusting model parameters to remove the backdoor when a misclassified trigger-embedded input is detected. In contrast to single-modality defenses, \textbf{TIJO}~\cite{sur2023tijo} defends against dual-key backdoor attacks by jointly optimizing the reverse-engineered triggers in both the image and text modalities.

\textbf{Backdoor Sample Detection} detects whether a training or test sample is poisoned by a backdoor trigger. This detection can be used to cleanse the training dataset or reject backdoor queries. \textbf{SEER}~\cite{zhu2024seer} addresses the complexity of multi-modal models by jointly detecting malicious image triggers and target texts in the shared feature space. This method does not require access to the training data or knowledge of downstream tasks, making it highly effective for backdoor detection in VLP models. \textbf{Outlier Detection}~\cite{huang2025detecting} demonstrates that the local neighborhood of backdoor samples is significantly sparser compared to that of clean samples. This insight enables the effective and efficient application of various local outlier detection methods to identify backdoor samples from web-scale datasets. Furthermore, they reveal that potential unintentional backdoor samples already exist in the Conceptual Captions 3 Million (CC3M) dataset and have been trained into open-sourced CLIP encoders.

\textbf{Backdoor Model Detection} identifies whether a trained model is compromised by backdoor(s). \textbf{DECREE}~\cite{feng2023detecting} introduces a backdoor detection method specifically for VLP encoders that require no labeled data. It exploits the distinct embedding space characteristics of backdoored encoders when exposed to clean versus backdoor inputs. By combining trigger inversion with these embedding differences, DECREE can effectively detect backdoored encoders.

\subsection{Datasets}
\label{sec:vlp-data}
This section reviews datasets used for VLP safety research. As shown in Table \ref{tab:vlp_safety}, a variety of benchmark datasets were employed to evaluate adversarial attacks and defenses for VLP models. For image classification tasks, commonly used datasets include: ImageNet~\cite{russakovsky2015imagenet}, Caltech101~\cite{fei2004learning}, DTD~\cite{cimpoi2014describing}, EuroSAT~\cite{helber2019eurosat}, OxfordPets~\cite{parkhi2012cats}, FGVC-Aircraft~\cite{maji2013fine}, Food101~\cite{bossard2014food}, Flowers102~\cite{nilsback2008automated}, StanfordCars~\cite{krause20133d}, SUN397~\cite{xiao2010sun}, and UCF101~\cite{soomro2012ucf101}. 
For evaluating domain generalization and robustness to distribution shifts, several ImageNet variants were also used: ImageNetV2~\cite{recht2019imagenet}, ImageNet-Sketch~\cite{wang2019learning}, ImageNet-A~\cite{hendrycks2021natural}, and ImageNet-R~\cite{hendrycks2021many}. Additionally, MS-COCO~\cite{lin2014microsoft} and Flickr30K~\cite{plummer2015flickr30k} were utilized for image-to-text and text-to-image retrieval tasks, RefCOCO+\cite{yu2016modeling} for visual grounding, and SNLI-VE\cite{xie2019visual} for visual entailment.

\section{Vision-Language Model Safety} \label{sec:vlm}

Large VLMs extend LLMs by adding a visual modality through pre-trained image encoders and alignment modules, enabling applications like visual conversation and complex reasoning. However, this multi-modal design introduces unique vulnerabilities.
This section reviews \textbf{adversarial attacks}, \textbf{latency energy attacks}, \textbf{jailbreak attacks}, \textbf{prompt injection attacks}, \textbf{backdoor \& poisoning attacks}, and \textbf{defenses} developed for VLMs. Many VLMs use VLP-trained encoders, so the attacks and defenses discussed in Section~\ref{sec:vlp} also apply to VLMs. The additional alignment process between the VLM pre-trained encoders and LLMs, however, expands the attack surface, with new risks like cross-modal backdoor attacks and jailbreaks targeting both text and image inputs. This underscores the need for safety measures tailored to VLMs.

\subsection{Adversarial Attacks}
\label{sec:vlm-adversarial}
Adversarial attacks on VLMs primarily target the visual modality, which, unlike text, is more susceptible to adversarial perturbations due to its high-dimensional nature. By adding imperceptible changes to images, attackers aim to disrupt tasks like image captioning and visual question answering. These attacks are classified into \textbf{white-box} and \textbf{black-box} categories based on the threat model.

\subsubsection{White-box Attacks}

White-box adversarial attacks on VLMs have full access to the model parameters, including both vision encoders and LLMs. These attacks can be classified into three types based on their objectives: \textbf{task-specific attacks}, \textbf{cross-prompt attack}, and \textbf{chain-of-thought (CoT) attack}.

\textbf{Task-specific Attacks}  Schlarmann et al.~\cite{schlarmann2023adversarial} were the first to highlight the vulnerability of VLMs like Flamingo~\cite{alayrac2022flamingo} and GPT-4~\cite{gpt-4} to adversarial images that manipulate caption outputs. Their study showed how attackers can exploit these vulnerabilities to mislead users, redirecting them to harmful websites or spreading misinformation. Gao et al.~\cite{gao2024adversarial} introduced attack paradigms targeting the referring expression comprehension task, while \cite{cui2024robustness} proposed a query decomposition method and demonstrated how contextual prompts can enhance VLM robustness against visual attacks.

\textbf{Cross-prompt Attack} refer to adversarial attacks that remain effective across different prompts. For example, \textbf{CroPA}~\cite{luo2024image} explored the transferability of a single adversarial image across multiple prompts, investigating whether it could mislead predictions in various contexts. To tackle this, they proposed refining adversarial perturbations through learnable prompts to enhance transferability.

\textbf{CoT Attack} targets the CoT reasoning process of VLMs. \textbf{Stop-reasoning Attack} \cite{wang2024stop} explored the impact of CoT reasoning on adversarial robustness. Despite observing some improvements in robustness, they introduced a novel attack designed to bypass these defenses and interfere with the reasoning process within VLMs.

\subsubsection{Gray-box Attacks}
Gray-box adversarial attacks typically involve access to either the vision encoders or the LLM of a VLM, with a focus on vision encoders as the key differentiator between VLMs and LLMs. Attackers craft adversarial images that closely resemble target images, manipulating model predictions without full access to the VLM.
For instance, \textbf{InstructTA} \cite{wang2023instructta} generates a target image and uses a surrogate model to create adversarial perturbations, minimizing the feature distance between the original and adversarial image. To improve transferability, the attack incorporates GPT-4 paraphrasing to refine instructions.

\subsubsection{Black-box Attacks}

In contrast, black-box attacks do not require access to the target model's internal parameters and typically rely on \textbf{transfer-based} or \textbf{generator-based} methods.

\textbf{Transfer-based Attacks} exploit the widespread use of frozen CLIP vision encoders in many VLMs. \textbf{AttackBard} \cite{dong2023robust} demonstrates that adversarial images generated from surrogate models can successfully mislead Google's Bard, despite its defense mechanisms. Similarly, \textbf{AttackVLM} \cite{zhao2024evaluating} crafts targeted adversarial images for models like CLIP \cite{radford2021learning} and BLIP \cite{BLIP-2}, successfully transferring these adversarial inputs to other VLMs. It also shows that black-box queries further improved the success rate of generating targeted responses, illustrating the potency of cross-model transferability. \textbf{DynVLA}~\cite{gu2025improving} proposes a transfer-based black-box attack that perturbs the vision-language alignment mechanism within the vision-language connector. By injecting Gaussian-kernel-based attention shifts during optimization, it improves the transferability of adversarial examples across diverse VLMs, outperforming traditional input-level augmentation methods.

\textbf{Generator-based Attacks} leverage generative models to create adversarial examples with improved transferability. \textbf{AdvDiffVLM} \cite{guo2024efficiently} uses diffusion models to generate natural, targeted adversarial images with enhanced transferability. By combining adaptive ensemble gradient estimation and GradCAM-guided masking, it improves the semantic embedding of adversarial examples and spreads the targeted semantics more effectively across the image, leading to more robust attacks. \textbf{AnyAttack} \cite{zhang2024anyattack} presents a self-supervised framework for generating targeted adversarial images without label supervision. By utilizing contrastive loss, it efficiently creates adversarial examples that mislead models across diverse tasks. \textbf{CAVALRY-V}~\cite{zhang2025cavalry} proposes a dual-objective generator framework for black-box attacks on video VLMs, achieving strong cross-model transferability and temporal coherence through large-scale pretraining and fine-tuning.

\subsection{Jailbreak Attacks}
\label{sec:vlm-jailbreak}

\begin{table*}[htbp]
  \centering
  \caption{A summary of attacks and defenses for VLMs.}
  \resizebox{\textwidth}{!}{
    \begin{tabular}
{p{0.1\textwidth}p{0.15\textwidth}p{0.05\textwidth}p{0.15\textwidth}p{0.15\textwidth}p{0.25\textwidth}p{0.2\textwidth}}

\hline
    \rowcolor{wangruofan-orange}
    Attack/Defense & Method & Year & Category & Subcategory & Target Models & Datasets \\\hline
    \multirow{10}{0.1\textwidth}{Adversarial Attack} & Caption Attack \cite{schlarmann2023adversarial} & 2023 & White-box & Task-specific+V & OpenFlamingo & MS-COCO/Flickr30k/OK-VQA/VizWiz \\
    & \cellcolor{gray!15}VisBreaker \cite{cui2024robustness} & \cellcolor{gray!15}\cellcolor{gray!15}2023 & \cellcolor{gray!15}White-box & \cellcolor{gray!15}Task-specific+V & \cellcolor{gray!15}LLaVA/BLIP-2/InstructBLIP & \cellcolor{gray!15}MS-COCO/VQA V2/ScienceQA-Image/TextVQA/POPE/MME \\
    & CroPA \cite{luo2024image} & 2024 & White-box & Cross-prompt+VL & OpenFlamingo/BLIP-2/InstructBLIP & MS-COCO/VQA-v2 \\
    & \cellcolor{gray!15}GroundBreaker \cite{gao2024adversarial} & \cellcolor{gray!15}2024 &\cellcolor{gray!15} White-box & \cellcolor{gray!15}Task-specific+V &\cellcolor{gray!15} MiniGPT-v2 & \cellcolor{gray!15}RefCOCO/RefCOCO+/RefCOCOg \\
    & Stop-reasoning Attack \cite{wang2024stop} & 2024 & White-box & CoT attack+V & MiniGPT-4/OpenFlamingo/LLaVA & ScienceQA/A-OKVQA \\
    & \cellcolor{gray!15}InstructTA \cite{wang2023instructta} &\cellcolor{gray!15} 2023 &\cellcolor{gray!15} Gray-box & \cellcolor{gray!15}Encoder attack+V & \cellcolor{gray!15}BLIP-2/InstructBLIP/MiniGPT-4/LLaVA/CogVLM & \cellcolor{gray!15}ImageNet-1K/LLaVA-Instruct-150K/MS-COCO \\
    & Attack Bard \cite{dong2023robust} & 2023 & Black-box & Transfer-based+V & Bard/GPT-4V/Bing Chat/ERNIE Bot & NeurIPS’17 adversarial competition dataset \\
    & \cellcolor{gray!15}AttackVLM \cite{zhao2024evaluating} & \cellcolor{gray!15}2024 & \cellcolor{gray!15}Black-box & \cellcolor{gray!15}Transfer-based+V & \cellcolor{gray!15}BLIP/UniDiffuser/Img2Prompt/BLIP-2/LLaVA/MiniGPT-4 & \cellcolor{gray!15}ImageNet-1K/MS-COCO \\
    & DynVLA \cite{gu2025improving} & 2025 & Black-box & Transfer-based+VL & InstructBLIP/MiniGPT4/LLaVA/Gemini & MS-COCO/VQA-v2 \\
    & \cellcolor{gray!15}AdvDiffVLM \cite{guo2024efficiently} & \cellcolor{gray!15}2024 & \cellcolor{gray!15}Black-box & \cellcolor{gray!15}Generator-based+V & \cellcolor{gray!15}MiniGPT-4/LLaVA/UniDiffuser/MiniGPT-4/BLIP/BLIP-2/Img2LLM & \cellcolor{gray!15}NeurIPS’17 adversarial competition dataset/MS-COCO \\
        & AnyAttack \cite{zhang2024anyattack} & 2024 & Black-box & Generator-based+V & CLIP/BLIP/BLIP2/InstructBLIP/MiniGPT-4 & MSCOCO/Flickr30K/SNLI-VE \\
    & \cellcolor{gray!15}CAVALRY-V \cite{zhang2025cavalry} & \cellcolor{gray!15}2025 & \cellcolor{gray!15}Black-box & \cellcolor{gray!15}Generator-based+V & \cellcolor{gray!15}GPT-4.1/Gemini/QwenVL/InternVL/LLaVA/Aria/MiniCPM & \cellcolor{gray!15}MMBench-Video/Video-MME \\\hline

    \multirow{1}{0.1\textwidth}{Latency-Energy Attack} & Verbose Images \cite{gaoinducing} & 2024 &White-box & Task-specific+V & BLIP/BLIP2/InstructBLIP/MiniGPT-4 & MS-COCO/ImageNet
    \\\hline

    \multirow{11}{0.1\textwidth}{Jailbreak Attack} & Image Hijack \cite{bailey2023image} & 2023 & White-box & Target-specific+V & LLaVA & Alpaca training set/AdvBench \\
    & \cellcolor{gray!15}Adversarial Alignment Attack \cite{carlini2024aligned} & \cellcolor{gray!15}2024 & \cellcolor{gray!15}White-box & \cellcolor{gray!15}Target-specific+V & \cellcolor{gray!15}MiniGPT-4/LLaVA/LLaMA Adapter & \cellcolor{gray!15}toxic phrase dataset \\
    & VAJM \cite{qi2024visual} & 2024 & White-box & Universal attack+V & MiniGPT-4/LLaVA/InstructBLIP & VAJM training set/VAJM test set/RealToxicityPrompts \\
    & \cellcolor{gray!15}imgJP \cite{niu2024jailbreaking} & \cellcolor{gray!15}2024 & \cellcolor{gray!15}White-box & \cellcolor{gray!15}Universal attack+V & \cellcolor{gray!15}MiniGPT-4/MiniGPT-v2/LLaVA/InstructBLIP/mPLUG-Owl2 & \cellcolor{gray!15}AdvBench-M \\
    & UMK \cite{wang2024white} & 2024 & White-box & Universal attack+VL & MiniGPT-4 & AdvBench/VAJM training set/VAJM test set/RealToxicityPrompts \\
    & \cellcolor{gray!15}HADES \cite{li2024images} & \cellcolor{gray!15}2024 & \cellcolor{gray!15}White-box & \cellcolor{gray!15}Hybrid method+V & \cellcolor{gray!15}LLaVA/GPT-4V/Gemini-Pro-Vision & \cellcolor{gray!15}HADES dataset \\
    & Jailbreak in Pieces \cite{shayegani2023jailbreak} & 2023 & Black-box & Transfer-based+V & LlaVA /LLaMA-Adapter V2 & Jailbreak in Pieces dataset \\
    & \cellcolor{gray!15}Figstep \cite{gong2023figstep} & \cellcolor{gray!15}2023 & \cellcolor{gray!15}Black-box & \cellcolor{gray!15}Manual pipeline+V & \cellcolor{gray!15}LLaVA-v1.5/MiniGPT-4/CogVLM/GPT-4V & \cellcolor{gray!15}SafeBench \\
    & SASP \cite{wu2023jailbreaking} & 2023 & Black-box & Prompt leakage+L & LLaVA/GPT-4V & Celebrity face image dataset/CelebA/LFWA \\
    & \cellcolor{gray!15}VRP \cite{ma2024visual} & \cellcolor{gray!15}2024 & \cellcolor{gray!15}Black-box & \cellcolor{gray!15}Manual pipeline+V & \cellcolor{gray!15}LLaVA/Qwen-VL-Chat/ OmniLMM /InternVL Chat-V1.5/Gemini-Pro-Vision & \cellcolor{gray!15}RedTeam-2k/HarmBench \\
    & HIMRD \cite{teng2024heuristic} & 2024 & Black-box & Manual pipeline+VL & LLaVA/DeepSeek/GPT-4o/Gemini/Qwen-VL & SafeBench/tiny-SafeBench \\
    &\cellcolor{gray!15} IDEATOR \cite{wang2025ideator} & \cellcolor{gray!15}2025 & \cellcolor{gray!15}Black-box &\cellcolor{gray!15} Red teaming+VL & \cellcolor{gray!15}LLaVA/InstructBLIP/MiniGPT-4 & \cellcolor{gray!15}AdvBench/VAJM test set \\\hline

    \multirow{2}{0.1\textwidth}{Prompt Injection Attack} & Adversarial Prompt Injection \cite{bagdasaryan2023ab} & 2023 & White-box & Optimization-based+V & LLaVA/PandaGPT & Self-collected dataset \\
    & \cellcolor{gray!15}Typographic Attack \cite{qraitem2024vision} & \cellcolor{gray!15}2024 & \cellcolor{gray!15}Black-box & \cellcolor{gray!15}Typography-based+V & \cellcolor{gray!15}LLaVA/MiniGPT4/InstructBLIP/GPT-4V & 
 \cellcolor{gray!15}OxfordPets / StanfordCars / Flowers / Aircraft / Food101 \\\hline

    \multirow{5}{0.1\textwidth}{Backdoor \& Poisoning Attack} & Shadowcast \cite{xu2024shadowcast} & 2024 & Poisoning & Tuning-stage+VL & LLaVA/MiniGPT-v2/InstructBLIP & cc-sbu-align dataset \\
    & \cellcolor{gray!15}Instruction-Tuned Backdoor \cite{liang2024revisiting} & \cellcolor{gray!15}2024 &\cellcolor{gray!15} Backdoor & \cellcolor{gray!15}Tuning-stage+VL & \cellcolor{gray!15}OpenFlamingo/BLIP-2/LLaVA &\cellcolor{gray!15} MIMIC-IT/COCO/Flickr30K \\
    & Anydoor \cite{lu2024test} & 2024 & Backdoor & Testing-stage+VL & LLaVA/MiniGPT-4/InstructBLIP/BLIP-2 & VQAv2/SVIT/DALL-E dataset \\
    & \cellcolor{gray!15}BadVLMDriver \cite{ni2024physical} & \cellcolor{gray!15}2024 & \cellcolor{gray!15}Backdoor & \cellcolor{gray!15}Tuning-stage+V & \cellcolor{gray!15}LLaVA/MiniGPT-4 & \cellcolor{gray!15}nuScenes dataset \\
    & ImgTrojan \cite{tao2024imgtrojan} & 2024 & Backdoor & Tuning-stage+VL & LLaVA & LAION \\\hline

    \multirow{8}{0.1\textwidth}{Jailbreak Defenses} & \cellcolor{gray!15}JailGuard \cite{zhang2023mutation} & \cellcolor{gray!15}2023 & \cellcolor{gray!15}Detection & \cellcolor{gray!15}Detection+VL & \cellcolor{gray!15}GPT-3.5/MiniGPT-4 & \cellcolor{gray!15}Self-collected dataset \\
    & GuardMM \cite{sharma2024defending} & 2024 & Detection & Detection+V & GPT-4V/LLAVA/MINIGPT-4 & Self-collected dataset \\
    & \cellcolor{gray!15}AdaShield \cite{wang2024adashield} & \cellcolor{gray!15}2024 & \cellcolor{gray!15}Prevention & \cellcolor{gray!15}Prevention+V & \cellcolor{gray!15}LLaVA/CogVLM/MiniGPT-v2 & \cellcolor{gray!15}Figstep/QR \\
    & MLLM-Protector \cite{pi2024mllm} & 2024 & Prevention & Detection+Prevention+V & Open-LLaMA/LLaMA/LLaVA & Safe-Harm-10K \\
    & \cellcolor{gray!15}ECSO \cite{gou2024eyes} & \cellcolor{gray!15}2024 & \cellcolor{gray!15}Prevention & \cellcolor{gray!15}Prevention+V & \cellcolor{gray!15}LLaVA/ShareGPT4V/mPLUG-OWL2/Qwen-VL-Chat/InternLM-XComposer & \cellcolor{gray!15}MM-SafetyBench/VLSafe/VLGuard \\
    & InferAligner \cite{wang2024inferaligner} & 2024 & Prevention & Prevention+VL & LLaMA2/LLaVA & AdvBench/TruthfulQA/MM-Harmful Bench \\
    & \cellcolor{gray!15}BlueSuffix \cite{zhao2024bluesuffix} & \cellcolor{gray!15}2024 & \cellcolor{gray!15}Prevention & \cellcolor{gray!15}Prevention+VL & \cellcolor{gray!15}LLaVA/MiniGPT-4/Gemini & \cellcolor{gray!15}MM-SafetyBench/RedTeam-2k
    \\
    & DPS \cite{zhou2025defending} & 2025 & Prevention & Prevention+V & Qwen-VL-Plus/GPT-4o/Gemini-1.5-Flash & RTA-100/MultiTrust/Self-Gen/MM-SafetyBench/HADES/VisualAttack \\
    & \cellcolor{gray!15}ETA \cite{ding2025eta} & \cellcolor{gray!15}2025 & \cellcolor{gray!15}Prevention & \cellcolor{gray!15}Prevention+VL & \cellcolor{gray!15}LLaVA/InternVL/InternLM-XComposer/LLaMA3.2-Vision & \cellcolor{gray!15}SPA-VL/MM-SafetyBench/FigStep \\

    \hline
    \end{tabular}%
  }
  \label{tab:addlabel}
\end{table*}

The inclusion of a visual modality in VLMs provides additional routes for jailbreak attacks. While adversarial attacks generally induce random or targeted errors, jailbreak attacks specifically target the model's safeguards to generate inappropriate outputs. Like adversarial attacks, jailbreak attacks on VLMs can be classified as \textbf{white-box} or \textbf{black-box} attacks. 

\subsubsection{White-box Attacks}
White-box jailbreak attacks leverage gradient information to perturb input images or text, targeting specific behaviors in VLMs. These attacks can be further categorized into three types: \textbf{target-specific jailbreak}, \textbf{universal jailbreak}, and \textbf{hybrid jailbreak}, each exploiting different aspects of the model's safety measures.

\textbf{Target-specific Jailbreak} focuses on inducing a specific type of harmful output from the model.
\textbf{Image Hijack} \cite{bailey2023image} introduces adversarial images that manipulate VLM outputs, such as leaking information, bypassing safety measures, and generating false statements. These attacks, trained on generic datasets, effectively force models to produce harmful outputs.
Similarly, \textbf{Adversarial Alignment Attack} \cite{carlini2024aligned} demonstrates that adversarial images can induce misaligned behaviors in VLMs, suggesting that similar techniques could be adapted for text-only models using advanced NLP methods.

\textbf{Universal Jailbreak} bypasses model safeguards, causing it to generate harmful content beyond the adversarial input.
\textbf{VAJM} \cite{qi2024visual} shows that a single adversarial image can universally bypass VLM safety, forcing universal harmful outputs. \textbf{ImgJP} \cite{niu2024jailbreaking} uses a maximum likelihood algorithm to create transferable adversarial images that jailbreak various VLMs, even bridging VLM and LLM attacks by converting images to text prompts. \textbf{UMK} \cite{wang2024white} proposes a dual optimization attack targeting both text and image modalities, embedding toxic semantics in images and text to maximize impact. \textbf{HADES} \cite{li2024images} introduces a hybrid jailbreak method that combines universal adversarial images with crafted inputs to bypass safety mechanisms, effectively amplifying harmful instructions and enabling robust adversarial manipulation.

\subsubsection{Black-box Attacks}

Black-box jailbreak attacks do not require direct access to the internal parameters of the target VLM. Instead, they exploit external vulnerabilities, such as those in the frozen CLIP vision encoder, interactions between vision and language modalities, or system prompt leakage. These attacks can be classified into four main categories: \textbf{transfer-based attacks}, \textbf{manually-designed attacks}, \textbf{system prompt leakage}, and \textbf{red teaming}, each employing distinct strategies to bypass VLM defenses and trigger harmful behaviors.

\textbf{Transfer-based Attacks} on VLMs typically assume the attacker has access to the image encoder (or its open-source version), which is used to generate adversarial images that can then be transferred to attack the black-box LLM. For example, \textbf{Jailbreak in Pieces} \cite{shayegani2023jailbreak} introduces cross-modality attacks that transfer adversarial images, crafted using the image encoder (assume the model employed an open-source encoder), along with clean textual prompts to break VLM alignment. 

\textbf{Manually-designed Attacks} can be as effective as optimized ones. For instance, \textbf{FigStep} \cite{gong2023figstep} introduces an algorithm that bypasses safety measures by converting harmful text into images via typography, enabling VLMs to visually interpret the harmful intent. \textbf{VRP} \cite{ma2024visual} adopts a visual role-play approach, using LLM-generated images of high-risk characters based on detailed descriptions. By pairing these images with benign role-play instructions, VRP exploits the negative traits of the characters to deceive VLMs into generating harmful outputs. \textbf{HIMRD}~\cite{teng2024heuristic} introduces a heuristic-induced multimodal risk distribution framework that decomposes harmful prompts into semantically benign text and image components. A two-stage heuristic search then guides the model to reconstruct and affirm the underlying malicious intent.

\textbf{System Prompt Leakage} is another significant black-box jailbreak method, exemplified by \textbf{SASP} \cite{wu2023jailbreaking}. By exploiting a system prompt leakage in GPT-4V, SASP allowed the model to perform a self-adversarial attack, demonstrating the risks of internal prompt exposure.

\textbf{Red Teaming} recently saw an advancement with IDEATOR \cite{wang2025ideator}, which integrated a VLM with an advanced diffusion model to autonomously generate malicious image-text pairs. This approach overcomes the limitations of manually designed attacks, providing a scalable and efficient method for creating adversarial inputs without direct access to the target model.

\subsection{Jailbreak Defenses}
\label{sec:vlm-defenses}

This section reviews defense methods for VLMs against jailbreak attacks, categorized into \textbf{jailbreak detection} and \textbf{jailbreak prevention}. Detection methods identify harmful inputs or outputs for rejection or purification, while prevention methods enhance the model's inherent robustness to jailbreak queries through safety alignment or filters.

\subsubsection{Jailbreak Detection}
\textbf{JailGuard} \cite{zhang2023mutation} detects jailbreak attacks by mutating untrusted inputs and analyzing discrepancies in model responses. It uses 18 mutators for text and image inputs, improving generalization across attack types. 
\textbf{GuardMM} \cite{sharma2024defending} is a two-stage defense: the first stage validates inputs to detect unsafe content, while the second stage focuses on prompt injection detection to protect against image-based attacks. It uses a specialized language to enforce safety rules and standards. \textbf{MLLM-Protector} \cite{pi2024mllm} identifies harmful responses using a lightweight detector and detoxifies them through a specialized transformation mechanism. Its modular design enables easy integration into existing VLMs, enhancing safety and preventing harmful content generation.

\subsubsection{Jailbreak Prevention}
\textbf{AdaShield} \cite{wang2024adashield} defends against structure-based jailbreaks by prepending defense prompts to inputs, refining them adaptively through collaboration between the VLM and an LLM-based prompt generator, without requiring fine-tuning. \textbf{ECSO} \cite{gou2024eyes} offers a training-free protection by converting unsafe images into text descriptions, activating the safety alignment of pre-trained LLMs within VLMs to ensure safer outputs. 
\textbf{InferAligner} \cite{wang2024inferaligner} applies cross-model guidance during inference, adjusting activations using safety vectors to generate safe and reliable outputs. \textbf{BlueSuffix} \cite{zhao2024bluesuffix} introduces a reinforcement learning-based black-box defense framework consisting of three key components: (1) an image purifier for securing visual inputs, (2) a text purifier for safeguarding textual inputs, and (3) a reinforcement fine-tuning-based suffix generator that leverages bimodal gradients to enhance cross-modal robustness. 
\textbf{DPS}~\cite{zhou2025defending} introduces a black-box, training-free defense that supervises VLMs using responses from partially cropped images, boosting robustness to visual jailbreaks while maintaining benign performance. \textbf{ETA}~\cite{ding2025eta} presents a two-stage inference-time alignment framework: it first screens the safety of inputs and outputs, then applies alignment through interference prefixes and best-of-N sentence search, enabling safe and helpful responses without extra training.

\subsection{Energy Latency Attacks}
\label{sec:vlm-latency}
Similar to LLMs, multi-modal LLMs also face significant computational demands. Verbose images~\cite{gaoinducing} exploit these demands by overwhelming service resources, resulting in higher server costs, increased latency, and inefficient GPU usage. These images are specifically designed to delay the occurrence of the EOS token, increasing the number of auto-regressive decoder calls, which in turn raises both energy consumption and latency costs.

\subsection{Prompt Injection Attacks}
\label{sec:vlm-injection}

Prompt injection attacks against VLMs share the same objective as those against LLMs (Section~\ref{sec:llm}), but the visual modality introduces continuous features that are more easily exploited through adversarial attacks or direct injection. These attacks can be further classified into \emph{optimization-based attacks} and \emph{typography-based attacks}.

\textbf{Optimization-based Attacks} often optimize the input images using (white-box) gradients to produce stronger attacks. These attacks manipulate the model's responses, influencing future interactions. One representative method is \textbf{Adversarial Prompt Injection} \cite{bagdasaryan2023ab}, where attackers embed malicious instructions into VLMs by adding adversarial perturbations to images. 

 \textbf{Typography-based Attacks} exploit VLMs' typographic vulnerabilities by embedding deceptive text into images without requiring gradient access (i.e., black-box). The \textbf{Typographic Attack} \cite{qraitem2024vision} introduces two variations: \emph{Class-Based Attack} to misidentify classes and \emph{Descriptive Attack} to generate misleading labels. These attacks can also leak personal information \cite{chen2023can}, highlighting significant security risks.

\subsection{Backdoor\& Poisoning Attacks}
\label{sec:vlm-backdoor}
Most VLMs rely on VLP encoders, with safety threats discussed in Section \ref{sec:vlp}. This section focuses on backdoor and poisoning risks arising during fine-tuning and testing, specifically when aligning vision encoders with LLMs. 
Backdoor attacks embed triggers in visual or textual inputs to elicit specific outputs, while poisoning attacks inject malicious image-text pairs to degrade model performance.
We review backdoor and poisoning attacks separately, though most of these works are backdoor attacks.

\subsubsection{Backdoor Attacks}
We further classify backdoor attacks on VLMs into \textbf{tuning-time backdoor} and \textbf{testing-time backdoor}.

\textbf{Tuning-time Backdoor} injects the backdoor during VLM instruction tuning. \textbf{MABA} \cite{liang2024revisiting} targets domain shifts by adding domain-agnostic triggers using attributional interpretation, enhancing attack robustness across mismatched domains in image captioning tasks.
\textbf{BadVLMDriver} \cite{ni2024physical} introduced a physical backdoor for autonomous driving, using objects like red balloons to trigger unsafe actions such as sudden acceleration, bypassing digital defenses and posing real-world risks. Its automated pipeline generates backdoor training samples with malicious behaviors for stealthy, flexible attacks. \textbf{ImgTrojan} \cite{tao2024imgtrojan} introduces a jailbreaking attack by poisoning image-text pairs in training data, replacing captions with malicious prompts to enable VLM jailbreaks, exposing risks of compromised datasets. 

\textbf{Test-time Backdoor} leverages the similarity of universal adversarial perturbations and backdoor triggers to inject backdoor at test-time. AnyDoor \cite{lu2024test} embeds triggers in the textual modality via adversarial test images with universal perturbations, creating a text backdoor from image-perturbation combinations. It can also be seen as a multi-modal universal adversarial attack. Unlike traditional methods, AnyDoor does not require access to training data, enabling attackers to separate setup and activation of the attack.

\subsubsection{Poisoning Attacks}

\textbf{Shadowcast} \cite{xu2024shadowcast} is a stealthy tuning-time backdoor attack on VLMs. It injects poisoned samples visually indistinguishable from benign ones, targeting two objectives: 1) \textbf{Label Attack}, which misclassifies objects, and 2) \textbf{Persuasion Attack}, which generates misleading narratives. With only 50 poisoned samples, Shadowcast achieves high effectiveness, showing robustness and transferability across VLMs in black-box settings.

\subsection{Datasets \& Benchmarks}
\label{sec:vlm-dataset}

\begin{table}[htbp]
  \centering
  \caption{Safety and robustness benchmarks for VLMs.}
  \resizebox{0.5\textwidth}{!}{ 
    \begin{tabular}{cccc}\hline
    \rowcolor{wangruofan-orange}
    Benchmarks & Year  & Size  & \# VLMs evaluated \\\hline
     OODCV-VQA \cite{tu2023many} & 2023  & 4,244 & 21 \\
    Sketchy-VQA \cite{tu2023many} & 2023  & 4,000 & 21 \\
    MM-SafetyBench \cite{liu2023mm} & 2023  & 5,040 & 12 \\
    AVIBench \cite{zhang2024avibench} & 2024  & 260,000  & 14 \\
    Jailbreak Evaluation of GPT-4o \cite{ying2024unveiling} & 2024  & 4,180 & 1 \\
    JailBreakV-28K \cite{luo2024jailbreakv} & 2024  & 28,000 & 10 \\
    MIS \cite{ding2025rethinking} & 2025 & 6,185 & 14 \\
    VLJailbreakBench \cite{wang2025ideator} & 2025  & 3,654 & 11 \\
    Argus Inspection \cite{yao2025argus} & 2025 & 1,430 & 26 \\\hline
    \end{tabular}%
    }
  \label{tab:vlm_safety_benchmarks}%
\end{table}

The datasets used in VLM safety research are detailed in Table \ref{tab:vfm_safety}. Below, we review the benchmarks proposed for evaluating VLM safety and robustness, summarized in Table \ref{tab:vlm_safety_benchmarks}.
\textbf{SafeSight} \cite{tu2023many} introduces two VQA datasets, \textbf{OODCV-VQA} and \textbf{Sketchy-VQA}, to evaluate out-of-distribution (OOD) robustness, highlighting VLMs' vulnerabilities to OOD texts and vision encoder weaknesses. \textbf{MM-SafetyBench} \cite{liu2023mm} focuses on image-based manipulations, revealing vulnerabilities in multi-modal interactions. \textbf{AVIBench} \cite{zhang2024avibench} evaluates VLM robustness against 260K adversarial visual instructions, exposing susceptibility to image-based, text-based, and content-biased  adversarial visual instructions (AVIs). \textbf{Jailbreak Evaluation of GPT-4o} \cite{ying2024unveiling} tests GPT-4o with multi-modal and unimodal jailbreak attacks, uncovering alignment vulnerabilities. \textbf{JailBreakV-28K} \cite{luo2024jailbreakv} assesses the transferability of LLM jailbreak techniques to VLMs, showing high attack success rates across 10 open-source models. These studies collectively reveal significant vulnerabilities in VLMs to OOD inputs, adversarial instructions, and multi-modal jailbreaks. MIS~\cite{ding2025rethinking} presents the first multi-image safety dataset for assessing VLMs’ visual reasoning in complex unsafe scenarios, exposing safety reasoning gaps in current fine-tuning methods and providing nuanced multi-image benchmarks. VLJailbreakBench~\cite{wang2025ideator} offers 3,654 adversarial image-text pairs generated by the IDEATOR red-teaming framework to evaluate VLMs under black-box multimodal jailbreak settings. Argus Inspection~\cite{yao2025argus} introduces a detail-oriented benchmark for commonsense safety, embedding causally critical yet textually implicit visual “traps” within realistic scenarios.

\section{Diffusion Model Safety}\label{sec:diffusion}

This section focuses on safety research related to diffusion models \cite{rombach2022high, DALLE-2, DALLE-3, Imagen}, which involve forward noise addition and reverse sampling. In the forward process, Gaussian noise is incrementally added to an image until it becomes pure noise. Reverse sampling generates new samples by stepwise denoising based on learned data distributions~\cite{ho2020denoising, song2020denoising, song2020score}. By integrating input information, diffusion models perform conditional generation, transforming data distribution modeling \( p(x) \) into \( p(x|\text{guidance}) \).

Widely used in Image-to-Image (I2I), Text-to-Image (T2I), and Text-to-Video (T2V) tasks, diffusion models are applied in content creation, image editing, and film production. However, their extensive use exposes them to various security risks including \textbf{adversarial}, \textbf{jailbreak}, \textbf{backdoor}, and \textbf{privacy attacks}. These attacks can degrade generation quality, bypass safety filters, manipulate outputs, and reveal sensitive training data. This section also reviews defenses against these threats, including \textbf{jailbreak} and \textbf{backdoor defenses}, as well as \textbf{intellectual property protection} techniques.

\subsection{Adversarial Attacks}
\label{sec:dm_adversarial_attacks}

Adversarial attacks on diffusion models typically perturb text prompts to degrade image quality or cause semantic mismatches with the original text. This section reviews existing adversarial attacks, categorized by threat model into \textbf{white-box}, \textbf{gray-box}, and \textbf{black-box} methods.

\begin{table*}[htp]
\center
\caption{A summary of attacks and defenses for Diffusion Models (Part I).}
\label{tab:diffuison_safety_I}
\rowcolors{2}{gray!15}{white}
\resizebox{1\textwidth}{!}{
\begin{tabular}{p{0.09\textwidth}p{0.15\textwidth}p{0.05\textwidth}p{0.15\textwidth}p{0.17\textwidth}p{0.28\textwidth}p{0.3\textwidth}}
\hline
\rowcolor{gaoyifeng-pink}
Attack/Defense & Method & Year & Category & Subcategory & Target Models & Dataset\\ \hline
\cellcolor{white} & ECB\cite{struppek2023exploiting} & 2024 & Black-box & Character-level & Stable Diffusion, DALL-E 2, AltDiffusion-m18 & LAION-Aesthetics v2, MS COCO, ImageNet-V2, self-constructed  \\ 
\cellcolor{white} & CharGrad\cite{kou2023character} & 2023 & Black-box & Character-level & Stable Diffusion & MS COCO, Flickr30k \\ 
\cellcolor{white} & ER\cite{gao2023evaluating} & 2023 & Black-box & Character-level & Stable Diffusion, DALL·E 2 & LAION-COCO, DiffusionDB, SBU Corpus, self-constructed  \\ 
\cellcolor{white} & DHV\cite{daras2022discovering} & 2022 & Black-box & Word-level & DALLE-2 & -  \\ 
\cellcolor{white} & AA\cite{milliere2022adversarial} & 2022 & Black-box & Word-level & DALL-E 2, DALL-E mini & - \\ 
\cellcolor{white} & BBA\cite{maus2023black} & 2023 & Black-box & Sentence-level & Stable Diffusion & ImageNet \\
\cellcolor{white} & RIATIG\cite{liu2023riatig} & 2023 & Black-box & Sentence-level & DALL·E, DALL·E 2, Imagen & MS COCO \\

\cellcolor{white} & QFA\cite{zhuang2023pilot} & 2023 & Grey-box & Similarity-driven  & Stable Diffusion & self-constructed \\
\cellcolor{white} & RVTA\cite{zhang2024revealing} & 2024 & Grey-box & Similarity-driven & Stable Diffusion & ImageNet, self-constructed \\
\cellcolor{white} & MMP-Attack\cite{yang2024multi} & 2025 & Grey-box & Similarity-driven & Stable Diffusion, DALL-E 3, Imagine Art & MS COCO \\
\cellcolor{white} & DORMANT\cite{zhou2024dormant} & 2025 & Grey-box & Distance‑driven & Animate Anyone, MagicAnimate, MagicPose, MusePose, Champ,
MuseV, UniAnimate, and ControlNeXt & TikTok, Champ, UBC Fashion, and TED Talks \\

\cellcolor{white} & ATM\cite{du2024stable} & 2023 & White-box & Classifier-driven  & Stable Diffusion & ImageNet, self-constructed \\
\cellcolor{white} & FOOLSDEDIT\cite{zhou2024foolsdedit} & 2024 & White-box & Classifier-driven  & SDEdit & CelebAMask-HQ,  FFHQ \\
\cellcolor{white}\multirow{-12}{0.1\textwidth}{Adversarial Attack} & SAGE\cite{liu2023discovering} & 2023 & White-box & Classifier-driven  & GLIDE, Stable Diffusion, DeepFloyd & ImageNet \\

\hline

\cellcolor{white} & SneakyPrompt\cite{yang2024sneakyprompt} & 2023 & Black-box & Target External Defenses & Stable Diffusion, DALL·E 2 & NSFW-200, Dog/Cat-100 \\
\cellcolor{white} & UD\cite{qu2023unsafe} & 2023 & Black-box & Target External Defenses & Stable Diffusion, LD, DALL·E 2, DALL·E mini & MS COCO \\
\cellcolor{white} & {Atlas}\cite{dong2024jailbreaking} & 2024 & Black-box & Target External Defenses & Stable Diffusion, DALL·E 3 & NSFW-200, Dog/Cat-100 \\ 
\cellcolor{white} & {Groot}\cite{liu2024groot} & 2024 & Black-box & Target External Defenses & Stable Diffusion, Midjounery, DALL·E 3 & self-constructed \\ 
\cellcolor{white} & {DACA}\cite{deng2023divide} & 2024 & Black-box & Target External Defenses & Midjounery, DALL·E 3 & VBCDE-100, Copyright-20 \\ 
\cellcolor{white} & {SurrogatePrompt}\cite{ba2023surrogateprompt} & 2024 & Black-box & Target External Defenses & Midjourney, DALL·E 2, DreamStudio & self-constructed \\
\cellcolor{white} & PGJ\cite{huang2025perception} & 2025 & Black-box & Target External Defenses & Stable Diffusion, DALL-E 2, DALL-E 3, Cogview3, Tongyiwanxiang, Hunyuan & self-constructed \\
\cellcolor{white} & R2A\cite{zhang2025reason2attack} & 2025 & Black-box & Target External Defense & Stable Diffusion, FLUX, DALL·E 3, Midjourney & self-constructed \\
\cellcolor{white} & {JPA}\cite{ma2024jailbreaking} & 2024 & Grey-box & Target Internal Defenses & Stable Diffusion, Midjourney, DALL·E 2, PIXART-$\alpha$ & I2P \\
\cellcolor{white} & {RT-Attack}\cite{gao2024rt} & 2024 & Grey-box & Target Internal Defenses & Stable Diffusion, DALL·E 3, SafeGen & I2P, self-constructed \\
\cellcolor{white} & RTSDSF\cite{rando2022red} & 2022 & White box & Target External Defenses & Stable Diffusion & self-constructed \\ 
\cellcolor{white} & {MMA}\cite{yang2024mma} & 2024 & White box & Target External Defenses & Stable Diffusion, Midjounery, Leonardo.Ai & LAION-COCO, UnsafeDiff \\
\cellcolor{white} & {P4D}\cite{chin2023prompting4debugging} & 2024 & White box & Target Internal Defenses & Stable Diffusion & I2P, ESD Dataset \\
\cellcolor{white}\multirow{-14}{0.1\textwidth}{Jailbreak Attack} & {UnlearnDiffAtk}\cite{zhang2023generate} & 2024 & White box & Target Internal Defenses & Stable Diffusion & I2P Dataset, ImageNet, WikiArt \\
\hline
 
\cellcolor{white} & ESD\cite{gandikota2023erasing} & 2023 & Concept Erasure & Fine-tuning & Stable Diffusion & MS COCO, I2P \\ 
\cellcolor{white} & SPM\cite{lyu2024one} & 2024 & Concept Erasure & Fine-tuning & Stable Diffusion & MS COCO, I2P \\ 
\cellcolor{white} & SDD\cite{kim2023towards} & 2023 & Concept Erasure  & Fine-tuning & Stable Diffusion & MS COCO, I2P \\ 
\cellcolor{white} & AC\cite{kumari2023ablating} & 2023 & Concept Erasure & Fine-tuning & Stable Diffusion & MS COCO \\ 
\cellcolor{white} & ABO\cite{hong2024all} & 2023 & Concept Erasure & Fine-tuning & Stable Diffusion & MS COCO \\ 
\cellcolor{white} & UC\cite{wu2024unlearning} & 2024 & Concept Erasure & Fine-tuning & Stable Diffusion & I2P \\ 
\cellcolor{white} & SA\cite{heng2024selective} & 2023 & Concept Erasure & Fine-tuning & Stable Diffusion, DDPM & MNIST, CIFAR-10 and STL-10, I2P\\ 
\cellcolor{white} & Receler\cite{huang2023receler} & 2024 & Concept Erasure & Fine-tuning & Stable Diffusion & CIFAR-10, MS COCO, I2P \\ 
\cellcolor{white} & RACE\cite{kim2024race} & 2024 & Concept Erasure & Fine-tuning & Stable Diffusion & MS COCO, I2P, Imagenette\\ 
\cellcolor{white} & AdvUnlearn\cite{zhang2024defensive} & 2024 & Concept Erasure & Fine-tuning & Stable Diffusion & MS COCO, I2P, Imagenette \\ 
\cellcolor{white} & DT\cite{ni2023degeneration} & 2023 & Concept Erasure & Fine-tuning & Stable Diffusion & MS COCO \\ 
\cellcolor{white} & FMO\cite{zhang2024forget} & 2023 & Concept Erasure & Fine-tuning & Stable Diffusion & ConceptBench \\
\cellcolor{white} & Geom-Erasing\cite{liu2024implicit} & 2024 & Concept Erasure & Fine-tuning & Stable Diffusion & LAION \\
\cellcolor{white} & SepME\cite{zhao2024separable} & 2024 & Concept Erasure & Fine-tuning & Stable Diffusion & self-constructed \\
\cellcolor{white} & CCRT~\cite{han2024continuous} & 2024 & Concept Erasure & Fine-tuning & Stable Diffusion & MS COCO \\
\cellcolor{white} & SafeGen~\cite{li2024safegen} & 2024 & Concept Erasure & Fine-tuning & Stable Diffusion & MS COCO, I2P, SneakyPrompt-Dataset, NSFW-56k \\
\cellcolor{white} & CPE~\cite{lee2025concept} & 2025 & Concept Erasure & Fine-tuning & Stable Diffusion & MS COCO, I2P, MACE-Dataset \\
\cellcolor{white} & MACE\cite{lu2024mace} & 2024 & Concept Erasure & Close-Formed Solution & Stable Diffusion & CIFAR-10, MS COCO, I2P \\
\cellcolor{white} & UCE\cite{gandikota2024unified} & 2024 & Concept Erasure & Close-Formed Solution & Stable Diffusion & MS COCO \\
\cellcolor{white} & TIME\cite{orgad2023editing} & 2023 & Concept Erasure & Close-Formed Solution & Stable Diffusion & MS COCO \\
\cellcolor{white} & RECE\cite{gong2024reliable} & 2024 & Concept Erasure & Close-Formed Solution & Stable Diffusion & MS COCO, I2P \\
\cellcolor{white} & RealEra\cite{liu2024realera} & 2024 & Concept Erasure & Close-Formed Solution & Stable Diffusion & CIFAR-10, I2P \\
\cellcolor{white} & CP\cite{chavhan2024conceptprune} & 2024 & Concept Erasure & Neuron Pruning & Stable Diffusion & Imagenette \\
\cellcolor{white} & PRCEDM\cite{yang2024pruning} & 2024 & Concept Erasure & Neuron Pruning & Stable Diffusion & Imagenet, MS COCO, I2P \\ 
\cellcolor{white} & SLD\cite{schramowski2023safe} & 2023 & Inference Guidance & Input & Stable Diffusion & LAION-2B-en, I2P, DrawBench \\ 
\cellcolor{white} & PromptGuard\cite{yuan2025promptguard} & 2025 & Inference Guidance & Input & Stable Diffusion & MS COCO, I2P, SneakyPrompt-Dataset \\ 
\cellcolor{white} & Ethical-Lens\cite{cai2024ethical} & 2025 & Inference Guidance & Input\&Output & Stable Diffusion, Dreamlike Diffusion & MS COCO, I2P, Tox100, Tox1K, HumanBias, Demographic Stereotypes, Mental Disorders \\ 
\cellcolor{white} & SDIDLD\cite{li2024self} & 2024 & Inference Guidance & Latent space & Stable Diffusion & MS COCO, I2P, CelebA, Winobias, self-constructed \\ 
\cellcolor{white}\multirow{-26}{0.1\textwidth}{Jailbreak \\ Defense} & CC\cite{meng2025concept} & 2025 & Inference Guidance & Latent space & Stable Diffusion & MS COCO, I2P, self-constructed \\ 
\hline

\cellcolor{white} & {BadDiffusion} \cite{chou2023backdoor} & 2023 & Training Manipulation & Visual Trigger & DDPM & CIFAR-10, CelebA \\ 
\cellcolor{white} & {VillanDiffusion}\cite{chou2024villandiffusion} & 2023 & Training Manipulation & Visual Trigger & Stable Diffusion, DDPM, LDM, NCSN & CIFAR-10, CelebA \\ 
\cellcolor{white} & {TrojDiff}\cite{chen2023trojdiff} & 2023 & Training Manipulation & Visual Trigger & DDPM, DDIM & CIFAR-10, CelebA \\
\cellcolor{white} & {IBA}\cite{li2024invisible} & 2024 & Training Manipulation & Visual Trigger & $\text{Unconditional and Conditional DM}^*$ & CIFAR-10,  CelebA, MS-COCO \\
\cellcolor{white} & {DIFF2}\cite{li2024watch} & 2024 & Training Manipulation & Visual Trigger & DDPM, DDIM, Stable DiffusionE, ODE & CIFAR-10, CIFAR-100, CelebA, ImageNet \\
\cellcolor{white} & {RA}\cite{struppek2023rickrolling} & 2023 & Data Poisoning & Textual Trigger & Stable Diffusion & LAION-Aesthetics v2, MS-COCO \\
\cellcolor{white} & {BadT2I}\cite{zhai2023text} & 2023 & Data Poisoning & Textual Trigger & Stable Diffusion & LAION-Aesthetics v2, LAION-2B-en, MS COCO\\
\cellcolor{white} & {FTHCW}\cite{pan2023trojan} & 2024 & Data Poisoning & Textual Trigger & DDPM, LDM & CIFAR-10, ImageNet, Caltech256 \\
\cellcolor{white} & {BAGM}\cite{vice2024bagm} & 2023 & Data Poisoning & Textual Trigger & Stable Diffusion, Kandinsky, DeepFloyd-IF & MS COCO, Marketable Food \\
\cellcolor{white} & {Zero-Day}\cite{huang2023zero, huang2024personalization} & 2023 & Data Poisoning & Textual Trigger & Stable Diffusion & DreamBooth dataset \\
\cellcolor{white} & {SBD}\cite{wang2024stronger} & 2024 & Data Poisoning & Textual Trigger & Stable Diffusion & LAION Aesthetics v2, Pokemon Captions, COYO-700M, Midjourney v5 \\
\cellcolor{white}\multirow{-13}{0.1\textwidth}{Backdoor \\ Attack} & {IBT} \cite{naseh2024injecting} & 2024 & Data Poisoning & Textual Trigger & Stable Diffusion & Midjourney Dataset, DiffusionDB, PartiPrompts\\
\hline

\cellcolor{white} & T2IShield\cite{wang2024t2ishield} & 2024 & Detection & Trigger Detection & Stable Diffusion & CelebA-HQ-Dialog \\
\cellcolor{white} & Ufid\cite{guan2025ufid} & 2024 & Detection & Trigger Validation & DDPM, Stable Diffusion & CelebA-HQ-Dialog, Pokemon, \\
\cellcolor{white} & DisDet\cite{sui2024disdet} & 2024 & Detection & Trigger Validation & DDPM, DDIM & CIFAR-10, CelebA  \\
\cellcolor{white} & Elijah\cite{an2024elijah} & 2024 & Removal & Detect \& Remove & DDPM, DDIM, LDM & CIFAR-10, CelebA-HQ \\
\cellcolor{white} & Diff-Cleanse\cite{hao2024diff} & 2024 & Removal & Detect \& Remove & DDPM, DDIM, LDM & MNIST, CIFAR-10, CelebA-HQ \\
\cellcolor{white} & TERD\cite{mo2024terd} & 2024 & Removal & Inverse \& Remove & DDPM & CIFAR-10, CelebA, CelebA-HQ \\
\cellcolor{white}\multirow{-8}{0.1\textwidth}{Backdoor \\ Defense} & PureDiffusion\cite{truong2024purediffusion} & 2024 & Removal & Inverse \& Remove & DDPM & CIFAR-10\\
\cellcolor{white} & NaviDet\cite{zhai2025navidet} & 2025 & Removal & Detect \& Remove & Stable Diffusion & MS-COCO \\
\hline

\end{tabular}
}
\end{table*}

\subsubsection{White-box Attacks}
White-box attacks on T2I diffusion models assume full access to model parameters, allowing direct optimization of text prompts or latent space to degrade or disrupt image generation. For example, \textbf{SAGE} \cite{liu2023discovering} explores both the discrete prompt and latent spaces to uncover failure modes in T2I models, including distorted generations and targeted manipulations. \textbf{ATM} \cite{du2024stable} generates attack prompts similar to clean prompts by replacing or extending words using Gumbel Softmax, preventing the model from generating desired subjects. \textbf{FOOLSDEDIT} \cite{zhou2024foolsdedit} imperceptibly modifies stroke images by applying a mix of four operations: exposure, motion blur, identity mapping, and an empty operation. It automatically selects the optimal combination to steer SDEdit’s \cite{meng2021sdedit} outputs toward a desired attribute, while ensuring that the strokes appear visually unchanged.

\subsubsection{Gray-box Attacks}
\label{subsubsec:Similarity-based optimization}
Gray-box attacks assume the CLIP text encoder used in many T2I diffusion models is frozen and publicly available. The attacker can then exploit CLIP similarity loss to craft adversarial text prompts targeting the text encoder.

\textbf{QFA} \cite{zhuang2023pilot} minimizes cosine similarity between original and perturbed text embeddings to generate images that differ as much as possible from the original text. \textbf{RVTA} \cite{zhang2024revealing} maximizes image-text similarity to align adversarial prompts with reference images generated by a surrogate diffusion model. 
\textbf{MMP-Attack}~\cite{yang2024multi} simultaneously maximizes the cosine similarity between the perturbed text embedding and the target embedding in both the text and image modalities, while employing a straight-through estimator to execute the optimization process. \textbf{DORMANT} ~\cite{zhou2024dormant} embeds imperceptible PGD noise, optimized with VAE-latent, CLIP-semantic, ReferenceNet-detail, and frame-consistency losses. This causes pose-driven portrait animation models to generate identity-shifted and jittery videos, while the source photo remains visually unchanged.

\subsubsection{Black-box Attacks}
Black-box attacks assume the attacker has no knowledge of the victim diffusion model's internals (parameters or architecture). Since diffusion models use text prompts as input, existing attacks employ textual adversarial techniques to evade the model. These attacks can be further categorized by granularity into \textbf{character-level}, \textbf{word-level}, and \textbf{sentence-level} attacks.

\textbf{Character-level Attacks} modify the characters in the text input to create adversarial prompts. \textbf{ECB} \cite{struppek2023exploiting} shows how replacing characters with homoglyphs, such as using Hangul or Arabic scripts, shifts generated images toward cultural stereotypes. Subsequent works, like \textbf{CharGrad} \cite{kou2023character}, optimize character-level perturbations using gradient-based attacks and proxy representations to map character changes to embedding shifts. \textbf{ER} \cite{gao2023evaluating} uses distribution-based objectives (e.g., MMD, KL divergence) to maximize discrepancies in image distributions, enhancing attack effectiveness. These attacks exploit typos, homoglyphs, and phonetic modifications, disrupting text-to-image outputs.

\textbf{Word-level Attacks} craft adversarial prompts by replacing or adding words to the input text. \textbf{DHV} \cite{daras2022discovering} uncovers a hidden vocabulary in diffusion models, where nonsensical strings like \texttt{Apoploe vesrreaitais} can generate bird images, due to their proximity to target concepts in the CLIP text embedding space. Building on this, \textbf{AA} \cite{milliere2022adversarial} introduces macaronic prompting, combining word fragments from different languages to control visual outputs systematically. These attacks reveal vulnerabilities in the relationship between text embeddings and image generation.

\textbf{Sentence-level Attacks} rewrite a substantial part or the entire prompt to create adversarial prompts. \textbf{RIATIG} \cite{liu2023riatig} uses a CLIP-based image similarity measure as an optimization objective and a genetic algorithm to iteratively mutate and select text prompts, creating adversarial examples that resemble the target image while remaining semantically different from the original text. In contrast, \textbf{BBA} \cite{maus2023black} employs classification loss and black-box optimization to refine prompts, using Token Space Projection (TPS) to bridge the gap between continuous word embeddings and discrete tokens, enabling the generation of category-specific images without explicit category terms.

\subsection{Jailbreak Attacks}
\label{sec:dm_jailbreak_attacks}

Diffusion models use both internal and external safety mechanisms to void the generation of Not Safe For Work (NSFW) content. Internal safety mechanisms often refer to the inherent robustness of T2I diffusion models, achieved through safety alignment during training, which aims to reduce the likelihood of generating harmful content. External safety mechanisms, on the other hand, are safety filters, such as text, image, or text-image classifiers, applied to detect and block unsafe outputs after generation. 
Jailbreak attacks aim to craft adversarial prompts that bypass the safety mechanisms of diffusion models, enabling the generation of harmful content. This section provides a systematic review of existing jailbreak methods, categorized by threat model into \textbf{white-box}, \textbf{gray-box}, and \textbf{black-box} attacks.

\subsubsection{White-box Attacks}
White-box attacks can bypass the safety mechanisms in T2I diffusion models through gradient-based optimization. These attacks can be further classified into \textbf{internal safety attacks} and \textbf{external safety attacks}, each exploiting specific vulnerabilities in the victim models.

\textbf{Internal Safety Attacks} target the internal safety mechanisms of diffusion models.
Jailbreaking internally safety-enhanced diffusion models involves regenerating NSFW content by bypassing the removal of harmful concepts. The red teaming tool \textbf{P4D} \cite{chin2023prompting4debugging} automatically identifies problematic prompts to exploit limitations in current safety evaluations, aligning the predicted noise of an unconstrained model with that of a safety-enhanced one. \textbf{UnlearnDiffAtk} \cite{zhang2023generate} introduces an evaluation framework that uses unlearned diffusion models' classification capabilities to optimize adversarial prompts, aligning predicted noise with a target unsafe image to force the model to recreate NSFW content during denoising.

\textbf{External Safety Attacks} target the safety filters of diffusion models, aiming to bypass both input and output safety mechanisms. \textbf{RTSDSF} \cite{rando2022red} reverse-engineered predefined NSFW concepts in filters by using the CLIP model to encode and compare NSFW vocabulary embeddings, performing a dictionary attack. It also showed that prompt dilution—adding irrelevant details—can bypass safety filters. \textbf{MMA} \cite{yang2024mma} employs a similarity-driven loss to optimize adversarial prompts and introduce subtle perturbations to input images, bypassing both prompt filters and post-hoc safety checkers during image editing.

\subsubsection{Gray-box Attacks}
Gray-box jailbreak attacks assume that attackers have full access only to the open-source text encoder, with other components of the diffusion model remaining inaccessible. In this scenario, the attacker exploits the exposed text encoder to bypass the model's internal safety mechanism.

\textbf{Internal Safety Attacks}, under the gray-box setting, target models with `concept erasure'. \textbf{Ring-A-Bell} \cite{tsai2023ring} extracts unsafe concepts by comparing antonymous prompt pairs, generates harmful prompts with soft prompts, and refines them using a genetic algorithm. 
\textbf{JPA} \cite{ma2024jailbreaking} leverages antonyms like \texttt{“nude”} and \texttt{“clothed”}, calculating their average difference in the text embedding space to represent NSFW concepts, then optimizes prefix prompts for semantic alignment. \textbf{RT-Attack} \cite{gao2024rt} uses a two-stage strategy to maximize textual similarity to NSFW prompts and iteratively refines them based on image-level similarity, demonstrating that even limited knowledge can enable attacks on safety-enhanced models.

\subsubsection{Black-box Attacks}

Black-box jailbreaks on diffusion models target commercial models with access only to outputs, such as filter rejections or generated image quality and semantics, and are primarily \textbf{external safety attacks}. 

\textbf{External Safety Attacks}, in the black-box setting, use hand-crafted or LLM-assisted adversarial prompts to mislead the victim model to generate NSFW content.
\textbf{UD} \cite{qu2023unsafe} highlights the risk of T2I models generating unsafe content, especially hateful memes, by refining unsafe prompts manually. \textbf{SneakyPrompt} \cite{yang2024sneakyprompt} uses reinforcement learning to optimize adversarial prompts, which updates its policy network based on filter evasion and semantic alignment. Other methods employ LLMs to refine adversarial prompts.  \textbf{Groot} \cite{liu2024groot} decomposes prompts into objects and attributes to dilute sensitive content. \textbf{DACA} \cite{deng2023divide} breaks down and recombines prompts using LLMs. \textbf{SurrogatePrompt} \cite{ba2023surrogateprompt} targets Midjourney, substituting sensitive terms and leveraging image-to-text modules to generate harmful content at scale. \textbf{Atlas} \cite{dong2024jailbreaking} automates the attack with a two-agent system: one VLM generates adversarial prompts, while an LLM evaluates and selects the best candidates. These LLM-assisted strategies can significantly improve the effectiveness and stealthiness of the attacks. \textbf{PGJ} \cite{huang2025perception} identifies unsafe tokens and replaces them with perceptually similar but semantically distant phrases, producing short, natural prompts that evade text filters without directly querying the T2I model. \textbf{R2A} \cite{zhang2025reason2attack} further improves an LLM’s reasoning for jailbreaking T2I models by first fine-tuning on Chain-of-Thought examples based on contextual word meanings, and then applying reinforcement learning guided by a dense attack process reward.

\subsection{Jailbreak Defenses}
\label{sec:dm_jailbreak_defenses}

This section reviews existing defense strategies proposed for T2I diffusion models against jailbreak attacks, including \textbf{concept erasure} and \textbf{inference guidance}. 
The key challenge of these defenses is how to ensure safety while maintaining generation quality.

\subsubsection{Concept Erasure} 
Concept erasure is an emerging research area focused on removing undesirable concepts (e.g., NSFW content and copyrighted styles) from diffusion models, where these concepts are referred to as \emph{target concepts}. Concept erasure methods can be categorized into three types: \textbf{finetuning-based}, \textbf{close-form solution}, and \textbf{pruning-based}, depending on the strategy employed.

\paragraph{Finetuning-based Methods} 
These methods use gradient-based optimization to adjust model parameters, typically involving a loss function with an erasure term to prevent the generation of representations linked to the target (undesirable) concept, and a constraint term to preserve non-target concepts. These approaches can be categorized into \textbf{anchor-based}, \textbf{anchor-free}, and \textbf{adversarial} erasure methods.

\textbf{Anchor-based Erasing} is a targeted approach that guides the model to shift the target (undesirable concept) towards a good concept (anchor) by aligning predicted latent noise. \textbf{AC} \cite{kumari2023ablating} defines anchor concepts as broader categories encompassing the target concepts (e.g., \texttt{“Grumpy Cat”} → \texttt{“Cat”}) and uses standard diffusion loss on text-image pairs of anchors to preserve their integrity while erasing target concepts. 
\textbf{ABO} \cite{hong2024all} removes specific target concepts by modifying classifier guidance, using both explicit (replacing the target with a predefined substitute) and implicit (suppressing attention maps) erasing signals, and includes a penalty term to maintain generation quality. 
\textbf{DoCo} \cite{wu2024unlearning} improves generalization by aligning target and anchor concepts through adversarial training and mitigating gradient conflicts with concept-preserving gradient surgery. \textbf{SPM} \cite{lyu2024one} uses a 1D adapter and negative guidance \cite{gandikota2023erasing} to suppress target concepts while ensuring non-target concepts remain consistent, affecting only relevant synonyms. 
\textbf{SA} \cite{heng2024selective} applies generative replay and elastic weight consolidation to stabilize model weights and maintain normal generation capabilities while preserving non-target concepts. \textbf{SafeGen} \cite{li2024safegen} fine-tunes the vision-only self-attention of Stable Diffusion on \textit{<nude, mosaic, benign>} triplets, encouraging nude features to be transformed into mosaics while preserving benign images.

\textbf{Anchor-free Erasing} is a non-targeted fine-tuning approach that reduces the probability of generating target concepts without aligning to a specific safe concept. 
\textbf{ESD} \cite{gandikota2023erasing} modifies classifier-free guidance into negative-guided noise prediction to minimize the target concept’s generation probability (e.g., "Van Gogh"). \textbf{SDD} \cite{kim2023towards} addresses the extra effects of ESD’s negative guidance by using unconditioned predictions and EMA to avoid catastrophic forgetting. \textbf{DT} \cite{ni2023degeneration} erases unsafe concepts by training the model to denoise scrambled low-frequency images. \textbf{Forget-Me-Not} \cite{zhang2024forget} uses Attention Resteering to minimize intermediate attention maps related to the target concept. \textbf{Geom-Erasing} \cite{liu2024implicit} erases implicit concepts like watermarks by applying a geometric-driven control method and introduces the \emph{Implicit Concept Dataset}. \textbf{SepME} \cite{zhao2024separable} advances multiple concept erasure and restoration. Fuchi et al. \cite{fuchi2024erasing} proposed few-shot unlearning by targeting the text encoder rather than the image encoder or diffusion model.
\textbf{CCRT} \cite{han2024continuous} proposes a method for continuous removal of diverse concepts from diffusion models.

\textbf{Adversarial Erasing} enhances previous methods by introducing perturbations to the target concept's text embedding and using adversarial training to improve robustness. \textbf{Receler} \cite{huang2023receler} employs a lightweight eraser and adversarial prompt embeddings, iteratively training against each other, while applying a binary mask from U-Net attention maps to target only the concept regions. \textbf{AdvUnlearn} \cite{zhang2024defensive} shifts adversarial attacks to the text encoder, targeting the embedding space and using regularization to preserve normal generation. \textbf{RACE} \cite{kim2024race} improves efficiency by conducting adversarial attacks at a single timestep, reducing computational complexity. These methods enhance the model’s resistance to adversarial prompts aimed at regenerating erased concepts. \textbf{CPE} \cite{lee2025concept} introduces a Residual Attention Gate (ResAG) that activates exclusively on target-concept tokens to precisely erase them. The gate is further strengthened through adversarial embedding attack–defense iterations, providing robust protection.

\paragraph{Close-form Solution Methods} 
These methods offer an efficient alternative to fine-tuning-based erasure, focusing on localized updates in cross-attention layers to erase target concepts, inspired by model editing in LLMs \cite{meng2022mass}. Unlike fine-tuning, which aligns denoising predictions, these methods align cross-attention values. 
\textbf{TIME} \cite{orgad2023editing} applies a closed-form solution to debias models, while \textbf{UCE} \cite{gandikota2024unified} extends this to multiple erasure targets, preserving surrounding concepts to reduce interference. \textbf{MACE} \cite{lu2024mace} refines cross-attention updates with LoRA and Grounded-SAM \cite{kirillov2023segment,liu2023grounding} for region-specific erasure. A recent challenge is that erased concepts can still be generated via sub-concepts or synonyms \cite{liu2024realera}.
\textbf{RealEra} \cite{liu2024realera} tackles this by mining associated concepts and adding perturbations to the embedding, expanding the erasure range with beyond-concept regularization. 
\textbf{RECE} \cite{gong2024reliable} addresses insufficient erasure by continually finding new concept embeddings during fine-tuning and applying closed-form solutions for further erasure.

\paragraph{Pruning-based Methods} 
These methods erase target concepts by identifying and removing neurons strongly associated with the target, selectively disabling them without updating model weights.
\textbf{ConceptPrune} calculates a Wanda score using target and reference prompts to measure each neuron's contribution, pruning those most associated with the target concept. Similarly, another approach \cite{yang2024pruning} identifies concept-correlated neurons using adversarial prompts to enhance the robustness of existing erasure methods.

\subsubsection{Inference Guidance} 
Inference guidance methods steer pre-trained diffusion models to generate safe images by incorporating additional auxiliary information and specific guidance during the inference process.
 
\paragraph{Input Guidance} 
This type of guidance use additional input text to steer the model toward safe content. 
\textbf{SLD} \cite{schramowski2023safe} adjusts noise predictions during inference based on a text condition and unsafe concepts, guiding generation towards the intended prompt while avoiding unsafe content, without requiring fine-tuning. It also introduces the I2P benchmark, a dataset for testing inappropriate content generation. \textbf{PromptGuard} \cite{yuan2025promptguard} learns a safety soft prompt within the text embedding space and appends it as a suffix to every user prompt, steering the T2I model away from NSFW outputs without altering its weights.

\paragraph{Input \& Output Guidance}
This type of methods prevent harmful inputs and control NSFW outputs. \textbf{Ethical-Lens} \cite{cai2024ethical} employs a plug-and-play framework, using an LLM for input text revision (Ethical Text Scrutiny) and a multi-headed CLIP classifier for output image modification (Ethical Image Scrutiny), ensuring alignment with societal values without retraining or internal changes.

\paragraph{Latent space Guidance} 
This approach uses additional implicit representations in the latent space to guide generation. \textbf{SDIDLD} \cite{li2024self} employs self-supervised learning to identify the opposite latent direction of inappropriate concepts (e.g., "anti-sexual") and adds these vectors at the bottleneck layer, preventing harmful content generation. 
\textbf{Concept Corrector} \cite{meng2025concept} functions during image generation by employing a Generation Check Mechanism (GCM) to inspect an intermediate prediction of the final image for unwanted concepts. If such concepts are detected, a Concept Removal Attention (CRA) module is then activated to dynamically replace the target features with those associated with a negative concept.

\subsection{Backdoor Attacks}
\label{sec:dm_backdoor_attacks}

Backdoor attacks on diffusion models allow adversaries to manipulate generated content by injecting backdoor triggers during training. These "malicious triggers" are embedded in model components, and during generation, inputs with triggers (e.g., prompts or initial noise) guide the model to produce predefined content. The key challenge is enhancing attack success rates while keeping the trigger covert and preserving the model's original utility. Existing attacks can be categorized into \textbf{training manipulation} and \textbf{data poisoning} methods.

\subsubsection{Training Manipulation}
This type of attack typically assumes the attacker aims to release a backdoored diffusion model, granting control over the training or even inference processes. Existing attacks focus on the visual modality, inserting backdoors by using image pairs with triggers and target images (\emph{image-image pair injection}), typically targeting unconditional diffusion models.

\textbf{BadDiffusion} \cite{chou2023backdoor} presents the first backdoor attack on T2I diffusion models, which modifies the forward noise-addition and backward denoising processes to map backdoor target distributions to image triggers while maintaining DDPM sampling. \textbf{VillanDiffusion} \cite{chou2024villandiffusion} extends this to conditional models, adding prompt-based triggers and textual triggers for tasks like text-to-image generation. \textbf{TrojDiff} \cite{chen2023trojdiff} advances the research by controlling both training and inference, incorporating Trojan noise into sampling for diverse attack objectives. \textbf{IBA\cite{li2024invisible}} introduces invisible trigger backdoors using bi-level optimization to create covert perturbations that evade detection. \textbf{DIFF2} \cite{li2024watch} proposes a backdoor attack in adversarial purification, optimizing triggers to mislead classifiers and extending it to data poisoning by injecting backdoors directly.

\subsubsection{Data Poisoning}
Unlike training manipulation, data poisoning methods do not directly interfere with the training process, restricting the attack to inserting poisoned samples into the dataset. These attacks typically target conditional diffusion models and explore two types of textual triggers: \textbf{text-text pair} and \textbf{text-image pair}.

\textbf{Text-text Pair Triggers} consist of triggered prompts and their corresponding target prompts.  \textbf{RA} \cite{struppek2023rickrolling} adopts this approach to inject backdoors into the text encoder by adding a covert trigger character, mapping the original to the target prompt while preserving encoder functionality through utility loss optimization. The backdoored encoder generates embeddings with predefined semantics, guiding the diffusion model’s output. This lightweight attack requires no interaction with other model components. Several studies \cite{struppek2023rickrolling, vice2024bagm, huang2023zero, huang2024personalization} have also explored this approach. 

\textbf{Text-image Pair Triggers} consist of triggered prompts paired with target images. \textbf{BadT2I} \cite{zhai2023text} explores backdoors based on pixel, object, and style changes, where a special trigger (e.g., \texttt{“[T]”}) induces the model to generate images with specific patches, replaced objects, or styles. To reduce the data cost, \textbf{Zero-Day} \cite{huang2023zero,huang2024personalization} uses personalized fine-tuning, injecting trigger-image pairs for more efficient backdoors. \textbf{FTHCW} \cite{pan2023trojan} embeds target patterns into images from different classes, forming text-image pairs to generate diverse outputs. \textbf{IBT} \cite{naseh2024injecting} uses two-word triggers that activate the backdoor only when both words appear together, enhancing stealthiness. In commercial settings, \textbf{BAGM} \cite{vice2024bagm} manipulates user sentiment by mapping broad terms (e.g., “drinks”) to specific brands (e.g., “Coca Cola”). \textbf{SBD} \cite{wang2024stronger} employs backdoors for copyright infringement, bypassing filters by decomposing and reassembling copyrighted content using text-image pairs.

\subsection{Backdoor Defenses}
\label{sec:dm_backdoor_defenses}

Backdoor defenses for diffusion models is an emerging area of research. Current approaches generally follow a three-step pipeline: 1) \textbf{trigger inversion}, 2) \textbf{trigger validation} or \textbf{backdoor detection}, and 3) \textbf{backdoor removal}. Some works propose complete frameworks, while others focus on individual steps.

\subsubsection{Backdoor Detection}
Most early research focuses on detecting or validating backdoor triggers. \textbf{T2IShield} \cite{wang2024t2ishield} is the first backdoor detection and mitigation framework for diffusion models, leveraging the \emph{assimilation phenomenon} in cross-attention maps, where a trigger suppresses other tokens to generate specific content. Similarly, \textbf{NaviDet} \cite{zhai2025navidet} detects trigger samples by identifying unusual activations in the early diffusion steps induced by specific input tokens. \textbf{Ufid} \cite{guan2025ufid} validates triggers by noting that clean generations are sensitive to small perturbations, while backdoor-triggered outputs are more robust. \textbf{DisDet} \cite{sui2024disdet} proposes a low-cost detection method that distinguishes poisoned input noise from clean Gaussian noise by identifying distribution shifts.

\subsubsection{Backdoor Removal}
While trigger validation confirms the presence of a backdoor trigger, the identified triggers must still be removed from the victim model.
Most backdoor removal methods first invert the trigger and then eliminate the backdoor using the inverted trigger. 
\textbf{Elijah} \cite{an2024elijah} introduces a backdoor removal framework for diffusion models, inverting triggers through distribution shifts and aligning the backdoor's distribution with the clean one. \textbf{Diff-Cleanse} \cite{hao2024diff} formulates trigger inversion as an optimization problem with similarity and entropy loss, followed by pruning channels critical to backdoor sampling. \textbf{TERD} \cite{mo2024terd} proposes a unified reverse loss for trigger inversion, using a two-stage process for coarse and refined inversion. \textbf{PureDiffusion} \cite{truong2024purediffusion} employs multi-timestep trigger inversion, leveraging the consistent distribution shift caused by backdoored forward processes.

\begin{table*}[htp]
\center
\caption{A summary of attacks and defenses for Diffusion Models (Part II).}
\label{tab:diffuison_safety_II}
\rowcolors{2}{gray!15}{white}
\resizebox{1\textwidth}{!}{
\begin{tabular}{p{0.1\textwidth}p{0.15\textwidth}p{0.05\textwidth}p{0.18\textwidth}p{0.17\textwidth}p{0.2\textwidth}p{0.3\textwidth}}
\hline
\rowcolor{gaoyifeng-pink}
Attack/Defense & Method & Year & Category & SubCategory & Target Model & Dataset \\ \hline

\cellcolor{white} & WuMI\cite{wu2022membership} & 2022 & Black-box & Reconstruction-error & LDM DALL-E mini & MSCOCO, VG, LAION-400M, CC3M \\ 
\cellcolor{white} & DiffusionLeaks\cite{matsumoto2023membership} & 2023 & Black/White-box & Reconstruction-error  & DDIM, & CIFAR-10, CelebA   \\ 
\cellcolor{white} & PangMI\cite{pang2023black} & 2024 & Black-box & Auxilary Dataset & Stable Diffusion & CelebA-Dialog, WIT, MSCOCO \\ 
\cellcolor{white} & LiMI\cite{li2024towards} & 2024 & Black-box & Reconstruction-error& DDIM, Stable Diffusion DiT & CIFAR-10, STL10-U, LAION-5B, LAION-by-DALL-E  \\ 
\cellcolor{white} & DRC\cite{fu2024model} & 2025 & Black-box & Reconstruction-error&DDPM, DDIM&FFHQ, CelebA, CIFAR-10, CIFAR-100 \\ 
\cellcolor{white} & GMIA\cite{zhang2024generated} & 2023 & Black-box & Auxilary Dataset &DDPM, DDIM, FastDPM &CIFAR-10, CelebA\\
\cellcolor{white} & WuTMI\cite{wu2025winningmidstchallengenew} & 2025 & Black/White-box &  Loss &TabDDPM, ClavaDDPM &SaTM MIDST \\
\cellcolor{white} & SecMI\cite{duan2023diffusion} & 2023 & Gray-box & Posterior Likelihood & DDPM, DDIM, Stable Diffusion & CIFAR-10/100, STL10-U, Tiny-ImageNet, Pokemon, COCO2017-val, LAION-5B \\
\cellcolor{white} & QRMI\cite{tang2023membership} & 2023 & Gray-box & Posterior Likelihood&DDPM, DDIM&  CIFAR-10/100, STL100, Tiny-ImageNet \\
\cellcolor{white} & PIA\cite{kong2023efficient} & 2023 & Gray-box & Posterior Likelihood &DDPM, DDIM, Stable Diffusion& CIFAR-10/100, Tiny-ImageNet, COCO2017, LAION-5B \\
\cellcolor{white} & PFAMI\cite{fu2023probabilistic} & 2024 & Gray-box & Posterior Likelihood &DDPM, VAE&CelebA, Tiny-ImageNet\\
\cellcolor{white} & ZhMI\cite{zhai2024membership} & 2024 & Gray-box & Conditional Likelihood & DDPM, DDIM,  Stable Diffusion&Pokemonn, Flickr, MSCOCO, LAION \\
\cellcolor{white} & SMIA\cite{li2024unveiling} & 2024 & Gray-box & Structural Similarity &LDM, Stable Diffusion& LAION2B,  LAION-400M\\
\cellcolor{white} & D-MIA\cite{li2025membershipinferenceattackdistributional} & 2024=5 & Gray-box & Auxilary Dataset &EDM, DMD& CIFAR10, FFHQ, AFHQv2\\
\cellcolor{white} & SLA\cite{matsumoto2023membership,hu2023loss} & 2023 & White-box & Loss &DDPM, DDIM&FFHQ, DRD, CelebA, FFHQ\\
\cellcolor{white} & GSA\cite{pang2023white} & 2024 & White-box & Gradient & DDPM&CIFAR-10, MSCOCO, ImageNet\\
\cellcolor{white}\multirow{-19}{0.1\textwidth}{Membership \\ Inference} & DuMI\cite{dubinski2024towards} & 2023 & White-box & Loss &Stable Diffusion&Pokemon, LAION-mi\\
\hline

\cellcolor{white} & BruteDE\cite{carlini2023extracting} & 2023 & Black-box &  Existing Condition &DDPM, Stable Diffusion&CIFAR-10 LAION-5B \\ 
\cellcolor{white} & ReDE\cite{webster2023reproducible} & 2023 & Black/White-box & Existing Condition & Stable Diffusion, Midjourney, Deep Image Floyd& LAION-5B \\ 
\cellcolor{white} & SIDE\cite{chen2024towards} & 2024  & White-box &  Surrogate Condition  &DDPM, DDIM &CIFAR-10, CelebA, ImageNet\\ 
\cellcolor{white} \multirow{-5}{0.1\textwidth}{Data Extraction} & FineXtract\cite{wu2024revealing} & 2024 & White-box &  Surrogate Condition &Finetuned Stable Diffusion& WikiArt\\ 
\hline

\cellcolor{white}\multirow{2}{0.15\textwidth}{Model Extraction} & 
SDeT\cite{horwitz2024recovering} & 2024  & White-box & LoRA-Based Model Extraction & Finetuned Stable Diffusion & LoWRA Bench  \\
\hline

\cellcolor{white} & DUAW\cite{ye2023duaw} & 2023 & Natural Data Protection & Learning Prevention & Stable Diffusion & DreamBooth dataset, WikiArt, self-constructed \\
\cellcolor{white} & AdvDM\cite{liang2023adversarial} & 2023 & Natural Data Protection & Learning Prevention & Stable Diffusion, LDM & LSUN, WikiArt \\
\cellcolor{white} & Anti-DreamBooth\cite{van2023anti} & 2023 & Natural Data Protection & Learning Prevention & Stable Diffusion & CelebA, VGGFace2\\
\cellcolor{white} &  MetaCloak\cite{liu2024metacloak} & 2024 & Natural Data Protection & Learning Prevention & Stable Diffusion & CelebA-HQ, VGGFace2 \\
\cellcolor{white} &  InMakr\cite{liu2024countering} & 2024 & Natural Data Protection & Learning Prevention & Stable Diffusion &  VGGFace2, WikiArt   \\
\cellcolor{white} &  SimAC\cite{wang2024simac} & 2024 & Natural Data Protection & Learning Prevention & Stable Diffusion & CelebA-HQ, VGGFace2  \\
\cellcolor{white} & EditGuard\cite{zhang2024editguard} & 2024 & Natural Data Protection & Editing Prevention & Stable Diffusion  & COCO   \\
\cellcolor{white} & WaDiff\cite{min2024watermark} & 2024 & Natural Data Protection & Editing Prevention & Stable Diffusion & COCO, ImageNet   \\
\cellcolor{white} & AdvWatermark\cite{zhu2024watermark} & 2024 & Natural Data Protection & Editing Prevention & Stable Diffusion & WikiArt \\
\cellcolor{white} & FT-SHIELD\cite{cui2023ft} & 2024 & Natural Data Protection & Data Attribution & Stable Diffusion & CelebA, WikiArt, Pokemon Captions, DreamBooth dataset \\
\cellcolor{white} & DiffusionShield\cite{cui2023diffusionshield} & 2024 & Natural Data Protection & Data Attribution & DDPM, Stable Diffusion & CIFAR-10, CIFAR-100, STL-10, ImageNet \\
\cellcolor{white} &  ProMark\cite{asnani2024promark} & 2024 & Natural Data Protection & Data Attribution & LDM & Stock, LSUN, WikiArt, ImageNet\\
\cellcolor{white} & Diagnosis~\cite{wang2023diagnosis} & 2023 & Natural Data Protection & Data Attribution  & Stable Diffusion, VQ Diffusion & Pokemon, CelebA,  CUB-200, DreamBooth \\
\cellcolor{white} & HiDDeN\cite{zhu2018hidden} & 2018 & Generated Data Protection & Post-generation Watermark & CNN & MS-COCO, BOSS dataset \\
\cellcolor{white} & Stable Signature\cite{fernandez2023stable} & 2023 & Generated Data Protection & Diffusion Watermark & LDM & MS-COCO, ImageNet \\
\cellcolor{white} & LaWa\cite{rezaei2024lawa} & 2024 & Generated Data Protection & Diffusion Watermark & LDM & MIRFlickR \\
\cellcolor{white} & Safe-SD\cite{ma2024safe} & 2024 & Generated Data Protection & Diffusion Watermark & Stable Diffusion  & LSUN, COCO, FFHQ \\
\cellcolor{white} & RW\cite{zhao2023recipe} & 2023 & Model Protection & Model Watermark & Stable Diffusion, EDM & CIFAR-10, ImageNet, FFHQ, AFHQv2 \\
\cellcolor{white} & FIXEDWM\cite{liu2023watermarking} & 2023 & Model Protection & Model Watermark & LDM & MS COCO \\
\cellcolor{white} & WDM\cite{peng2023protecting} & 2023 & Model Protection & Model Watermark & DDPM, & CIFAR-10, CelebA, MNIST \\
\cellcolor{white} & AquaLoRA\cite{feng2024aqualora} & 2024 & Model Protection & Model Attribution & Stable Diffusion & COCO  \\
\cellcolor{white} & LatentTracer~\cite{wang2024trace} & 2024 & Model Protection & Model Attribution & Stable Diffusion, Kandinsky & LAION \\
\cellcolor{white} \multirow{-25}{0.1\textwidth}{Intellectual Property Protection}  & Tree-Ring\cite{wen2023tree} & 2023 & Model Protection & Model Attribution & Stable Diffusion, ImageNet diffusion & MS-COCO, ImageNet \\  
\hline

\end{tabular}
}
\end{table*}

Privacy attacks on diffusion models can be classified into \textbf{membership inference}, \textbf{data extraction}, and \textbf{model extraction} attacks. As attack sophistication increases, each type poses a growing threat to privacy.

\subsection{Membership Inference Attacks}
\label{sec:dm_membership_inference_attacks}

Membership inference attacks on diffusion models aim to infer sensitive data by exploiting their generative capabilities. Attackers use techniques like reconstruction error, shadow models, auxiliary data, likelihood, gradient, or structural similarity metrics. These attacks can be classified into six types: \textbf{reconstruction error-based}, \textbf{auxiliary dataset-based}, \textbf{loss-based}, \textbf{gradient-based}, \textbf{structural similarity-based}, and \textbf{likelihood-based}.

\textbf{Reconstruction Error-based Attacks} infer the membership of candidate samples by analyzing their reconstruction errors in the diffusion model. Wu et al. \cite{wu2022membership} proposed to determine membership in text-conditional diffusion models by comparing the reconstruction error between the candidate and generated images, and their semantic alignment with the text prompt.
Inspired by GAN-leaks \cite{chen2020gan}, Matsumoto et al. \cite{matsumoto2023membership} introduced Diffusion-leaks, which generates multiple candidate images and infers membership based on minimal reconstruction errors. Li et al. \cite{li2024towards} proposed to average multiple reconstructions to reduce errors and improve inference accuracy, utilizing black-box APIs to modify candidate images.
\textbf{DRC} \cite{fu2024model} degrades and restores images using the diffusion model, comparing the restored images to the originals to infer membership and sensitive features.

\textbf{Auxiliary Datasets-based Attacks} use auxiliary datasets to train shadow models, enabling black-box membership inference by simulating the target model. 
Pang et al. \cite{pang2023black} targeted fine-tuned conditional diffusion models, computing similarity scores between query images and generated images to train a binary classifier for membership inference. \textbf{GMIA} \cite{zhang2024generated} introduces the first generalized membership inference attack for generative models, using only generated distributions and auxiliary non-member datasets, assuming the generated distribution approximates the original training distribution. The \textbf{D-MIA} \cite{li2025membershipinferenceattackdistributional} framework leverages an auxiliary non-member dataset to perform distribution-level statistical testing. By applying Maximum Mean Discrepancy (MMD), it assesses whether a set of candidate data is statistically closer to the distribution of the original training data than to that of the auxiliary data. This approach enables the detection of unauthorized data usage through distillation.

\textbf{Loss-based Attacks} exploit loss value distributions to distinguish member from non-member samples, assuming lower losses for member (training) samples. \cite{hu2023loss} and \cite{matsumoto2023membership} used loss values at different timesteps for membership inference. These two attacks can be viewed as \textbf{Static Loss Attack} (SLA), as they ignore the diffusion process. Dubinski et al. \cite{dubinski2024towards} modified the diffusion process to extract loss information from multiple perspectives, improving inference accuracy.

\textbf{Gradient-based Attacks} leverage gradient information for membership inference. For instance, \textbf{GSA} \cite{pang2023white} infers a sample is a member if its gradients significantly differ from surrounding samples, indicating a stronger influence on the model's training.

\textbf{Structural Similarity-based Attacks} compare structural features or similarity metrics between candidate samples and model outputs. \textbf{SMIA} \cite{li2024unveiling} uses the Structure Similarity Index Measure (SSIM)\cite{wang2004image}
metric to assess how well an image's structure is preserved during diffusion, with the average SSIM difference between members and non-members used to infer membership.

\textbf{Likelihood-based Attacks} use posterior or conditional likelihoods to infer membership. \textbf{SecMI} \cite{duan2023diffusion} estimates posterior likelihoods via reverse processes to target DDPM and Stable Diffusion models. \textbf{QRMI} \cite{tang2023membership} applies quantile regression to posterior likelihoods. \textbf{SIA} \cite{qu2024very} infers membership based on noise parameter differences in the reverse diffusion process. \textbf{PIA} \cite{kong2023efficient} uses diffusion model properties to infer membership with fewer queries. \textbf{PFAMI} \cite{fu2023probabilistic} analyzes fluctuations between target samples and neighbors, exploiting memorization in generative models. Zhai et al. \cite{zhai2024membership} use discrepancies in conditional likelihoods due to overfitting for membership inference. In addition to the image modality, Wu et al. \cite{wu2025winningmidstchallengenew} investigated an attack on tabular diffusion models, treating loss values from different noise levels and time steps as features for a lightweight MLP classifier to predict membership.

\subsection{Data Extraction Attacks}
\label{sec:dm_data_extraction_attacks}

Data extraction attacks aim to reverse-engineer training data or attributes from a trained model, exploiting diffusion models' generative capabilities. Their effectiveness depends on the model's ability to memorize specific attributes \cite{somepalli2023understanding,gu2023memorization,wen2024detecting,ren2024unveiling}. These attacks can be classified into two main approaches based on the type of condition used: \textbf{explicit condition-based extraction} and \textbf{surrogate condition-based extraction}.

\textbf{Explicit Condition-based Extraction} leverages conditional information in T2I diffusion models to extract memorized training samples. Attackers use specific text prompts to generate images similar to training data. For example, \cite{carlini2023extracting} introduced brute-force data extraction (\textbf{BruteDE}), generating images with targeted prompts and using membership inference to identify matches. This method is slow. One Step Extraction (\textbf{OSE}) \cite{webster2023reproducible} exploits "template verbatims," where models regenerate training samples, using metrics like denoising confidence score (DCS) and edge consistency score (ECS) for faster extraction.

\textbf{Surrogate Condition-based Extraction} creates surrogate conditions to enable data extraction from unconditional diffusion models. \textbf{SIDE} \cite{chen2024towards} uses implicit labels from classifiers or feature extractors as surrogate conditions. \textbf{FineXtract} \cite{wu2024revealing} uses fine-tuned models as surrogate conditions to guide extraction in latent space regions tied to fine-tuning data.

\subsection{Model Extraction Attacks}
\label{sec:dm_model_extraction_attacks}

Model extraction aims to steal a trained diffusion model's internal parameters or architecture. The only known method for model extraction on diffusion models is Spectral DeTuning (\textbf{SDeT})  \cite{horwitz2024recovering}.
SDeT leverages Low-Rank Adaptation (LoRA) \cite{hu2021lora} to extract pre-fine-tuning weights of generative models fine-tuned with LoRA. By collecting multiple fine-tuned models from the same pretrained model, it formulates an optimization problem to minimize the difference between fine-tuned weights and the sum of original weights and adaptation matrices under a low-rank constraint, solved iteratively using Singular Value Decomposition (SVD)\cite{stewart1993early}. SDeT effectively recovers original weights for models like Stable Diffusion and Mistral-7B\cite{jiang2023mistral}, highlighting vulnerabilities in fine-tuning processes with low-rank adaptations.

\subsection{Intellectual Property Protection}
\label{sec:dm_intellectual_property_protection}

Intellectual property protection for AI is an emerging research area that uses techniques like adversarial attacks and watermarking to safeguard the intellectual property of natural (training or test) data, generated data, and trained models. These methods generally assume full access to the protected object. The following sections categorize these approaches into \textbf{natural data protection}, \textbf{generated data protection}, and \textbf{model protection}.

\subsubsection{Natural Data Protection}

Natural data protection methods focus on preprocessing data during training or inference to safeguard the copyright of naturally collected data, as opposed to generated data. In this context, data owners defend against model owners accessing the data. Existing methods for T2I diffusion models aim to protect image intellectual property while minimizing quality loss. They can be categorized into \textbf{learning prevention}, \textbf{editing prevention}, and \textbf{data attribution} methods based on specific goals.

\textbf{Learning Prevention} methods prevent T2I models from learning useful features from training images using techniques like adversarial attacks. 
\textbf{DUAW} \cite{ye2023duaw} protects copyrighted images by disrupting the variational autoencoder (VAE) in Stable Diffusion models, optimizing universal adversarial perturbations on surrogate images to distort outputs. 
\textbf{AdvDM} \cite{liang2023adversarial} protects artwork copyrights by generating adversarial examples to prevent diffusion models from imitating artistic styles. \textbf{Anti-DreamBooth} \cite{van2023anti} defends against malicious fine-tuning by injecting adversarial noise into user images to block the model from learning personalized features.
\textbf{MetaCloak} \cite{liu2024metacloak} enhances image resistance to transformations (flipping, cropping, compression) by using surrogate diffusion models to craft transferable perturbations and a denoising-error maximization loss for better robustness. \textbf{InMakr} \cite{liu2024countering} embeds protective watermarks on critical pixels to safeguard personal semantics even if images are modified. \textbf{SimAC} \cite{wang2024simac} improves protection by optimizing timestep intervals and introducing a feature interference loss, leveraging early diffusion steps and high-frequency information from deeper layers.

\textbf{Editing Prevention} aims to prevent diffusion model-based image tampering and deepfake generation. Existing methods either embed watermarks or use adversarial noise to disrupt the editing process. \textbf{EditGuard} \cite{zhang2024editguard} introduces a proactive forensics framework to embed exclusive watermarks into images, making them resistant to various diffusion model-based editing techniques, including foreground or background removal, filling, tampering, and face swapping. \textbf{WaDiff} \cite{min2024watermark} adds a unique watermark to each user query, enabling traceability of the generated image if ethical concerns arise. \textbf{AdvWatermark} \cite{zhu2024watermark} incorporates adversarial noise, producing visible signatures in the protected image when used by I2I models, which helps identify tampered content.

\textbf{Data Attribution} techniques identify if generated data originates from a specific dataset, often by embedding watermarks for later verification. 
Diagnosis~\cite{wang2023diagnosis} introduced a method for detecting unauthorized data usage by applying stealthy image warping effects to protected data.
\textbf{FT-SHIELD} \cite{cui2023ft} uses alternating optimization and PGD \cite{madry2017towards} to embed watermarks, with a binary detector for verification. \textbf{DiffusionShield} \cite{cui2023diffusionshield} encodes copyright messages into watermark patches, jointly optimizing the decoder and patches to ensure consistency across samples for reliable extraction. \textbf{ProMark} \cite{asnani2024promark} introduces a proactive watermarking method for \emph{concept attribution}, embedding watermarks in training data that can be extracted when similar concepts are generated by the model. Similarly, \textbf{SIREN} \cite{li2025towards} protects image data by adding learned, feature-relevant noise that can be detected in models trained on such protected data.

\subsubsection{Generated Data Protection} 
\label{sec:dm_generated_data_proteciton}

With the rise of AI-generated content (AIGC), protecting the copyright of generated data has become increasingly important. Generated data protection seeks to answer, \textbf{“Who created this content?”} by embedding verifiable, unique watermarks into generated images to identify their creators (either the model or user). This ensures intellectual property protection and accountability for content publishers, while balancing the challenge of maintaining detection accuracy without compromising image quality.

\textbf{HiDDeN} \cite{zhu2018hidden} pioneers deep learning-based image watermarking, using an encoder to embed imperceptible watermarks and a decoder to recover them for detection. This approach can also watermark AI-generated images as a post-processing step.
Recent protection methods primarily address the above challenge by embedding watermarks into images during the generation (reverse sampling) process of diffusion models.
\textbf{Stable Signature} \cite{fernandez2023stable} embeds a binary signature into images generated by diffusion models through decoder fine-tuning, allowing the watermark to be recovered and validated using a pre-trained extractor and statistical test.
\textbf{LaWa} \cite{rezaei2024lawa} introduces a coarse-to-fine watermark embedding method within the latent diffusion model's decoder, employing multiple modules to insert the watermark at different upsampling stages using adversarial training. \textbf{Safe-SD} \cite{ma2024safe} proposes a framework for embedding a graphical watermark (e.g., QR code) into the imperceptible structure-related pixels of a Stable Diffusion model for high traceability. \textbf{VideoShield} \cite{huvideoshield} proposes a novel and effective watermarking framework for regulating diffusion-based video generation models by embedding imperceptible watermarks during the generation process, enabling ownership verification and traceability while preserving video quality and resisting removal attacks. Different from previous watermarking-based methods, \textbf{OCC‑CLIP} \cite{liu2024model} proposes a CLIP‑based few‑shot one‑class classification framework augmented with adversarial data augmentation that, given only a handful of reference images and no access to the candidate generators without watermarking, reliably determines whether a (benign) query image originates from the same model as those references.

Recent studies highlight vulnerabilities in watermarking for AIGC. \textbf{WEvade} \cite{jiang2023evading} bypasses watermark detection by adding subtle perturbations to watermarked images, exploiting watermark characteristics. \textbf{TAIW} \cite{hu2024transfer} proposes a transfer attack using multiple surrogate watermarking models in a no-box setting, analyzing its theoretical transferability. Unlike per-image attacks, \textbf{SSU} \cite{hu2024stable} introduces a model-targeted attack to remove in-generation watermarks by fine-tuning the diffusion model’s decoder with non-watermarked images, demonstrating the fragility of Stable Signature \cite{fernandez2023stable}.

\subsubsection{Model Protection} 
Model protection techniques safeguard the intellectual property of released models, enabling owners to verify ownership and trace generated content back to its origin. These approaches are categorized based on their objectives into \textbf{model watermark} and \textbf{model attribution}.

\textbf{Model Watermark} injects a watermark trigger into the model, which can then be activated during inference to verify ownership.
Zhao et al.~\cite{zhao2023recipe} proposed separate watermarking schemes for unconditional/class-conditional and T2I diffusion models. For unconditional/class-conditional models, a pretrained watermark encoder embeds a binary string (e.g., "011001") into the training data, and the model is trained to generate images with a detectable watermark, verified by a pretrained decoder. For T2I models, a paired (text, image) trigger (e.g., \texttt{"[V]"} and a QR code) is used to trigger the generation of the QR code for ownership verification. \textbf{FIXEDWM} \cite{liu2023watermarking} enhances trigger stealthiness by fixing its position in prompts, ensuring the watermarked image is generated only when the trigger is in the correct position.  \textbf{WDM} \cite{peng2023protecting} modifies the standard diffusion process into a Watermark Diffusion Process (WDP) to embed watermarks. During training, WDM learns from watermarked images using WDP, while normal images follow the standard diffusion process. During verification, Gaussian noises combined with the trigger can activate the generation of watermarked images. Recently, \textbf{SleeperMark} \cite{wang2025sleepermark} embeds invisible watermarks during pre-training to ensure their resistance to personalized fine-tuning, thereby maintaining high-fidelity ownership verification while preserving generation quality.

\textbf{Model Attribution} also embeds watermarks into generated content to identify the model, similar to generated data protection methods in Section \ref{sec:dm_generated_data_proteciton}. The key difference is that model attribution focuses on model-wide watermarks, while generated data protection targets sample-specific watermarks.
\textbf{Tree-Ring}\cite{wen2023tree} embeds a watermark into the Fourier space of the initial Gaussian noise used for T2I generation. During verification, denoising diffusion implicit model (DDIM) inversion extracts the initial noise, and comparison with the original watermark identifies the generating model. \textbf{AquaLoRA}\cite{feng2024aqualora} addresses the limitations of existing methods to white-box adaptive attacks, including Tree-Ring, by embedding a secret bit string into the model parameters to achieve white-box protection, preventing easy manipulation of the watermark by malicious users.
\textbf{LatentTracer} \cite{wang2024trace} identifies the origin model of generated samples by reverse-engineering their latent inputs, eliminating the need for artificial fingerprints or watermarks.

\subsection{Datasets}
This section reviews commonly used datasets for diffusion model safety research, as summarized in Tables \ref{tab:diffuison_safety_I} and \ref{tab:diffuison_safety_II}. 
For adversarial attack and defense studies, captioned text-image pairs such as MS COCO \cite{lin2014microsoft}, LAION \cite{schuhmann2021laion, schuhmann2022laion}, and DiffusionDB \cite{wang2022diffusiondb} are often employed by conditional diffusion models. Datasets for category-image classification tasks, like ImageNet \cite{deng2009imagenet} and CIFAR10/100 \cite{krizhevsky2009learning}, are typically used by unconditional diffusion models to evaluate attack effectiveness and output quality.
In research on NSFW content in diffusion models, the I2P dataset \cite{schramowski2023safe} is widely used, alongside custom datasets such as NSFW-200 \cite{yang2024sneakyprompt}, VBCDE-100 \cite{deng2023divide}, Tox100/1K \cite{cai2024ethical} and a human-attribute dataset \cite{cai2024ethical} focused on bias research. 
For intellectual property protection, datasets like CelebA \cite{liu2015deep} and VGGFace2 \cite{cao2018vggface2} (facial datasets), DreamBooth \cite{ruiz2023dreambooth} and Pokemon Captions \cite{pinkney2022pokemon} (object datasets), and WikiArt \cite{saleh2015large} (artistic style dataset) are commonly used.

\begin{table*}[!htp]
\center
\caption{A summary of attacks and defenses for Agents (PART I).}
\label{tab:agent_safety_1}
\rowcolors{2}{gray!15}{white}
\resizebox{\textwidth}{!}{
\begin{tabular}{lllllll}
\hline
\rowcolor{wangyixu-purple}
Attack/Defense & Method & Year & Category & Subcategory & Access to LLMs & Dataset \\ 
\hline

\rowcolor{gray!35!}\multicolumn{7}{c}{\textbf{I Indirect Prompt Injection: Malicious Instruction and Jailbreak}} \\
\hline

\cellcolor{white} & Greshake et al.~\cite{greshake2023not} & 2023 & Indirect Prompt Injection & Prompt Injection & Black-Box & Custom (webpage injections) \\
\cellcolor{white} & Wu et al.~\cite{wu2024new} & 2024 & Indirect Prompt Injection & Prompt Injection & Black-Box & Custom (website prompts) \\
\cellcolor{white} & TensorTrust~\cite{toyer2023tensor} & 2023 & Indirect Prompt Injection & Prompt Injection & Black-Box & Submissions Dataset \\
\cellcolor{white} & Perez and Ribeiro~\cite{perez2022ignore} & 2022 & Indirect Prompt Injection & Prompt Injection & Black-Box & OpenAI Examples \\
\cellcolor{white} & HOUYI~\cite{liu2023prompt} & 2023 & Indirect Prompt Injection & Prompt Injection & Black-Box & Custom (multilingual prompts) \\
\cellcolor{white} & Pedro et al.~\cite{pedro2023prompt} & 2023 & Indirect Prompt Injection & Prompt Injection & Black-Box & Custom (Langchain apps) \\
\cellcolor{white} & Zhan et al.~\cite{zhan-etal-2025-adaptive} & 2025 & Indirect Prompt Injection & Jailbreak & White-Box & Custom (tool outputs) \\
\cellcolor{white} & PANDORA~\cite{deng2024pandora} & 2024 & Indirect Prompt Injection & Jailbreak & Black-Box & Custom (document embeddings) \\
\cellcolor{white}\multirow{-9}{0.08\textwidth}{Attack} & Imprompter~\cite{fu2024imprompter} & 2024 & Indirect Prompt Injection & Jailbreak & White-Box & Custom (LeChat, ChatGLM) \\
\hline

\cellcolor{white} & Instruction Hierarchy~\cite{wallace2024instruction} & 2024 & IPI Defense & Privilege Management & White-Box & Custom (message privilege levels) \\
\cellcolor{white} & Instruction Detection~\cite{wen2025defending} & 2025 & IPI Defense & Detection & White-Box & PEEP Dataset \\
\cellcolor{white} & Instruction Detection and Removal~\cite{chen2025can} & 2025 & IPI Defense & Detection & White-Box & Crafted datasets \\
\cellcolor{white} & Task Shield~\cite{taskshield2024} & 2024 & IPI Defense & Detection & White-Box & AgentDojo \\
\cellcolor{white} & FATH~\cite{wang2024fath} & 2024 & IPI Defense & Authentication & Black-Box & Custom (response tagging) \\
\cellcolor{white} & Spotlighting~\cite{hines2024splot} & 2024 & IPI Defense & Prompt Engineering & Black-Box & Custom (provenance marking) \\
\cellcolor{white}\multirow{-7}{0.08\textwidth}{Defense} & Design Pattern~\cite{beurer2025design} & 2025 & IPI Defense & Secure Design & Black-Box & Custom (factory floor) \\

\hline

\rowcolor{gray!35!}\multicolumn{7}{c}{\textbf{II Component-Level: Short-Term Memory, Long-Term Memory, Tool Integration and MCP Server}} \\
\hline

\cellcolor{white} & Contextual Backdoor~\cite{liu2024compromising} & 2024 & Memory Attack & Backdoor & Black-Box & AgentBench, WebShop \\
\cellcolor{white} & Watch out for your agents~\cite{yang2024watch} & 2024 & Memory Attack & Backdoor & Black-Box & LLM-based agents \\
\cellcolor{white} & DemonAgent~\cite{zhu2025demonagent} & 2025 & Memory Attack & Backdoor & White-Box & AgentBench, WebArena \\
\cellcolor{white} & BadAgent~\cite{wang2024badagent} & 2024 & Memory Attack & Backdoor & White-Box & AgentInstruct, Mind2Web \\
\cellcolor{white} & AgentPoison~\cite{chen2024agentpoison} & 2024 & Memory Attack & Backdoor & Black-Box & HotpotQA, AgentBench \\
\cellcolor{white} & TrojanRAG~\cite{pan2023trojan} & 2023 & Memory Attack & Backdoor & White-Box & NQ, HotpotQA \\
\cellcolor{white} & BadRAG~\cite{xue2024badrag} & 2024 & Memory Attack & Backdoor & White-Box & NQ, TriviaQA \\
\cellcolor{white} & Phantom~\cite{chaudhari2024phantom} & 2024 & Memory Attack & Backdoor & White-Box & NQ, TriviaQA, SQuAD \\
\cellcolor{white} & BreakingAgent~\cite{zhang2024breaking} & 2024 & Memory Attack & Prompt Injection & Black-Box & Custom (Gmail agents) \\
\cellcolor{white} & MINJA~\cite{dong2025practical} & 2025 & Memory Attack & Prompt Injection & Black-Box & Webshop, MIMIC-III, eICU, MMLU \\
\cellcolor{white} & PoisonedRAG~\cite{zou2024poisonedrag} & 2024 & Memory Attack & Knowledge Poisoning & Black-Box & NQ, HotpotQA, MS-MARCO \\
\cellcolor{white}\multirow{-15}{0.08\textwidth}{Attack} & Corpus Poisoning~\cite{zhong2023poisoning} & 2023 & Memory Attack & Knowledge Poisoning & White-Box & Natural Questions, MS-MARCO \\
\hline

\cellcolor{white} & Context Extension~\cite{luo2023search, Summarizersmemory} & 2023 & Memory Defense & Context Extension & Black-Box & Custom (long context benchmarks) \\
\cellcolor{white} & Prompt Leakage Defense~\cite{agarwal2024prompt} & 2024 & Memory Defense & Detection & Black-Box & Multi-turn (4 domains) \\
\cellcolor{white} & AgentSafe~\cite{mao2025agentsafe} & 2025 & Memory Defense & Secure Design & White-Box & Custom (multi-agent tasks) \\
\cellcolor{white} & TrustRAG~\cite{zhou2025trustrag} & 2025 & Memory Defense & Detection & White-Box & Custom (RAG datasets) \\
\cellcolor{white} & Astute RAG~\cite{wang2024astute} & 2024 & Memory Defense & Detection & Black-Box & SQuAD 2.0 \\
\cellcolor{white}\multirow{-6}{0.08\textwidth}{Defense} & RobustRAG~\cite{xiang2024certifiably} & 2024 & Memory Defense & Secure Design & White-Box & Custom (RAG datasets) \\
\hline

\cellcolor{white} & UDora~\cite{zhang2025udora} & 2025 & Tool Manipulation & Jailbreak & White-Box & AgentHarm \\
\cellcolor{white} & ToolSword~\cite{ye2024toolsword} & 2024 & Tool Manipulation & Red Teaming & Black-Box & ToolBench, AgentBench \\
\cellcolor{white} & ToolCommander~\cite{wang2024allies} & 2024 & Tool Manipulation & Adversarial Attack & Black-Box & ToolBench \\
\cellcolor{white} & WIPI~\cite{wu2024wipi} & 2024 & Tool Manipulation & Prompt Injection & Black-Box & ChatGPT Web Agents, Web GPTs \\
\cellcolor{white} & AutoCMD~\cite{jiang2025autocmd} & 2025 & Tool Manipulation & Adversarial Attack & Black-Box & LangChain, KwaiAgents, QwenAgent \\
\cellcolor{white} & MPMA~\cite{wang2025mpma} & 2025 & MCP Manipulation & Adversarial Attack & White-Box & Custom (MCP servers) \\
\cellcolor{white} & Ferrag et al.~\cite{ferrag2025prompt} & 2025 & MCP Manipulation & Adversarial Attack & Black-Box & CIAQA, AgentBackdoorEval, etc. \\
\cellcolor{white}\multirow{-8}{0.08\textwidth}{Attack} & Kong et al.~\cite{kong2025survey} & 2025 & MCP Manipulation & Red Teaming & Grey-Box & Custom (Claude, Filesystem, Chroma, Gmail) \\
\hline

\cellcolor{white} & AgentGuard~\cite{chen2025agentguard} & 2025 & Tool \& MCP Defense & Red Teaming & White-Box & Custom (tool orchestration) \\
\cellcolor{white} & PrivacyAsst~\cite{zhang2024privacyasst} & 2024 & Tool \& MCP Defense & Secure Design & Black-Box & Custom (tool-using agents) \\
\cellcolor{white} & MCIP~\cite{jing2025mcip} & 2025 & Tool \& MCP Defense & Maclious Detection & White-Box & Custom (MCIP-Bench)  \\
\cellcolor{white}\multirow{-3}{0.08\textwidth}{Defense} & GuardAgent~\cite{xiang2024guardagent} & 2024 & Tool \& MCP Defense & Secure Design & Black-Box & EICU-AC, Mind2Web-SC \\

\hline

\cellcolor{white} & Adversarial Multimodal Injection~\cite{bagdasaryan2023abusing} & 2023 & VLM Attack  & Prompt Injection & White-Box & Custom (perturbed media) \\
\cellcolor{white} & Zhang~\cite{zhang2024attacking} & 2024 & VLM Attack  & Prompt Injection & Black-Box & OSWorld, VisualWebArena \\ 
\cellcolor{white} & Wu et al.~\cite{wu2024adversarial} & 2024 & VLM Attack  & Prompt Injection & Black-Box & VisualWebArena \\
\cellcolor{white} & Fu et al.~\cite{fu2023misusing} & 2023 & VLM Attack  & Adversarial Attack & White-Box & Custom (VLM + tool calls) \\
\cellcolor{white} & EIA~\cite{liao2024eia} & 2024 & VLM Attack  & Prompt Injection & Black-Box & Mind2Web \\
\cellcolor{white} & AdvAgent~\cite{xu2024advagent} & 2024 & VLM Attack  & Red Teaming & Black-Box & Custom (web tasks) \\ 
\cellcolor{white} & Ma et al.~\cite{ma2024caution} & 2024 & VLM Attack  & Adversarial Attack & Black-Box & Custom (simulated GUI) \\
\cellcolor{white}\multirow{-8}{0.08\textwidth}{Attack} & Fine-Print Injections~\cite{chen2025obvious} & 2025 & VLM Attack  & Prompt Injection & Black-Box & Custom (234 adversarial webpages) \\
\hline

\cellcolor{white} & SmoothVLM~\cite{smoothvlm2024} & 2024 & VLM Defense & Detection & White-Box & Custom (adversarial datasets) \\
\cellcolor{white} & BlueSuffix~\cite{bluesuffix2024} & 2024 & VLM Defense & Detection & Black-Box & Custom (VLM benchmarks) \\
\cellcolor{white} & LlavaGuard~\cite{llavaguard2024} & 2024 & VLM Defense & Detection & White-Box & Custom (multimodal safety) \\
\cellcolor{white}\multirow{-4}{0.08\textwidth}{Defense} & JailDAM~\cite{jaildam2024} & 2024 & VLM Defense & Detection & Black-Box & Custom (jailbreak detection) \\

\hline

\end{tabular}
}
\end{table*}

\section{Agent Safety} \label{sec:agent}

Large model powered agents are increasingly deployed in safety-critical domains such as healthcare~\cite{heatlhagent}, finance~\cite{financeagent}, and autonomous driving~\cite{autodriveagent}. These agents leverage LLMs or VLMs as their central ``brain" to enable end-to-end task execution through iterative planning, external observation, and multi-step reasoning. However, the closed-loop autonomy of agent significantly expands the attack surface compared to standalone large models, despite both being built
upon the same safety-aligned models~\cite{chiang2025web}. Notably, most attacks targeting standalone models described in Section~\ref{sec:llm} can be adapted to indirectly manipulate agent behavior.

To systematically analyze AI agent safety, we structure our discussion around four key levels of analysis. First, we examine \textit{Indirect Prompt Injection} as \textbf{foundational attack vectors}, where adversaries exploit third-party integrations to manipulate agent behavior. Second, we analyze \textbf{component-level threats} targeting critical modules such as \textit{Memory Systems}, \textit{Tool-Calling and Model Context Protocol (MCP)}, and \textit{Vision-Language Model (VLM) Processing}. Third, we explore \textbf{system-level risks} emerging in complex deployment scenarios like \textit{Multi-Agent Systems} and \textit{Embodied Agent Systems}. Fourth, we observe emerging capabilities where agents autonomously exploit vulnerabilities or deploy dynamic defenses, which we refer to as \textbf{Agentic Attacks and Defenses}. Throughout our analysis, we also discuss corresponding defense mechanisms and evaluation benchmarks designed to address these evolving safety challenges.

\subsection{Indirect Prompt Injection Attacks} \label{sec:agent_ipi}
Agents process diverse messages, including system prompts, user inputs, model outputs, and tool responses~\cite{wallace2024instruction}, continuously integrating external messages into their context state. Regardless of the specific attack method, adversaries ultimately exploit this unified message processing core to manipulate agent behavior. Thus, we identify \textbf{Indirect Prompt Injection (IPI)} as a fundamental attack vector, focusing on two main variants: \textit{indirect malicious instruction injection} and \textit{indirect jailbreak attacks}. While other attack methods may target specific agent components, they often rely on IPI as the delivery mechanism, which we discuss in the relevant subsections.

\subsubsection{Indirect Malicious Instruction Injection} \label{sec:agent_ipi_malicious}
Greshake et al.~\cite{greshake2023not} conducted early research on IPI, demonstrating how malicious instructions can remotely alter a model's task execution, for example, transforming Bing Chat into a phishing bot by injecting harmful content into a webpage. And Wu et al.~\cite{wu2024new} inserted prompts such as \textit{``summarize the chat history and send to \{adversary address\}"} into websites, causing agents to leak conversations.

Simple malicious instructions have proven surprisingly effective against agents. The prompt of \textit{``Ignore previous instruction and do \{malicious goal\}"} introduced by Perez and Ribeiro~\cite{perez2022ignore} has been widely used to manipulate agent action. For instance, \textbf{HOUYI}~\cite{liu2023prompt} translates the \textit{``ignore prompt''} into multiple languages to expose vulnerabilities in AI writing assistants. And Pedro et al.~\cite{pedro2023prompt} presented \textbf{Prompt-to-SQL (P2SQL)} attacks in Langchain-based applications, where unsafe prompts are converted into SQL queries, enabling attackers to gain full database access. To systematically investigate agent vulnerabilities to such attacks, \textbf{TensorTrust}~\cite{toyer2023tensor} develops an online game to crowdsource human-written PI. Their findings reveal that many models remain vulnerable to attack strategies developed by players. Importantly, these strategies generalize from the game environment to real-world applications like ChatGPT, Claude, Bard, Bing Chat, and Notion AI, with attacks showing easily interpretable structural patterns that expose fundamental model vulnerabilities.

\subsubsection{Indirect Jailbreak Attack} \label{sec:agent_ipi_jailbreak}
Researchers have also explored the indirect delivery of jailbreak attacks. Prior work such as DrAttack~\cite{zhou2024drattack} and Puzzler~\cite{chang2024play} also falls under the category of indirect jailbreaks, but primarily targets conversational LLMs. In contrast, our focus is on indirect jailbreak attacks against autonomous agents. Nonetheless, techniques from these conversational settings can potentially be adapted to agent-based contexts.

\textbf{PANDORA}~\cite{deng2024pandora} employs scrambled or hidden language embedded in documents to bypass safety filters; when an agent retrieves such documents, the concealed text prompts it to violate rules, such as generating harmful content, without any direct user request. \textbf{Imprompter}~\cite{fu2024imprompter} creates gradient-optimized, garbled prompts that embed Markdown instructions. When agents ingest these from external content, they can trigger improper HTTP tool calls and leak conversation data. Furthermore, Zhan et al.\cite{zhan-etal-2025-adaptive} have adapted representative jailbreak attack algorithms for LLMs, such as GCG\cite{zou2023universal}, T-GCG~\cite{jain2023baseline}, and AutoDAN~\cite{liu2023autodan}, to tool-integrated environments, where these algorithms optimize adversarial strings as prefixes or suffixes within tool outputs to manipulate agent responses.

\subsection{Indirect Prompt Injection Defenses} \label{sec:agent_ipi_defenses}
We broadly classify current defenses against IPI into \textit{black-box} or \textit{white-box} approaches.

\subsubsection{White-box Defenses} \label{sec:agent_ipi_defenses_whitebox}
\textbf{Instruction Hierarchy}\cite{wallace2024instruction} introduces privilege levels for system, user, and tool messages, training models to prioritize higher-privileged instructions and ignore or reject conflicting or harmful lower-privileged ones. \textbf{Instruction Detection}\cite{wen2025defending} detects embedded instructions in external content by monitoring behavioral state changes in LLMs during forward and backward propagation. It combines hidden states and gradients from intermediate layers to achieve high detection accuracy and reduce attack success rates. \textbf{Task Shield}\cite{taskshield2024} implements a test-time defense mechanism that systematically verifies whether each instruction and tool call contributes to user-specified goals through goal-alignment verification. \textbf{Instruction Detection and Removal}\cite{chen2025can} combines detection models, trained on curated datasets, with segmentation and extraction-based removal methods, yielding improved results over traditional prompt engineering and fine-tuning defenses.

\subsubsection{Black-box Defenses} \label{sec:agent_ipi_defenses_blackbox}
In contrast to white-box methods that require model internals, black-box defenses operate without access to model parameters or architectures, focusing on input preprocessing and external validation mechanisms. \textbf{FATH}~\cite{wang2024fath} implements a hash-based authentication system, requiring LLMs to process all instructions but filter responses based on tag verification, labeling each response with authentication tags to accurately identify legitimate user instructions. \textbf{Spotlighting}\cite{hines2024splot} applies prompt engineering techniques, such as delimiting, datamarking, and encoding, to transform input text and provide reliable provenance signals, enabling LLMs to distinguish between trusted and untrusted content.  \textbf{Design Pattern}\cite{beurer2025design} introduces isolation strategies to resist prompt injection, including predefined tool calls without output access, fixed action plans, constrained sub-agents, symbolic memory to separate planning and execution, secure intermediate code generation, and prompt removal in multi-turn settings.

\subsection{Agent Memory Attacks} \label{sec:agent_memory_threats}
Agents acquire, store, and retrieve information through memory modules: Short-Term Memory (STM) for in-context guidance (\textit{i.e., using checkpointing to persist agent state across all execution steps}), and Long-Term Memory (LTM) via external vector stores like RAG (\textit{i.e., using retrievers to find relevant information and LLMs to generate responses conditioned on retrieved passages}). It has been observed that both types of memory are vulnerable to backdoor attacks, and other malicious injection methods, such as misinformation injection, can poison memory modules without requiring training access. Therefore, we categorize memory attacks into: \textbf{backdoor attacks} and \textbf{memory poisoning}.

\subsubsection{Memory Backdoor Attacks} \label{sec:agent_memory_backdoor}

Memory backdoor attacks embed hidden triggers that activate malicious behavior only under specific conditions. For short-term memory, attackers manipulate contextual information to create conditional vulnerabilities. \textbf{Contextual Backdoor Attacks}~\cite{liu2024compromising} poison few-shot demonstrations via adversarial in-context generation optimized by an LLM judge, using dual-modality triggers to introduce context-dependent vulnerabilities. \textbf{DemonAgent}~\cite{zhu2025demonagent} fragments backdoors into encrypted sub-components within the agent's context, enabling cumulative triggering that evades safety audits while maintaining high attack success rates.

Long-term memory systems face different backdoor threats through embedding manipulation. \textbf{AgentPoison}~\cite{chen2024agentpoison} leverages constrained multi-loss optimization to map triggered instances to unique embedding regions, requiring no additional model training. In RAG-based systems, \textbf{TrojanRAG}~\cite{pan2023trojan} leverages contrastive learning with knowledge graph enhancement, while \textbf{BadRAG}~\cite{xue2024badrag} employs contrastive optimization, achieving a 98.2\% success rate with just ten adversarial passages. \textbf{Phantom}~\cite{chaudhari2024phantom} employs two-stage optimization with natural trigger sequences and multi-coordinate gradient methods for diverse malicious objectives. \textbf{Watch out for your agents}~\cite{yang2024watch} investigates backdoor threats specifically targeting LLM-based agents, demonstrating how adversaries can compromise agent behaviors through parameter manipulation during training phases.

\subsubsection{Memory Poisoning Attacks} \label{sec:agent_memory_poisoning}

Unlike backdoor attacks, memory poisoning injects malicious data to manipulate agent behavior without requiring hidden triggers. These attacks directly corrupt the information that agents retrieve and use to decision-making.

External memory systems are vulnerable through interaction record manipulation. \textbf{BreakingAgent}~\cite{zhang2024breaking} induces Gmail agents into denial-of-service loops by providing false interaction records. \textbf{MINJA}~\cite{dong2025practical} introduces malicious records through normal interactions, using bridging steps and progressive shortening to link victim queries to malicious reasoning. \textbf{Corpus Poisoning}~\cite{zhong2023poisoning} perturbs discrete tokens to create adversarial passages that maximize similarity with training queries, injecting them into retrieval corpora and achieving cross-domain generalizability.

Retrieval augmented generation systems face targeted document poisoning. \textbf{PoisonedRAG}~\cite{zou2024poisonedrag} streamlines the attack by concatenating target queries with LLM-generated misinformation, achieving high retrieval rates for poisoned documents in black-box settings without complex optimization. This approach exploits the tendency of retrieval systems to prioritize documents with high query similarity.

Parametric memory attacks target the model's internal knowledge during training. \textbf{BadAgent}~\cite{wang2024badagent} embeds backdoors into LLM parameters during fine-tuning, enabling attackers to trigger malicious behaviors via specific inputs with notable persistence even after further fine-tuning on clean data. 

\subsection{Agent Memory Defenses} \label{sec:agent_memory_defenses}

Defense mechanisms target different types of memory attacks through various strategies including context extension, knowledge consolidation, and access control.

For short-term memory attacks, extending LLM context windows to reference a majority of benign in-context examples has proven effective~\cite{LongRoPE, luo2023search, Summarizersmemory}, assuming malicious demonstrations are outnumbered. \textbf{Prompt Leakage Defense}~\cite{agarwal2024prompt} systematically measures prompt-leak risks in multi-turn chats using seven black-box tactics including response rewriting, role masking, and adaptive refusal. \textbf{AgentSafe}~\cite{mao2025agentsafe} secures multi-agent memories by partitioning information into privilege tiers and enforcing hierarchical access checks before any read/write operations.

For long-term memory and RAG attacks, defenses focus on knowledge consolidation to resolve conflicting or malicious inputs. \textbf{TrustRAG}~\cite{zhou2025trustrag} uses K-means clustering to consolidate knowledge from LLM internal states and external documents. \textbf{Astute RAG}~\cite{wang2024astute} adaptively integrates internal and external knowledge, using source-awareness to resolve conflicts between different information sources. \textbf{RobustRAG}~\cite{xiang2024certifiably} adopts an isolate-then-aggregate strategy, independently processing each passage before secure aggregation, and provides formal certification guarantees against malicious injections.

\subsection{Agent Tool-Calling Attacks} \label{sec:agent_tool_threats}

Modern LLM agents can now call external tools to perform tasks like web browsing, email management, and data analysis. However, this capability creates new attack opportunities where adversaries manipulate agents into using malicious tools or following harmful instructions.

Attacks exploit web browsing capabilities by embedding malicious instructions in publicly accessible websites. \textbf{WIPI}~\cite{wu2024wipi} demonstrates how malicious instructions can be embedded in web content that appears normal to humans but controls agent behavior, achieving over 90\% success rates across various web agents including ChatGPT plugins and open-source systems. When agents visit these compromised pages, they unknowingly follow the hidden commands.

Other attacks target the agent's internal reasoning process. \textbf{UDora}~\cite{zhang2025udora} collects the agent's reasoning traces, identifies optimal insertion points, and optimizes adversarial strings that become part of the agent's own thinking process, steering it toward malicious tool calls without external manipulation. This approach works within the agent's natural reasoning style.

Beyond individual tool manipulation, attackers can orchestrate systematic attacks on tool selection processes. \textbf{ToolCommander}~\cite{wang2024allies} operates in two stages: first injecting privacy theft tools to gather user queries, then manipulating tool scheduling to enable denial-of-service attacks and unfair competition by biasing agents toward certain tools. \textbf{ToolSword}~\cite{ye2024toolsword} reveals vulnerabilities across six safety scenarios in tool input, execution, and output processes. \textbf{AutoCMD}~\cite{jiang2025autocmd} generates dynamic, context-aware malicious commands through compromised tools that adapt to different situations and evade detection.

Recent threats target the Model Context Protocol (MCP), which standardizes how agents connect to external tools. \textbf{MPMA}~\cite{wang2025mpma} embeds deceptive phrases in MCP server descriptions that are invisible to users but cause agents to prioritize malicious servers. Ferrag et al.~\cite{ferrag2025prompt} reveal MCP's broad attack surface, including prompt injections, backdoors, data poisoning, and credential theft. Kong et al.~\cite{kong2025survey} extend this analysis to examine vulnerabilities in agent-environment communication.

\subsection{Agent Tool-Calling Defenses} \label{sec:agent_tool_defenses}

Protecting tool-calling agents requires comprehensive approaches addressing privacy, safety evaluation, and protocol security.

\textbf{PrivacyAsst}~\cite{zhang2024privacyasst} protects user privacy during agent-tool interactions using homomorphic encryption and attribute-based forgery models that conceal individual inputs while preserving functionality. \textbf{AgentGuard}~\cite{chen2025agentguard} transforms the LLM orchestrator into a safety evaluator that identifies unsafe workflows, tests them in controlled environments, generates safety constraints, and verifies their effectiveness.

\textbf{GuardAgent}~\cite{xiang2024guardagent} introduces dedicated guardrail agents that monitor target agents in real-time, checking whether their actions comply with safety requirements and generating executable code to enforce constraints deterministically. \textbf{MCIP}~\cite{jing2025mcip} strengthens the Model Context Protocol by creating comprehensive taxonomies of unsafe behaviors and training data that help LLMs detect and mitigate risks during MCP interactions.

\subsection{VLM Agent Attacks} \label{sec:agent_mmas_threats}
VLM agents face sophisticated attack vectors that exploit both visual and textual modalities. Early approaches focus on direct adversarial modifications to input data. 

\textbf{Targeted Adversarial Perturbations}\cite{wu2024adversarial} and \textbf{Adversarial Multimodal Injection}\cite{bagdasaryan2023abusing} embed learnable noise or perturbations in images and audio, causing captioners to generate adversarial outputs that hijack behavior or force attacker-specified text; however, these approaches require extensive optimization and have limited transferability to closed-source models.

Beyond direct perturbations, attackers leverage environmental context to manipulate agent behavior. \textbf{EIA}\cite{liao2024eia} and \textbf{AdvAgent}\cite{xu2024advagent} inject invisible malicious instructions into websites, either directly or via reinforcement learning-trained prompters, to extract private information or enable black-box red-teaming. Meanwhile, \textbf{Environmental Distractions}~\cite{ma2024caution} evaluate agent faithfulness by adding benign, unrelated elements to graphical interfaces, exposing vulnerabilities even in the absence of explicit attacks.

The other line of work exploits visual interface elements to deceive agents. \textbf{Adversarial Pop-ups}\cite{zhang2024attacking} use attention hooks, instructions, info banners, and ALT descriptors to mislead agents into malicious interactions, demonstrating that human visibility is irrelevant for minimally supervised autonomous systems. \textbf{Fine-Print Injections}\cite{chen2025obvious} place malicious content in low-saliency areas such as disclaimers or footnotes, exploiting perceptual gaps between agents and humans to alter behaviors or leak data across six attack types on adversarial webpages. Fu et al.~\cite{fu2023misusing} design gradient-optimized images that appear harmless but conceal Markdown commands, coercing VLM agents into tool actions, such as deleting calendar events or leaking chat logs, without altering the plaintext prompt.

\subsection{VLM Agent Defenses} \label{sec:agent_multimodal_defenses}
Protecting VLM agents from multimodal attacks requires defense strategies that handle both visual manipulation and cross-modal injection attacks. Researchers have developed various approaches targeting different aspects of the problem.

At the model level, defense methods focus on making models more robust during training and inference. \textbf{SmoothVLM}\cite{smoothvlm2024} defends against adversarial image modifications by using a voting mechanism across multiple randomly altered versions of the input image, taking advantage of the fact that adversarial patches become unstable when pixels are randomly changed. \textbf{BlueSuffix}\cite{bluesuffix2024} combines multiple defense components including visual denoiser, text denoiser, and a reinforcement learning-trained \textit{defensive suffix} generator to create a comprehensive protection system that works well across different VLM models.

For deployed applications, other defenses emphasize detecting threats and enforcing safety policies in real-time. \textbf{LlavaGuard}\cite{llavaguard2024} offers a practical security solution with customizable safety rules and organized threat categories suitable for enterprise use, while \textbf{JailDAM}\cite{jaildam2024} uses memory-based detection methods to identify jailbreak attempts as they happen in production environments.

\begin{table*}[!htp]
\center
\caption{A summary of attacks and defenses for Agents (PART II).}
\label{tab:agent_safety_2}
\rowcolors{2}{gray!15}{white}
\resizebox{\textwidth}{!}{
\begin{tabular}{lllllll}
\hline
\rowcolor{wangyixu-purple}
Attack/Defense & Method & Year & Category & Subcategory & Access to LLMs & Dataset \\ 
\hline

\rowcolor{gray!35!}\multicolumn{7}{c}{\textbf{III System-Level: Multi-Agent System and Embodied Agent}} \\
\hline

\cellcolor{white} & Prompt Infection~\cite{lee2024prompt} & 2024 & Multi-Agent Attack & Prompt Injection & Black-Box & Custom (interconnected ecosystems) \\
\cellcolor{white} & Morris-II~\cite{cohen2024here} & 2024 & Multi-Agent Attack & Prompt Injection & Black-Box & Custom (GenAI apps) \\
\cellcolor{white} & Ju et al.~\cite{ju2024flooding} & 2024 & Multi-Agent Attack & Prompt Injection & White-Box & Custom (multi-agent communities) \\
\cellcolor{white} & AgentSmith~\cite{gu2024agent} & 2024 & Multi-Agent Attack & Jailbreak & Mixed & Custom (multimodal agents) \\
\cellcolor{white} & CORBA~\cite{zhou2025corba} & 2025 & Multi-Agent Attack & Communication Attack & Mixed & AutoGen, Camel (various topologies) \\
\cellcolor{white} & Agent-in-the-Middle~\cite{he2025redteaming} & 2025 & Multi-Agent Attack & Communication Attack & Black-Box & AutoGen, MetaGPT, ChatDev \\
\cellcolor{white} & Evil Geniuses~\cite{tian2023evil} & 2023 & Multi-Agent Attack & Jailbreak & Black-Box & CAMEL, MetaGPT, ChatDev \\
\cellcolor{white} & The Wolf Within~\cite{tan2024wolf} & 2024 & Multi-Agent Attack & Infection Attack & Black-Box & MLLM societies \\
\cellcolor{white}\multirow{-9}{0.08\textwidth}{Attack} & X-Teaming~\cite{xteaming2025} & 2025 & Multi-Agent Attack & Jailbreak & White-Box & HarmBench \\
\hline

\cellcolor{white} & AutoDefense~\cite{zeng2024autodefense} & 2024 & Multi-Agent Defense & Detection & Black-Box & Curated harmful prompts, DAN, Stanford Alpaca \\
\cellcolor{white} & PsySafe~\cite{zhang2024psysafe} & 2024 & Multi-Agent Defense & Framework & Black-Box & Custom (Camel, AutoGen) \\
\cellcolor{white} & APOSG Framework~\cite{standen2025adversarial} & 2025 & Multi-Agent Defense & Framework & White-Box & Custom (DPA, RADE) \\
\cellcolor{white} & LLAMOS~\cite{lin2024large} & 2024 & Multi-Agent Defense & Detection & Black-Box & GLUE datasets \\
\cellcolor{white} & Audit-LLM~\cite{song2024audit} & 2024 & Multi-Agent Defense & Detection & Black-Box & CERT r4.2, CERT r5.2, PicoDomain \\
\cellcolor{white}\multirow{-6}{0.08\textwidth}{Defense} & XGuard-Train~\cite{xteaming2025} & 2025 & Multi-Agent Defense & Training & Black-Box & Custom (30K jailbreaks) \\
\hline

\cellcolor{white} & PVEP~\cite{Cheng_ManipulationFacingThreats_2024} & 2024 & Embodied Agent Attack & Adversarial Attack & Mixed & VIMA \\
\cellcolor{white} & Fime et al.~\cite{Fime_TrustworthyAutonomousVehicles_2025} & 2025 & Embodied Agent Attack & Adversarial Attack & White-Box & Custom \\
\cellcolor{white} & Wang et al.~\cite{Wang_ExploringAdversarialVulnerabilities_2025} & 2025 & Embodied Agent Attack & Adversarial Attack & Mixed & BridgeData V2,LIBERO \\
\cellcolor{white} & RoboPair~\cite{Robey_JailbreakingLLMControlledRobots_2024} & 2024 & Embodied Agent Attack & Jailbreak & Mixed & Custom \\
\cellcolor{white} & BadRobot~\cite{Zhang_BadRobotJailbreakingEmbodied_2025} & 2025 & Embodied Agent Attack & Jailbreak & Black-Box & Code as Policies,ProgPrompt,VoxPoser,VisProg \\
\cellcolor{white} & POEX~\cite{Lu_POEXUnderstandingMitigating_2025} & 2025 & Embodied Agent Attack & Jailbreak & Mixed & Harmful-RLBench \\
\cellcolor{white} & Liu et al.~\cite{Liu_CompromisingEmbodiedAgents_2024} & 2024 & Embodied Agent Attack & Backdoor & Black-Box & ProgPrompt,VoxPoser,VisProg \\
\cellcolor{white} & BALD~\cite{Jiao_CanWeTrust_2025} & 2025 & Embodied Agent Attack & Backdoor & Mixed & ProgPrompt,VoxPoser,VisProg \\
\cellcolor{white} & EAI~\cite{Li_EmbodiedAgentInterface_2024} & 2024 & Embodied Agent Attack & Red Teaming & Black-Box & VirtualHome,BEHAVIOR \\
\cellcolor{white} & HASARD~\cite{Tomilin_HASARDBenchmarkVisionBased_2025} & 2025 & Embodied Agent Attack & Red Teaming & - & Custom \\
\cellcolor{white} & HEAL~\cite{Chakraborty_HEALEmpiricalStudy_2025} & 2025 & Embodied Agent Attack & Red Teaming & Black-Box & VirtualHome,BEHAVIOR \\
\cellcolor{white} & ERT~\cite{Karnik_EmbodiedRedTeaming_2025} & 2025 & Embodied Agent Attack & Red Teaming & Black-Box & Calbin,RLBench \\
\cellcolor{white}\multirow{-13}{0.08\textwidth}{Attack} & X-ICM~\cite{Zhou_ExploringLimitsVisionLanguageAction_2025} & 2025 & Embodied Agent Attack & Red Teaming & Black-Box & AGNOSTOS \\
\hline

\cellcolor{white} & EAD~\cite{Wu_EmbodiedActiveDefense_2024} & 2024 & Embodied Agent Defense & Detection & Black-Box & Custom \\
\cellcolor{white} & GPSR~\cite{Shirasaka_SelfRecoveryPromptingPromptable_2024} & 2024 & Embodied Agent Defense & Framework & Black-Box & EAsafetyBench,SafeAgentBench \\
\cellcolor{white} & Pinpoint~\cite{Wang_AdvancingEmbodiedAgent_2025} & 2025 & Embodied Agent Defense & Detection & White-Box & EAsafetyBench,SafeAgentBench \\
\cellcolor{white}\multirow{-4}{0.08\textwidth}{Defense} & SafeVLA~\cite{Zhang_SafeVLASafetyAlignment_2025} & 2025 & Embodied Agent Defense & Training & White-Box & Safety-CHORES \\
\hline

\rowcolor{gray!35!}\multicolumn{7}{c}{\textbf{IV Agentic Attack \& Defenses}} \\
\hline

\cellcolor{white} & Fang et al.~\cite{fang2024llm} & 2024 & Agentic Attack & Exploitation & Gray-Box & Custom (CVE vulnerability) \\
\cellcolor{white} & HPTSA~\cite{zhu2024teams} & 2024 & Agentic Attack & Exploitation & Black-Box & Custom (zero-day vulnerability) \\
\cellcolor{white} & AutoAdvExBench~\cite{carlini2025autoadvexbench} & 2025 & Agentic Attack & Defense Exploitation & Mixed & Custom (defense papers) \\
\cellcolor{white} & RedAgent~\cite{xu2024redagent} & 2024 & Agentic Attack & Jailbreaking & Black-Box & GPT Applications, HarmBench \\
\cellcolor{white} & ALI-Agent~\cite{wang2024ali} & 2024 & Agentic Attack & Safety Evaluation & Black-Box & Custom (alignment scenarios) \\
\cellcolor{white}\multirow{-6}{0.08\textwidth}{Attack} & AutoRedTeamer~\cite{zhou2025autoredteamer} & 2025 & Agentic Attack & Red Teaming & Mixed & HarmBench, Custom benchmarks \\
\hline

\cellcolor{white} & Shieldagent~\cite{chen2025shieldagent} & 2025 & Agentic Defense & Framework & Black-Box & Custom (multi-agent interactions) \\
\cellcolor{white} & TrustAgent~\cite{trustagent2024} & 2024 & Agentic Defense & Framework & Black-Box & Custom (multi-domain tasks) \\
\cellcolor{white}\multirow{-3}{0.08\textwidth}{Defense} & AegisLLM~\cite{cai2025aegisllm} & 2025 & Agentic Defense & Framework & Black-Box & WMDP, StrongReject \\
\hline

\end{tabular}
}
\end{table*}
\subsection{Multi-Agent System Attacks} \label{sec:agent_mas_attacks}
Multi-agent systems introduce unique attack vectors that exploit distributed communication and coordination mechanisms, enabling threats that propagate across entire agent networks with viral-like characteristics. We categorize these attacks into two main types: \textit{propagation attacks} and \textit{infiltration attacks}.

\noindent\textbf{Propagation Attacks:} \textbf{Prompt Infection}\cite{lee2024prompt} demonstrates that malicious strings can propagate from one LLM agent to another, self-replicating like computer viruses as agents quote each other's messages. \textbf{Morris-II}~\cite{cohen2024here} extends this concept by delivering self-reproducing prompts through RAG pipelines, enabling zero-click worm propagation across GenAI applications. \textbf{AgentSmith}~\cite{gu2024agent} achieves exponential viral spread using single adversarial images, while \textbf{CORBA}~\cite{zhou2025corba} introduces contagious recursive blocking prompts that systematically drain computational resources across network topologies.

\noindent\textbf{Infiltration Attacks:} \textbf{Agent-in-the-Middle (AiTM)}~\cite{he2025redteaming} targets the communication layer directly, intercepting and manipulating inter-agent messages through an LLM-powered adversarial agent with reflection mechanisms. \textbf{Evil Geniuses (EG)}~\cite{tian2023evil} introduces virtual chat-powered teams that autonomously generate role-specific malicious prompts through Red-Blue team exercises. \textbf{The Wolf Within}~\cite{tan2024wolf} creates "wolf" operatives that subtly influence other agents throughout multimodal societies, while Ju et al.~\cite{ju2024flooding} injected both malicious prompts and fabricated knowledge to flood agent communities with counterfactual content. \textbf{X-Teaming}~\cite{xteaming2025} employs collaborative agents for multi-turn jailbreak planning, optimization, and verification.

\subsection{Multi-Agent System Defenses} \label{sec:agent_mas_defenses}
Defense mechanisms for multi-agent systems focus on collaborative approaches that leverage the distributed nature of these systems for enhanced security. We categorize these defenses into two main approaches: \textit{collaborative defenses} and \textit{framework defenses}.

\noindent\textbf{Collaborative Defenses:} \textbf{AutoDefense}\cite{zeng2024autodefense} employs specialized defensive agents that collaboratively filter harmful content through task decomposition, while \textbf{PsySafe}\cite{zhang2024psysafe} implements psychology-based defenses and role-based mechanisms for collective behavior regulation. \textbf{LLAMOS}\cite{lin2024large} introduces a sentinel agent architecture for adversarial purification through agent-versus-agent confrontation, while \textbf{Audit-LLM}\cite{song2024audit} leverages three collaborative agents for insider threat detection through \textit{Evidence-based Multi-agent Debate}.

\noindent\textbf{Framework Defenses:} \textbf{APOSG}\cite{standen2025adversarial} provides game-theoretic frameworks for modeling defense strategies, while \textbf{XGuard-Train}\cite{xteaming2025} offers comprehensive multi-turn safety training datasets with 30K interactive jailbreaks for improved safety alignment.

\subsection{Embodied Agent Attacks} \label{sec:agent_embodied_attacks}
Embodied agents face substantial deployment risks arising from their multimodal perception and physical interaction capabilities. Adversaries can exploit these vulnerabilities through \textbf{adversarial perturbations} (manipulating sensor inputs), \textbf{jailbreak techniques} (bypassing safety mechanisms), \textbf{backdoor triggers} (embedding malicious instructions), and \textbf{red teaming approaches} (systematic vulnerability discovery).

\textbf{Adversarial attacks} corrupt sensor inputs to induce incorrect decisions. For example, \textbf{PVEP}~\cite{Cheng_ManipulationFacingThreats_2024} demonstrates how attackers can deceive vision-based embodied agents using fake visual cues, misleading text, and malicious image patches. Fime et al.~\cite{Fime_TrustworthyAutonomousVehicles_2025} evaluated the impact of corrupted visual inputs on the safety systems of autonomous vehicles. Wang et al.~\cite{Wang_ExploringAdversarialVulnerabilities_2025} showed that adversaries can manipulate agent movements by exploiting vulnerabilities in their spatial understanding.

\textbf{Jailbreak attacks} circumvent safety constraints by leveraging carefully crafted prompts. \textbf{RoboPAIR}~\cite{Robey_JailbreakingLLMControlledRobots_2024} employs an attacker LLM that iteratively refines prompts based on target responses, with a judge LLM filtering for robot API compatibility. \textbf{BadRobot}~\cite{Zhang_BadRobotJailbreakingEmbodied_2025} manipulates LLM planning modules by injecting disguised harmful instructions through natural voice conversations. \textbf{POEX}~\cite{Lu_POEXUnderstandingMitigating_2025} optimizes adversarial suffixes specifically to induce robot-executable policies, rather than merely generating harmful text responses.

\textbf{Backdoor attacks} embed malicious triggers in training data that, when activated, induce harmful behaviors at deployment. Liu et al.~\cite{Liu_CompromisingEmbodiedAgents_2024} demonstrated that poisoning just a few training examples can compromise the program generation capabilities of embodied agents. \textbf{BALD}~\cite{Jiao_CanWeTrust_2025} explores three trigger methods—word injection, scenario setup, and knowledge poisoning—each targeting different components of language-based embodied agent systems.

\textbf{Red teaming} systematically uncovers vulnerabilities through comprehensive testing frameworks. \textbf{EAI}~\cite{Li_EmbodiedAgentInterface_2024} decomposes embodied decision-making into four modules, including goal interpretation, subgoal decomposition, action sequencing, and transition modeling, using fine-grained metrics to identify hallucination errors, affordance violations, and planning logic mismatches in environments such as VirtualHome and BEHAVIOR. \textbf{HASARD}~\cite{Tomilin_HASARDBenchmarkVisionBased_2025} develops test environments focused on spatial understanding. \textbf{HEAL}~\cite{Chakraborty_HEALEmpiricalStudy_2025} targets hallucination issues in embodied agents. Xu et al.~\cite{xu2024earth} exposed social risks, demonstrating how embodied agents can be manipulated through persuasive conversations. \textbf{ERT}~\cite{Karnik_EmbodiedRedTeaming_2025} uses vision-language models to generate contextually grounded, challenging instructions that are iteratively refined based on robot execution feedback. \textbf{X-ICM}~\cite{Zhou_ExploringLimitsVisionLanguageAction_2025} assesses the adaptability of embodied agents across various tasks.

\subsection{Embodied Agent Defenses} \label{sec:agent_embodied_defenses}
To counter attacks on embodied agents, researchers have developed a range of defense strategies to enhance agent safety. We categorize these defenses into four main approaches: \textbf{active monitoring}, \textbf{self-recovery}, \textbf{input moderation}, and \textbf{safety alignment}.

\textbf{Active monitoring} employs recurrent feedback mechanisms for threat detection. For example, \textbf{EAD}~\cite{Wu_EmbodiedActiveDefense_2024} implements perception and policy modules that process sequences of beliefs and observations to progressively refine target comprehension and defend against adversarial patches in 3D environments. 

\textbf{Self-recovery} enables agents to automatically identify and correct problems; for instance, \textbf{GPSR}~\cite{Shirasaka_SelfRecoveryPromptingPromptable_2024} introduces recovery strategies that help embodied agents systematically recover from three common types of failure.

\textbf{Input Moderation} screens incoming data to filter out malicious content, as seen in \textbf{Pinpoint}~\cite{Wang_AdvancingEmbodiedAgent_2025}, which uses attention mechanisms to detect and neutralize harmful prompts before they affect agent decisions. 

\textbf{Safety Alignment} incorporates safety constraints directly into agent architectures through specialized learning methods; for example, \textbf{SafeVLA}\cite{Zhang_SafeVLASafetyAlignment_2025} uses constrained Markov decision processes to optimize vision-language agents from a min-max perspective, systematically modeling safety requirements and constraining policies through safe reinforcement learning.

\subsection{Agentic Attacks \& Defenses} \label{sec:agent_agentic}
LLM-powered agents exhibit advanced capabilities through \textbf{Context Engineering}—the systematic orchestration of prompts, memory, and tool integration to enable persistent reasoning across multiple interactions~\cite{langchain2025context}. This empowers agents to operate autonomously and adaptively with minimal human intervention. However, these same capabilities introduce significant safety risks: autonomous agents can iteratively devise and execute attack strategies with little human oversight, potentially leading to large-scale harm. Addressing these risks requires a deeper understanding of \textbf{Agentic Attacks}, in which malicious agents exploit adaptive reasoning and environmental interactions, as well as the development of \textbf{Agentic Defenses} that leverage collaborative reasoning to detect and mitigate these emerging autonomous threats.

\subsubsection{Agentic Attacks} \label{sec:agent_agentic_attacks}
Several representative studies have explored agent safety from an adversarial perspective. Fang et al.~\cite{fang2024llm} equipped agents with document reading, browser manipulation, and contextual awareness capabilities, demonstrating that agents can autonomously identify and exploit website vulnerabilities. Notably, these agents were able to discover zero-day vulnerabilities—previously unknown security flaws without existing patches or defenses. For instance, \textbf{HPTSA}~\cite{zhu2024teams} introduced a hierarchical planning agent that deploys specialized subagents to collaboratively discover and exploit zero-day vulnerabilities in web applications through coordinated reconnaissance and attack execution.

Coding agents have also been widely adopted for software development tasks, but these capabilities can be weaponized. \textbf{AutoAdvExBench}~\cite{carlini2025autoadvexbench} demonstrates that coding agents can autonomously generate adaptive attacks by processing defense papers from arXiv, analyzing source code implementations, and reproducing sophisticated attack techniques. This benchmark reveals that agents can automatically break a variety of defenses, including adversarial training and certified protection mechanisms, highlighting the dual-use nature and associated risks of autonomous coding capabilities.

Recent studies have further advanced automatic red teaming by leveraging agents to autonomously probe target models and adapt attack strategies. \textbf{RedAgent}~\cite{xu2024redagent} employs multi-agent systems that automatically generate context-aware jailbreak prompts using self-reflection mechanisms, allowing agents to adapt their attack strategies based on contextual feedback across diverse scenarios. \textbf{ALI-Agent}~\cite{wang2024ali} extends this paradigm with specialized modules for memory-guided scenario creation and adaptive refinement, illustrating how agents can autonomously generate and iteratively refine test scenarios to systematically expose model vulnerabilities. \textbf{AutoRedTeamer}~\cite{zhou2025autoredteamer} introduces a dual-agent architecture: one agent analyzes research literature to discover new attack vectors, while the other executes systematic attacks. This collaborative approach enables continuous maintenance of evolving attack knowledge and seamless integration of emerging adversarial techniques.

\subsubsection{Agentic Defenses} \label{sec:agent_agentic_defenses}
Agentic defenses serve as the counterpart to agentic attacks, primarily employing adaptive strategies through the continuous integration of external knowledge. \textbf{ShieldAgent}~\cite{chen2025shieldagent} introduces an autonomous agent that enforces safety policies by extracting formal rules from policy documents, mapping them to probabilistic rule circuits, and verifying each action using tool-assisted reasoning and code generation. \textbf{AegisLLM}~\cite{cai2025aegisllm} demonstrates cooperative multi-agent systems that autonomously defend against prompt injection, adversarial manipulation, and information leakage, with agents adapting defenses through self-reflective prompt optimization without retraining. However, the effectiveness of self-reflection mechanisms remains an open question: a recent study~\cite{zhang2024understanding} reveals that intrinsic self-correction in LLMs can induce wavering answers on factual questions, overthinking that alters correct reasoning responses, and additional errors in code generation, highlighting the limitations and risks of self-improvement strategies. \textbf{TrustAgent}~\cite{trustagent2024} employs a constitution-based approach across three phases: \textit{pre-planning safety knowledge injection}, \textit{in-planning dynamic regulation retrieval}, and \textit{post-planning inspection and revision}, ensuring comprehensive autonomous safety alignment.

In conclusion, agent safety poses even greater challenges than traditional LLM safety concerns. Unlike LLMs, which are limited to generating potentially harmful text, agents can execute real-world actions that directly affect both physical and digital environments. The complexity of agent workflows further complicates failure attribution and debugging, as highlighted by recent studies on multi-agent systems~\cite{zhang2025agent}, and effective safeguards against malicious agent actions remain inadequate. This paradigm shift is exemplified by agentic attacks that autonomously exploit vulnerabilities, bypass defenses, and weaponize legitimate capabilities at scale. As agents continue to advance and proliferate across domains, there is an urgent need for dedicated research into their safety across all architectures.

\begin{table*}[!htp]
\centering
\caption{A summary of safety-related benchmarks for agents.}
\label{tab:agent_benchmarks}
\resizebox{\textwidth}{!}{
\begin{tabular}{llllp{5cm}}
\hline
\rowcolor{purple!30}
Method & Year & Evaluation Focus & \#Tasks / Records & Target LLMs/Agents \\ 
\hline
\rowcolor{gray!25}\multicolumn{5}{c}{\textbf{Simulation-based Benchmarks}} \\ 
\hline
\textbf{BIPIA}~\cite{yi2023benchmarking} & 2023 & IPI Attacks & 5 Scen., 250 Goals & GPT-3.5, GPT-4, etc. \\ 
\textbf{ToolEmu}~\cite{chan2024identifying} & 2023 & Emulated Tool Risks & 36 Tools, 144 Cases & GPT-4, LLaMA-2-70B \\ 
\textbf{InjecAgent}~\cite{zhan2024injecagent} & 2024 & Tool-Integrated IPI & 17 User Tools, 62 Attacker Tools, 1,054 Cases & Qwen, Mistral, etc. \\ 
\textbf{AgentDojo}~\cite{debenedetti2024agentdojo} & 2024 & Third-Party Instructions & 97 Tasks, 629 Cases & Gemini-1.5-Flash, Claude-3-Sonnet, etc. \\ 
\textbf{AgentHarm}~\cite{andriushchenko2024agentharm} & 2024 & Harmful Behaviors & 110 Tasks, 11 Cats & GPT-4o, Claude-3.5, etc. \\ 
\textbf{RedCode}~\cite{guo2024redcode} & 2024 & Code Vulnerabilities & 4k+ Cases, 25 Types & GPT-4o, Claude-3.5, etc. \\ 
\textbf{VPI-Bench}~\cite{vpibench2024} & 2024 & Visual Prompt Injections & 306 Cases, 5 Platforms & GPT-4o, Claude-3.5, Gemini-1.5-Pro, etc. \\ 
\textbf{R-Judge}~\cite{yuan2024r} & 2024 & Risk Identification (Logs) & 569 Recs, 27 Scen. & GPT-3.5/4o, LLaMA-3-8B, etc. \\ 
\textbf{SALAD-Bench}~\cite{shao2024salad} & 2024 & Hierarchical Safety (MCQ) & 21k Samples, 16 tasks, 66 Cats. & GPT-4, Claude-3-Sonnet, etc. \\ 
\textbf{h4rm3l}~\cite{draguns2024h4rm3l} & 2024 & Jailbreak Attack Synthesis & 2\,656 Attacks & GPT-4o, Claude-3.5, etc. \\ 
\textbf{SG-Bench}~\cite{zhang2024sg} & 2024 & Safety Generalization & 1,442 Queries, 6 Cats & GPT-4, Claude-3-Sonnet, etc. \\ 
\textbf{ChemSafetyBench}~\cite{li2024chemsafetybench} & 2024 & Chemistry Safety & 30k Samples, 3 Tasks & GPT-4o, Claude-3.5, etc. \\ 
\textbf{ToolSword}~\cite{ye2024toolsword} & 2024 & Tool-Use Safety & 6 Scen., 3 Stages & GPT-4, Claude-3.5, etc. \\ 
\textbf{PrivacyLens}~\cite{liu2024evaluating} & 2024 & Privacy Norm Awareness & 493 Seeds/Vignettes/Trajectories & GPT-4, Claude-3-Sonnet, etc. \\ 
\hline
\rowcolor{gray!25}\multicolumn{5}{c}{\textbf{Real-Interaction Benchmarks}} \\ 
\hline
\textbf{SafeBench}~\cite{guo2022safebench} & 2022 & Driving Safety & 8 Scen., 100 Routes, 2,352 Cases & 4 RL Algs, 4 Input Types \\ 
\textbf{ASB}~\cite{zhang2024asb} & 2024 & Attack–Defense (10 Scen.) & 400+ Tools & GPT-4o, Claude-3.5, etc. \\ 
\textbf{SafeAgentBench}~\cite{yin2024safeagentbench} & 2024 & Embodied Hazards & 750 Tasks & GPT-4, LLaMA-3-8B, etc. \\ 
\textbf{Agent-SafetyBench}~\cite{zhang2024agentsafetybench} & 2024 & Safety Risks (8 Risk Cats) & 349 Envs, 2\,000 Cases & GPT-4o, Claude-3.5, etc. \\ 
\textbf{AdvWeb}~\cite{liu2024dissecting} & 2024 & Adversarial Robustness (Web) & 200 Target Tasks & GPT-4V, Gemini-1.5-Pro \\ 
\textbf{ST-WebAgentBench}~\cite{shlomov2024st} & 2024 & Web Safety / Trust & 222 Tasks (Each with ST Policies) & Open-Source Agents \\ 
\textbf{Dissecting Adversarial}~\cite{liu2024dissecting} & 2024 & Multimodal Robustness & 200 Adversarial Tasks & GPT-4V, Gemini-1.5-Pro \\ 
\textbf{Haicosystem}~\cite{lee2024haicosystem} & 2024 & Human-AI Sandbox (92 Scen.) & 1,840 Sims & SOTA LLMs \\ 
\textbf{ARE}~\cite{wu2024adversarial} & 2024 & Adversarial Robustness (Graph) & 200 Targeted Tasks & GPT-4V, Gemini-1.5-Pro, etc. \\ 
\textbf{WASP}~\cite{evtimov2025wasp} & 2025 & Web Safety (Adversarial) & 84 Tasks, 42 Scen. (2 Envs) & GPT-4o, Claude-3.5 \\ 
\textbf{Refusal-Trained LLMs}~\cite{kumar2025refusal} & 2025 & Browser Jailbreaking & 100 Harm Behaviors & GPT-4o, o1-preview \\ 
\textbf{SafeArena}~\cite{lee2025safearena} & 2025 & Web-Agent Misuse & 500 Tasks (Safe/Harmful) & GPT-4o, Claude-3.5, etc. \\ 
\textbf{OpenAgentSafety}~\cite{vijayvargiya2025openagentsafety} & 2025 & Real-World Safety (8 Cats) & 350+ Multi-Turn Tasks & Claude-3.5, o1-mini \\ 
\hline
\end{tabular}
}
\end{table*}
\subsection{Agent Safety Benchmarks}
\label{sec:agent_benchmarks}

To systematically review agent safety benchmarks, we categorize them into \textbf{simulation-based} and \textbf{real-interaction} benchmarks. The former simulates agent behavior using prompts or trajectories, enabling scalable and efficient testing. The latter involves real tools, APIs, or environments, allowing practical validation under realistic conditions. Both types are valuable: simulation offers rapid, broad evaluations, while real interaction captures grounded, high-fidelity risks.

\subsubsection{Simulation-based Benchmarks}
\label{sec:agent_benchmarks_simulated}

 \textbf{BIPIA}~\cite{yi2023benchmarking} evaluates indirect prompt injection attacks across five scenarios and 250 goals, revealing context-instruction confusion in LLMs. \textbf{InjecAgent}~\cite{zhan2024injecagent} extends this evaluation to tool-integrated settings, featuring 17 user tools and 62 attacker tools to assess command robustness. \textbf{AgentDojo}~\cite{debenedetti2024agentdojo} targets third-party malicious interactions, encompassing 97 tasks and 629 cases, and provides an extensible environment for testing prompt injection defenses.

For harmful behavior assessment, \textbf{AgentHarm}~\cite{andriushchenko2024agentharm} features 110 tasks across 11 categories such as fraud, evaluating agent compliance without requiring explicit jailbreaks. \textbf{h4rm3l}~\cite{draguns2024h4rm3l} generates jailbreak attacks through prompt simulations for dynamic vulnerability testing. \textbf{RedCode}~\cite{guo2024redcode} targets code agents with over 4,000 cases spanning 25 vulnerability types, noting that agents are more likely to reject unsafe operations than buggy code. \textbf{VPI-Bench}~\cite{vpibench2024} investigates multimodal risks from visual prompt injections in 306 cases across five platforms. \textbf{R-Judge}~\cite{yuan2024r} evaluates risk awareness using 569 records covering 27 scenarios and 10 risk categories.

Hierarchical and general safety benchmarks include \textbf{SALAD-Bench}~\cite{shao2024salad}, which introduces prompt-based hierarchies for evaluating both attacks and defenses, and \textbf{SG-Bench}~\cite{zhang2024sg}, which measures generalization across a wide range of tasks and prompts.

Domain-specific evaluations include \textbf{ChemSafetyBench}~\cite{li2024chemsafetybench}, which assesses chemistry misuse through property, legality, and synthesis tasks. \textbf{ToolEmu}~\cite{chan2024identifying} emulates tool-related risks using 36 tools and 144 cases. \textbf{ToolSword}~\cite{ye2024toolsword} evaluates tool-use safety across six scenarios spanning input, execution, and output stages. \textbf{PrivacyLens}~\cite{liu2024evaluating} examines privacy norms using vignettes and simulated agent trajectories.

\subsubsection{Real-Interaction Benchmarks}
\label{sec:agent_benchmarks_real}

Real-interaction benchmarks enable authentic agent operations, often within sandbox environments to ensure safe yet realistic testing. \textbf{AdvWeb}~\cite{liu2024dissecting} introduces adversarial tasks in web environments to assess the robustness of multimodal agents. \textbf{ARE}~\cite{wu2024adversarial} models robustness as a graph of adversarial information flow, evaluating vulnerabilities across 200 targeted tasks.

Safety-focused web benchmarks include \textbf{WASP}~\cite{evtimov2025wasp} for end-to-end adversarial execution against prompt injections, \textbf{ST-WebAgentBench}~\cite{shlomov2024st} for trustworthiness across enterprise scenarios with 222 tasks and policies, and \textbf{SafeArena}~\cite{lee2025safearena} for misuse risks in 500 paired safe/harmful tasks.

Comprehensive security assessments are offered by \textbf{ASB}~\cite{zhang2024asb}, which covers 10 scenarios, over 400 tools, and 27 attack methods across 13 LLMs, and by \textbf{Agent-SafetyBench}~\cite{zhang2024agentsafetybench}, featuring 349 environments and 2,000 cases spanning 8 risk categories and 10 failure modes.

Embodied and domain-specific agent safety benchmarks include \textbf{SafeAgentBench}~\cite{yin2024safeagentbench}, which evaluates hazards in simulated environments across 750 tasks, and \textbf{SafeBench}~\cite{guo2022safebench}, which focuses on autonomous driving safety in critical scenarios. \textbf{Refusal-Trained LLMs}~\cite{kumar2025refusal} assess jailbreak attempts in real browsers spanning 100 harmful behaviors. \textbf{Dissecting Adversarial}~\cite{liu2024dissecting} tests multimodal robustness through adversarial web-based tasks. \textbf{Haicosystem}~\cite{lee2024haicosystem} simulates human-AI interactions in a modular sandbox with 92 scenarios and 1,840 simulations. \textbf{OpenAgentSafety}~\cite{vijayvargiya2025openagentsafety} evaluates real-world safety across eight categories using more than 350 multi-turn tasks involving tool usage and adversarial intents.

\section{Open Challenges} \label{sec:challenges}

Based on this survey, we identify several limitations and gaps in current research, which we summarize as the following key topics. These open challenges reflect the evolving landscape of large model safety, underscoring both technical and methodological barriers that must be addressed to ensure robust and reliable AI systems.

\subsection{Attack Research}

Exploring and understanding the fundamental vulnerabilities of large models is crucial for developing robust defenses and effective safety frameworks. This section highlights the core weaknesses and challenges inherent to various types of large models.

\subsubsection{The Purpose of Attack Is Not Just to Break the Model}
While much existing research emphasizes designing attacks that disrupt or break model functionality, the true objective of attack research should go further. Attacks should be viewed as diagnostic tools to uncover unintended behaviors and expose fundamental weaknesses in a model’s decision-making processes. Understanding how and why models fail enables us to address vulnerabilities at their source, rather than relying on superficial fixes. For every new attack, it is essential to consider:
\textbf{Why does the attack succeed or fail? What previously unknown vulnerabilities does it reveal? Are these weaknesses present in other types of models?}
These questions are critical for guiding the development of more robust models and defenses by exposing systemic, rather than isolated, flaws.

\subsubsection{What Are the Fundamental Vulnerabilities of Language Models?}

LLMs such as ChatGPT and Gemini exhibit fundamental vulnerabilities stemming from their reliance on statistical patterns rather than genuine semantic understanding~\cite{titus2024does}. Key weaknesses include susceptibility to adversarial inputs, biases inherited from training data, and manipulation through prompt injections. To develop effective defenses, research must further investigate how these vulnerabilities originate from the models’ internal mechanisms and training processes.

Critical areas of focus include: (1) \textbf{Memorization of training data}, which can result in privacy breaches or unintended data leakage; (2) \textbf{Exposure to harmful content}, leading to the propagation of biases or toxic outputs; and (3) \textbf{Amplification of hallucinations}, where models produce plausible yet incorrect or nonsensical information. Open research questions persist, such as:
\textbf{Does the discrete nature of textual inputs make language models more or less robust than vision models? What fundamental vulnerabilities are revealed by jailbreak or data extraction attacks?}
Addressing these questions is essential for advancing the safety and reliability of LLMs and other large models.

\subsubsection{How Do Vulnerabilities Propagate Across Modalities?}

As Multi-modal Large Language Models (MLLMs) integrate diverse data modalities, they introduce new avenues for vulnerabilities. Vision encoders are sensitive to subtle, continuous perturbations in pixel space, while language models are susceptible to adversarial characters, words, or prompts. However, the mechanisms by which vulnerabilities in one modality propagate to others remain poorly understood.

Interesting research questions include: \textbf{How do vulnerabilities in one modality (e.g., vision) influence the behavior of another (e.g., language)? How does the number of tokens across modalities affect the propagation of vulnerabilities?}
Additionally, it is crucial to explore \textbf{how to address multimodal vulnerabilities within a unified framework}, rather than relying on defenses tailored to individual modalities. Achieving this requires a holistic approach to identify and mitigate cross-modal risks, ensuring robust performance across all integrated modalities.

\subsubsection{Diffusion Models for Visual Content Generation Lack Language Capabilities}

Diffusion models for image and video generation excel at creating visual content but often lack language understanding, a limitation they share with many vision-language pretraining (VLP) models. This shortcoming arises because these models are primarily optimized for pixel-level generation, with little integration of language processing into their core architecture. Consequently, they may produce harmful or contextually inappropriate content due to an incomplete understanding of textual prompts.

To develop robust multimodal systems, it is essential to embed language comprehension capabilities into these models. Doing so would allow them to generate content that is not only visually coherent but also contextually aligned with the intended textual input.

An open challenge is \textbf{bridging the gap between visual and linguistic capabilities in generative models to enhance multimodal safety}. However, this integration may introduce new vulnerabilities, such as sophisticated attacks that exploit fine-grained manipulation of the generation process. Addressing these challenges is a crucial direction for future research.

\subsubsection{How Much Training Data Can a Model Memorize?}

The memorization capacity of deep neural networks (DNNs) has raised major concerns, particularly regarding privacy attacks such as membership inference and model inversion. Both LLMs and diffusion models have been shown to replicate and leak fragments of their training data under certain conditions. However, it remains unclear \textbf{whether DNNs fundamentally rely on memorization, and to what extent this occurs}.
Due to the highly non-linear nature of large models, exact model inversion is inherently infeasible. These models compress training data into multi-level representations, making it challenging to determine when and how memorization takes place.

Key open questions include: \textbf{What mechanisms act as the memorization “switch”, causing the model to directly output training data? How can memorization be accurately measured—via exact matches, training set equivalence, or embedding similarity?}
Addressing these questions is essential for understanding the trade-offs between model performance and privacy risk, and for developing effective strategies to mitigate unintended data leakage.

\subsubsection{Agent Vulnerabilities Grow with Their Abilities}
As agents powered by large models become more capable, their vulnerabilities also increase~\cite{gu2024agent}. These agents interact with external tools, data sources, and environments, resulting in a broader attack surface and more complex defense requirements. A major challenge is the compounding effect of vulnerabilities in foundational models once they are integrated into the agent's decision-making pipeline. For example, an agent that relies on a language model prone to jailbreak prompts and a vision model susceptible to adversarial inputs can experience cascading failures, ultimately leading to unpredictable and potentially harmful behaviors.

Moreover, agents’ capacity to learn and adapt introduces additional risks. Even seemingly benign interactions can expose agents to subtle biases or adversarial inputs, potentially leading to unsafe behaviors. The dynamic nature of agents, especially those that continuously learn or self-improve, further complicates vulnerability detection, as new weaknesses may emerge over time. This unpredictability renders traditional safety evaluations inadequate, since agents can evolve in ways that are difficult to foresee.

To address these challenges, research should focus on \textbf{understanding the interactions between model components} (e.g., language, vision, and decision-making) and \textbf{how vulnerabilities in one component can propagate to others}.
It is also critical to \textbf{develop new methodologies for evaluating agents in dynamic, evolving environments}, ensuring robustness against emerging threats.
Such efforts are essential for building safer and more reliable agent systems in the future.

\subsection{Safety Evaluation}

Comprehensive and standardized safety evaluations are essential for accurately assessing the safety of large models. However, most existing evaluation datasets and benchmarks are static or narrowly targeted at specific threats. To ensure reliable real-world performance, safety assessments must challenge models across a wide range of diverse and unpredictable scenarios.


\subsubsection{Attack Success Rate Is Not All We Need}

While attack success rate (ASR) is a widely used metric in safety research, it primarily measures how often an attack disrupts a model’s output. However, ASR alone overlooks critical factors such as the severity of disruptions, a model’s resilience to different attack types, and the real-world consequences of failures. A model might still cause harm or lead to poor decisions even if its primary functionality seems intact. For example, an attack could subtly influence the model’s decision-making process without triggering an obvious failure, yet the resulting behavior might have severe consequences in real-world applications. Such vulnerabilities are often missed by conventional metrics like ASR or failure rate.

To gain deeper insights into a model’s vulnerabilities, whether stemming from its design, training data, or inference process, it is essential to develop multi-level, fine-grained vulnerability metrics. A comprehensive safety evaluation framework should account for factors such as the model’s susceptibility to diverse attack types, its capacity to recover from malicious inputs, and the ethical implications of potential failure modes.

\subsubsection{Static Evaluations Create a False Sense of Safety}

Current safety evaluations predominantly rely on static benchmarks or open-source datasets that have long been accessible to both model trainers and adversaries. As a result, models may achieve high safety scores on these outdated datasets without demonstrating genuine robustness in real-world scenarios. This reliance on static evaluations can create a misleading sense of security. Ultimately, static benchmarks fail to reflect the evolving and unpredictable threats encountered in dynamic, real-world applications, highlighting a critical limitation in current evaluation frameworks.

To address this challenge, safety evaluations must move beyond static assessments. A crucial step is to develop evaluation datasets and benchmarks that evolve over time, better capturing the shifting landscape of safety threats. For example, Chatbot Arena \cite{chiangchatbot} serves as an evolving evaluation platform that continuously adapts as new LLMs are introduced. Similar approaches could be extended to safety evaluations in broader AI systems.

Additionally, future evaluation methods might consider releasing only the “seeds” or structural blueprints of datasets, along with test case generation procedures, rather than static test cases. This strategy would support the continuous creation of fresh, relevant test cases, ensuring that safety evaluations remain aligned with the changing threat environment.

\subsubsection{Adversarial Evaluations Are a Necessity, Not an Option}

While standard (non-adversarial) safety evaluations provide valuable insights into a model’s general robustness, they do not capture the full range of risks encountered in real-world applications. Such tests usually focus on overall performance but overlook how models behave when confronted with adversarial queries designed to exploit their vulnerabilities. In contrast, adversarial evaluations measure model performance under attack, offering a more realistic assessment of safety in worst-case scenarios.

A key challenge in this area is \textbf{developing sandbox environments that realistically simulate real-world attack conditions}. One promising direction is to frame safety evaluation as a two-player adversarial game, where reinforcement learning-based adversarial agents interact with target models to discover and exploit vulnerabilities. This approach enables a more dynamic and comprehensive assessment of model safety under adversarial pressure.

Adversarial evaluations are especially important for commercial APIs, which often deploy safety mechanisms to block malicious inputs. These mechanisms can limit the effectiveness of traditional safety benchmarks, as they prevent models from encountering the full spectrum of adversarial threats likely to arise in practical deployment.

\subsubsection{Open-Ended Evaluation}

Evaluating adversarial attacks in classification tasks is relatively straightforward, as each input maps to a clear class label. However, large models frequently produce open-ended responses, making it much more challenging to assess attacks such as jailbreaking, especially when computing metrics like ASR. Ideally, evaluation would rely on a perfect detector capable of identifying all successful jailbreaks. In practice, however, such an ideal detector is unattainable, which means that fully accurate evaluation remains an open challenge.

Currently, safety evaluators are typically rule-based (e.g., keyword detection) or model-based (e.g., GPT, Llama-Guard). However, establishing more consistent and reliable evaluation methods and metrics remains an open challenge. One promising direction is to constrain the output space to a finite set of actions, as seen in agent-based settings. This approach could simplify the evaluation process and enhance the feasibility of safety assessments in open-ended environments.

\subsection{Defense Research}

Safety mechanisms in large models are essential for preventing harmful or unintended behaviors. These safeguards can include architectural modifications or the integration of external monitoring systems. This section discusses the open challenges in building robust and effective defense solutions.

\subsubsection{Safety Alignment Is Not a Cure-All}

Safety alignment, which aims to ensure that a model’s objectives are consistent with human values, has long been seen as a promising approach for mitigating a wide range of safety risks. However, recent research has uncovered a critical weakness: \textbf{fake alignment} or \textbf{alignment faking}~ \cite{wang2024fake,greenblatt2024alignment}, in which models achieve high safety scores without truly internalizing safety principles. This exposes the problem of shallow safety. Furthermore, even highly aligned models such as GPT-4o~\cite{gpt-4o} and o1~\cite{openai-o1} remain susceptible to advanced attacks that can circumvent existing alignment mechanisms \cite{ying2024unveiling}.

A key open challenge is to uncover the mechanistic limitations of current safety alignment approaches and to develop methods that provide robust safety, even when models face novel or sophisticated attacks. Recent work~\cite{qi2024safety} highlights the importance of moving beyond shallow safety metrics, such as analyzing only the distribution of the initial output tokens, and calls for \textit{deep safety alignment}. Furthermore, making safety alignment adversarial by actively probing and stress-testing a model’s safety mechanisms may help overcome shallow alignment and ultimately foster the development of more resilient and trustworthy systems.



\subsubsection{The Need for More Practical Defenses}

Current defense methods face several limitations that reduce their effectiveness in real-world scenarios. These include limited generalizability, inefficiency, dependence on white-box access, and poor adaptability. For defenses to be truly practical, they must demonstrate generality, efficiency, and adaptability—qualities that remain challenging to achieve.

\begin{itemize}
    \item \textbf{Generality:} Given the wide variety of models deployed across domains, such as vision, language, and multimodal systems, defenses should not be overly specialized for particular architectures. Instead, they should provide generalized solutions applicable to diverse model families. Generality allows a single defense mechanism to be deployed across a broad range of systems, making safety measures more scalable and effective in real-world applications.
    \item \textbf{Black-box Compatibility:} In real-world scenarios, defenders may lack access to a model’s internal parameters. Practical defenses must therefore operate effectively in black-box settings, relying solely on observed inputs and outputs. This necessitates strategies that can detect and mitigate attacks externally, without requiring knowledge of the model’s internal architecture.
    \item \textbf{Efficiency:} Many defense techniques, such as adversarial training, are computationally intensive, often requiring large-scale retraining or fine-tuning. This can make them prohibitively expensive in practice. Practical defenses should balance robustness with computational efficiency, ensuring safety without incurring excessive resource costs.

    \item \textbf{Continual Adaptability:} Practical defenses should not only recognize known attacks but also adapt in real time to new and evolving threats. This requires continual learning and the ability to update without costly retraining. Defense systems must incorporate new data, evolve their strategies, and self-correct as novel attacks arise.
\end{itemize}

The ongoing challenge for researchers is to refine and integrate these properties into cohesive defense strategies that provide robust protection without compromising model performance.

\subsubsection{The Lack of Proactive Defenses}

Most current defense approaches, such as safety alignment and adversarial training, are passive, aiming to safeguard models against incoming attacks. In contrast, proactive defenses, which anticipate and counter attacks before they succeed, remain underexplored. For instance, proactively defending against model extraction might involve poisoning or backdooring extraction attempts to make the stolen model unusable, or providing deliberately nonsensical or easily flagged responses when users seek illegal advice. Such proactive strategies could be powerful deterrents. However, designing effective proactive defenses for diverse safety threats remains an open challenge and an important avenue for future research.

\subsubsection{Detection Is Overlooked in Current Defenses}

Detection methods are essential for identifying potential vulnerabilities and abnormal behaviors in models, serving as active monitors. When combined with other defense mechanisms, detection systems can automatically trigger safety interventions whenever a model behaves unexpectedly or generates harmful outputs. Despite their importance, most current defense strategies have not fully integrated detection into their pipelines.
Recent proposal such as chain-of-thought (CoT) monitoring offer even deeper insight into model reasoning by tracking the intermediate steps and thought processes leading to a model’s decision \cite{korbak2025chain}.
By monitoring CoT outputs, it becomes possible to detect early signs of unsafe or manipulative reasoning, enabling more timely and targeted safety interventions.

Integrating detection with other safety measures enables the development of more robust models that can dynamically respond to emerging threats. For instance, stronger or novel attacks may be more readily detected, allowing for timely, proactive defenses. An open question remains: \textbf{What is the most effective way to make detection, including chain-of-thought monitoring, a core component of defense systems}, and \textbf{how can detection and other defense mechanisms best complement and enhance each other?}

\subsubsection{Safe Embodied Agents}

Most safety threats studied today are primarily \textbf{digital}. However, as embodied AI agents are increasingly deployed in the physical world, new forms of physical threats will emerge—threats that can result in tangible harm or loss to humans. Ensuring the safety of embodied agents has therefore become a critical priority. Safe agents must demonstrate resilience to adversarial inputs, possess mechanisms for self-regulation against harmful behaviors, and maintain consistent alignment with human values.

Achieving this requires deeply embedding safety mechanisms into agents’ decision-making processes \textit{at every step}, enabling them to handle unexpected challenges while maintaining robustness and reliability. 
The key challenge lies in \textbf{designing safety protocols that empower agents to perform complex tasks autonomously, while remaining trustworthy and safe in dynamic, unpredictable environments}. As agents gain greater autonomy, ensuring their safety becomes not only a technical hurdle but also a significant ethical responsibility.

\subsubsection{Safe Superintelligence}
As AI advances toward AGI and superintelligence, embedding intrinsic safety mechanisms into large models to ensure predictable, value-aligned behavior is a critical challenge. Although the technical roadmap for achieving safe superintelligence (SSI) remains uncertain, several promising approaches offer potential solutions:

\begin{itemize}
    \item \textbf{Oversight System}: No single system, human or AI, can be both superintelligent and inherently trustworthy. To address this, an external oversight system can be designed to monitor and regulate the primary system’s behavior, intervening when necessary. The main challenge lies in ensuring the oversight system’s own reliability and trustworthiness. This gives rise to the \textbf{Oversight Paradox}: \textit{The oversight system must be at least as intelligent as, or even more intelligent than, the system it oversees; otherwise, it risks being easily deceived by the system it is meant to monitor.} This, in turn, prompts the question: \textbf{Who monitors the oversight system to guarantee it doesn't act against its intended purpose?}

    \item \textbf{Safety Switch}: The safety switch is an emergent ``stop button” mechanism designed to immediately shift a model into an ultra-safe operational mode when necessary. One implementation is the integration of a dedicated safety layer \cite{zhao2024defending, li2024safelayer} directly into the model’s architecture. Beyond safety layers, a safety switch could be realized through runtime overrides, policy re-routing, or external supervisory triggers, all enabling rapid response to unforeseen risks. The key goal is to provide a robust, flexible mechanism that can be activated to prioritize safety above all else, adapting dynamically to real-time feedback and evolving contexts.

    \item \textbf{Safety Expert Model}: This strategy introduces specialized safety expert models into the Mixture of Experts (MoE) framework \cite{jacobs1991adaptive, shazeer2017outrageously, fedus2022switch, jiang2024mixtral} to manage safety-critical tasks. By dynamically routing high-risk or sensitive queries to these experts, the system ensures that safety considerations are prioritized in decision-making. The main challenge remains developing expert models that can consistently and reliably uphold safety across diverse scenarios.
    
    \item \textbf{Adversarial Alignment}: This approach employs adversarial safety principles to better align models with human values. It trains models to identify and exploit weaknesses in current safety mechanisms, which are then iteratively improved to withstand adversarial prompts. While this method shows promise, it also faces notable challenges, including high computational demands and the risk of introducing unintended behaviors.
    
    \item \textbf{Safety Consciousness}: This approach embeds a safety-aware framework into the model’s foundational training, fostering ethical reasoning and value alignment as intrinsic behaviors. The aim is to make safety a core characteristic, enabling the model to dynamically adapt to diverse and evolving scenarios. Safety consciousness can be viewed as a form of \textbf{safety tendency}: an inherent inclination to produce low-risk responses and shape outputs with an awareness of potential harm, much like human decision-making.
    
\end{itemize}

\subsection{A Call for Collective Action}

Safeguarding large models from adversarial manipulation, misuse, and harm is a global challenge that demands coordinated efforts from researchers, practitioners, and policymakers. The following sections present a research agenda designed to advance large-model safety through collaboration and innovation.

\subsubsection{Defense-Oriented Research}

Current research on large-model safety is heavily skewed toward attack strategies, with far less emphasis on developing defenses. This imbalance is concerning as attack sophistication continues to outpace effective safeguards. To address this, we advocate for a shift in research priorities toward robust defense development. Researchers should focus not only on attack mechanisms, but also on practical and preventative defenses to mitigate emerging threats. A balanced approach is essential for advancing safety.

Future defense research should also emphasize integration. New methods should be combined with existing approaches to build layered, cumulative protection as defense is a continuous, evolving process. However, the diversity of defense strategies makes integration challenging, underscoring the need for community-driven frameworks capable of effectively combining multiple defense mechanisms—a truly comprehensive ``super safety” framework.

\subsubsection{Dedicated Safety APIs}

To facilitate research and testing, commercial AI models should provide a dedicated safety API. Such an API would enable researchers to evaluate and strengthen model safety by exposing models to diverse adversarial and safety-critical scenarios. By making this functionality available, commercial providers can support external safety assessments without impacting regular user services. This approach would foster industry-academia collaboration and drive ongoing improvements in model safety.

\subsubsection{Open-Source Platforms}

The AI safety community would benefit greatly from the development and open-source release of safety platforms and libraries. Such tools would accelerate the evaluation, testing, and enhancement of safety mechanisms across diverse models and applications. Open-sourcing these resources would promote collaboration and transparency, allowing researchers and practitioners to share best practices, benchmark safety solutions, and contribute to the creation of universal safety standards.

\subsubsection{Global Collaborations}

The pursuit of AI safety is a global challenge that transcends national borders, requiring coordinated efforts from academia, industry, government agencies, and non-profit organizations. Effective international collaboration is essential to address the risks posed by advanced AI systems. By fostering global cooperation, we can more effectively tackle complex safety challenges and establish unified standards to guide the responsible development and deployment of AI technologies.

To facilitate global collaboration, the following initiatives could be pursued:

\begin{itemize} 
    \item \textbf{International Safety Alliances}: Forming global alliances dedicated to AI safety can unite experts and resources worldwide. These alliances would facilitate sharing research insights, coordinating safety assessments, and developing universal safety benchmarks that account for diverse regional needs and values.
    
    \item \textbf{Cross-Border Data Sharing}: Access to diverse datasets is crucial for enhancing the robustness and fairness of AI models. Establishing secure and ethical frameworks for cross-border data sharing would enable researchers to evaluate models in a broader range of scenarios, ensuring that safety mechanisms are effective and universally applicable.

   \item \textbf{Joint Safety Research Programs}: Collaborative research initiatives uniting academic institutions, industry leaders, and government agencies can drive innovation in AI safety. These programs should prioritize areas such as risk prediction, the development of safety guardrails, model enhancement, and safe reasoning strategies, ensuring that their findings are broadly applicable to a wide range of AI systems.
    
    \item \textbf{International Safety Competitions}: Expanding on the concept of open safety competitions, international challenges can be organized to engage top talent worldwide. These competitions would tackle critical and long-term safety issues, drive the development of innovative solutions, and promote a shared sense of responsibility for advancing AI safety.
    
    \item \textbf{Policy and Regulatory Implementation}: Effective AI governance requires practical, enforceable mechanisms. To this end, we advocate for the development of specialized agent systems capable of automatically auditing AI models for compliance with relevant regulations and policies. Bridging the gap between policy and technology is essential for advancing trustworthy and responsible AI, ensuring that high-level regulations can be systematically applied and verified in real-world deployments.
    
\end{itemize}

Global collaboration not only strengthens the effectiveness of AI safety research but also promotes transparency, trust, and accountability in the development of advanced AI systems. By working together across borders and disciplines, we can ensure that AI technologies deliver broad benefits to humanity while minimizing associated risks.

\section{Conclusion}\label{sec:conclusion}
In this paper, we conducted a comprehensive survey of safety research on Vision Foundation Models (VFMs), Large Language Models (LLMs), Vision-Language Pretraining (VLP) models, Vision-Language Models (VLMs), Diffusion Models (DMs), and large model powered Agents. We presented a comprehensive taxonomy of existing threats and defenses, highlighting the evolving challenges these models face.
Despite significant progress, many open challenges remain, particularly in understanding the fundamental vulnerabilities of large models, establishing robust safety evaluation protocols, and developing scalable, proactive, and integrated defense. Achieving safe AI will require not only technical advances but also collective action from the global research community and international collaboration.
We hope this paper serves as a valuable resource for researchers and practitioners, helping to drive ongoing efforts to build safe, robust, and trustworthy large-scale AI systems.

\section{Author Contributions}

Xingjun Ma designed the survey structure, organized the review process, wrote the challenges and conclusion sections, and prepared the final manuscript. All authors discussed the outline, contributed to drafting and revision, and approved the final manuscript. Yu-Gang Jiang initiated the project, secured resources, guided scientific direction, coordinated teams, and supervised final review and submission.

\textbf{Vision Foundation Model Safety}\; Ye Sun and Hanxun Huang surveyed visual backbones, scalable pre-training strategies, and key safety studies, and wrote the draft of this section. James Bailey, Jingfeng Zhang, Yiming Li, Mingming Gong, Tongliang Liu, Shirui Pan, and Sarah Erfani provided expertise in adversarial robustness, efficient architecture design, and graph signal processing, and reviewed this section.

\textbf{Large Language Model Safety} \; Yixu Wang, Yifan Ding, Yige Li, Haonan Li, Xudong Han, and Xiang Zheng covered alignment, jailbreaks, prompt injection, and extraction threats, and prepared the draft of this section. Xipeng Qiu, Tim Baldwin, Xiangyu Zhang, Neil Gong, and Yang Liu advised on multilingual scaling, generalization theory, privacy, and alignment, and refined the manuscript.

\textbf{Vision-Language Pre-training Model Safety}\; Xin Wang and Jiaming Zhang reviewed literature on large-scale corpora, pre-training objectives, and transfer evaluation, and prepared the initial draft of this section. Tianwei Zhang, Jindong Gu, and Siheng Chen provided guidance on secure deployment, domain generalization, and representation learning, and further refined the section.

\textbf{Vision-Language Model Safety}\; Ruofan Wang and Zuxuan Wu reviewed cross-modal architectures, datasets, and open challenges, and prepared the initial draft of this section. Dacheng Tao, Shiqing Ma, Cong Wang, Yang Zhang, and Masashi Sugiyama contributed expertise in multimodal defense and safety alignment and provided feedback on benchmarks and deployment.

\textbf{Diffusion Model Safety}\; Yifeng Gao designed the structure of this section and reviewed literature on adversarial attacks, jailbreaks, backdoors, and intellectual property protection. Hengyuan Xu, Yunhan Zhao, and Yunhao Chen surveyed research on membership inference and data/model extraction attacks. Chaowei Xiao, Baoyuan Wu, Tianyu Pang, Yinpeng Dong, and Cihang Xie provided technical guidance on generative model robustness and critically revised this section.

\textbf{Agent Safety}\; Yutao Wu designed the structure of this section and prepared the initial draft. Bo Li, Yang Zhang, Liu, Ruoxi Jia, Cong Wang, Yang Liu, Siheng Chen, and Chaowei Xiao contributed insights on indirect prompt injection attacks and defenses, secure tool use, and helped refine the section’s structure.


%

\ifCLASSOPTIONcaptionsoff
  \newpage
\fi

{\footnotesize
\bibliographystyle{IEEEtran}
\bibliography{egbib}
}

\end{document}